# Earth ECS Upper Limit using Modified Energy Budget Methods and Trend Analyses v.4

**Michael D. Mill...March 2025**
Contact: m.d.mill.climate@gmail.com.....

**ABSTRACT:** Earth Global and regional effective thermal "conductance" $\mathbf{G_{eff}}$ (in $(W/m^2)$/°C and often labeled $\lambda$ in climate research) and the related Equilibrium Climate Sensitivity (**ECS**) are evaluated by applying a modified version of the Energy Budget method, and using data only after 1970. By removing Periodic Interfering temperature components (using a novel PIR process) and applying high frequency filtering, an extraordinarily near linear temperature response is revealed, enhancing accurate $G_{eff}$ calculation and avoiding the pre-1970 aerosol forcing and $E^*_{Ocean}$($\equiv$ Ocean Heat Content per area) absorption uncertainties. A formal/empirical method is used to determine more reliable values of $Q(t) \equiv d[E^*_{Ocean}(t)]/dt$ . Using NOAA[17,12] data, and after PIR, it is shown that: 1) The Energy Budget Method can be realistically applied to the Ocean and Land regions independently, 2) the "historical" 1980-2020 most likely $\mathbf{G_{eff}}$ values for Global, global Ocean, and global Land regions are ≥ **1.72, 2.21, 1.25** $(W/m^2)$/°C respectively, and the corresponding median $\mathbf{ECS_{eff}}$ values are ≤ **2.15, 1.67, 2.96** °C/2x$CO_2$ respectively; where the updated IPCC AR5 orthodox independent global Forcing value of 0.4 $(W/m^2)$/Decade and $\Delta F_{2xCO2}$= 3.7 $W/m^2$ were used. The Global average $\mathbf{ECS_{true}}$ evaluation of ≤ **2.09** °C is 70% of the IPCC AR6 ECS[15] estimate of **3.0** °C, but 126% of the $ECS_{eff}$ value reported by Lewis[1] (**1.66** °C) . The estimated Oceans average TCR/ECS ratio = **0.71**, and the global average TCR/ECS ratio = **0.83**, and $ECS_{Land}/ECS_{Ocean}$ = **1.77** . [Results using HADCRUT temperature data instead are similar, but 6% "cooler" over land, and 8% "warmer" over oceans.] A simplified physically realistic formal/empirical Coarse 2-D Global Climate Model is derived wherein variation of $G_{eff}(t)$ until equilibrium (i.e. "pattern effects") are proven to be negligible or "cooling", *using these Methods*. And so it is likely $ECS_{true} \leq ECS_{eff}$. Very long lagging warming effects (e.g. glacial melt) are not included in this ECS. Uncertainties (often speculative or subjective) are determined separately herein, based on the readers own preferences. However, an ECS 1$\sigma$ upper limit uncertainty of only **10%** is a coarse estimate. And for comparison, a more orthodox application of the Energy Budget Method was also employed using the more recent reanalyzed E*(t) data of Cheng[13]. *That* calculated $ECS_{eff.Global}$ = **1.90** to **2.15**°C, is equal or *smaller* than using the better "modified" method above.

---

**Sections**

---

## A) Introduction

Within this monograph the following is achieved:

**1**) Histories of Temperature, Global Forcing, and Ocean Energy are determined with <u>sharply improved accuracy</u> using novel/modified methods of data analysis (see sections **B, C, D** and **SS3,4,5**). As a consequence

**2**) the critical Global ECS (Equilibrium Climate Sensitivity) is determined within <u>sharply reduced upper limits</u> using a novel method of Trend Analysis, to an extent that has evaded Climatology for 40 years; specifically $ECS_{Global}$ = **2.09 °C** with an upper limit uncertainty of very approximately **10%** (see section **A**, **E** and **SS1**,**2**). This far exceeds current IPCC AR6 accuracy. And this is extended to global Ocean, Land, and local regions.

**3**) These uncertainties are evaluated in sections **E.2** and **G.1,2,3.** In particular a physically realistic 2-D Coarse GCM (Global Climate Model) is necessarily derived which demonstrates (along with more orthodox GCMs) the *insignificance* of the often over estimated "pattern effect" variation at equilibrium (see **G.2**, **SS6,7**). The IPCC AR6 misevaluation of ECS, and that large uncertainty, is explained in section **H**.

..............................................................................................................................................................................

The evaluation of ECS using Energy Budget methods has been addressed by Lewis[1] , Otto[16], Spencer[28] ,et al., and reviewed by Forster[2], Sherwood[3], and Lewis[27], but will be independently introduced here as follows.

A quantity defined as the effective thermal "conductance" $G_{eff}$ of the Earth or various regions of the Earth can be evaluated based on the Energy Conservation principle applied to the observed total ocean energy, the calculated total independent Forcings ($W/m^2$) (mostly anthropogenic), and the global averaged temperature all as functions of time ("G" and the concept of a "thermal" circuit conductance are taken from electrical engineering...see **SS6**). Regionally the term $G_{eff}$ (often called "λ" in climate research) is defined *under known conditions* as

e.1 $\Delta I(t) = G_{eff} \cdot \Delta \mathbf{T_A(t)}$ , and so $G_{eff} = \Delta I(t)/\Delta \mathbf{T_A(t)}$

;where ΔI is defined as the increase of the thermal power "conducted" (i.e. radiated or convected or transported or conducted) out of a region and divided by the area of that region, as is caused by an increase in the area weighted average of the surface temperature ΔT in that same region ($\Delta \mathbf{T_A}$ in **bold** indicates an **Area** weighted average of ΔT). Here, ΔI(t) does *NOT* include power flow into that region's stored thermal energy. Note, all "Δ" quantities are define to all equal zero at some identical time $t_o$. This is taken to be an essentially "instantaneous" relationship certainly exhibiting no lag for variable moving averages of a year or more. In **SS1** es1.32 and es1.35 **[Note "SSx" refers to Sub-Section x, located at the end of the monograph; "e.x" refers to an equation number x, and es1.x indicates an equation "x" in SS1 ]** it is shown that $G_{eff}$ can be considered a true constant independent of time only if

e.2 $\Delta T(x,y,t)=\Delta T_\alpha(x,y)\cdot\Delta T_\beta(t)$ ;where (x,y) *represents* any 2-D position on the Globe surface, and "t" is time.

That is to say, the relative proportional value of Δtemperatures across the globe does not change over time. This constraint on the Energy Budget method has been often overlooked, could be a source of error, and is now known as the "pattern effect". It will be seen that this simplifying proportionality does hold over the evaluation ranges used herein to good approximation, but the effect on $G_{eff}$ at equilibrium must be considered separately in sections **E.1**, **G.2** and **SS7** . We are also presuming the linearity of e.1 such that $G_{eff}$ is also independent of $\Delta \mathbf{T_A(t)}$ (excluding "pattern effects") over a small range of several °C (i.e. perturbation theory applies). And this *is* verified empirically and convincingly by the striking linearity of $\Delta \mathbf{T_A(t)}$, described in Sections **B**, **G.2**.

An energy balance equation is developed in **SS1** es1.36 (which holds for independent regions as well as the entire globe, and for time pre-filtered variables...see **SS1**) which yields the result

e.3 $\Delta \mathbf{T_A}(t')\cdot G_{eff} =\Delta F(t')-\{d(\Delta E^*(t'))/dt' - [d(\Delta E^*(t'))/dt']_{@t'=0}\}$ ;where

$t' \equiv (t-t_o)$
$G_{eff}$ = *effective* thermal conductance...a true constant for the given region ($=(W/m^2)/°C$)
$\Delta \mathbf{T_A}(t') \equiv \mathbf{T_A}(t) - \mathbf{T_A}(t_o)$ ≡ the area weighted average surface temperature (°C) *change* as a function of time t'
$\Delta F(t') \equiv F(t) - F(t_o)$ ≡ temperature *independent* (or external) Forcing *change* for a given region; Forcing ≡ net independent Δpower flow input divided by the regional area ($=W/m^2$), as a function of the time t'
$E^*(t')$ ≡ Total stored thermal energy (mostly into the Oceans) divided by the region area ($= W\cdot years/m^2$).
$\Delta E^*(t') \equiv E^*(t) - E^*(t_o)$ ≡ total *change* of stored energy (mostly in the oceans) divided by region area ($=(W\cdot years)/m^2$).

As noted in **SS1,2** all the time dependant variables above may be replaced by the running averages (or succession of running averages) of the original "instantaneous" or yearly averaged variables. And the equation e.3 will hold true as long as *all* variables are averaged in this same manor. This averaging will be specified for the particular case. The various running averages are defined as "operators" H1[], H4[], H5[], H9[] in **SS2** and are especially useful. **[These are also notated as HF4, HF5, etc. in some figures.]**

Now subtract e.3 at $t'=t'_2$ from e.3 at $t'=t'_1$ , defining a range $\Delta t=t'_2-t'_1$, and yielding

e.4 a) $\underline{\Delta}T\cdot G_{eff} = \underline{\Delta}F - \underline{\Delta}Q$ ;where

b) $\underline{\Delta}T \equiv (\Delta\mathbf{T_A}(t'_2)- \Delta\mathbf{T_A}(t'_1))$

c) $\underline{\Delta}F \equiv (\Delta F(t'_2)- \Delta F(t'_1))$

d) $Q(t') \equiv d(\Delta E^*(t'))/dt'$

e) $\Delta Q(t') \equiv Q(t') - Q(t'=0 \text{ or } t=t_o) = d(\Delta E^*(t'))/dt' - [d(\Delta E^*(t'))/dt']_{\mathbf{@t'=0}}$

f) $\underline{\Delta}Q \equiv \Delta Q(t'_2) - \Delta Q(t'_1) = [d(\Delta E^*(t'))/dt']_{\mathbf{@t'=t'2}} - [d(\Delta E^*(t'))/dt']_{\mathbf{@t'=t'1}}$

Note, the $[d(\Delta E^*(t'))/dt']_{\mathbf{@t'=0}}$ term has cancelled out of the $\underline{\Delta}Q$ equation, which is convenient.
Finally then rewrite e.4a, as

e.5 $G_{eff} =(\underline{\Delta}F-\underline{\Delta}Q)/\underline{\Delta}\mathbf{T}$ ; using definitions e.4b,c,f .

<u>It is also useful to integrate e.3 (or e.4b,c,f) over the time range (t'-a) to (t'+a), and then divide by 2·a, so as to create more reliable time averages of all the variables therein</u>. This will be further described in Section **F** . This is considered herein as improved but "orthodox" Energy Budget methodology.

So the Earth system effective thermal conductance $G_{eff}$ can be determined by a calculation using the independent forcing, temperature, and total stored thermal energy/Area (i.e. $W\cdot years/m^2$) as a function of time. This Area may be the entire globe or any contiguous subsection of it *IF* the described conditions of e.2 approximately hold over the time range t' = 0 to $t_2'$ (see **SS1**).

Once $G_{eff}$ has been determined we can determine the equilibrium temperature change for an applied independent (or external) step forcing $\Delta F$ at $t_o$. For a system that starts in equilibrium at $t_o(t'=0)$, and ends in equilibrium at $t'=\infty$ then

e.6 $[d(\Delta E^{\mathbf{*}}(t'))/dt']_{@t'=0} = [d(\Delta E^{\mathbf{*}}(t'))/dt']_{@t'=\infty} = 0$ .

Then apply e.6 to e.3 and let $t'\rightarrow \infty$ yields

e.7 a) $\Delta\mathbf{T_A}(\infty)\cdot G_{eff} =\Delta F(\infty)-\{0 - 0\}$ or

b) $\Delta\mathbf{T_A}(\infty)=\Delta\mathbf{T_A}(\text{at equilibrium})=\Delta F(\infty)/G_{eff} =\Delta F_{total}/G_{eff}$ .

This presumes the ΔTemperature profile across the Globe does not change for the equilibrium case. This is not exactly true and will be discussed in following sections **E.1**, **G.2** and **SS7** . Note "ECS" (Equilibrium Climate Sensitivity) is defined as the "at equilibrium" ΔTemperature increase following a doubling of atmospheric$CO_2$ . The doubling induces an effective Forcing "$\Delta F/(2xCO_2)$". Using e.7, this is equal to

e.7c $ECS \equiv \Delta\mathbf{T_A}(\infty) = \Delta F_{total}/G_{eff} = \Delta F_{2xCO2}/G_{eff}$ .

Note, in this <u>empirical</u> evaluation it is advantageous to use a definition of **direct** Green House Gas Radiative Forcing (DRF or RF) that is almost "instantaneous" but defined <u>after</u> equilibration of the stratospheric temperature has taken place, and of a $G_{eff}$ that is <u>*specific*</u> to Green House Gas (GHG) bulk atmospheric Forcing (i.e. $G_{eff.GHG}$). This is more fully described in section **C**(first paragraph), **G.1**, and **SS6** . The concept of Effective Radiative Forcing is a theoretical construct not used in this case, though it is discussed in **SS6**.

The evaluation ranges selected are informed by the following understandings. We wish to avoid ranges starting before 1970. This avoids the large and very uncertain aerosol forcing ramp that occurred between 1945 and 1970. And 1970 is the approximate start of a large greenhouse gas forcing ramp, which is the source of nearly all the ΔF in the evaluation range (excluding volcanic Forcing).

**Unless otherwise specified all variables are pre-filtered with the H5[ ] or H9[ ] operator; specifically, a 3 point symmetrical moving average of the yearly averages applied 4 times in succession [which is defined as an H4[ ] operator], followed by a symmetric 5 or 9 point moving average, respectively. The endpoints are averaged uniquely as described in SS2.**

---

**B) Evaluation of T, ΔT and $\underline{\Delta}$T, and PIR**

**<u>B.1</u>** The surface temperature original data is taken from NOAA[17] data set circa 2022.
**[The annually average surface temperature data for various area weighted regional averages can be found in a special "all data" spread sheet within the "TREND-ANALYSIS-NOAA-(etc).xls . The NOAA monthly grid (5°x5°) is obtained from the "air.mon.anom.nc" file. The NCDUMP text listing of this is available in "noaa-grid-T-1880-2021-source.txt". The annually averaged version is available in "noaa-grid-T-1880-2021-annual.txt", where the "zero" longitude is redefined at 180° West. NOAA Global, Land and Ocean annual averaged time series data are also available at: https://www.ncei.noaa.gov/access/monitoring/climate-at-a-glance/global/time-series/ .**
**These are all found within the "source-Temperature-data" folder.] [Note, most referenced papers are available in the folder "papers"] [Find all resources in the end page repositories]**

As discussed previously the annually averaged temperature data is first filtered with the H5[] or H9[] operators(filters). **[These operations are provided in the "TREND-ANALYSIS-NOAA-(etc).xls" files.]**

**PIR discussion**: Figures 1 and 2 illustrate two frequency analyses of global average temperature for over a century (see C.D Keeling[4], B. Copeland[5] respectively). These show strong or notable frequency components very near 21.2, 9.3, and 15.2 cycle periods. The first is likely due to the solar 21 year magnetic cycle. The second is almost certainly due to the 18.6 year lunar precession cycle in which the sun, moon, earth, and the earth equatorial plane are all nearly collinear every 9.3 years, resulting in periodic (and parametric) tidal effects in the oceans and atmosphere. Figure 3 shows the 9.3 year cyclical effects on a pressure ridge over Australia (I. R. Wilson[6]). [This can also be described as those times the plane of the earth solar orbit, the plane of the lunar earth orbit, and the equatorial plane of the earth are co-linear (but not co-planar)] .

The source of the 15.2 year component is unknown. However, it is not necessary that the physical causes are known. Whatever the cause, these definite components exist, but are not often acknowledged. And certainly the corresponding independent Forcings are not included in any of the DRF or ERF forcing data used herein. It is therefore appropriate to remove these components from the temperature record artificially using a process called Periodic Interference Removal, i.e. PIR. The results of this removal can be striking.

Generally, the amplitude and phase of a 9.3 and 21.2 year sinusoidal component are adjusted and then subtracted from the original complete temperature function $T_o(t)$ (forming a modified $T_m(t)$) until an error metric is minimized.

The metric is defined as the integration (or sum) of the absolute values of the LF15[ ] (or LF9[ ]) operator on the yearly averaged modified $T_m(t)$ over a given time evaluation range. The LF15[ ] operator is a band pass (low frequency blocking) filter defined specifically as the H5[$T_m(t)$] value minus the 15 year symmetrical moving average of H5[$T_m(t)$] for a given year, i.e.

e.8 $$LF15[T_m(t)]=H5[T_m(t)] - AVERAGE[\ H5[T_m(t)]\ ]_{15.years.symmetric}\ .$$

The LF9[] operator is the same except using a 9 year symmetrical moving average. Thus only the interfering periodic components mostly remain (after LF15[$T_m(t)$] filtering); and these components are removed from $T_o(t)$ by subtraction and the minimization of the metric sum, in an iterative process

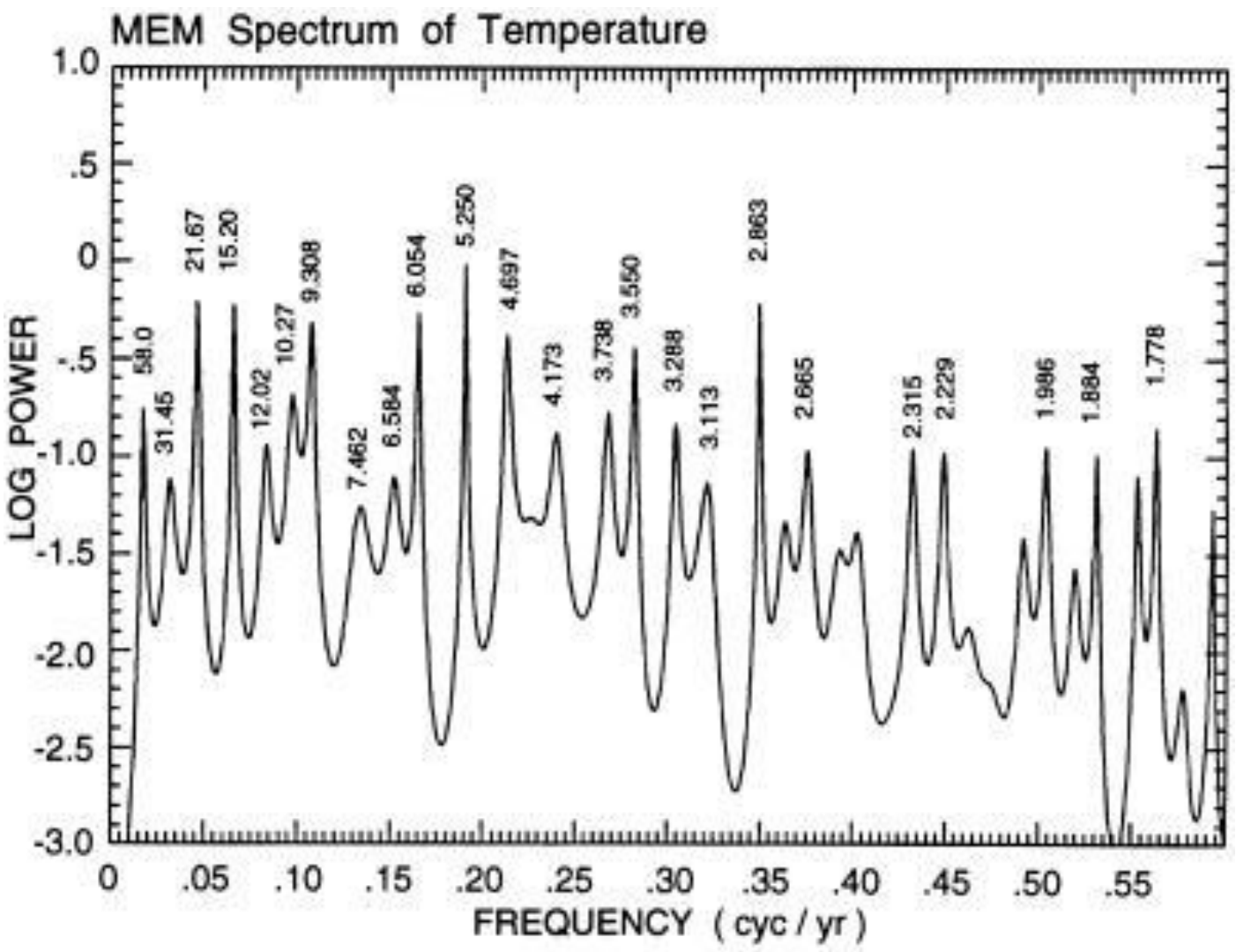

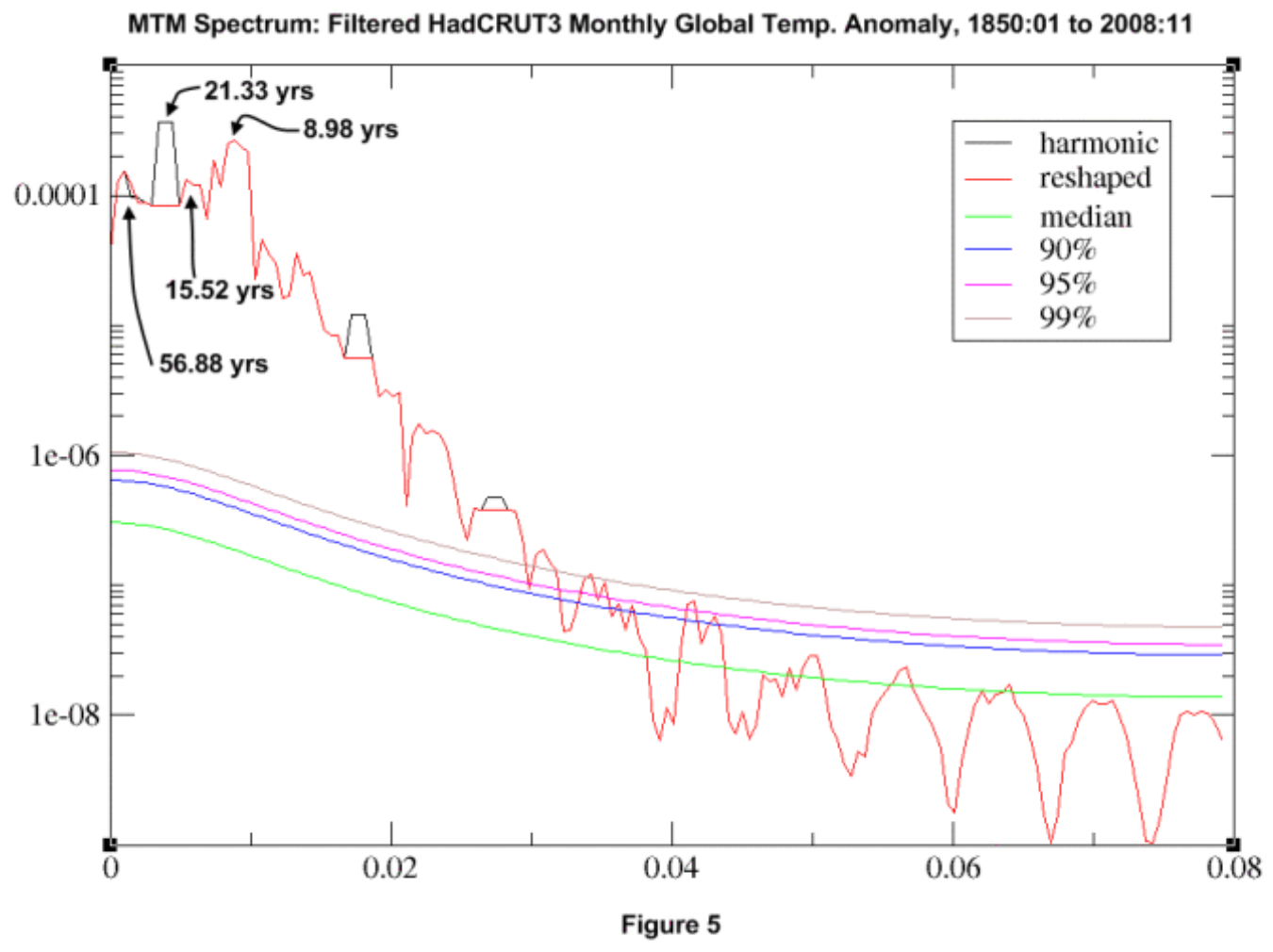

Fig.1 source: ref.[4] CC BY-NC-SA 4

Fig.2 source: ref.[5] CC BY-NC-SA 4

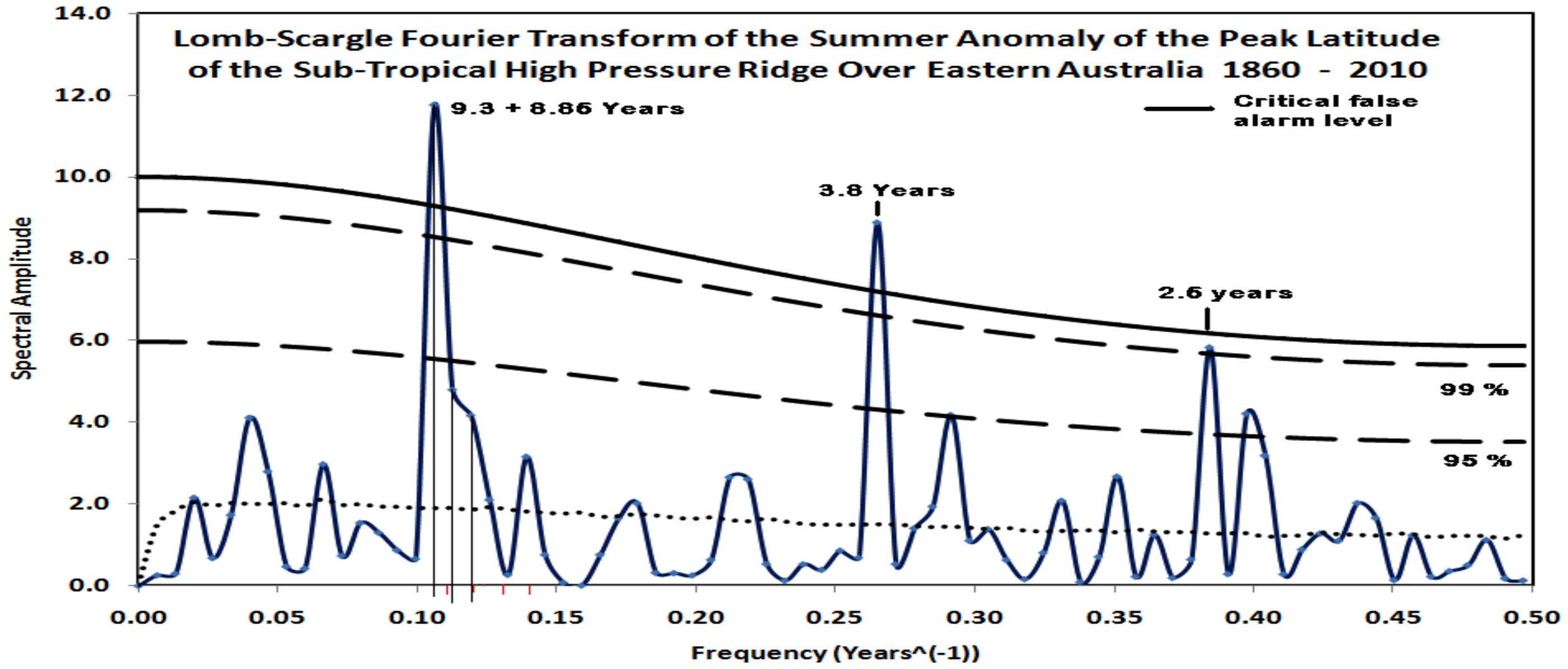

Fig.3 source: ref.[6] CC BY-NC-SA 4

(**using the Excel "Solver" tool**) that varies the phase and amplitude of the interfering signal. The final $T_m(t)$ is then the "uncontaminated" result. It is critical to use a sum of absolute values, and not a sum of the squares as the metric.

For 2 or 3 interfering components the specific process is as follows: The phase and amplitude of the first (9.3 year) component are varied iteratively to minimize the LF9[$T_m(t)$] metric using the Excel Solver tool, where the second and third (21.2 and 15.2 year) component amplitudes (which start at zero) and phase are held constant. Then the process is repeated for the second, and then third component using the LF15[$T_m(t)$] metric. And this is all repeated once again in a two pass process. It is important to use a summation range that is at least 3 full interfering component cycles, and use as few interfering components as possible. Note, PIR is performed as a part of the variable pre-filtering.

For PIR we choose to use data taken after 1950 for better quality (and avoiding the WWII disruptions). The solar and solar magnetic cycle timing and amplitude become notably non-uniform after 2007. Also a solar cycle maximum and minimum occur in 1948 and 2007, respectively. Therefore the PIR summation range of 1948-2007 is chosen (somewhat arbitrarily) for NOAA temperature data to accommodate the 21.2 year period (solar magnetic cycle) interference component...this being nearly 3 full cycles in length. [Because of the reduction in the solar-magnetic fields after 2007, the subtracted 21.2 year sinusoidal interference signal is multiplied by ½ after 2007 as a very rough approximation of this true interference reduction...although it makes virtually no difference in the final $G_{eff}$ calculation.] **[The working spread sheets to produce these $T_A(t)$ values and figures are "TREND-ANALYSIS-NOAA-(etc).xls" files.]** The resulting PIR regional temperature functions used are described below.

**1) NOAA Global exTROPAC Averaged Surface Temperature**

**[This refers to the Global region excluding the Tropical Pacific (i.e. exTROPAC), defined as the region from 20N - 20S latitude from the east coast of Australia to the west coast of South America. This region is excluded because extraordinary forcings or fluctuations (and temperature variations) occur there (results from unpublished work), that are unknown or unacknowledged in the orthodox ERF values used herein. Therefore, for comparison only, this unwanted complexity is removed to better reveal underling linearities. It is an acceptable procedure in any case. The longitudinal extent *could* better be reduced by 30° on both ends since these unusual forcings occur nearer the center of the pacific.]**

**Fig.4.**1**, 3 show global average surface temp after PIR**. Note the nearly linear response between 1975 and 2003, after H5 or H9 filtering and PIR processing of the temperature data. In this case PIR removes the 21.2, 9.3, and 15.2 yr periodic components, to produce a T(t) curve that is extraordinarily smooth and quite linear from 1975 to 2020 (excluding the small transient "dip" from 2005 to 2013), even using only the milder H4[] filter**. Fig.4.2 shows results of the LF15 and LF9 operators (filters) after PIR minimization**. **See Fig.4.1,2,3 and Fig.4.1B,2B,3B for the PIR and non PIR cases, respectively**. The interference components using LF9,15 evaluated over 1948-2007 are:

$T(t)=0.0631\cos([(t-1979.68)2\pi/9.3] + 0.0536\cos[(t-1981.81)2\pi/21.2] + 0.0234\cos[(t-1973.4)2\pi/15.1]$ °C.

The effects of the 21.2 yr and 9.3 yr periodic "interfering" components are clear and real. The resulting near linearity cannot realistically result from a simple coincidental random process but reveals a true underlying long term linear forcing trend, after high frequency noise is filtered out. The PIR process does not "force" a linearity on the output, as the transient dip and the upward ramp bend at 1970 illustrate. [Other random frequency pairs for the interference components were tested resulting in an error metric at least twice a large as for the much better pair at 21.2 and 9.3 year periodicities. Convincing evidence of the accuracy of the PIR method will be provided at the end of this section, in section **B.2**]

**Using the NOAA Global exTROPAC data, the H5[T(t)] best fit slope after PIR between 1980 and 2003 is 0.180 C/Decade.**

**2) NOAA Globally Averaged Surface Temperature**

Fig.4.4,**6** show global average surface temperature after PIR. Note again the nearly linear response between 1975 and 1995, after H5 or H9 filtering and PIR processing of the temp data. Note again the *larger* transient "dip" from 1995 to 2004 to 2015. It appears the TROPAC region adds much of the transient non-linear component. Fig.4.5 also shows results of the LF15 and LF9 operators (filters) after PIR minimization. The interference components using LF9,15 evaluated over 1948-2007 are:

$T(t)=0.0613\cos([(t-1979.72)2\pi/9.3] + 0.0547\cos[(t-1982.51)2\pi/21.2]$.

Figures 4.4B, 4.5B, 4.6B display the respective plots using no PIR, for comparison.

**Using the NOAA Global data, the H5[T(t)] best fit slope after PIR between 1980 and 1995 is 0.192 C/Decade.**

**3) NOAA Global (60N-60S Latitude) Averaged Surface Temperature**

**[This refers to the Global region between 60N and 60S Latitude]**

The comments here are similar to **2)** above. See Fig.4.7,8,**9** and Fig.4.7B,8B,9B for the PIR and non-PIR cases, respectively. The interference components using LF9,15 evaluated over 1948-2007 are:

$T(t)=0.0626\cos([(t-1979.69)2\pi/9.3] + 0.0515\cos[(t-1982.31)2\pi/21.2]$.

Note, the same transient "dip". **Using the NOAA Global (60N-60S) data, the H5[T(t)] best fit slope after PIR between 1980 and 1995 is 0.182 C/Decade.**

**4) NOAA Global Oceans exTROPAC Averaged Surface Temperature**

The comments here are similar to **2)** above including the linearity of the PIR temperature between 1970 and 2020 with clear small transient deviations from 1995-2013. See Fig.5.1,2,**3** and Fig.5.1B,2B,3B for the PIR and non-PIR cases, respectively. The interference components using LF9,15 evaluated over 1948-2007 are: $T(t)=0.0565\cos([(t-1979.7)2\pi/9.3] + 0.0423\cos[(t-1981.24)2\pi/21.2] + 0.0239\cos[(t-1972.29)2\pi/15.1]$.

**Using the NOAA Ocean exTROPAC data, the H5[T(t)] best fit slope after PIR between 1980 and 1995 is 0.127 C/Decade.**

**5) NOAA Global Oceans Averaged Surface Temperature**

The comments here are similar to **2)** above except the same transient deviation is more pronounced (due to the extraordinary forcings and/or fluctuations occurring within the TROPAC region?). See Fig.5.4,5,**6** and Fig.5.4B,5B,6B for the PIR and non-PIR cases, respectively.

The interference components using LF9,15 evaluated over 1948-2007 are:

$T(t)=0.0482\cos([(t-1979.52)2\pi/9.3] + 0.0428\cos[(t-1982.37)2\pi/21.2]$ .

**Using the NOAA Global Ocean data, the H5[T(t)] best fit slope after PIR between 1980 and 1995 is 0.137 C/Decade.**

**6) NOAA Oceans (60N-60S Latitude) Averaged Surface Temperature**

The comments here are similar to **2)** above. The same transient "dip" starts in 1995 (due to the extraordinary forcings and/or fluctuations occurring within the TROPAC region?). See Fig.5.7,8,**9** and Fig.5.7B,8B,9B for the PIR and non-PIR cases, respectively. The interference components using LF15,9 evaluated over 1948-2007 best include the 10.6 year solar cycle component, and are:

$T(t)=0.0434\cos([(t-1978.24)2\pi/9.3] + 0.0455\cos[(t-1982.21)2\pi/21.2] + 0.0255\cos[(t-1980.88)2\pi/10.6]$.

**Using the NOAA Ocean (60N-60S) data, the H5[T(t)] best fit slope after PIR between 1980 and 1995 is 0.138 C/Decade.**

**7) NOAA Global Land Averaged Surface Temperature**

The comments here are similar to **1)** above. In this case PIR uses the 21.2, 9.3 periodic components in the 2-pass process. Fig.6.3 shows again a nearly linear response over 1975-2020, with a small transient "dip" after 2003-2015 See Fig.6.1,2,**3** and Fig.6.1B,2B,3B for the PIR and non PIR cases, respectively. The interference components using LF9,15 evaluated over 1948-2007 are:

$T(t)=0.0888\cos([(t-1979.72)2\pi/9.3] + 0.067\cos[(t-1982.95)2\pi/21.2]$.

Volcanic cooling effects could have an effect on the temperature linearity from 1960-1995 (to be discussed in following sections)? **Using the NOAA Land data, the H5[T(t)] endpoint to endpoint slope after PIR between 1995 to 2020 (which avoids any volcanic effects) is 0.33 C/Decade.**

**8) NOAA Land (60N-60S Latitude) Averaged Surface Temperature**

In this case PIR uses the 21.2, 9.3, and 15.2 yr periodic components in the 2-pass process. Fig.6.6 shows a not as nearly linear response over 1970-1990 (compared to **7)** above), although nearly linear from 1980-2020, with the small transient "dip" after 2003-2015. See Fig.6.4,5,**6** and Fig.6.4B,5B,6B for the PIR and non PIR cases, respectively. The interference components using LF9,15 evaluated over 1948-2007 are:

$T(t)=0.086\cos([(t-1979.8)2\pi/9.3] + 0.0767\cos[(t-1982.21)2\pi/21.2] + 0.0192\cos[(t-1973.91)2\pi/15.1]$.

Volcanic cooling may have an even more visible effect in this case on the temperature linearity from 1960-1995 (to be discussed in following sections)? **Using the NOAA Land data, the H5[T(t)] endpoint to endpoint slope after PIR between 1995 to 2020 (which avoids any volcanic effects) is 0.31 C/Decade.**

**9) NOAA TROPAC Averaged Surface Temperature**

For comparison and insight, the T(t) for the Tropical Pacific is included. In this case PIR best uses the 21.2, 10.6 (the solar cycle period), and 15.2 yr periodic components in the 2-pass process. See Fig.7.1,2,**3** and Fig.7.1B,2B,3B for the PIR and non PIR cases, respectively. The interference components using LF15,9 evaluated over 1948-2007 are:

$T(t)=0.0913\cos([(t-1984.04)2\pi/21.2] + 0.0832\cos[(t-1980.56)2\pi/10.6] - .0713\cos[(t-1972.84)2\pi/15.1]$.

The 9.3 yr component does not seem to be needed in this minimization, and the 15.1 year component is of opposite sign as in the other regions(?). Fig.7.3B (using no PIR) shows much large irregularity in this region. However Fig.7.3 (after PIR) shows the possibility of piecewise linearity between 1950 and 1995, but with a very large transient "depression" in the trend after 1995, similar to but much larger than the deviation seen in the remainder of the globe. This seems consistent with the previous results; but the PIR results are certainly questionable here. The TROPAC region is considered to be the primary source of the temperature trend irregularities.

**Using the NOAA TROPAC data, the H5[T(t)] and H9[T(t)] best fit slope after PIR between 1975 and 1994 is 0.23 C/Decade.**

The ratio of NOAA Global Land/Ocean temperature rates is (.33/.137)= **2.41** . This compares to the Scafetta[26] reported values of 2.32 (80N-60S) for NOAA. And the NOAA ratio of **2.41** is 42% larger (i.e. **2.41**/1.7) than the average of nearly all current GCMs[26] (Global Climate Models). [Using HADCRUT T(t) data this becomes **2.41**·(.94/1.08)=**2.10**, see below.]

It is expected, and required, that $\Delta T_{global.ave}=0.3\cdot\Delta T_{land.ave}+0.7\cdot\Delta T_{ocean.ave}$ , where 0.3 and 0.7 are the global fractions of land and ocean areas respectively. And this does hold for the **PIR** NOAA temperature rates described in **2), 5), 7)** above within 2%, i.e.

e.9 NOAA:PIR: 0.7·**0.137**+0.3·**0.33**=.195=1.016·(**0.192**) (C/Decade) .

The accuracy of the PIR process in this regard, will be discussed in **B.2** below.

Most of the filtered and PIR processed T(t) plots indicate a nearly linear underlying Forcing (probably starting before 1970) and Temperature response observed from 1975 or 1980 extending to 2020. The most striking evidence being the Global average *both* excluding *and* including the Tropical Pacific region (see Fig.4.3,6), and the Global Land average (see Fig.6.3). Thus, this same underlying Forcing (presumed Global in nature) occurs over the Oceans also, even though there are more interfering transient forcings and fluctuations in effect there. This understanding allows us to separate (remove) the transient deviations from the underlying linear T(t) and F(t) component, thus simplifying the analysis.

The values of $\Delta\mathbf{T_A}$ or $\underline{\Delta}$T (as used in e.5, e.3) can be determined from these plots.

**[Figures 4.3,3B through 7.3,3B (the H5 and H9 filtered Temperature plots) are listed for easy comparison (PIR vs no PIR) below on the following pages...The rest of Figures 4.x through 7.x etc are available in repositories listed in the end page]**

**[!!Note: all figures below plot annually averaged discrete values linked by smooth curves for visualization. The curves must be translated + ½ year to represent their real time values!!]**

These figures follow:

Fig.4.3,3B Global exTROPAC
Fig.4.6,6B Global
Fig.4.9,9B Global (60N-60S)
Fig.5.3,3B Global Oceans exTROPAC
Fig.5.6,6B Global Oceans
Fig.5.9,9B Global Oceans (60N-60S)
Fig.6.3,3B Global Land
Fig.6.6,6B Global Land (60N-60S)
Fig.7.3,3B TROPAC

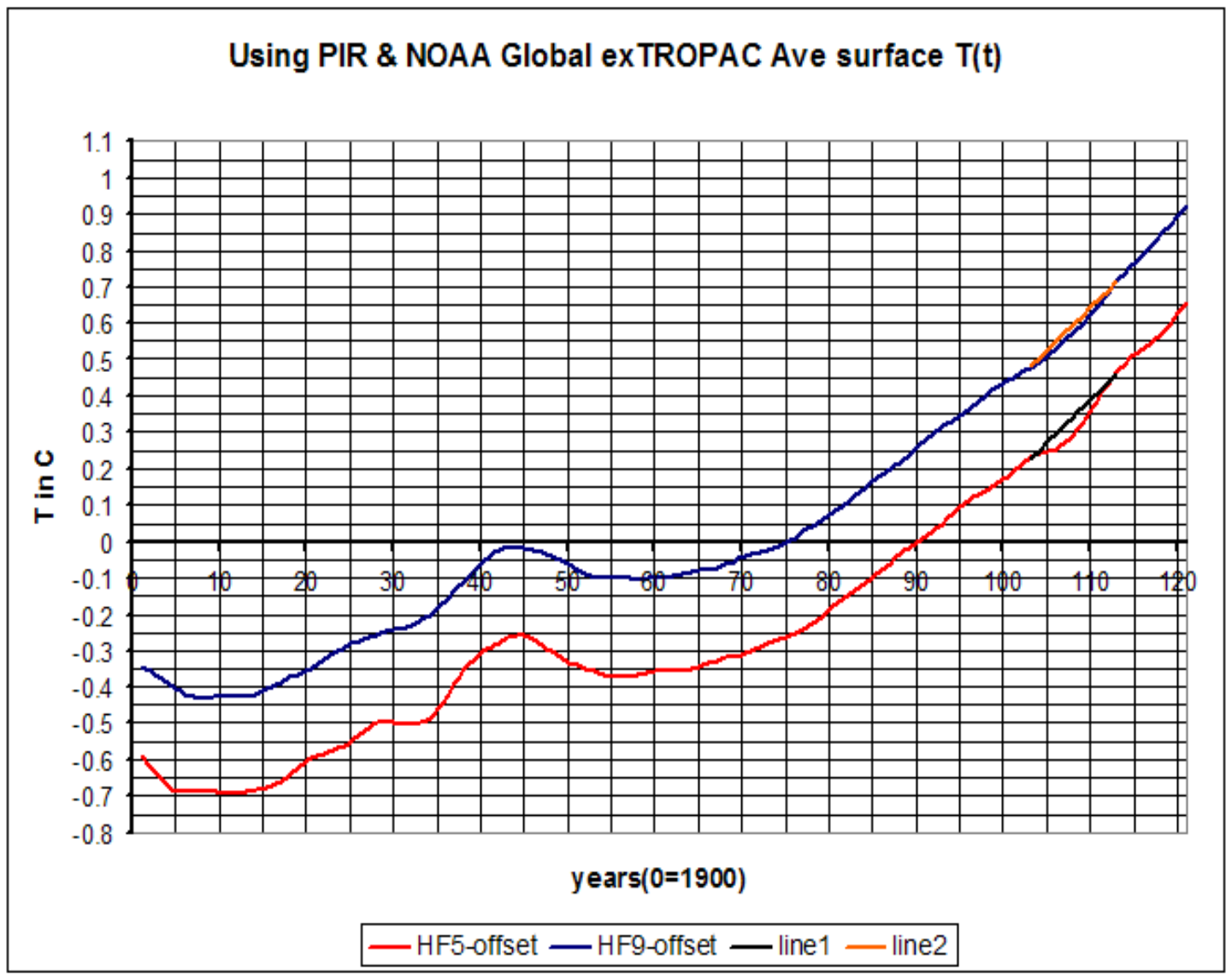

Fig.4.3(discrete annual ave. plots; curves for visualization)

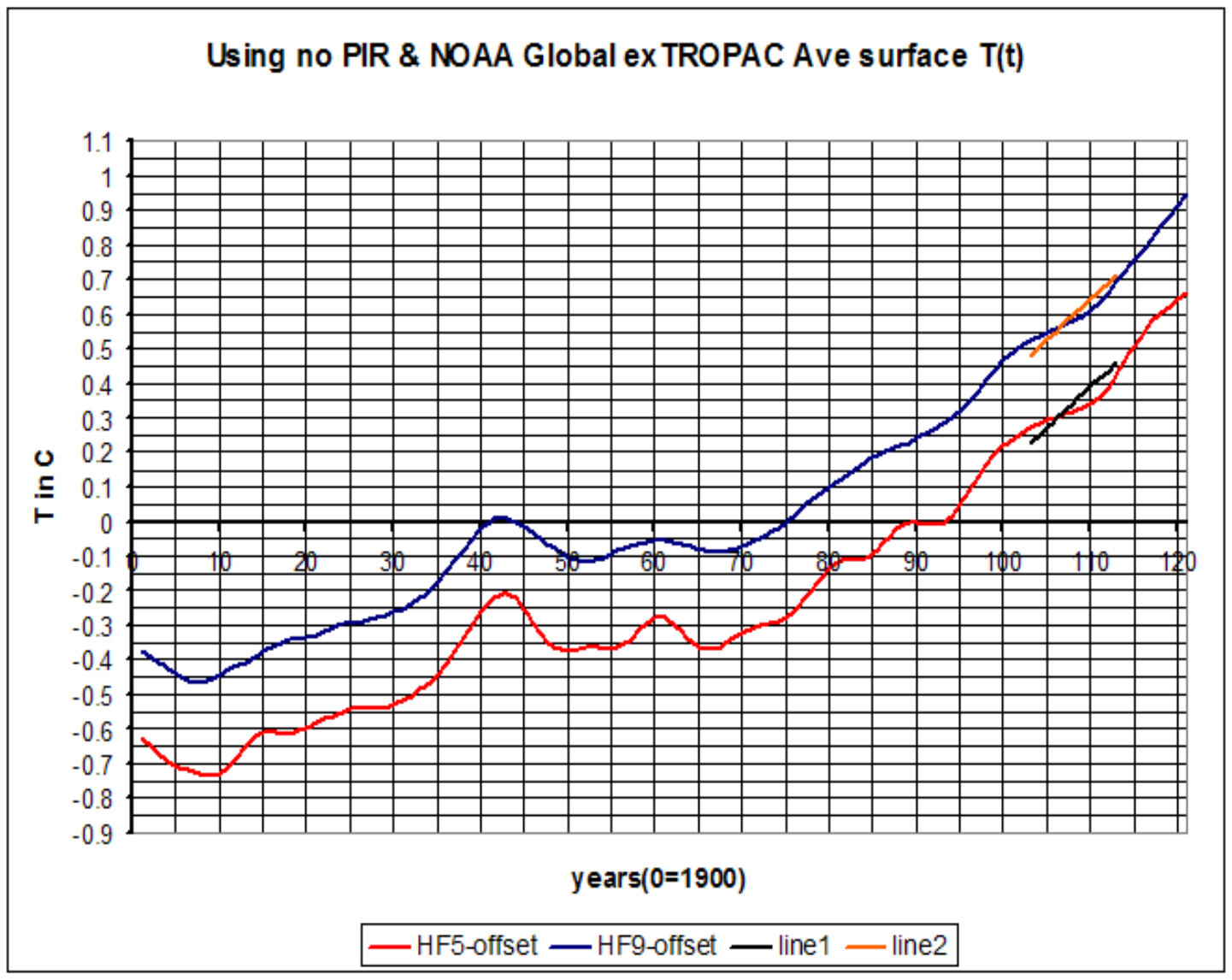

Fig.4.3B (discrete annual ave. plots; curves for visualization)

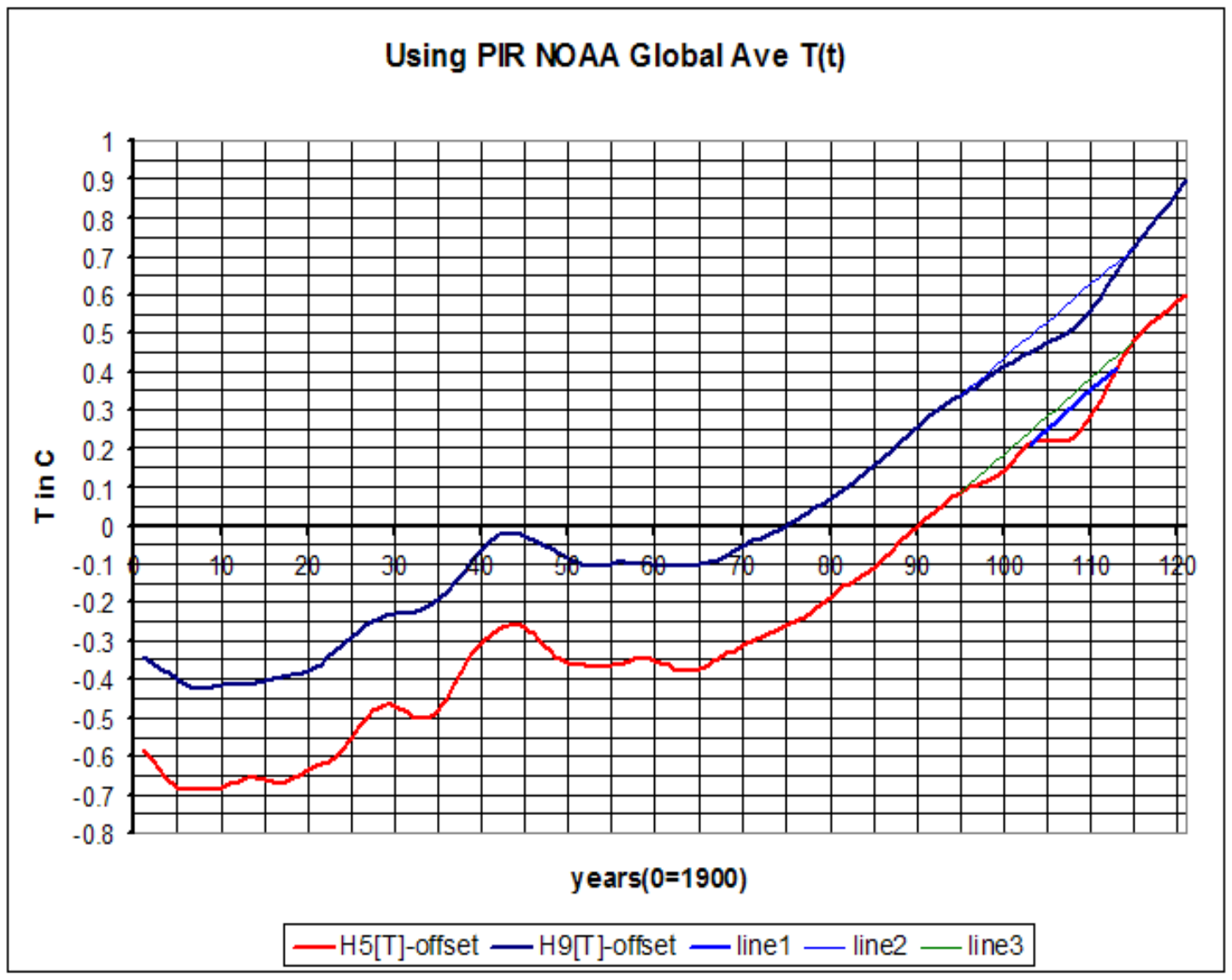

Fig.4.6(discrete annual ave. plots; curves for visualization)

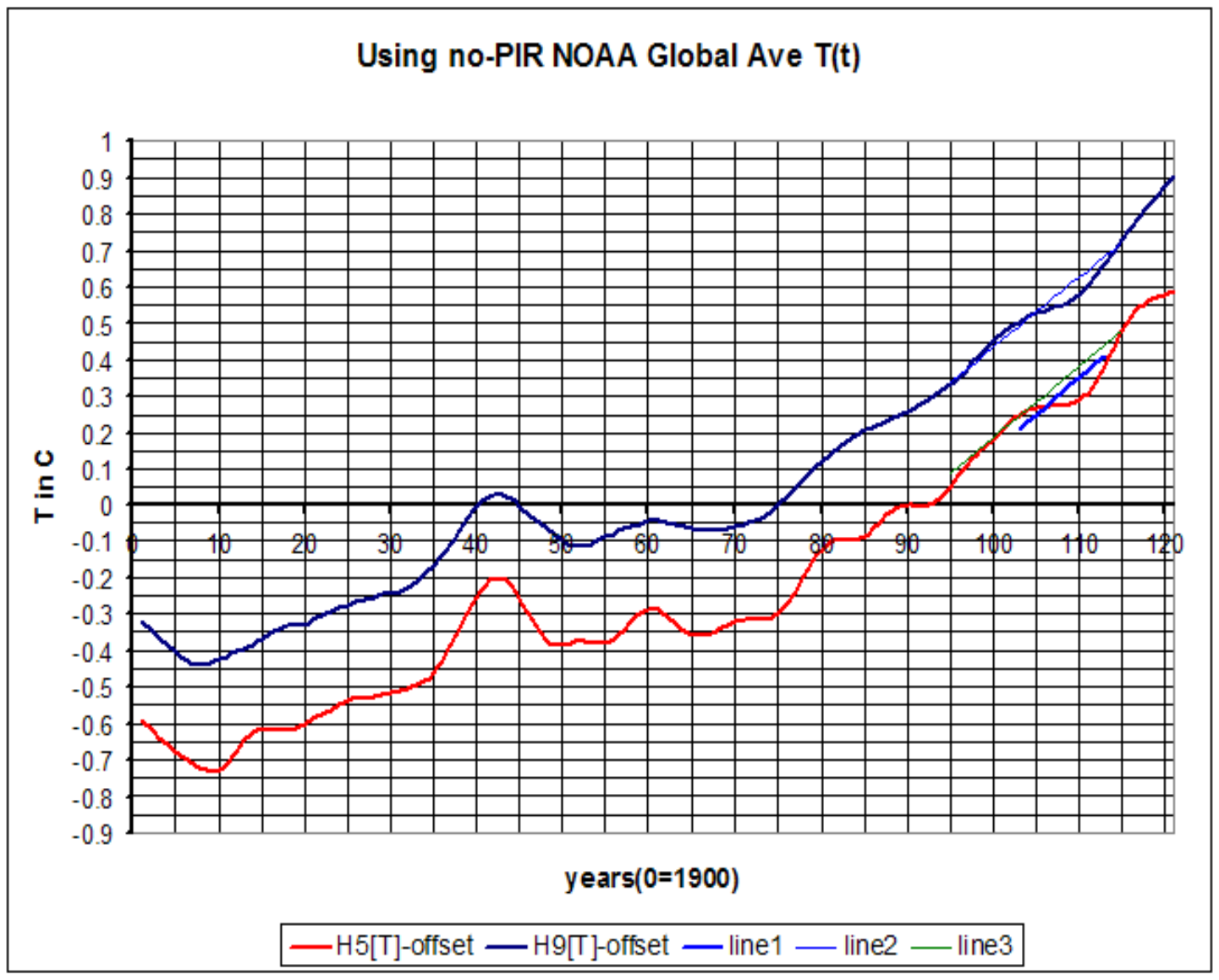

Fig.4.6B(discrete annual ave. plots; curves for visualization)

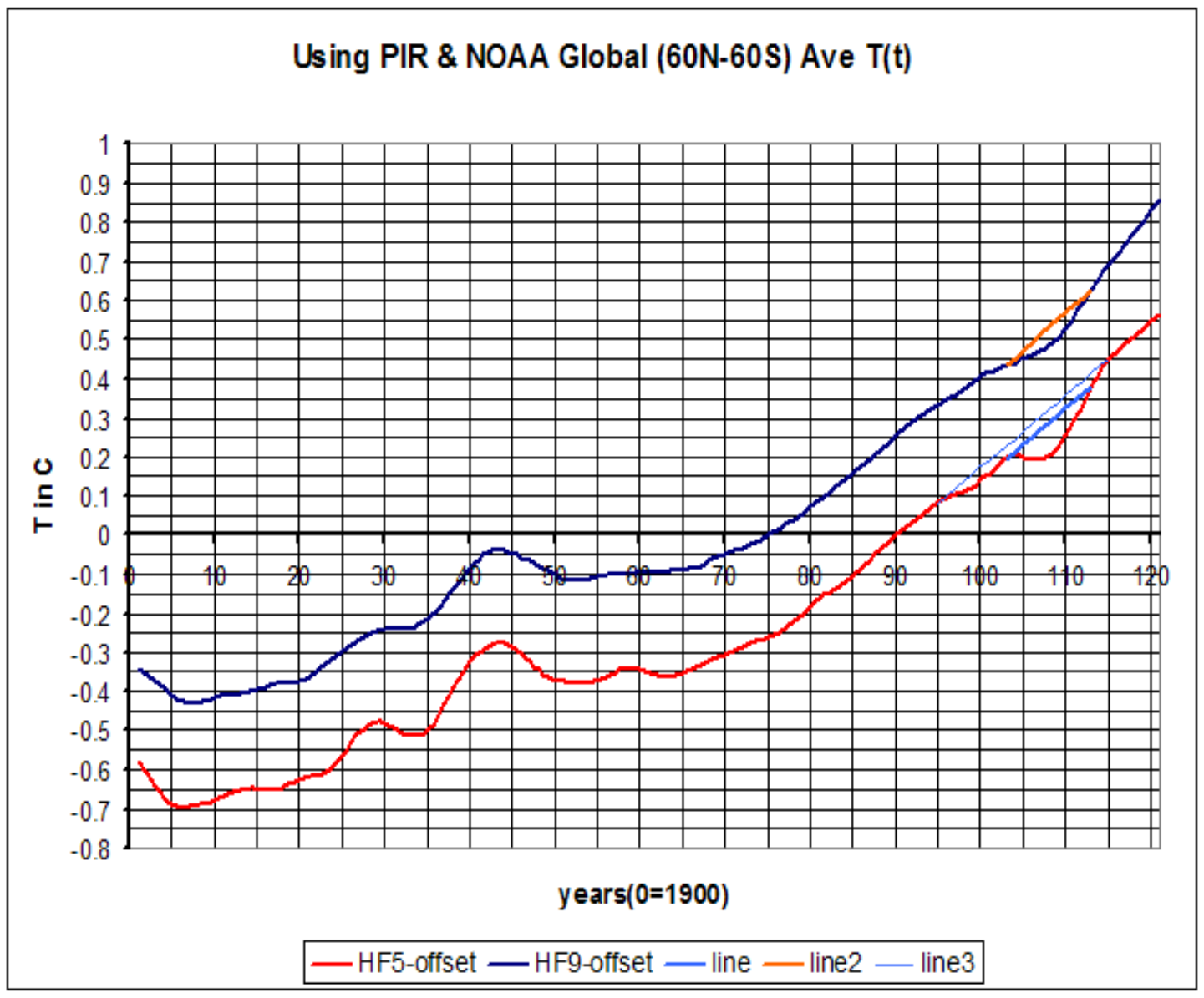

Fig.4.9(discrete annual ave. plots; curves for visualization)

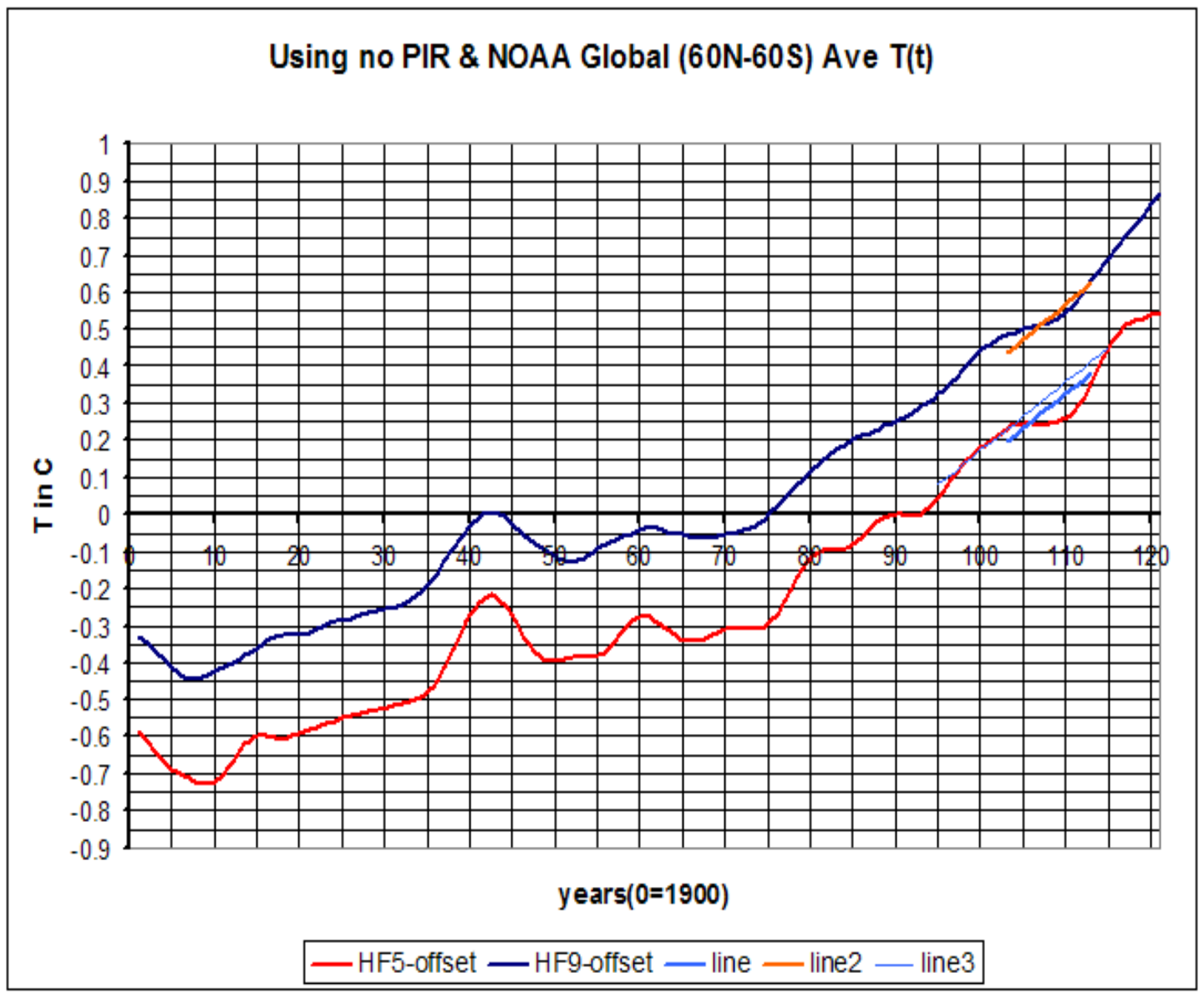

Fig.4.9B(discrete annual ave. plots; curves for visualization)

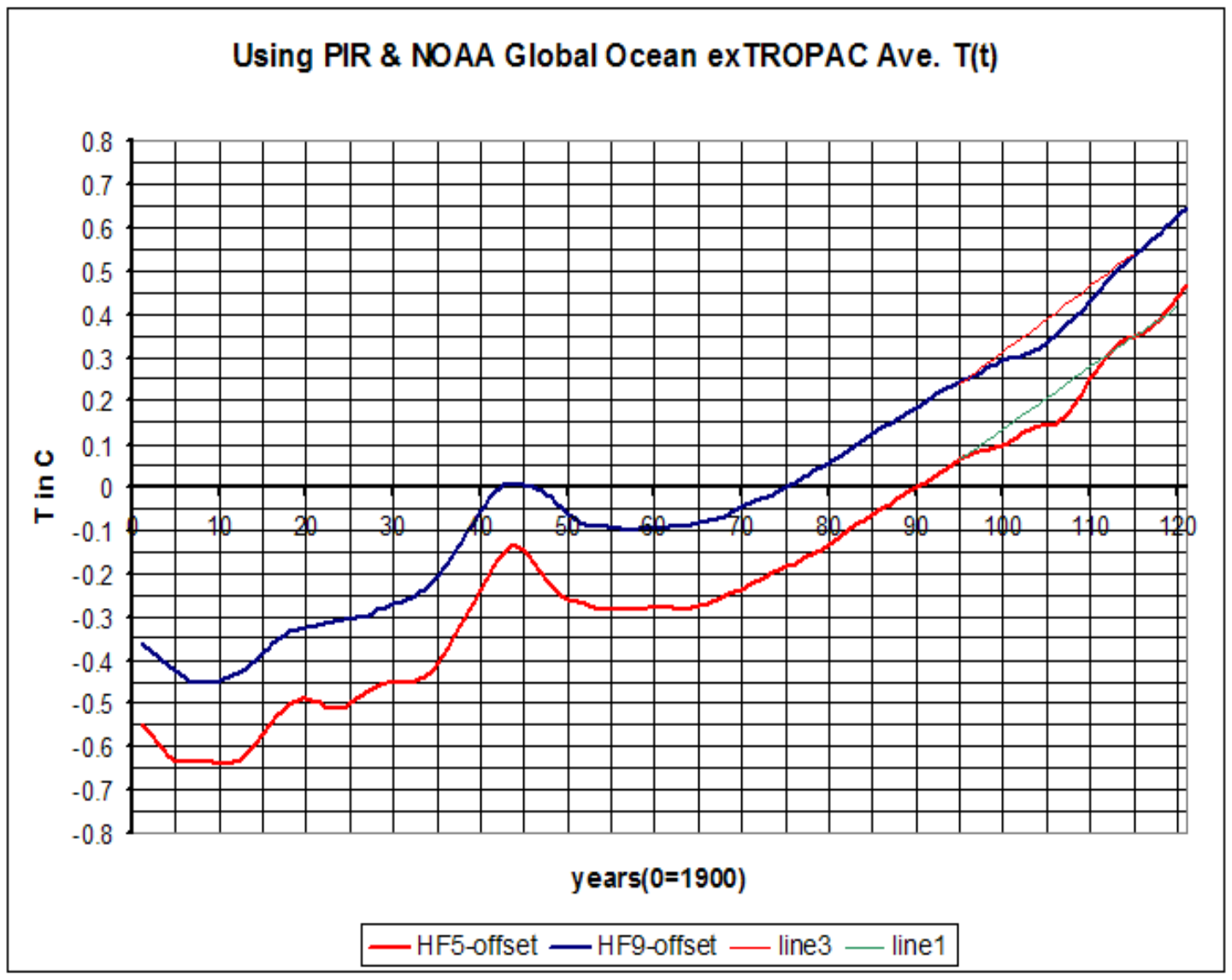

Fig.5.3(discrete annual ave. plots; curves for visualization)

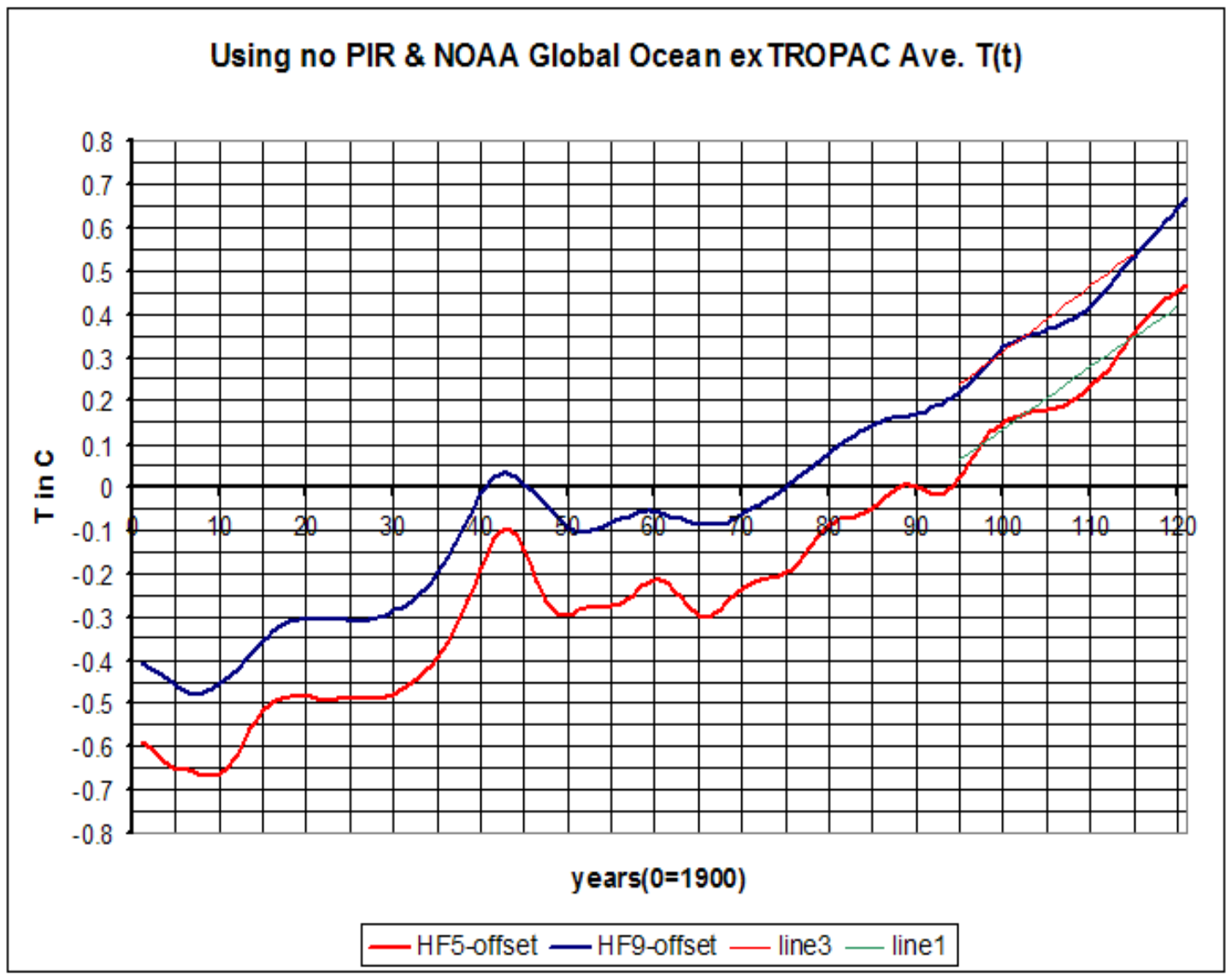

Fig.5.3B(discrete annual ave. plots; curves for visualization)

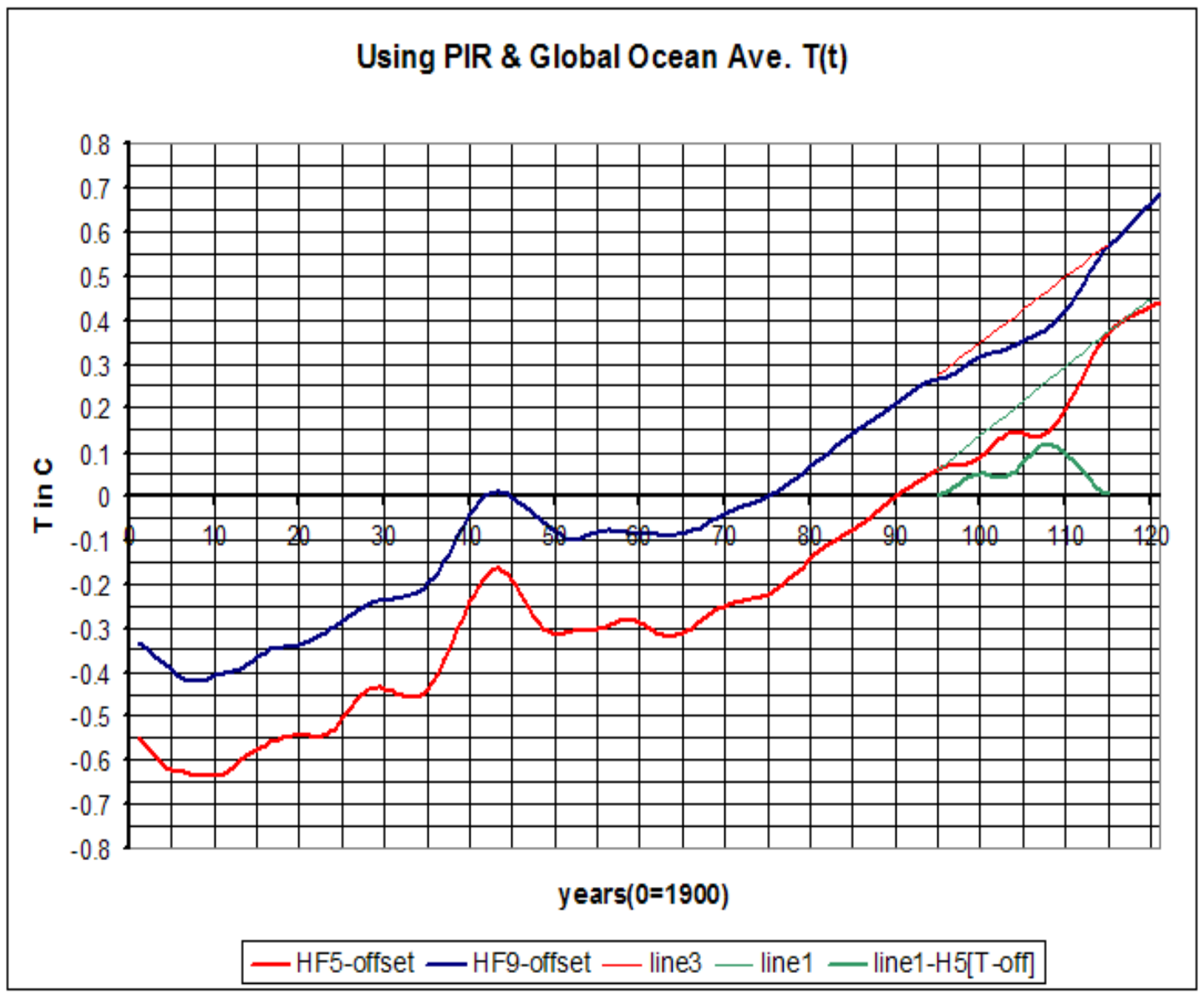

Fig.5.6(discrete annual ave. plots; curves for visualization)

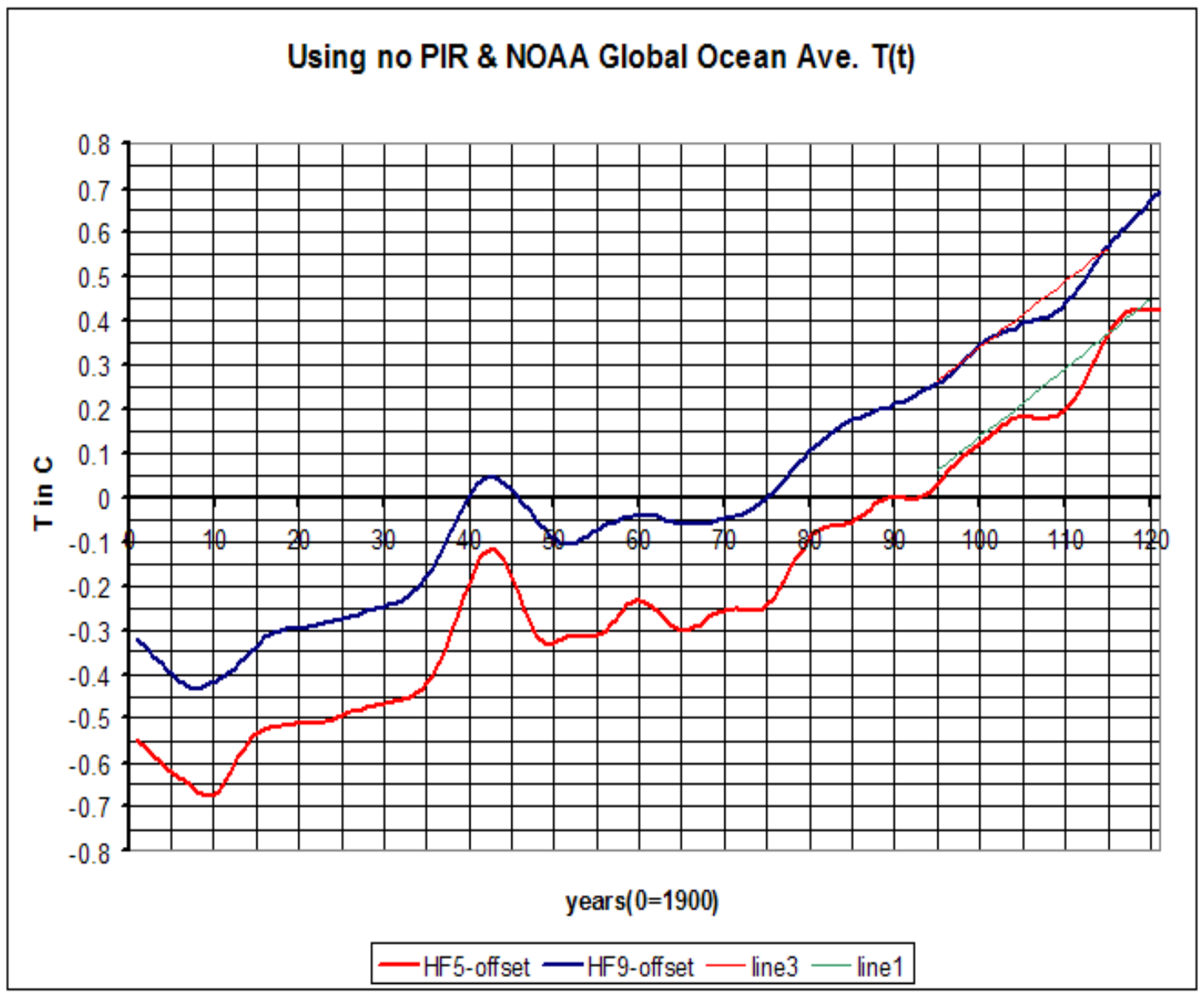

Fig.5.6B(discrete annual ave. plots; curves for visualization)

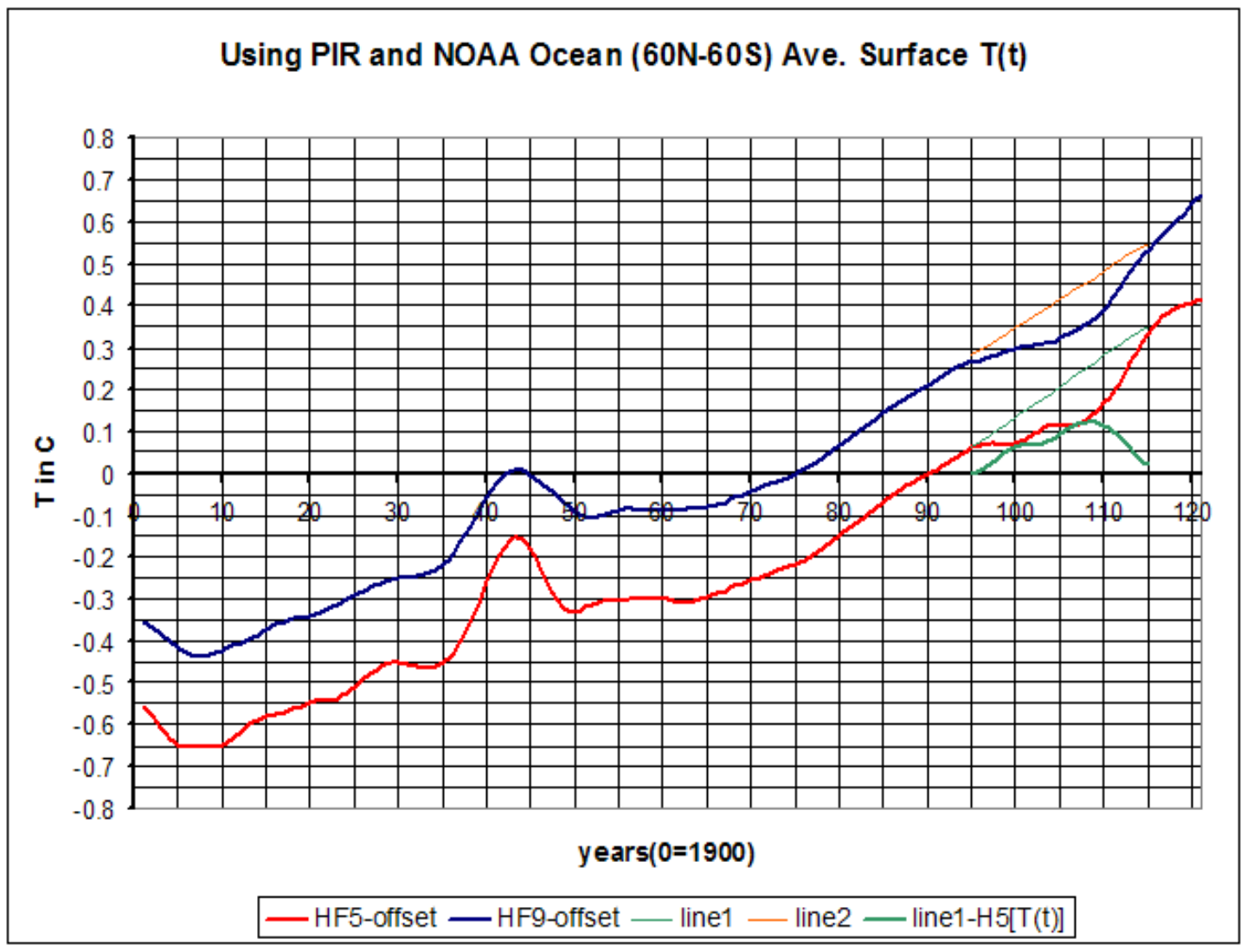

Fig.5.9(discrete annual ave. plots; curves for visualization)

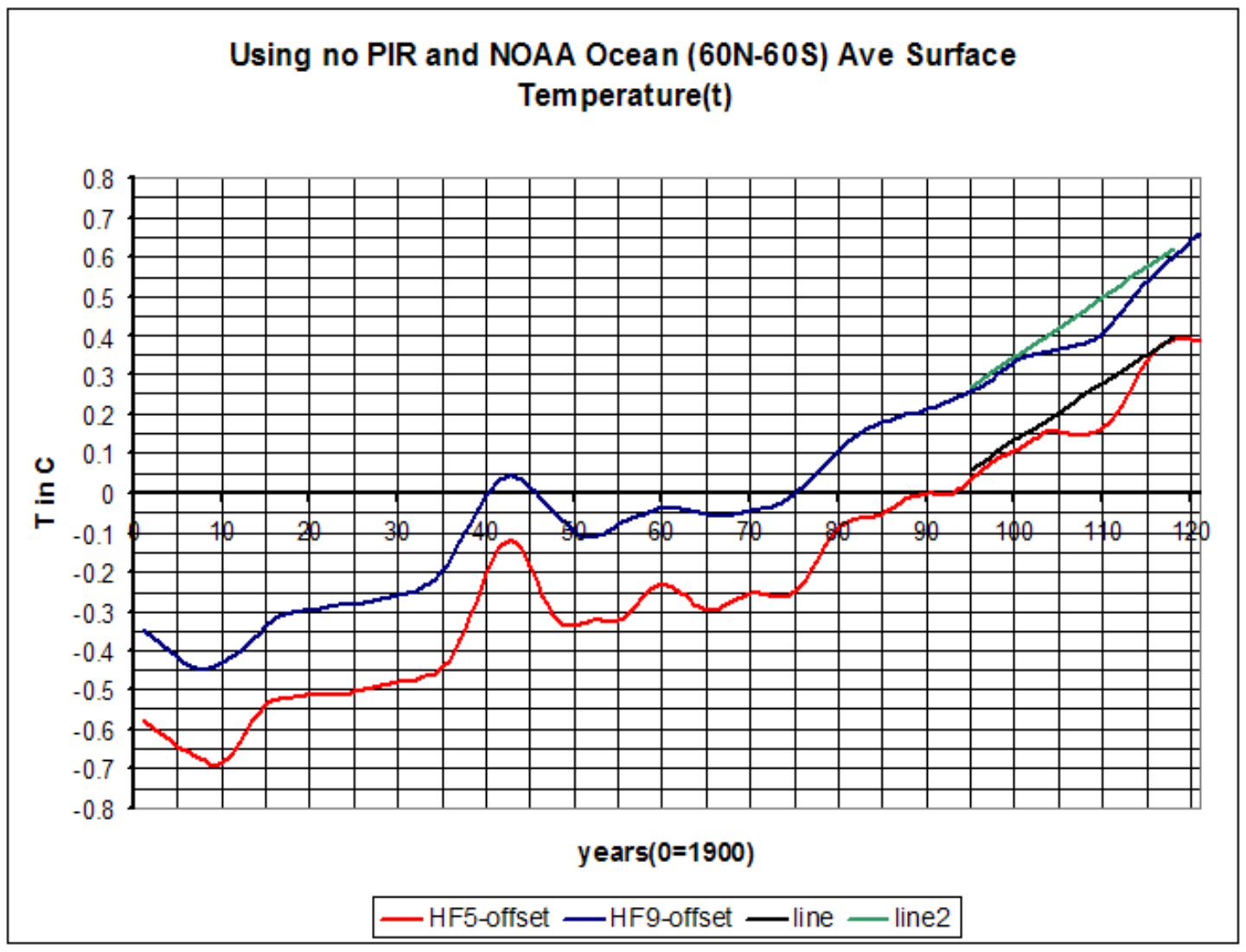

Fig.5.9B(discrete annual ave. plots; curves for visualization)

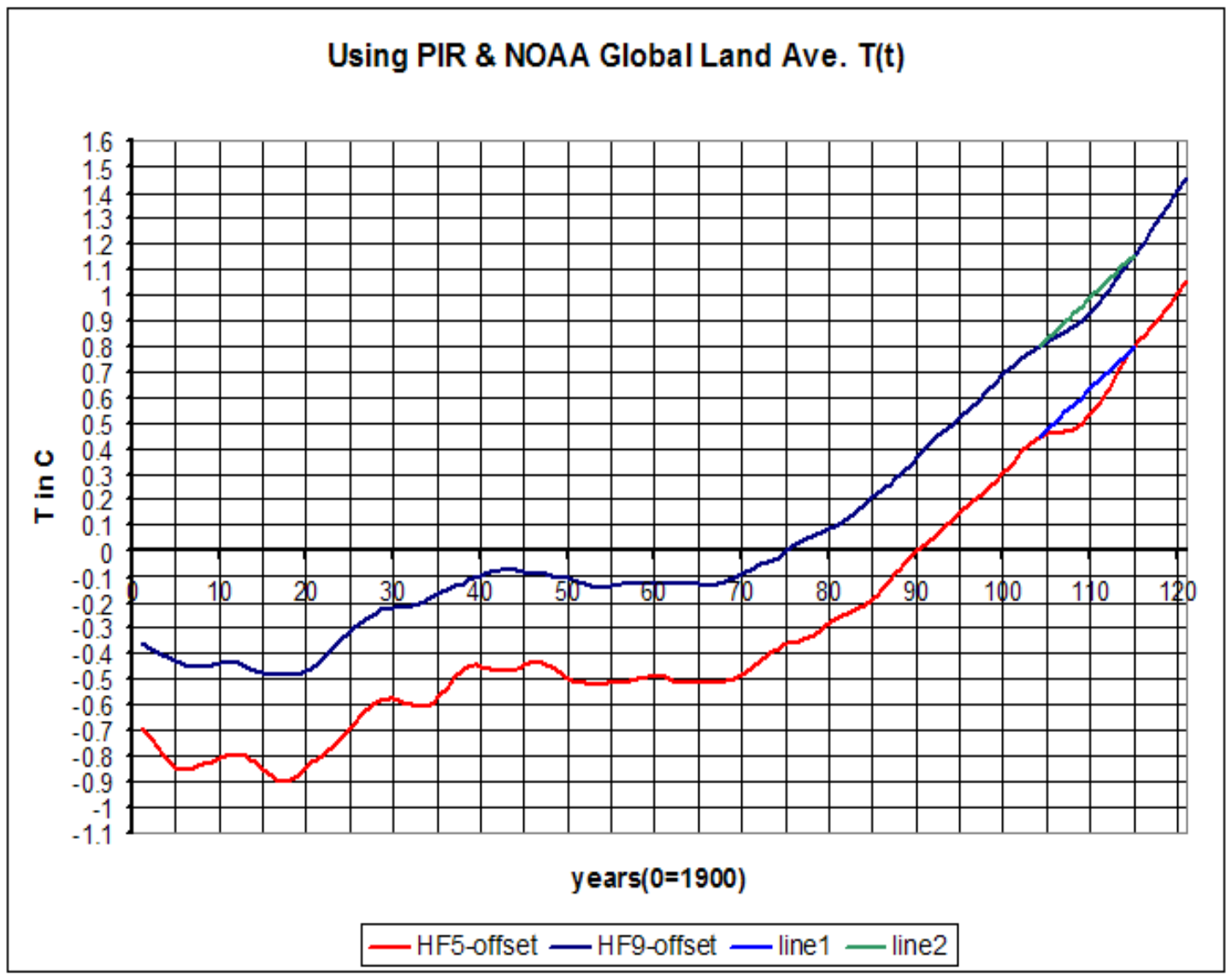

Fig.6.3(discrete annual ave. plots; curves for visualization)

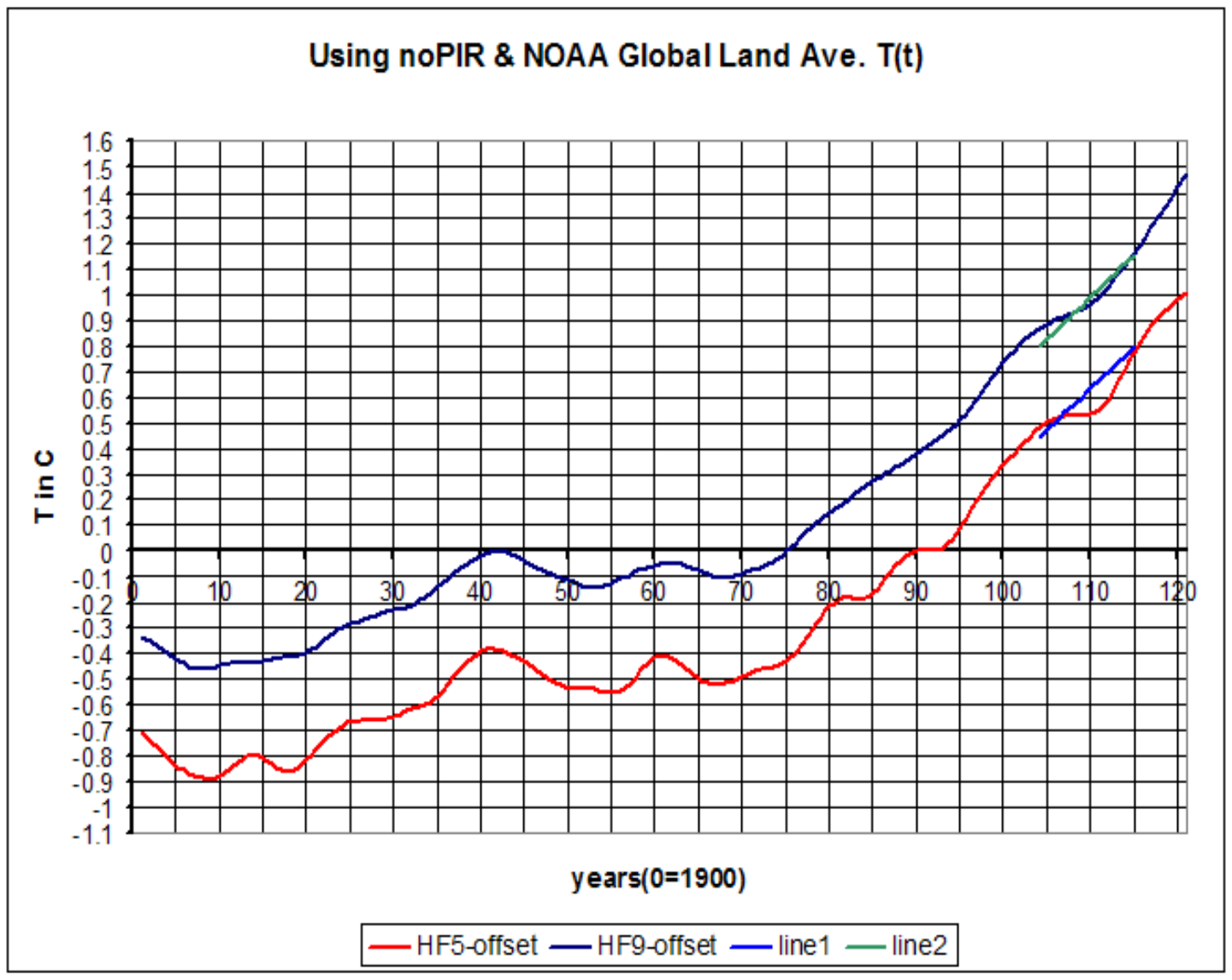

Fig.6.3B(discrete annual ave. plots; curves for visualization)

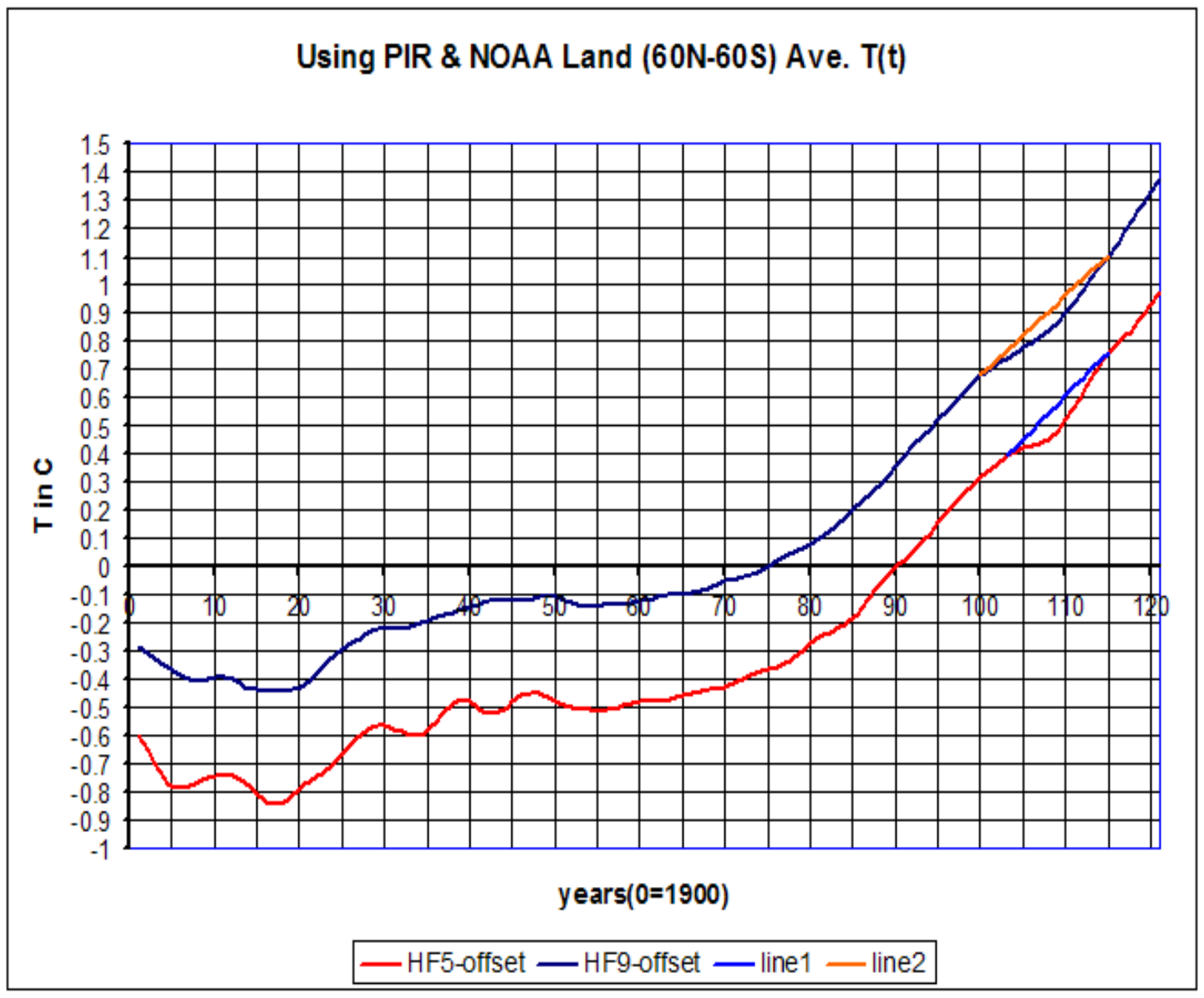

Fig.6.6(discrete annual ave. plots; curves for visualization)

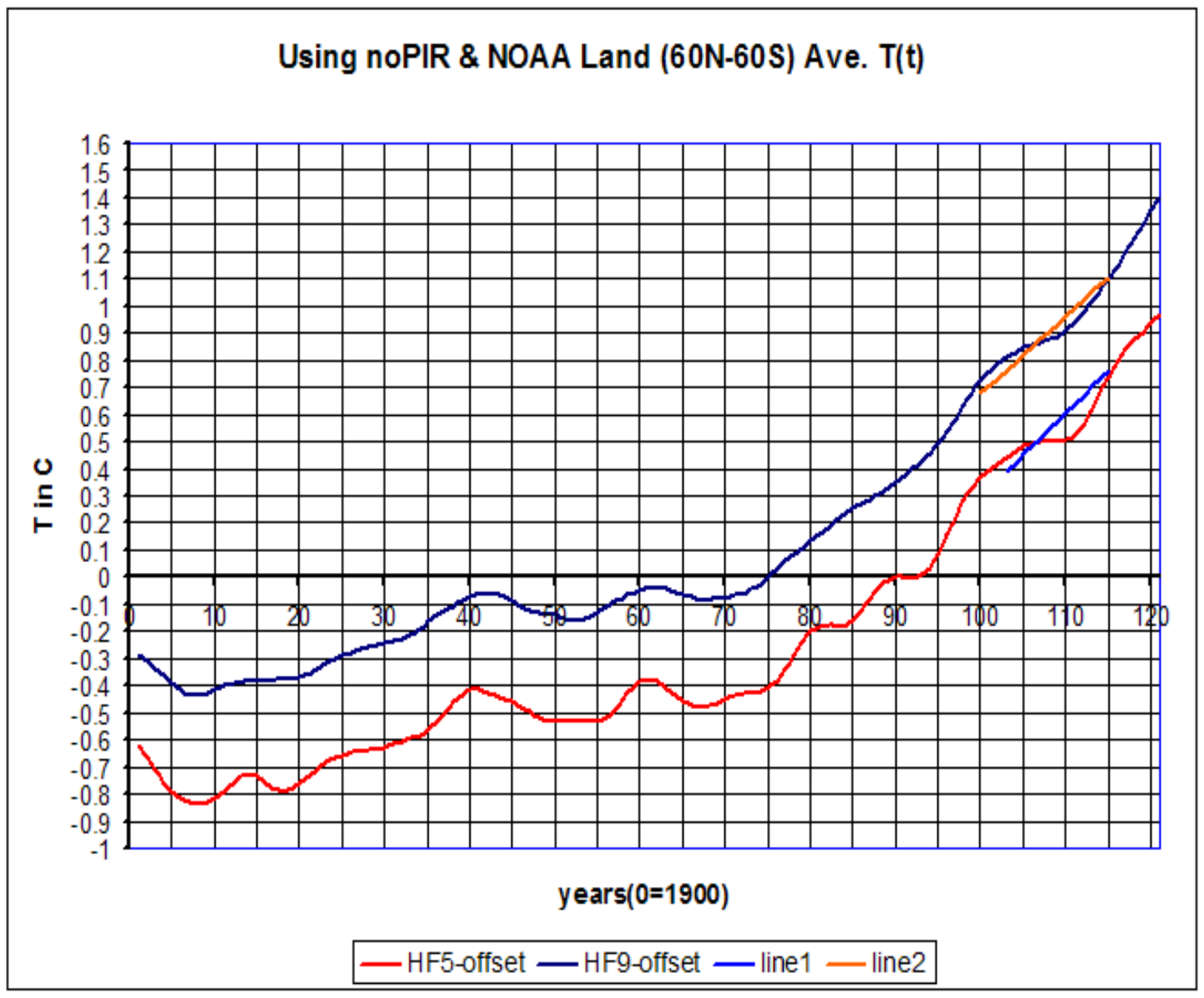

Fig.6.6B(discrete annual ave. plots; curves for visualization)

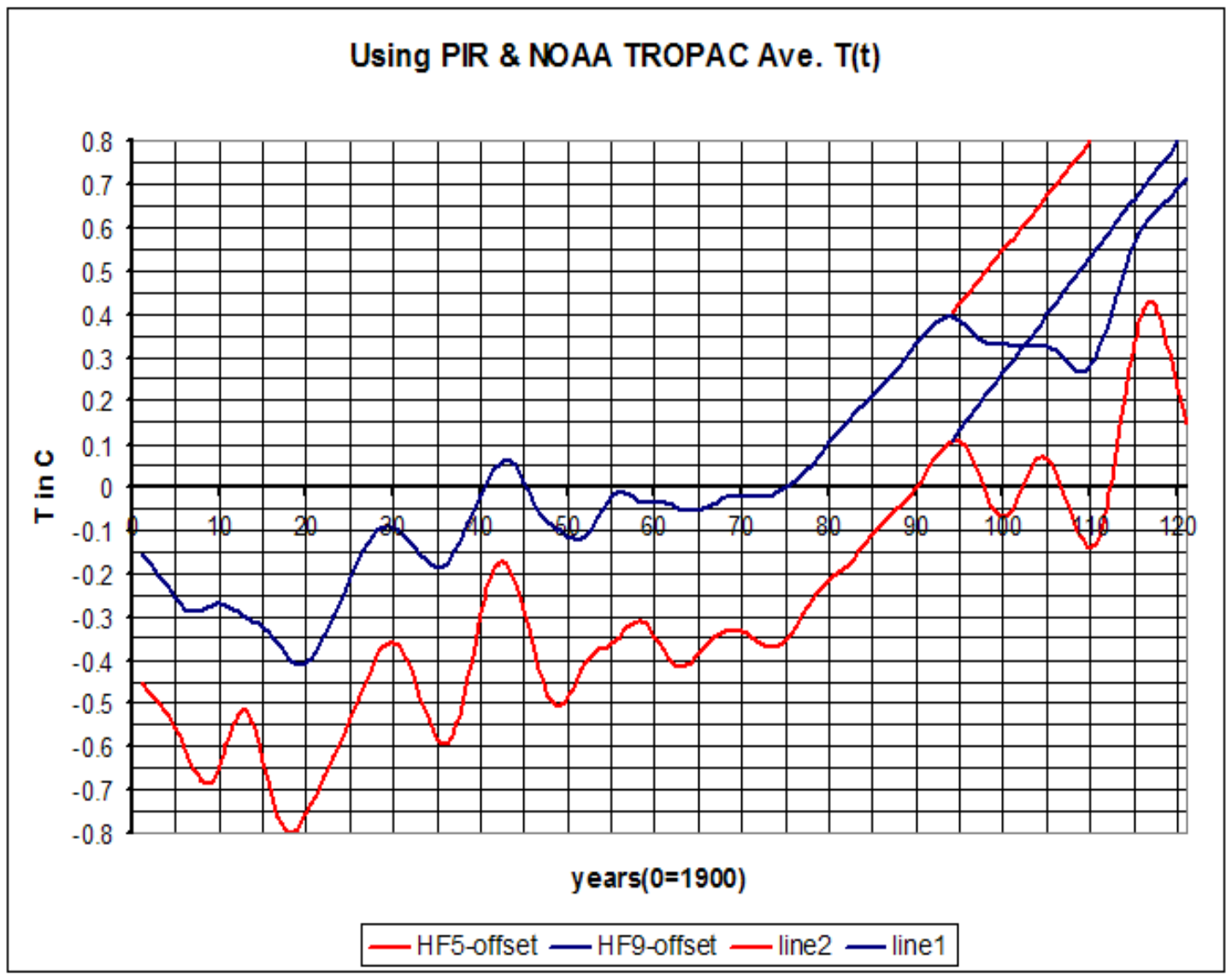

Fig.7.3(discrete annual ave. plots; curves for visualization)

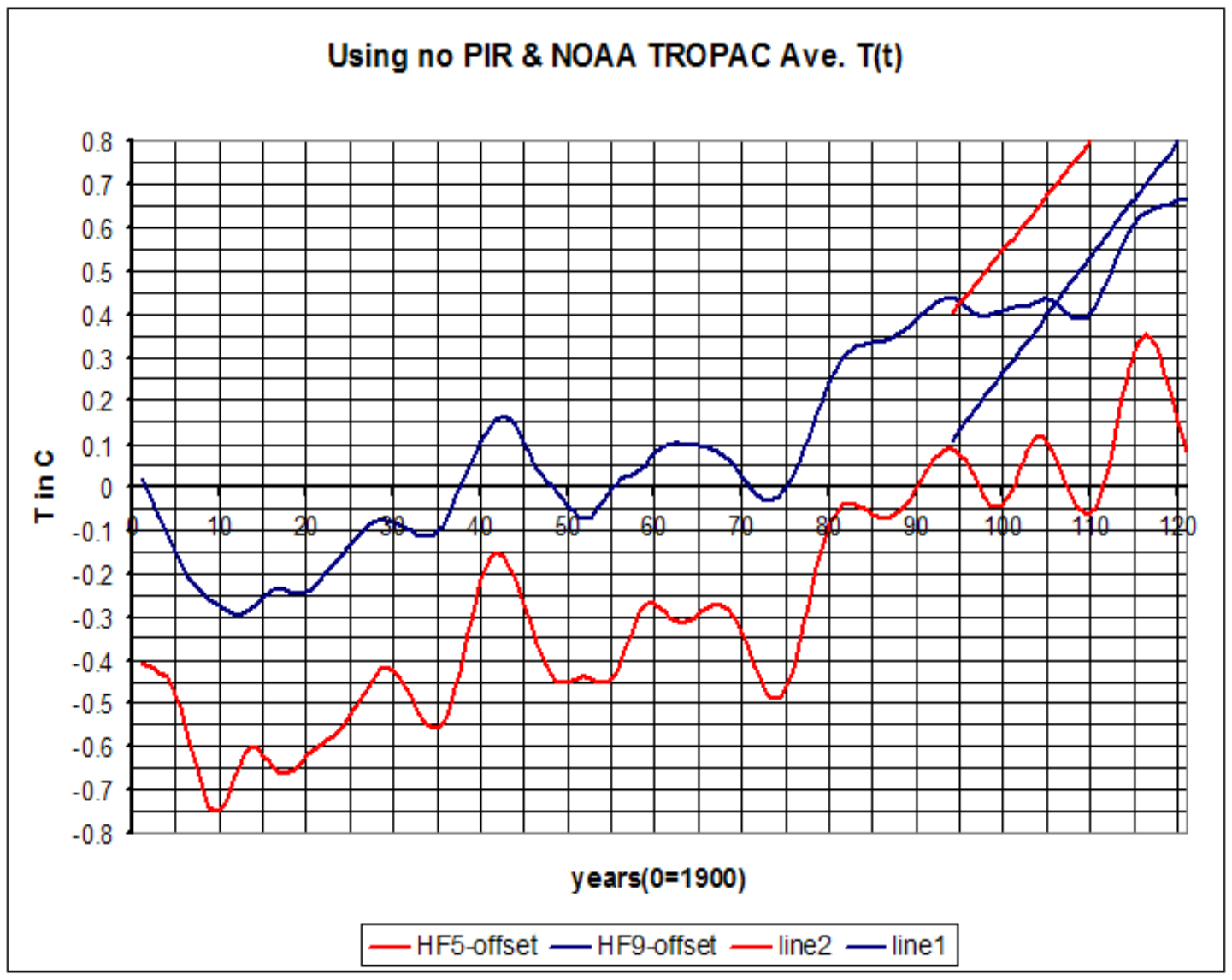

Fig.7.3B(discrete annual ave. plots; curves for visualization)

The cause of the oft observed transient temperature excursions from linearity (i.e. the "dip") occurring from about 1995,2000-2015 deserves comment (see Fig.5.6,5.9). If this were the response to an unknown independent external globally uniform radiative forcing then some adjustment to $\Delta F(t)$ would be required. However, the **land** average response (i.e. dip) to a transient global forcing (RF) should be very much larger than the similar **ocean** average response *because* the values of Ocean thermal capacitance and $G_{eff}$ are so much larger than those of over the land. Yet we observe they are smaller...compare Fig.6.3(land) with Fig.5.6(ocean).

But the observation above *is* consistent with a transient change in certain Ocean thermal transmission parameters and Ocean Energy distribution (*independent* of surface temperature variation); for example, a transient increase in the mixing layer, and/or deep ocean turbulent conduction, or transient vertical or horizontal circulation (especially over the Tropical Pacific?). Such a redistribution might reduce surface temperature (a dip), but would not in itself effect *total* E* in principle [see section **D** and **SS4,5** for details]. The land temperature "dip" could then only be a much weaker coupling (mixing) of the Ocean temperature into the land...which is as observed.

However, a transient radiative forcing might *only* exist over the Tropical Pacific (for unknown reasons), which would also account for the observation above. Comparisons of theoretical versus observed Q(t) in sect. **D** indicate this explanation is somewhat less likely, but not definitively.

Therefore it is possible no external independent RF $\Delta F(t)$ is required to account for this transient excursion (1995-2015). In any case, this forcing uncertainty will be avoided in section **E**.
However, these alternative explanations of the "dip" *do* effect the Q(t) evaluations of section **D**.

**[!Note: although not specifically shown above, the PIR temperature trends (and therefore ECS values) using HADCRUT temperature data instead are similar, but 6% "cooler" over land, and 8% "warmer" over oceans.!]**

### B.2) PIR Accuracy Verification

It is expected, and mathematically required, that

e.10 $\Delta T(t)_{global.ave}=0.3*\Delta T(t)_{land.ave}+0.7*\Delta T(t)_{ocean.ave}$

at all times t, where 0.3 and 0.7 are the global fractions of land and ocean areas respectively. This will be proven to be true for the NOAA yearly averaged ("raw") observed temperature data (Fig.8) and the highly filtered (H5[ ] in Fig.9) data for all years.

However, the non-linear PIR process is applied to the Global, Land and Ocean regions *independently*, AND there is no expectation the above relation should hold after PIR is performed. The relation could be expected to hold only if the PIR procedure *DOES* in fact properly remove only specific independent sinusoidal frequency components from all regions as intended. In that case the remaining *measured* temperature components should and would, by necessity, satisfy the relation.

It will be shown that after filtering and PIR the relation e.10 does indeed hold very closely (see Fig.10,11). It is unexpected that such an extensive approximate procedure could satisfy e.10 coincidentally, and this essentially verifies the validity of the PIR process.

In the following plots (Fig.8,9,10,11):
G ≡ Δ Global temperature average
O ≡ Δ Global Ocean temperature average
L ≡ Δ Global Land temperature average
G-O7L3 ≡ G - (O·0.7 + L·0.3)

In actuality the fractions of .72 and .28 produce the best result and are used across the 1900-2021 range, as opposed to .7 and .3 respectively, for some unknown reason. The value of (G-O7L3) is virtually zero in all cases verifying the relation e.10 in all cases. Note also that the simplifying proportionality e.2 does seem to hold over the evaluation ranges used herein to good approximation, specifically in Fig.10,11 where $t_o$ is set to1980. **[The working spread sheet used to produce these figures is "T-regional-comparisons-over-time.xls"]**

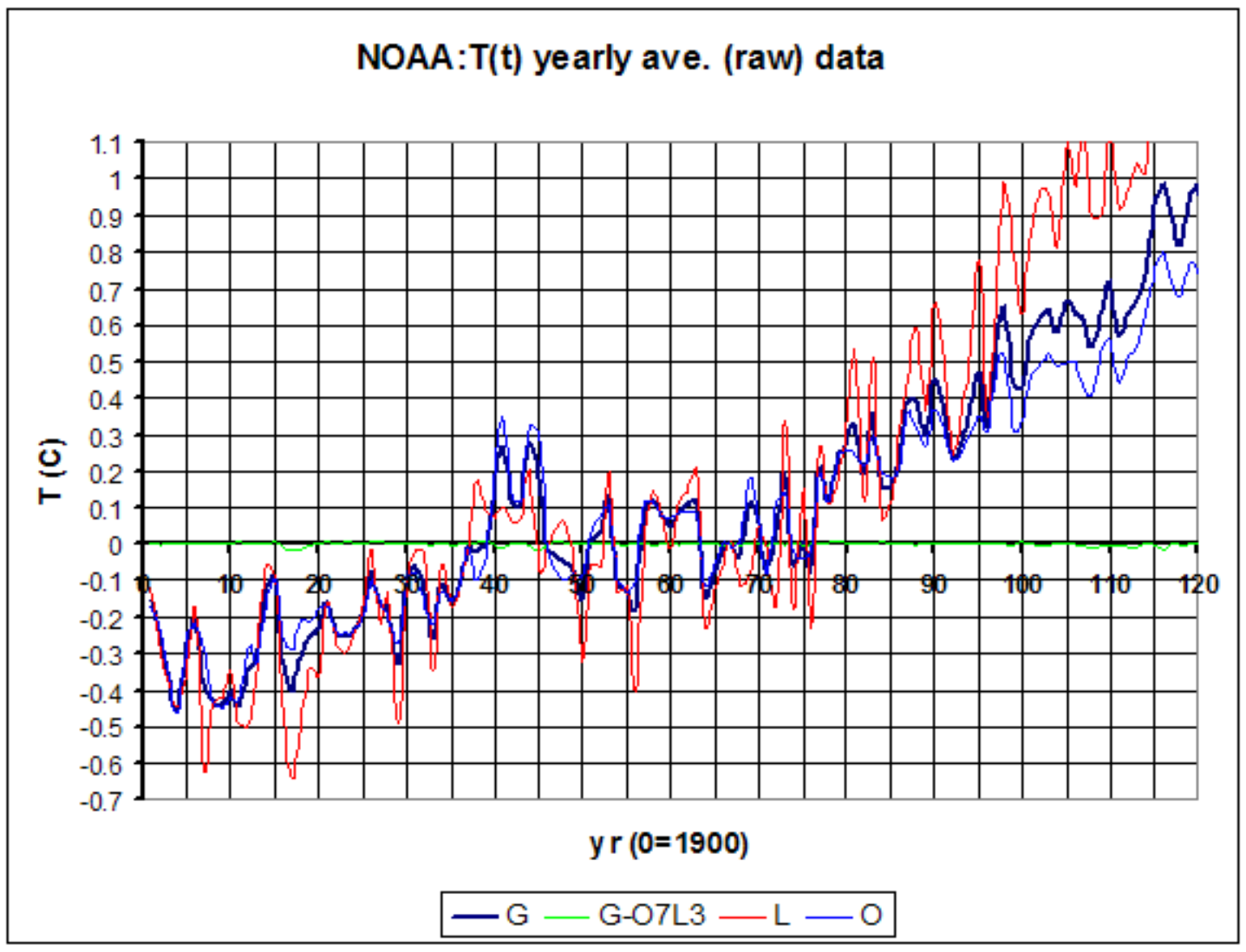

Fig.8(discrete annual ave. plots; curves for visualization)

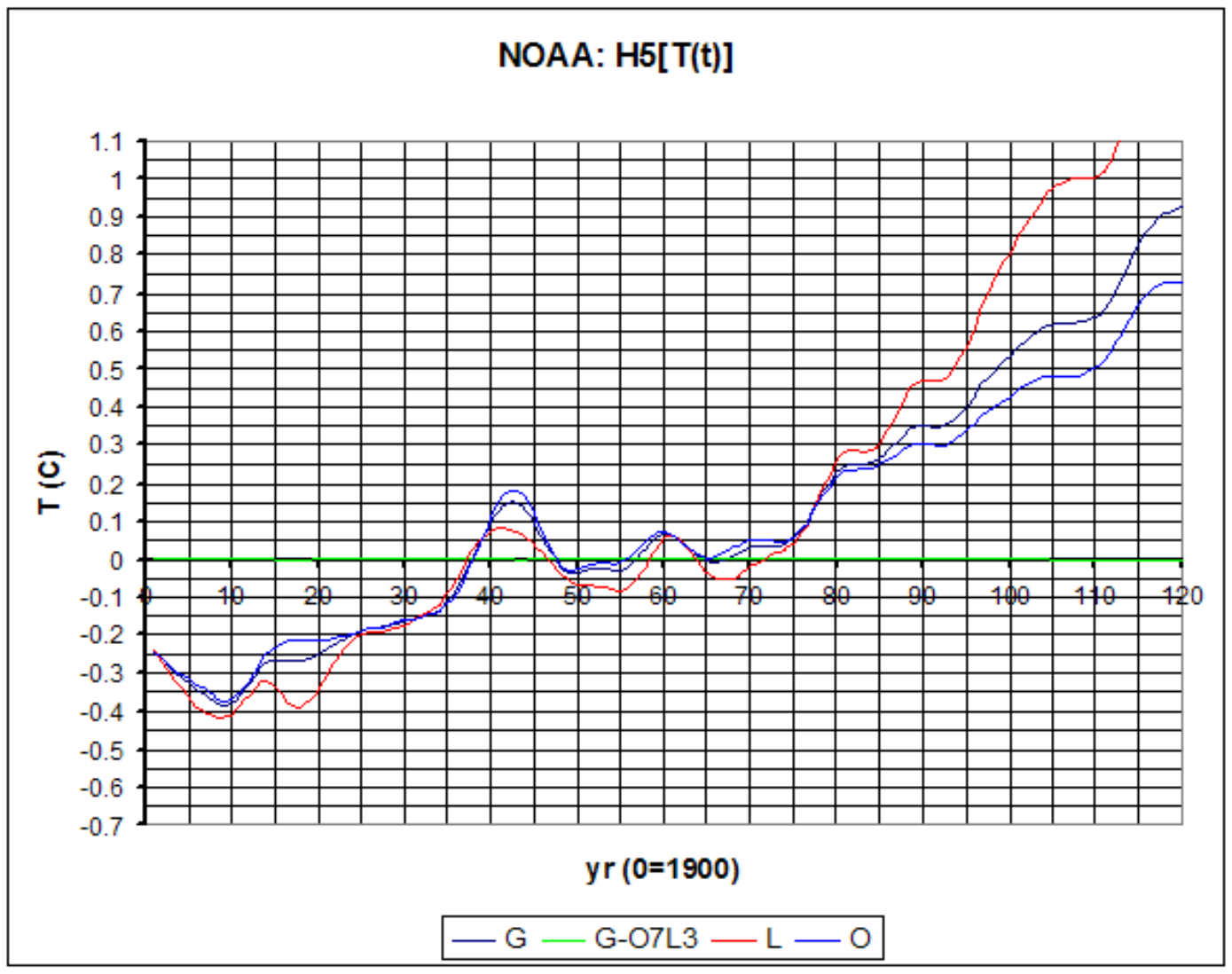

Fig.9(discrete annual ave. plots; curves for visualization)

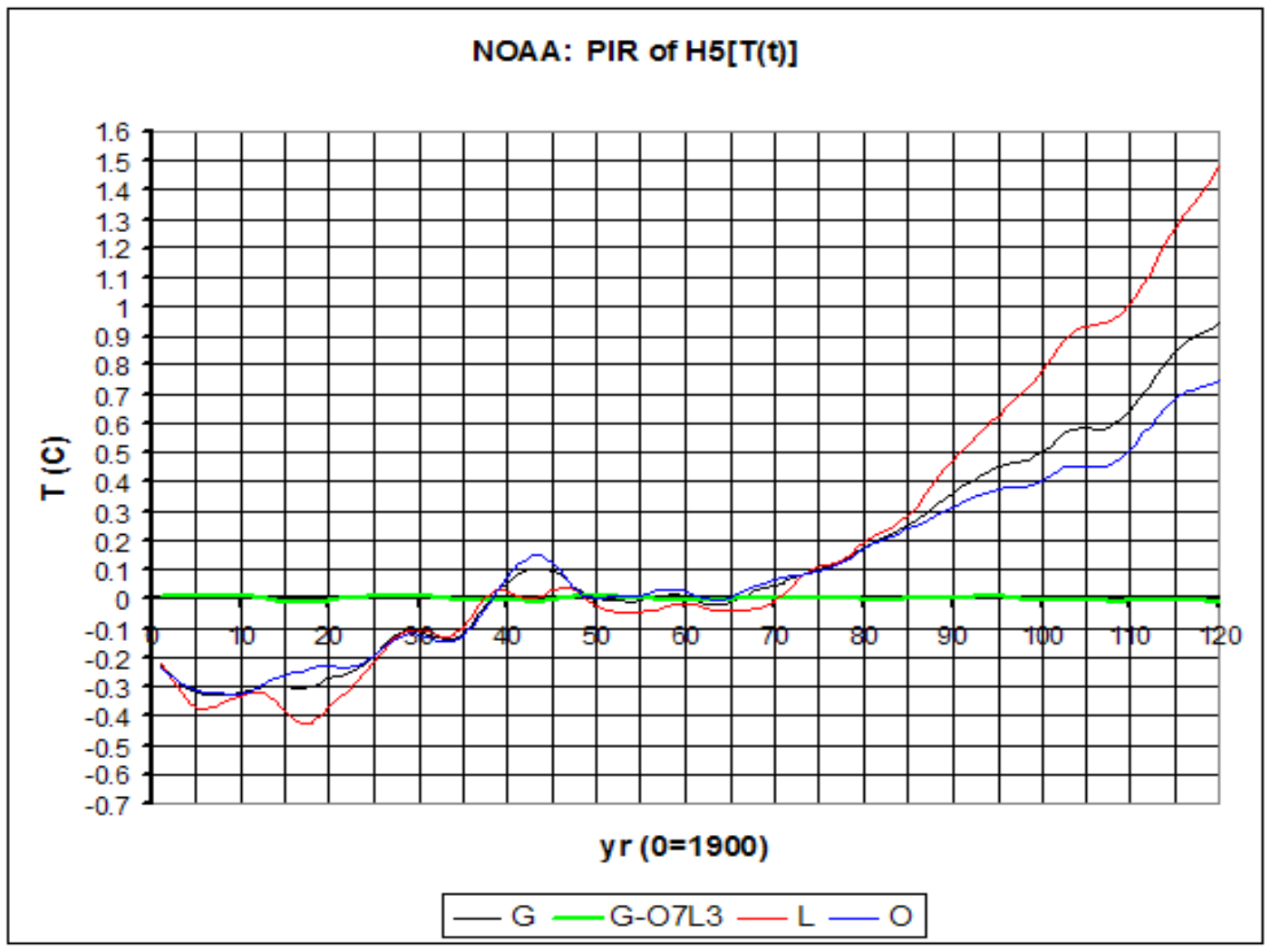

Fig.10(discrete annual ave. plots; curves for visualization)

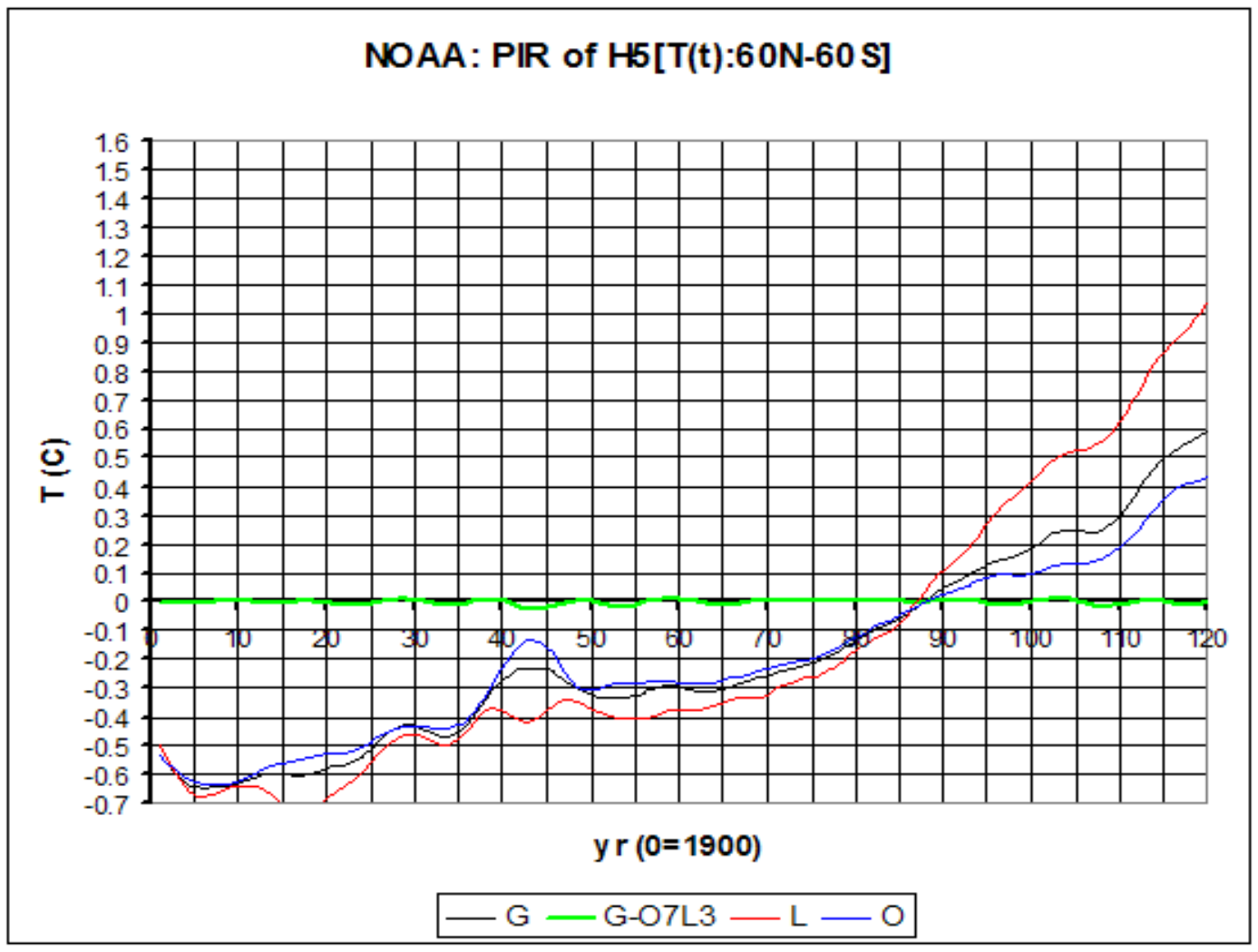

Fig.11(discrete annual ave. plots; curves for visualization)

---

## C) Evaluation of F, $\Delta F$ and $\underline{\Delta}F$

Starting with $\Delta F(t')$, as defined in e.3, presume nearly all anthropogenic forcing is due to globally well mixed gases and thus uniform with respect to land as compared to ocean. And presume volcanic and solar variations are similarly regionally uniform. It *may* be true that land-use forcing are slightly larger over the land, but that is omitted here. Note the IPCC AR5[7] definition of $CO_2$ and GHG **ERF** (effective radiative forcing) is the same as $CO_2$ **DRF** (direct radiative forcing) defined in Section **A** above and preferred in this analysis. Therefore effectively IPCC ERF ≈ DRF since nearly all **Δ**Forcing is GHG in the evaluation period (excluding Volcanic). The GHG 60N-60S $\Delta F$ average will differ from the global average, being about 4% larger.

Forcing estimates are taken from the Lewis[1] & Curry *updated* IPCC AR5 ERF and IPCC AR5[7] ERF plots shown in Fig.12 and 13 respectively, and specific volcanic forcing from Schmidt[8] et al. in Fig.14, and Gregory[9] et al. in Fig.15 . The total anthropogenic forcing function (Fig.12) is so nearly linear it is modeled here as linear after 1970 at 0.4 ($W/m^2$)/Decade as is shown in Fig.16 along with the H5, H9 filtered values. Note, the small periodic solar cycle forcing variation is omitted for simplification.

The yearly averaged volcanic forcing is estimated using Gregory[9] from 1968-1978. The yearly averages from 1979-2016 are taken directly from Schmidt[8] as shown in Fig.17(upper). This volcanic forcing can be multiplied by a volcanic scalar "VS" to produce an effective forcing.

The original aerosol cooling trend (IPCC AR5 ERF) is shown to be continuing after 1970 in Fig.12 . However results from PNNL[10] (Fig.20) and GISP2 ice cores[11] (Fig.21) indicate a constancy or reduction of the primary sulfate aerosols after 1975. Similarly, the IPCC AR6[15] WG1 TS:figTS.9 (see Fig.18) also indicates a constancy or reduction of the primary aerosol forcing *magnitude* after 1975. This is important because it indicates the aerosol $\Delta F$ component trend to be zero, or even positive; whereas otherwise this component has a very large uncertainty. **Thus evaluating data after 1975 removes most of the aerosol uncertainty**.

**[Since an upper limit on temperature change and ECS evaluations is considered most desired, the "*maximal* ECS" option is *defined* herein as that which produces a smaller value of $G_{eff}$ and a larger value of ECS]**. Thus the revised $\Delta F(t)$ of Fig.12 is a *maximal* value, since $\Delta F$ total is likely slightly larger than shown!

Thus, the total forcing is approximated as a linearly increasing value of 0.4 (W/m$^2$)/Decade after 1970, plus a possible Volcanic component. But calculate the **volcanic component** of $\underline{\Delta}F-\underline{\Delta}Q$ over **land** in e.4c,f . Let $t'_2$ occur well after 1995 (and thus the volcanic component $[\Delta F(t'_2)-\Delta Q(t'_2)_{Land}]$ is essentially zero, i.e. essentially no Volcanic activity), and set $t'_1$ before 1990 during volcanic cooling where $[\Delta F(t'_1)-\Delta Q(t'_2)_{Land}] \le 0$; where $\Delta Q/\Delta F \le 1$, always. Then $\underline{\Delta}F_{volc.}-\underline{\Delta}Q_{volc.} \equiv ([\Delta F_{volc.}(t'_2)-\Delta Q(t'_2)_{Land}] - [\Delta F_{volcanic}(t'_1)-\Delta Q(t'_1)_{Land}]) \ge 0$ . Thus any volcanic forcing (cooling) component increases $G_{eff}$ where $G_{eff} = (\underline{\Delta}F_{total} - \underline{\Delta}Q_{total})/\underline{\Delta}T$, and decreases ECS where $ECS=\Delta F_{2xCO2}/G_{eff}$ . Therefore setting the volcanic component of $[\Delta F(t'_2)-\Delta Q(t'_2)_{Land}]$ identically to zero over **land** results in a higher land ECS calculation, and is thus also always the *maximal* ECS option.

Because the volcanic components occur suddenly, the high frequency components over the **Lands** are mostly immediately absorbed into (or out of) the surprisingly non-trivial thermal capacitance of the atmosphere and extra-oceanic energy sinks. During the very large (-**2.4** W/m$^2$, see Fig.14,17) Mt. Pinatubo volcanic forcing of 1992, the yearly averaged "raw" temperature excursion *from* the highly filtered H9[ ] average (shown in Fig.23 (and Fig.6.4B) for land (60N-60S)) is only about **-0.23**°C and barely distinguishable from the noisy adjacent variation. Thus, over Land, the Volcanic *componen*t of the term $|\Delta F_{land}-\Delta Q_{Land}|$ is much smaller than $|\Delta F_{volcanic}|$ itself. Further, the H5[T(t)] "Pinatubo volcanic" excursion from the local linear trend (i.e. H9[T(t)], see Fig.23) is only ~**0.04**C . This is about 3.3% of the ~1.2°C linear $\Delta T$ land variation over a 40 year evaluation range; thus the volcanic component is a very small part of the $\Delta T$ evaluation.

The value of volcanic $(\Delta F_{land}-\Delta Q_{Land})$ can be estimated as $(\Delta F_{land}-\Delta Q_{Land}) = \Delta T \cdot G_{eff.Land} = \mathbf{0.04} \cdot 1.5 = 0.06$ W/m$^2$ ;where $G_{eff.Land} \le 1.5$ W/(m$^2\cdot$°C). This is only 0.06(Wm$^{-2}$)/1.6(Wm$^{-2}$) = 3.8% of the total $\Delta F_{Land}$ over a 40 year evaluation, and which occurs for only a few years. As a result the **volcanic component** of the slope of the "best linear fit" of $(\Delta F-\Delta Q)$ over the entire evaluation range over Land is essentially zero and need not be calculated in Section **E**, and is a *maximal* choice, as described above.

And the thermal capacitance of the Oceans is *much* larger than over the Land (14x). Thus, the high frequency components over the **Oceans** are immediately absorbed into (or out of) the large thermal capacitance of the Oceans, so that $\Delta F_{volcanic} \approx \Delta Q_{volcanic}$. As a result, the **volcanic component** of $(\Delta F-\Delta Q_{Ocean})$ over the oceans is virtually zero and need not be calculated in Section **E)**, and is also the *maximal* choice (just as with land regions described above)

**So, assuming this volcanic component ($\Delta F_{volc.} - \Delta Q_{volc.}$) to be zero Globally (i.e., VS=0) is the *maximal* ECS option, the simplest option, a very good approximation, and may be so assumed**. Thus *maximal* F(t), $\Delta F(t)$, and $\underline{\Delta}F$ can be calculated simply using Fig.16, VS=0, where $\Delta Q_{volc.}$ must also be set to zero artificially.

It should be noted that the Mt. Pinatubo, El Chikon, and Agung eruptions were separated by 9.3 year multiples (see Fig.15). And the Mt Pinatubo eruption of 1991 occurred just before the 1992 lunar tidal maximum shown in Fig.22 (Keeling[4]) . Similarly, the Mt. Katmai, Santa Maria, and Krakatau eruptions were separated by very nearly 9.3 year multiples, also. Recalling the 9.3 year lunar cycles discussed in Section **B**, this is surely not random coincidence? The eruptions seem in some way synchronized to the lunar-earth tidal cycle. If so, it can be difficult to separate the volcanic forcing effects from the completely separate lunar parametric tidal 9.3 year fluctuation cycle that is occurring simultaneously. This latter cycle is shown very convincingly in Fig.19 for NOAA Land H5[T(t)] data (60N-60S) after PIR (see Section **B 8)** ), but where the 9.3 year component is ***not*** compensated. Here the 9.3 year repetitive component is clearly observed from 1950-2005.

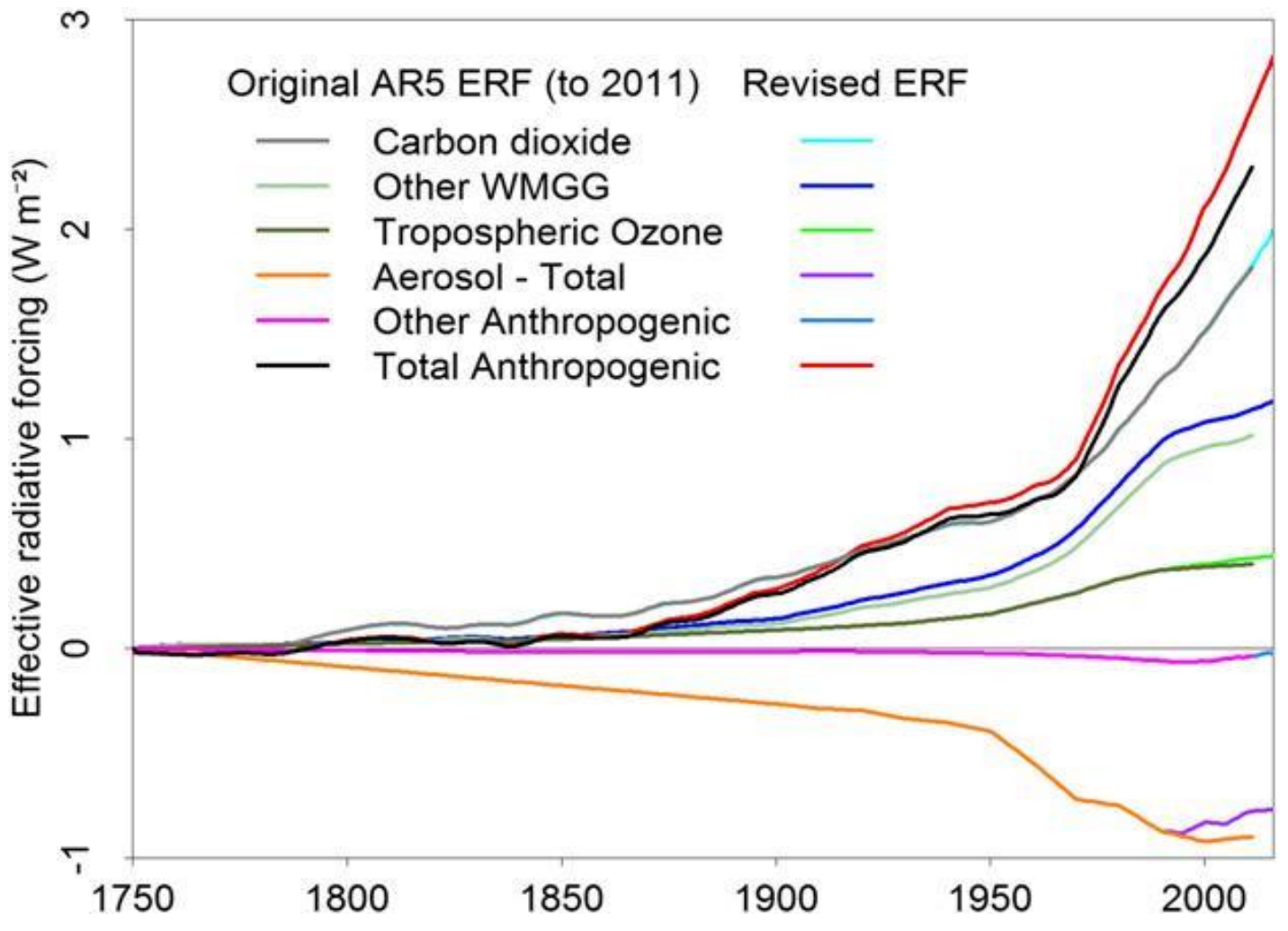

Fig.12 source: ref.[1] CC BY-NC-SA 4

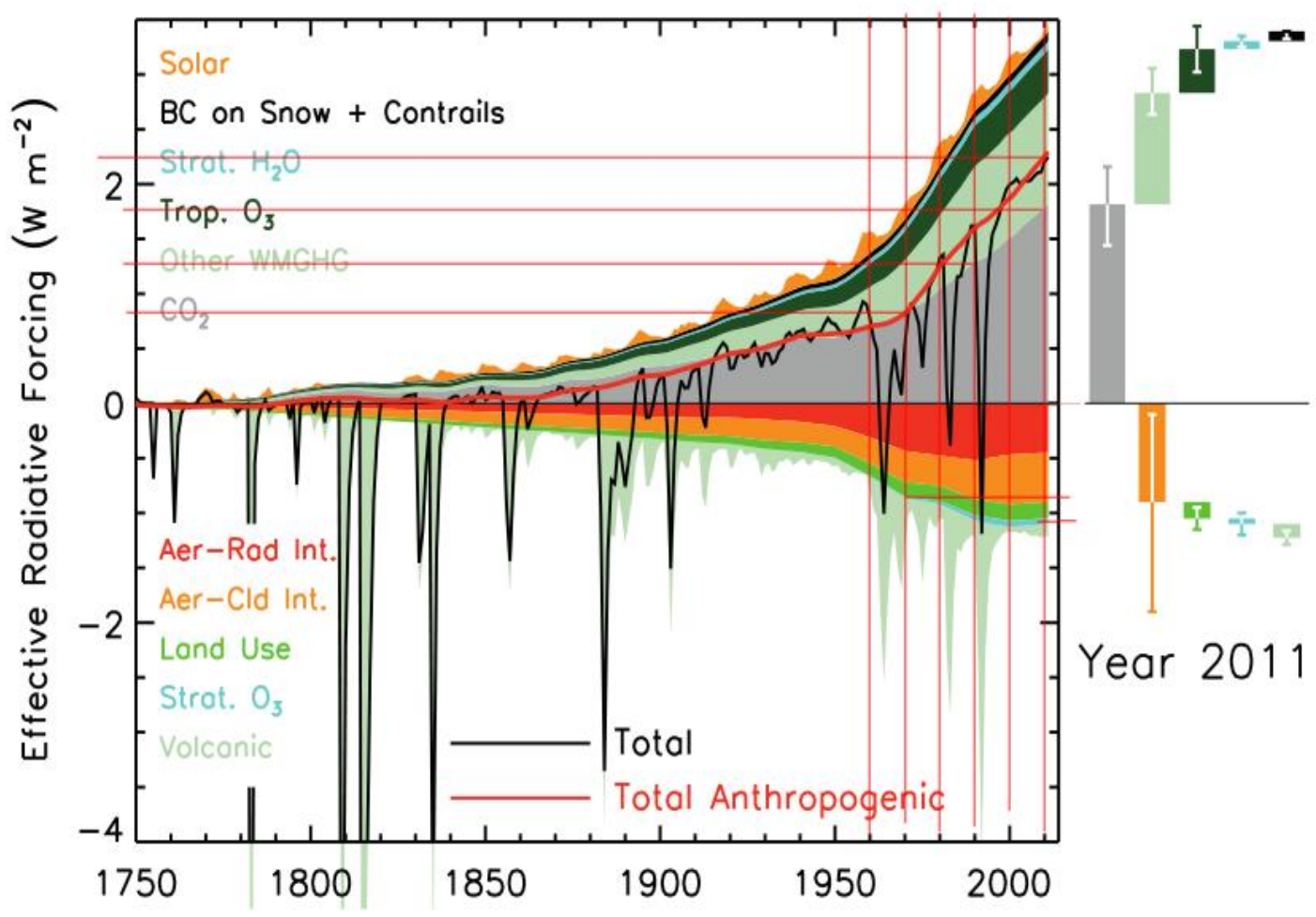

Fig.13 source: ref.[7] CC BY-NC-SA 4

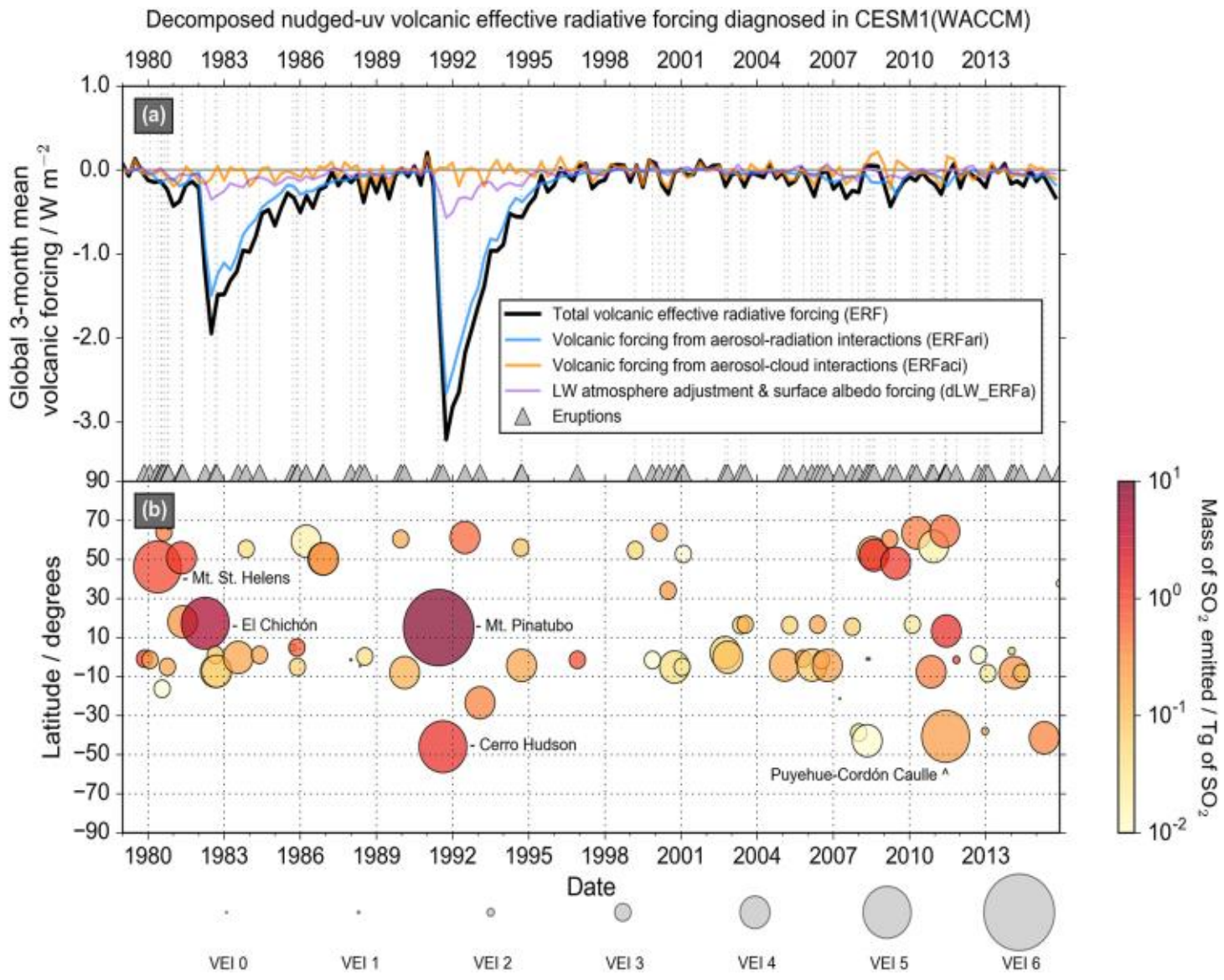

Fig.14 source: ref.[8] CC BY-NC-SA 4

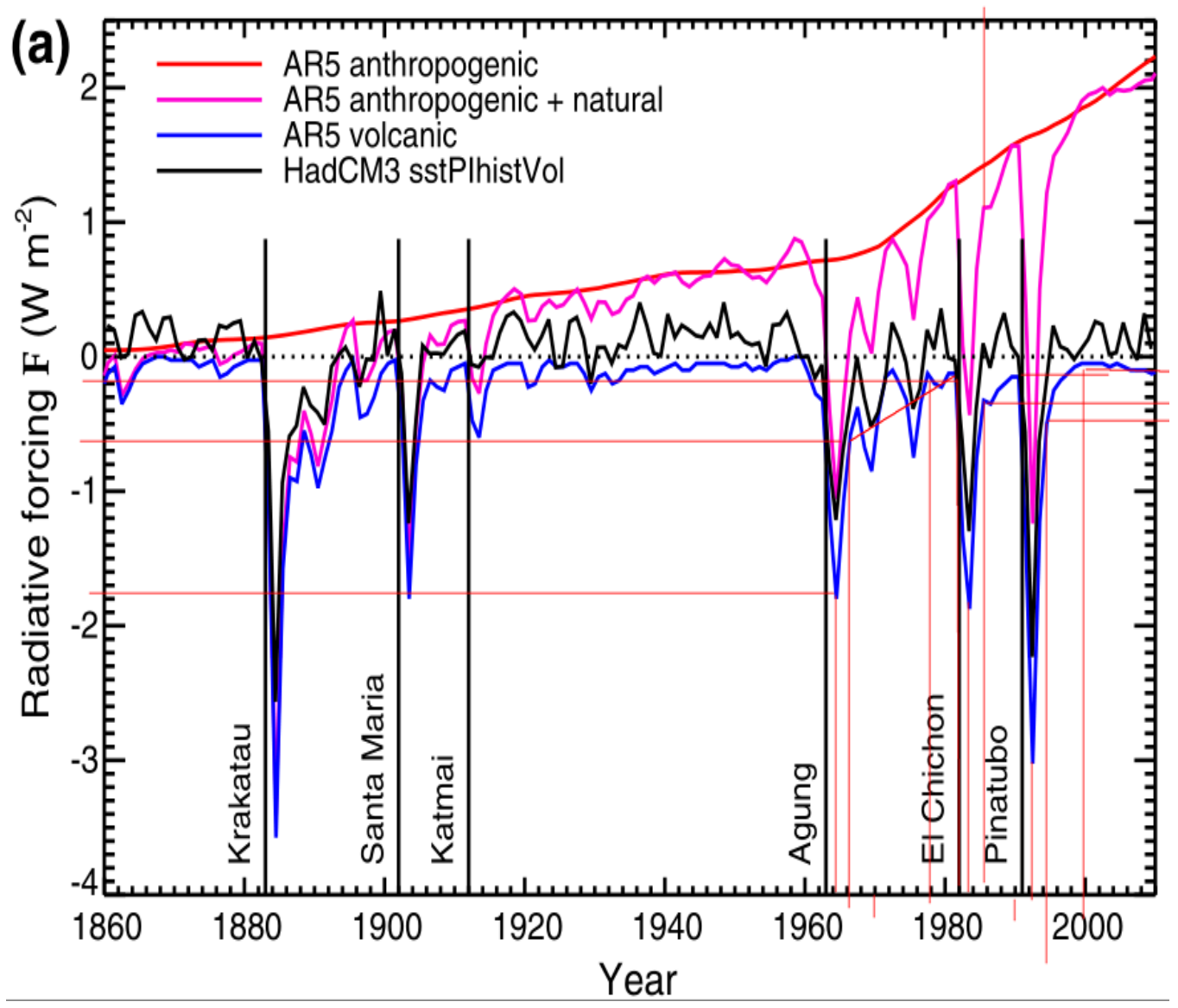

Fig.15 source: ref.[9] CC BY-NC-SA 4

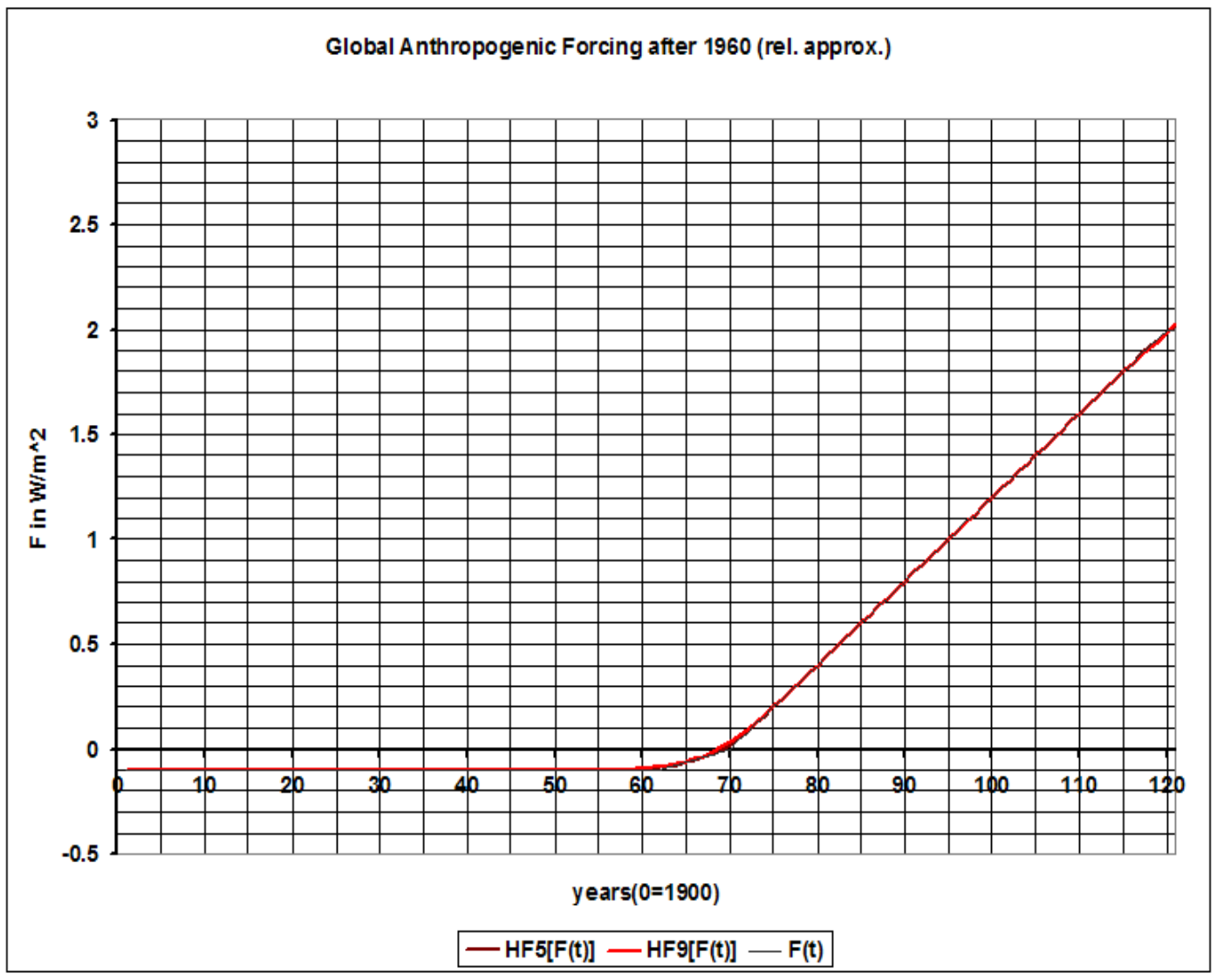

Fig.16 (discrete annual ave. plots;curves for visualization)

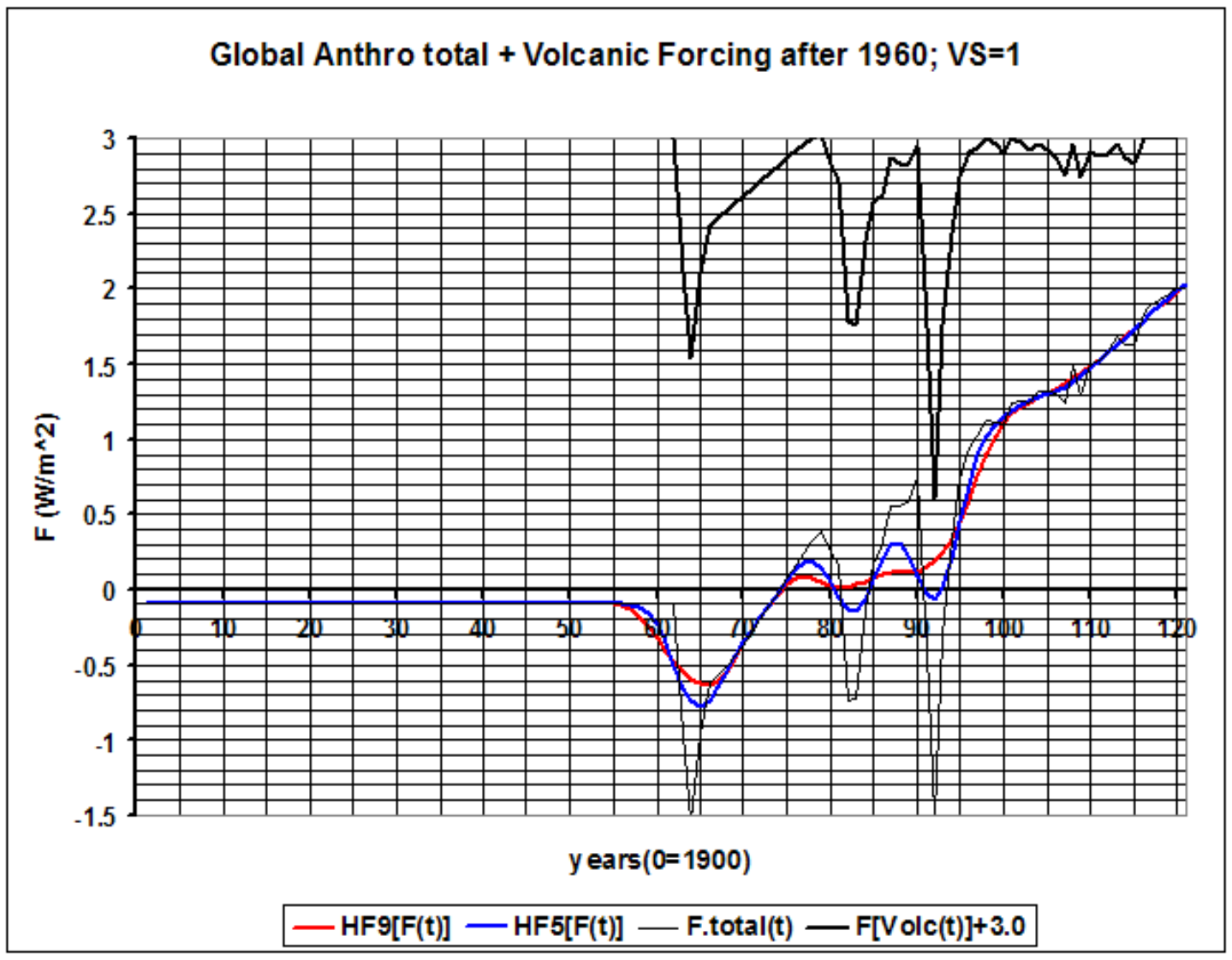

Fig.17 (discrete annual ave. plots;curves for visualization)

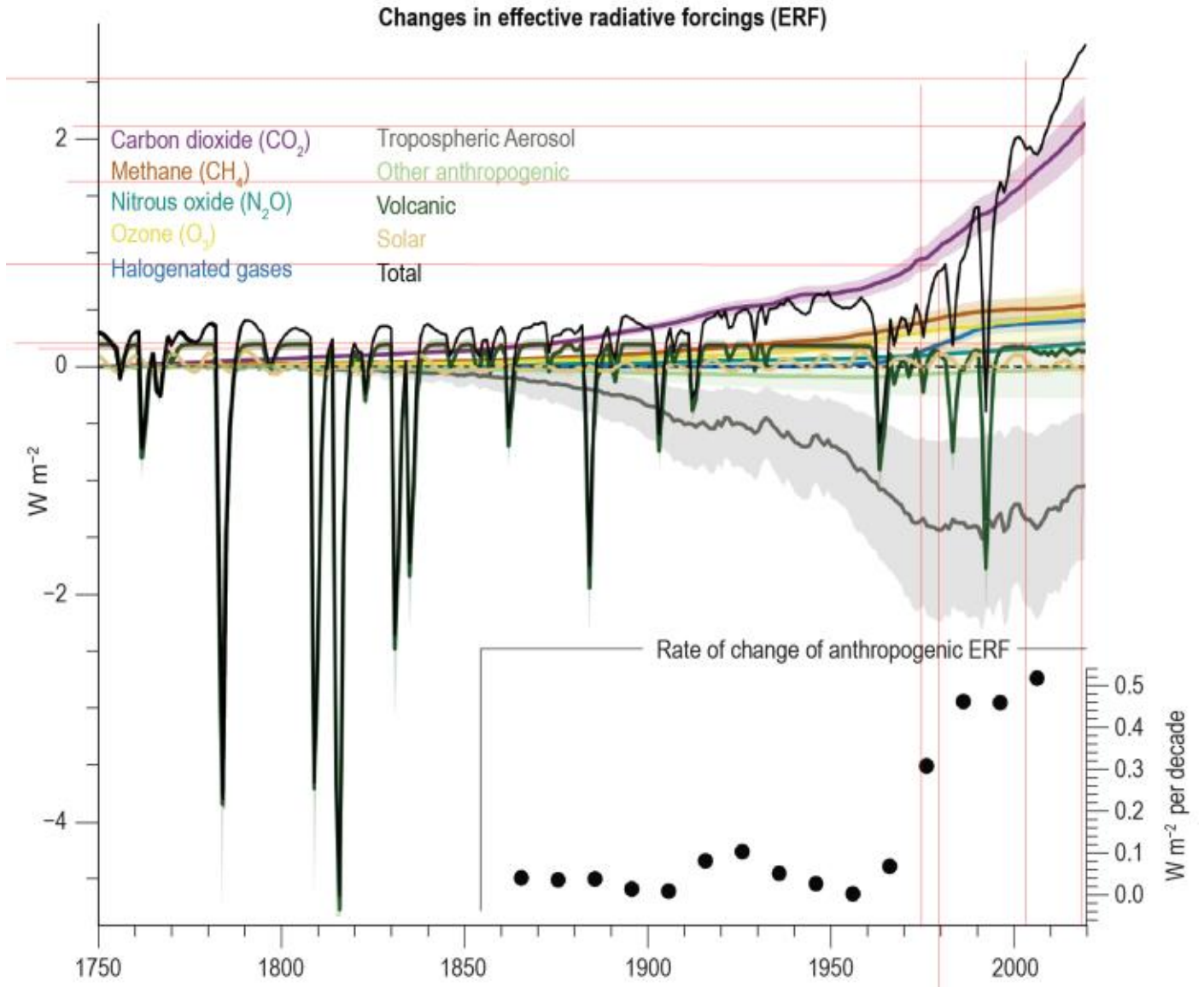

Fig.18 source: ref.[15] CC BY-NC-SA 4

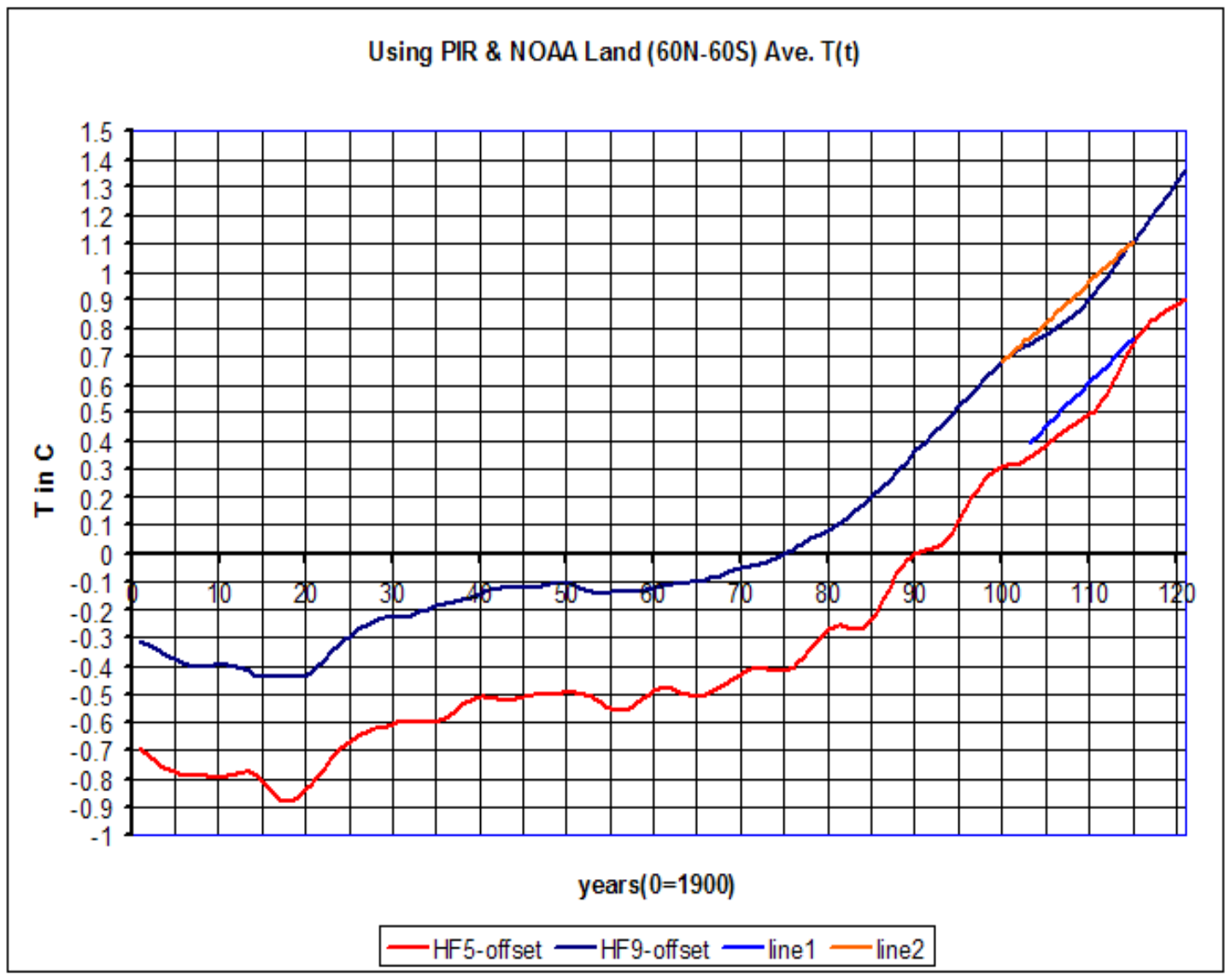

Fig.19 (discrete annual ave. plots; curves for visualization)

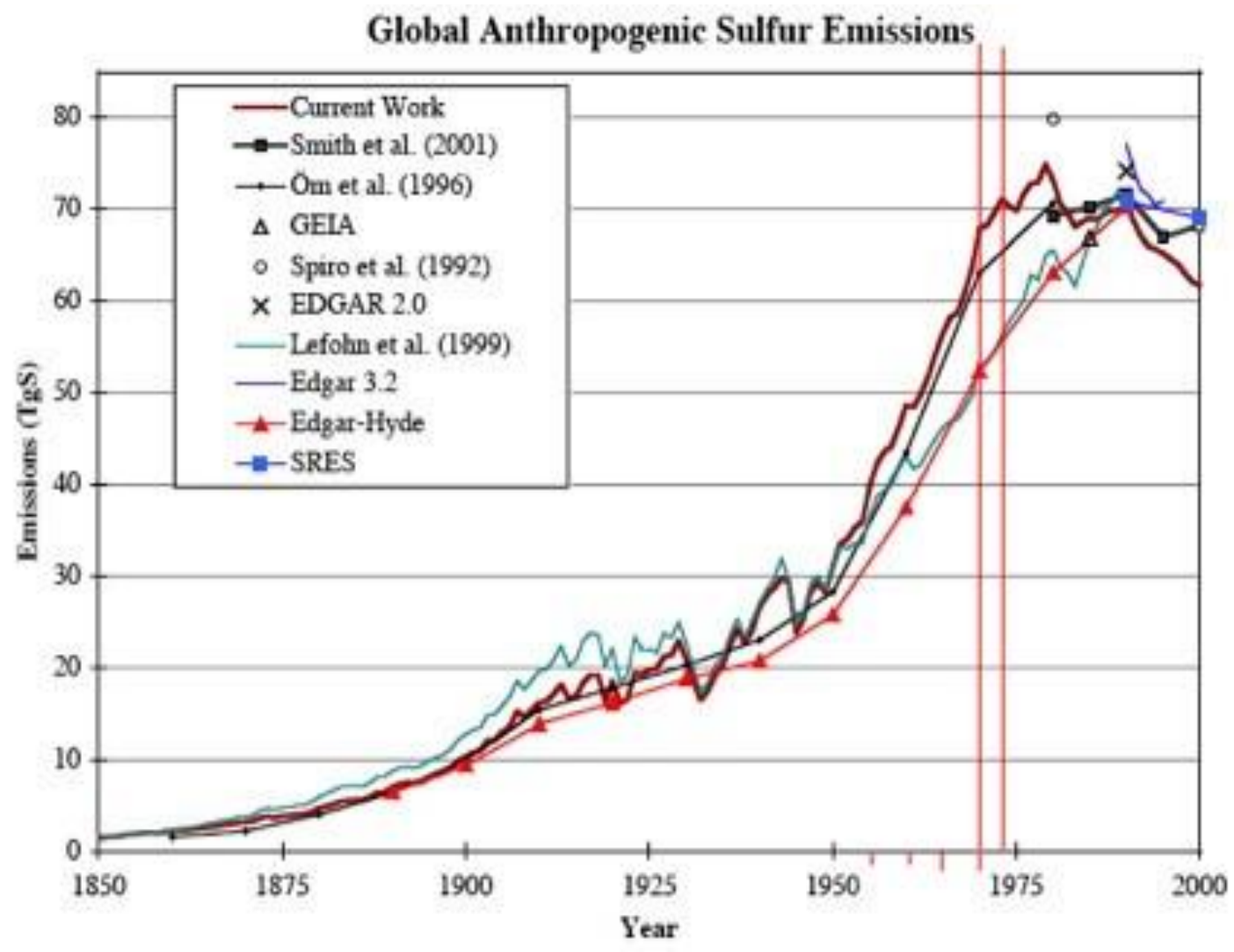

Fig.20 source: ref.[10] CC BY-NC-SA 4

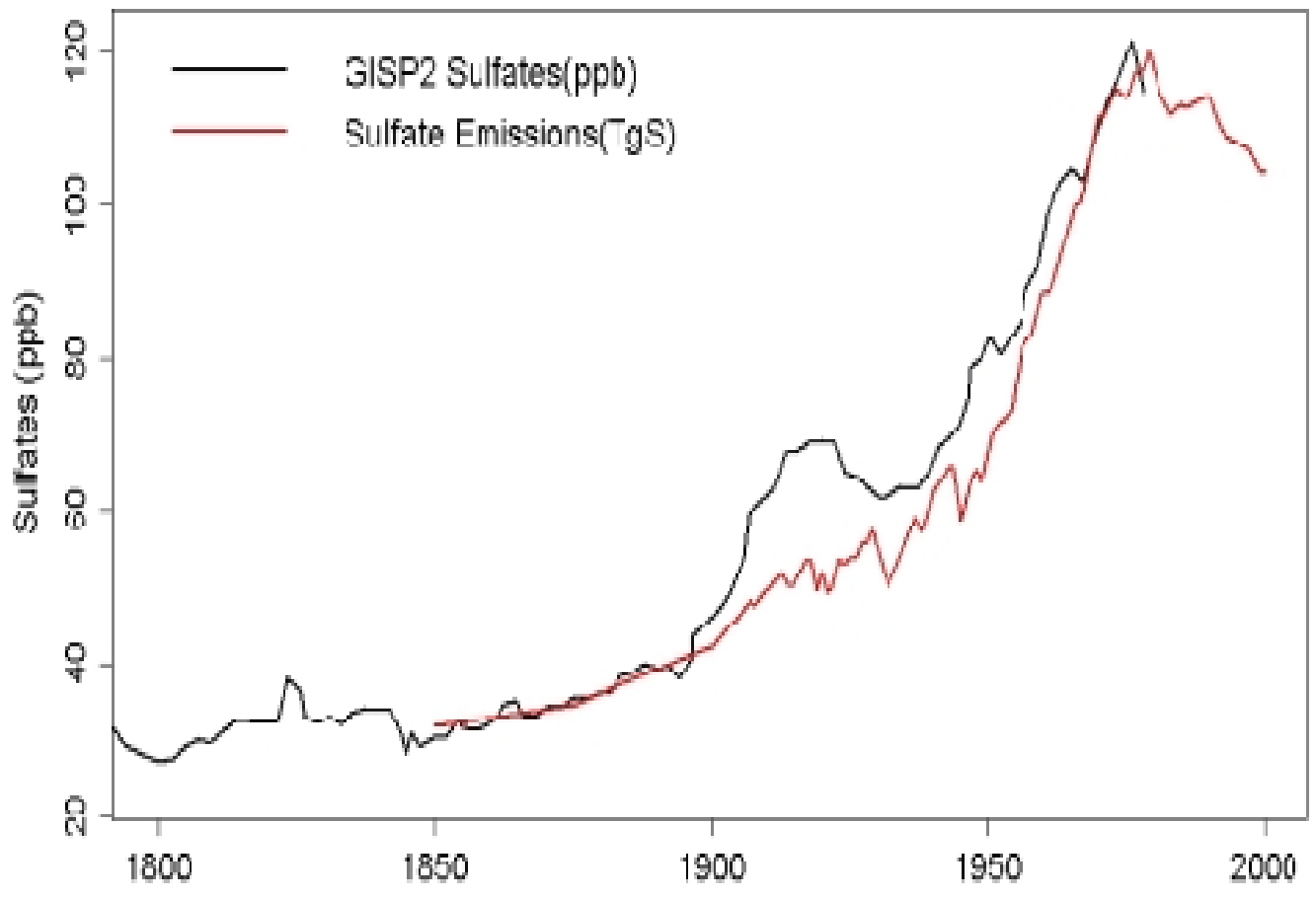

Fig.21 source: ref.[11] CC BY-NC-SA 4

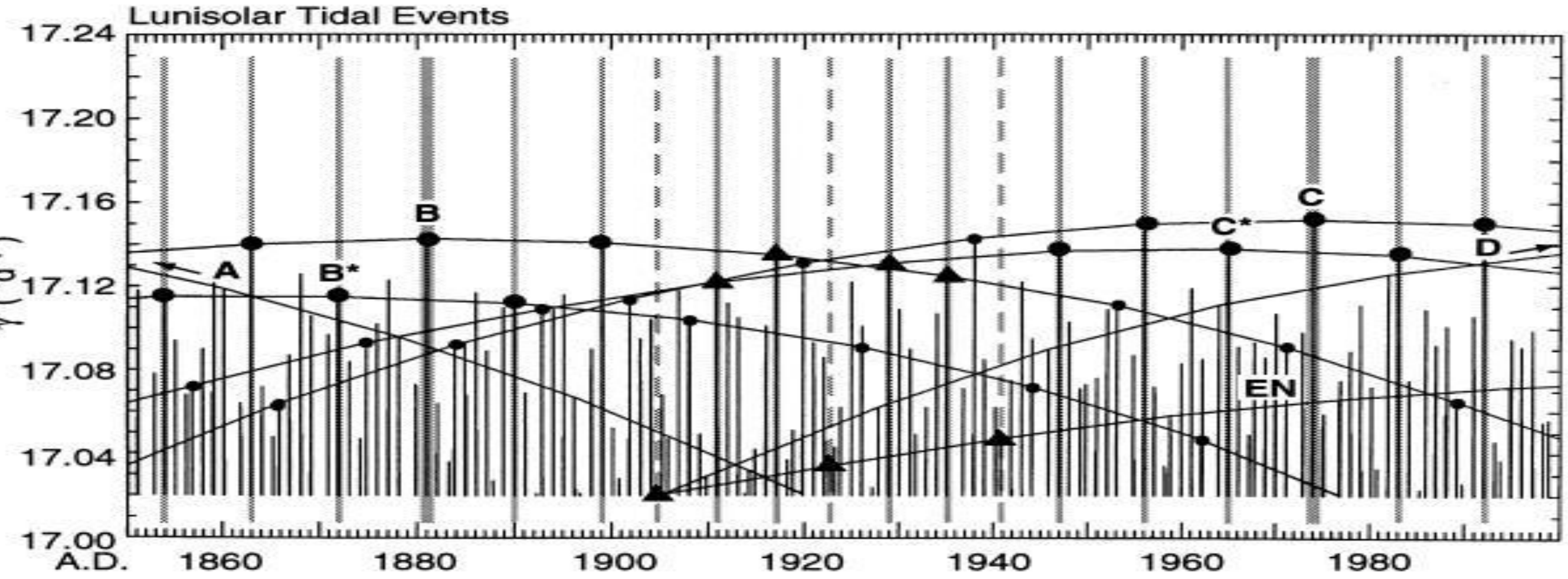

Fig.22 source: ref.[4] CC BY-NC-SA 4

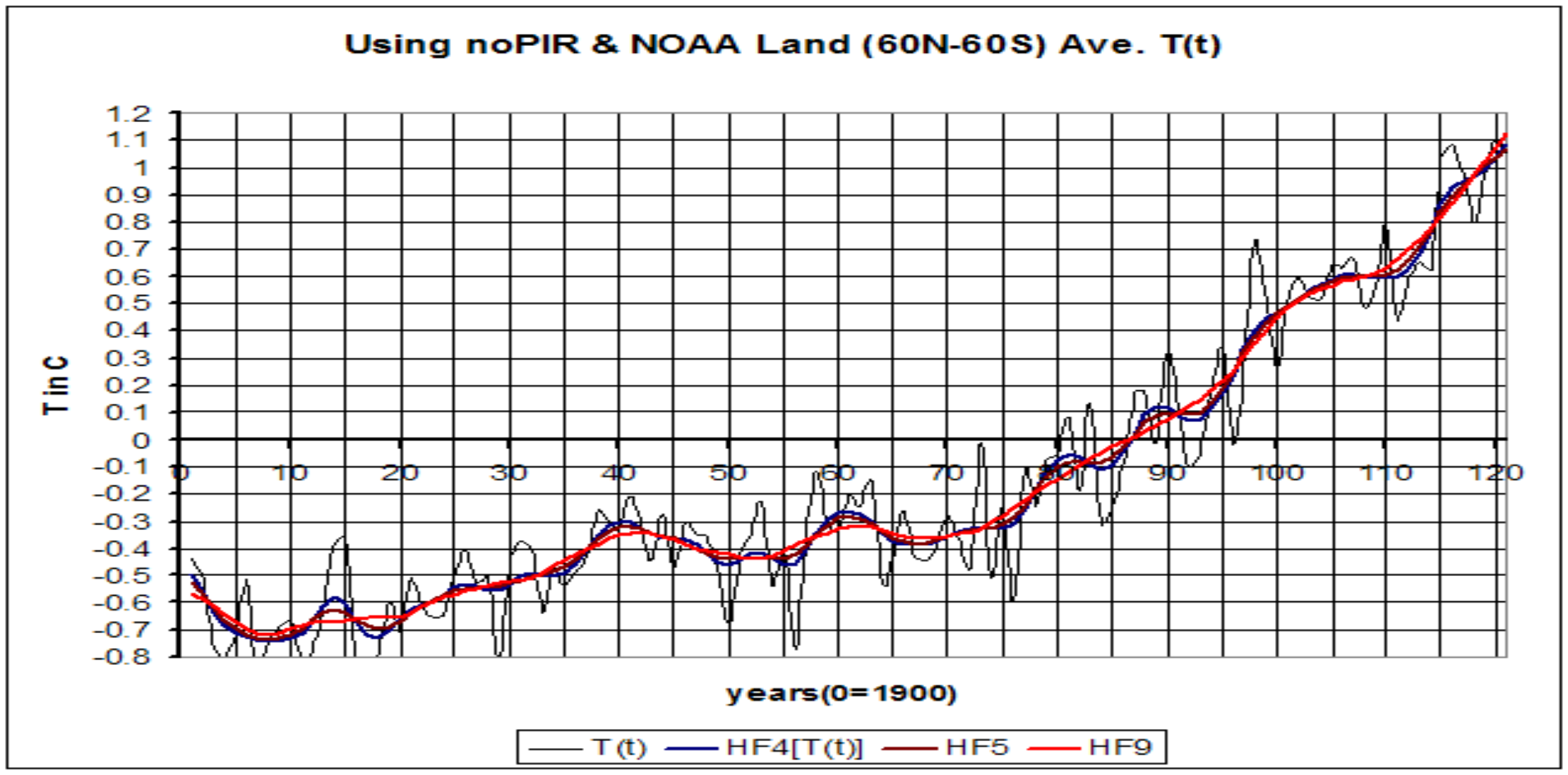

Fig.23(discrete annual ave. plots; curves for visualization)

The value of independent ΔF due to green house gas variation is smaller at higher colder latitudes. Thus the value of averaged ΔF 60N-60S is about 4% larger than that of the full 90N-90S average. This is similarly true for $\Delta F_{2xCO2}$. This small effect is included in the calculations for 60N-60S regions.

**[The working spread sheet used to produce these values and figures for Fig.16-Fig.17 is "TREND-ANALYSIS-AR5-ERF-E3.xls" for various values of "VS", and "volcanic-forcing.txt"]**

---

**D) Evaluation of Q(≡*d*[E*]/*dt*), ΔQ and $\underline{\Delta}$Q; Q Models**

Q, ΔQ and $\underline{\Delta}$Q are defined in e.4d,e,f, in terms of ΔE*(t') which is defined in e.3 and es1.11 (3), and where t'=(t-$t_o$), and $t_o$ is the selected reference year. The value of ΔE*(t) (5 yr moving averaged) is shown in Fig.24 for the Global Oceans top 2000 meters using NOAA Ocean Climate Laboratory[12] data extended from 1957-2021. The similar values for the Global Oceans (60N-60S) are shown in Fig.26 . And the reanalyzed one year Global averaged values from Cheng[13] are shown in Fig.28 **[see also Cheng-OHC2000m_monthly_timeseries.txt, and OE-Cheng-list-1940-2021.xls]**. There are other types of global thermal energy storage/absorption but these account for only about 7% of the Global total[13]. Therefore Global ΔOcean storage (0-2km) will be scaled by 1.07 as an approximation of the total value. The temperature of the abyssal oceans below 2km are self regulated and very close to freezing due to the continual upwelling of cold water sourced from the sinking cold polar waters, and are essentially independent of the surface temperature changes directly above. Thus, there is essentially no *dependant* ocean thermal energy change in this zone and it can be omitted, i.e. the 0-2km zone data is sufficient. [Reports in Cheng[13] (Figure 6) *do* indicate such abyssal ΔEnergy storage is less than 8% of the Ocean total, resulting in an ECS increase of about 1.6% (see Section **E.2**), i.e. insignificant in any case.]

The value of ΔE can be converted to ΔE* by noting the global ocean area is $360\cdot10^6$ $(km)^2 = 3.6(10^{14})m^2$. And ΔE is given as $X\cdot10^{22}$ Joules. So

$$\Delta E^* = \frac{X\cdot10^{22}\cdot J}{3.6(10^{14})m^2}\;\frac{W\sec}{J}\;\frac{year}{3.15(10^7)\sec} = X\cdot(0.88)\,(W\cdot yrs)/m^2$$

The resulting plots of NOAA ΔE*(t) and H5[ΔE*(t)] where $t_o$= 1957 or 1990 respectively is given in Fig.24 .

Using the data from Cheng[13] Fig.28, and setting $t_0$=1995

e.11 $Q_1 \equiv [d(H5[\Delta E^*(t')])/dt']_{@t'=0,t=1995} = 0.75\ W/m^2$ , $Q_2 \equiv [d(H5[\Delta E^*(t')])/dt']_{@t'=20,t=2015} = 0.81\ W/m^2$ .

Thus, using e.4f over the evaluation range of 1995 through 2015:

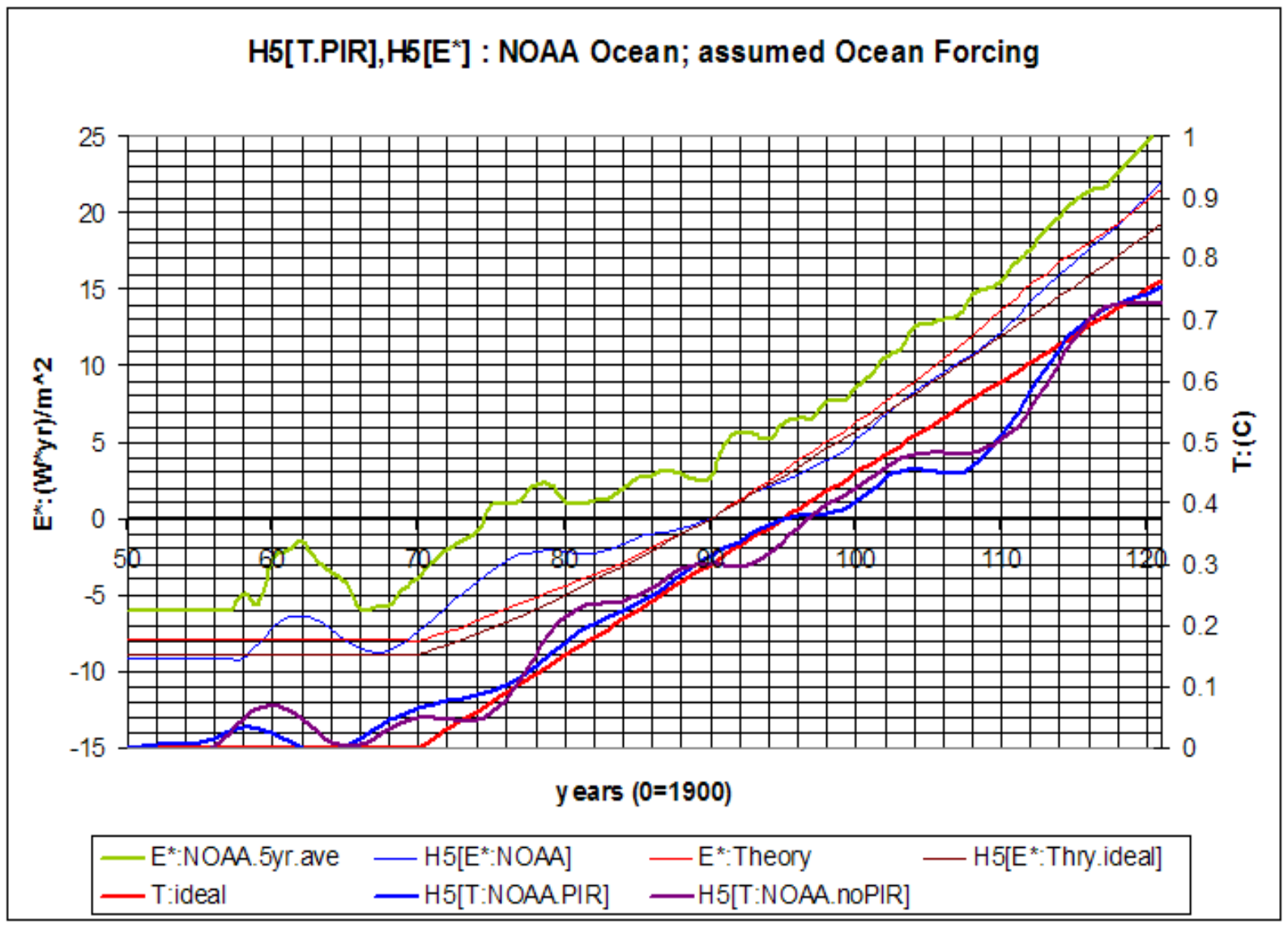

Fig.24 (discrete annual ave. plots; curves for visualization)

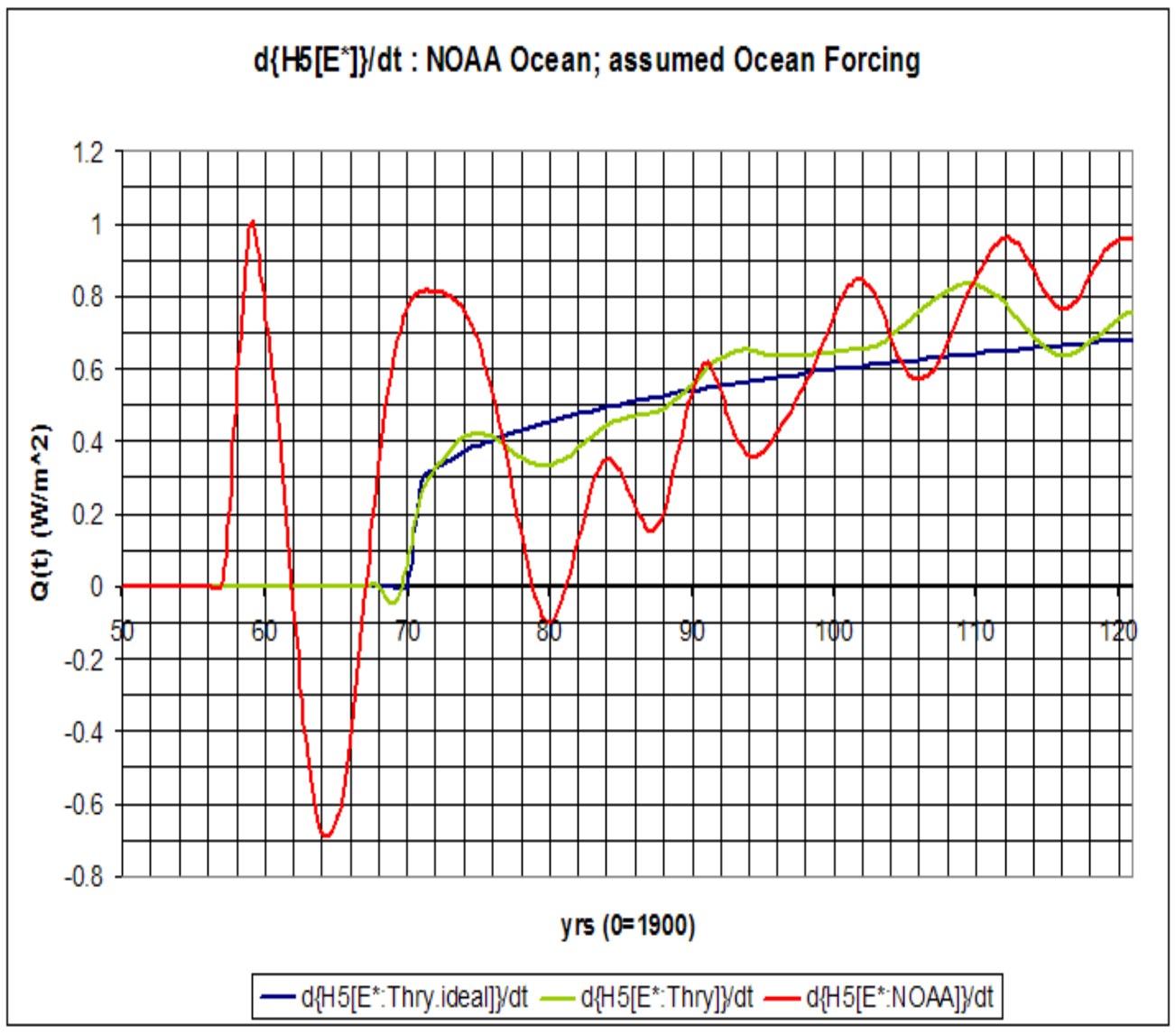

Fig.24B (discrete annual ave. plots; curves for visualization)

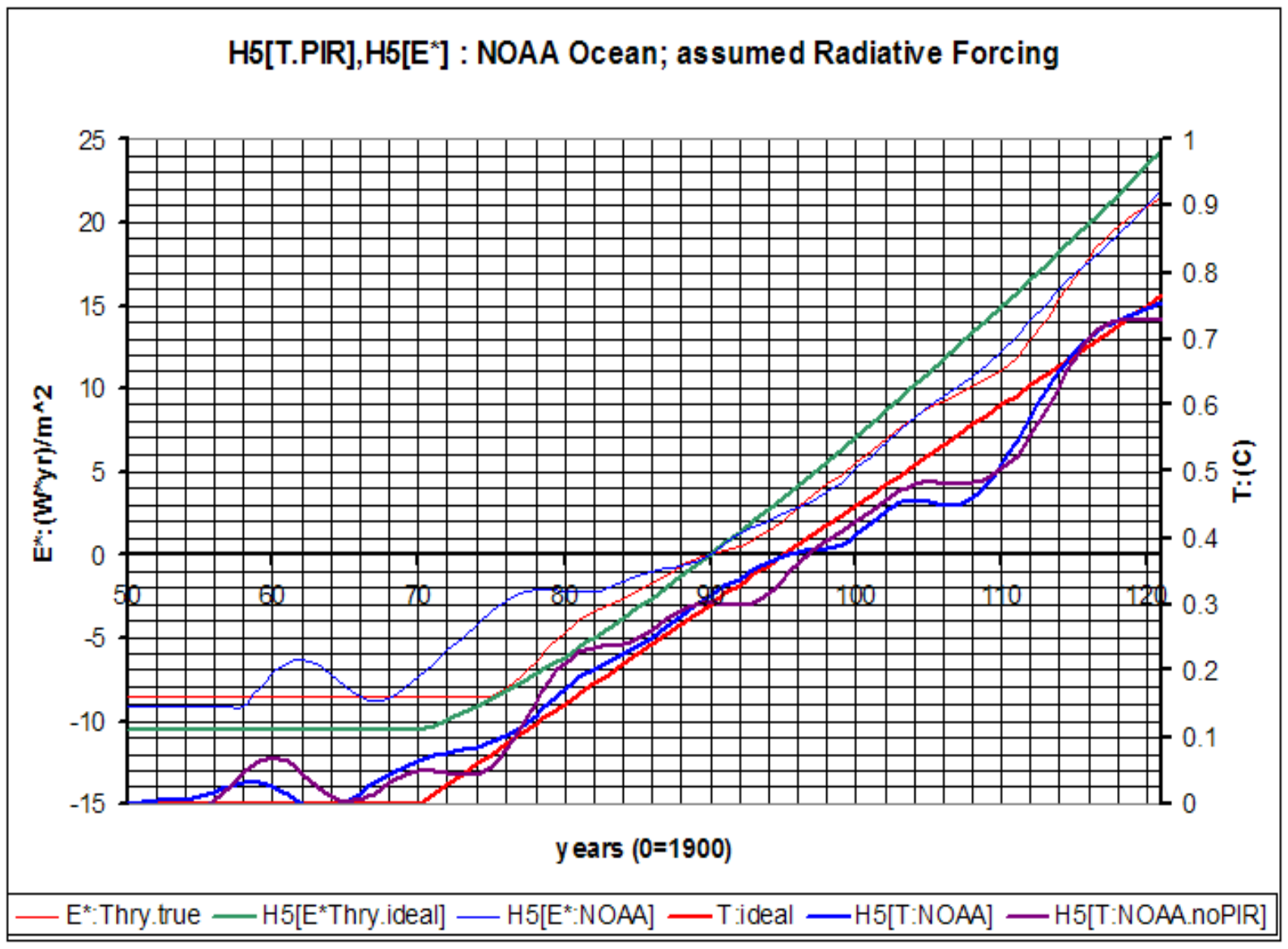

Fig.25 (discrete annual ave. plots; curves for visualization)

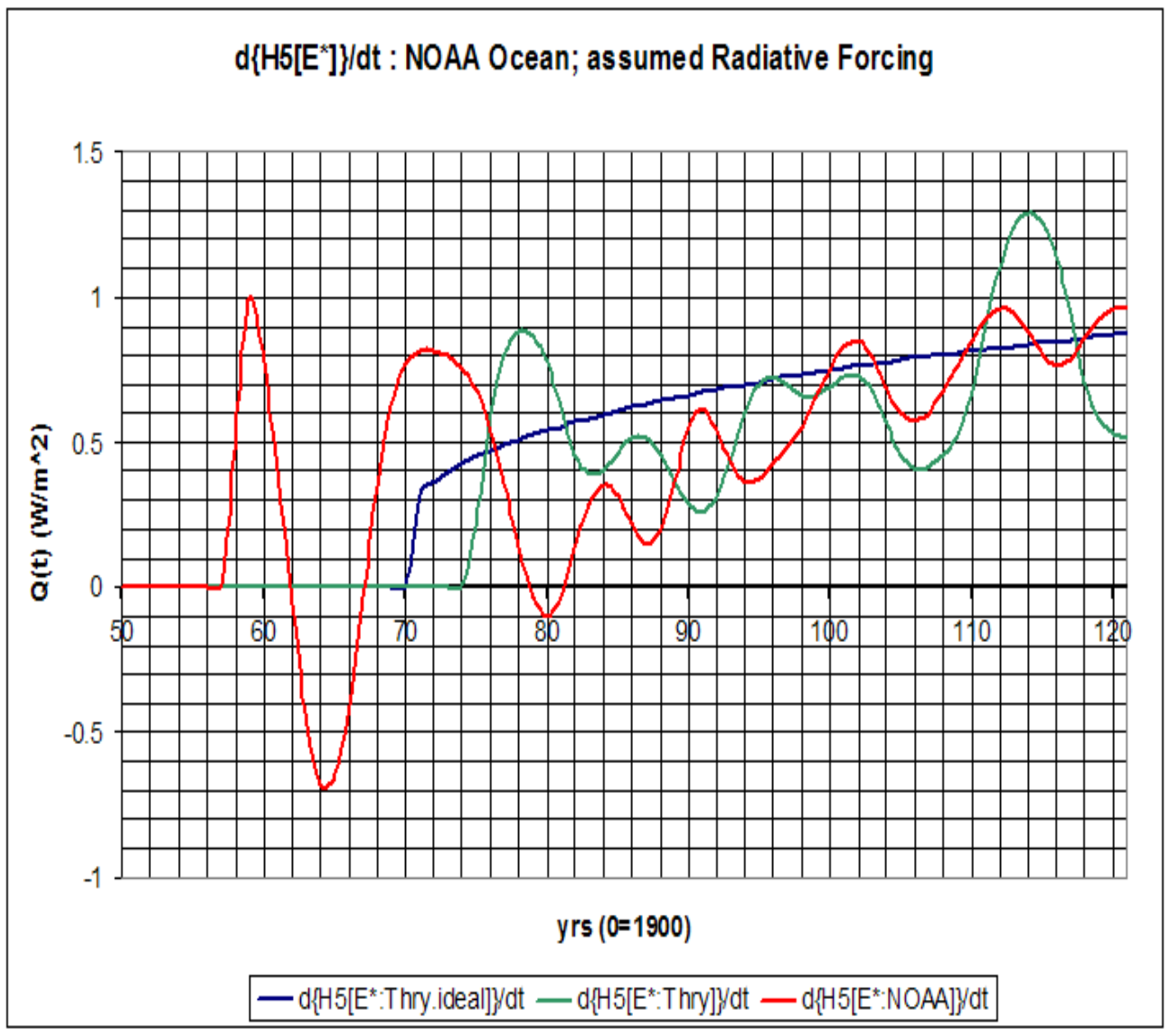

Fig.25B (discrete annual ave. plots; curves for visualization)

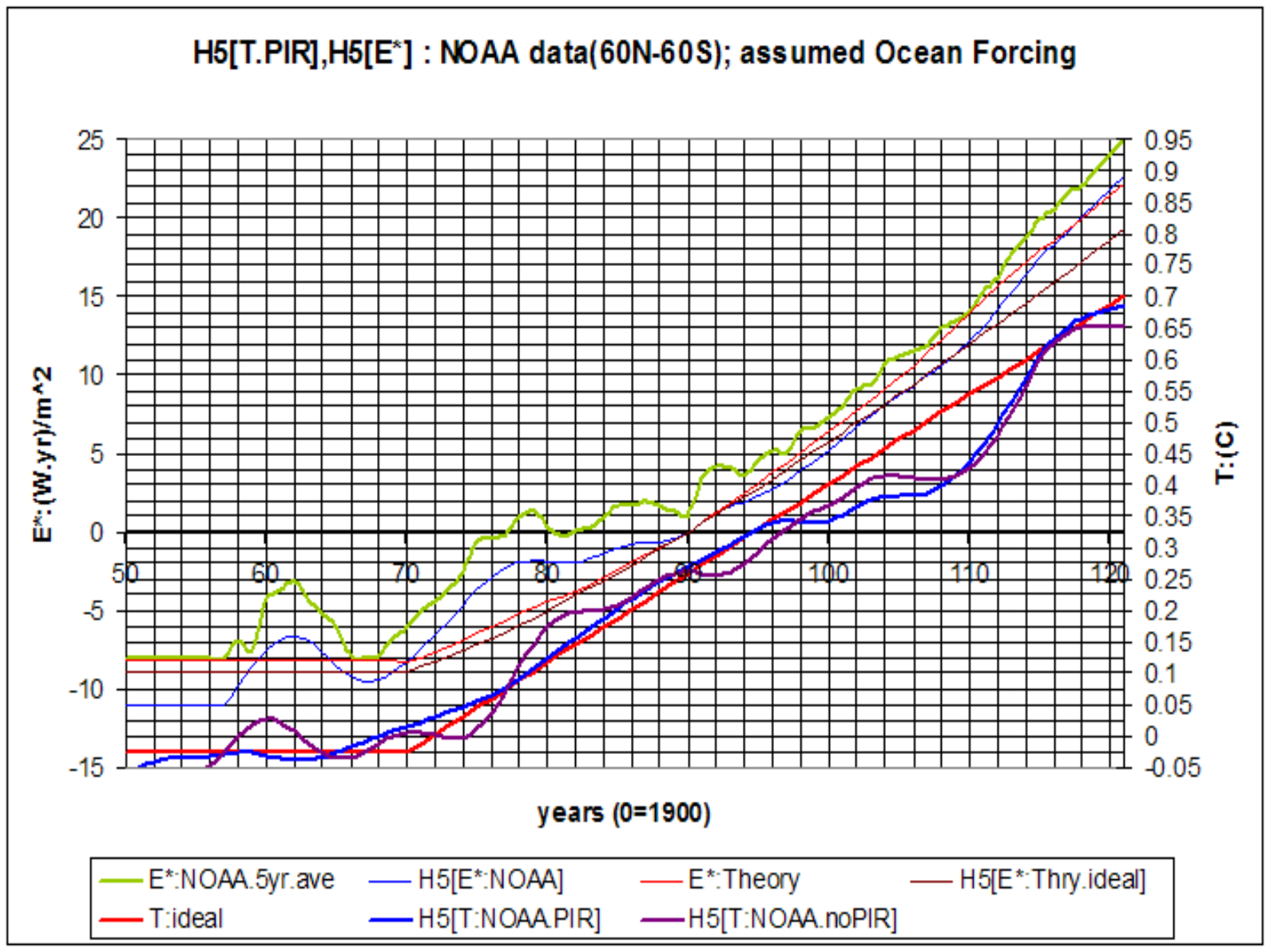

Fig.26 (discrete annual ave. plots; curves for visualization)

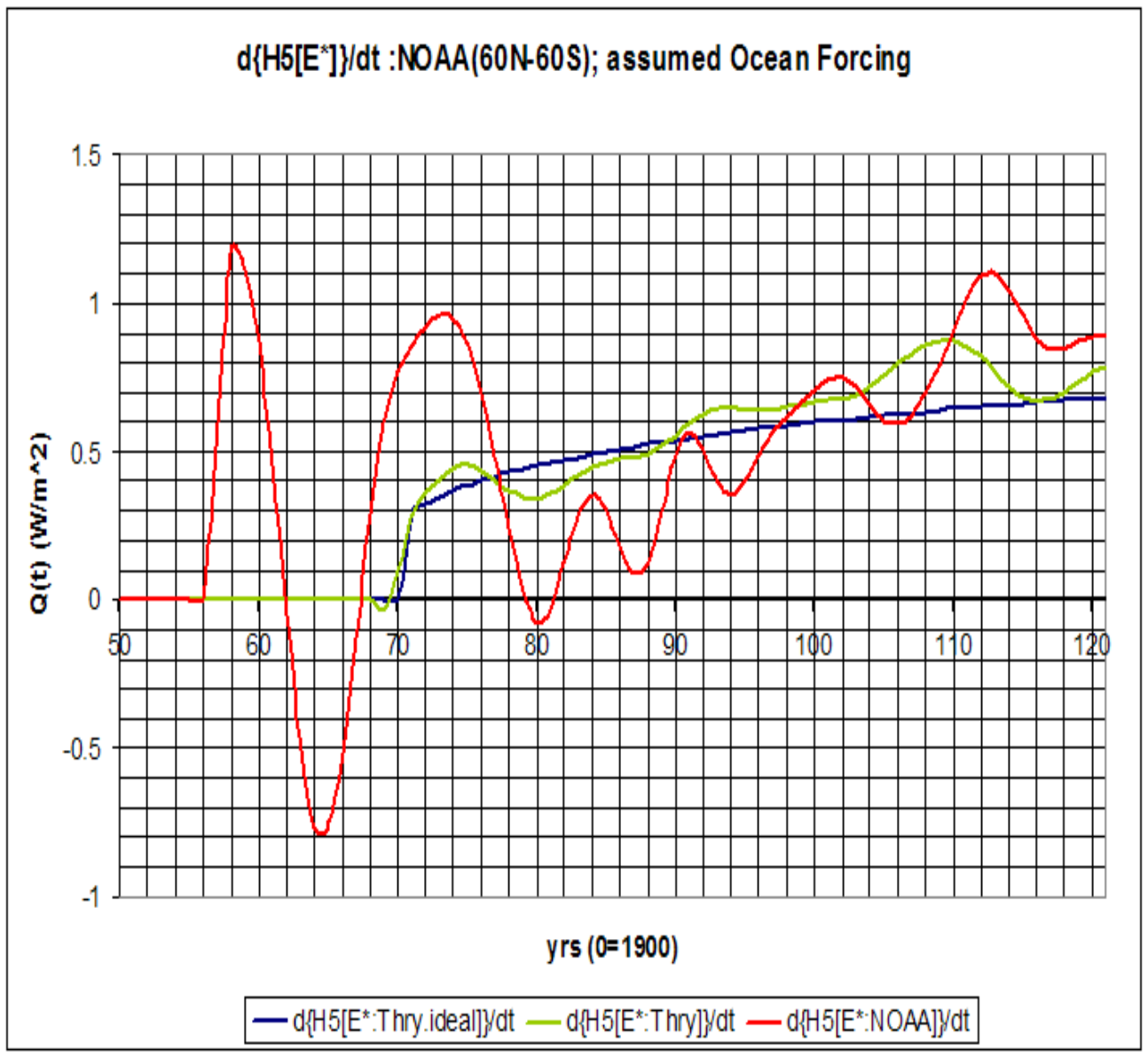

Fig.26B (discrete annual ave. plots; curves for visualization)

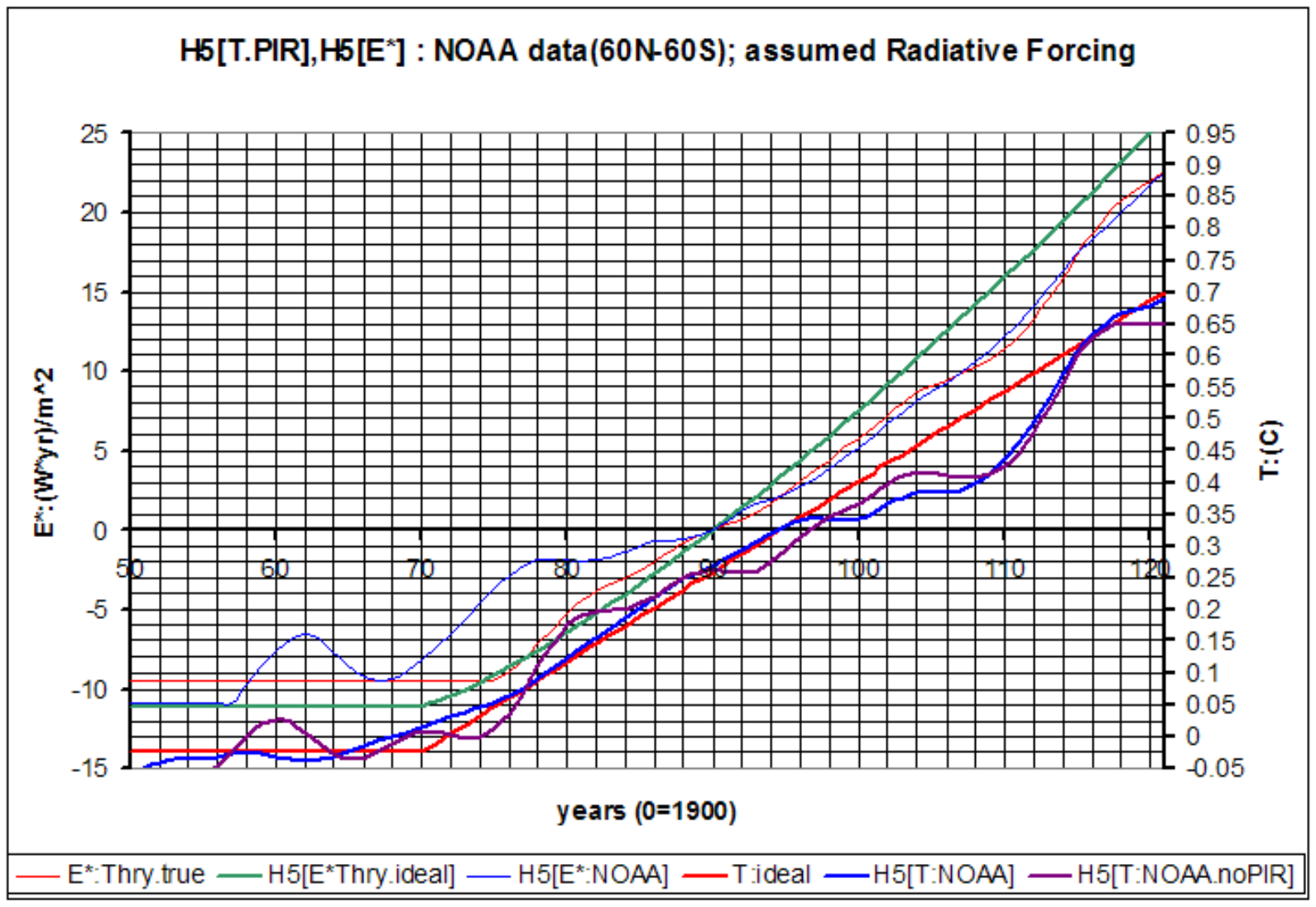

Fig.27 (discrete annual ave. plots; curves for visualization)

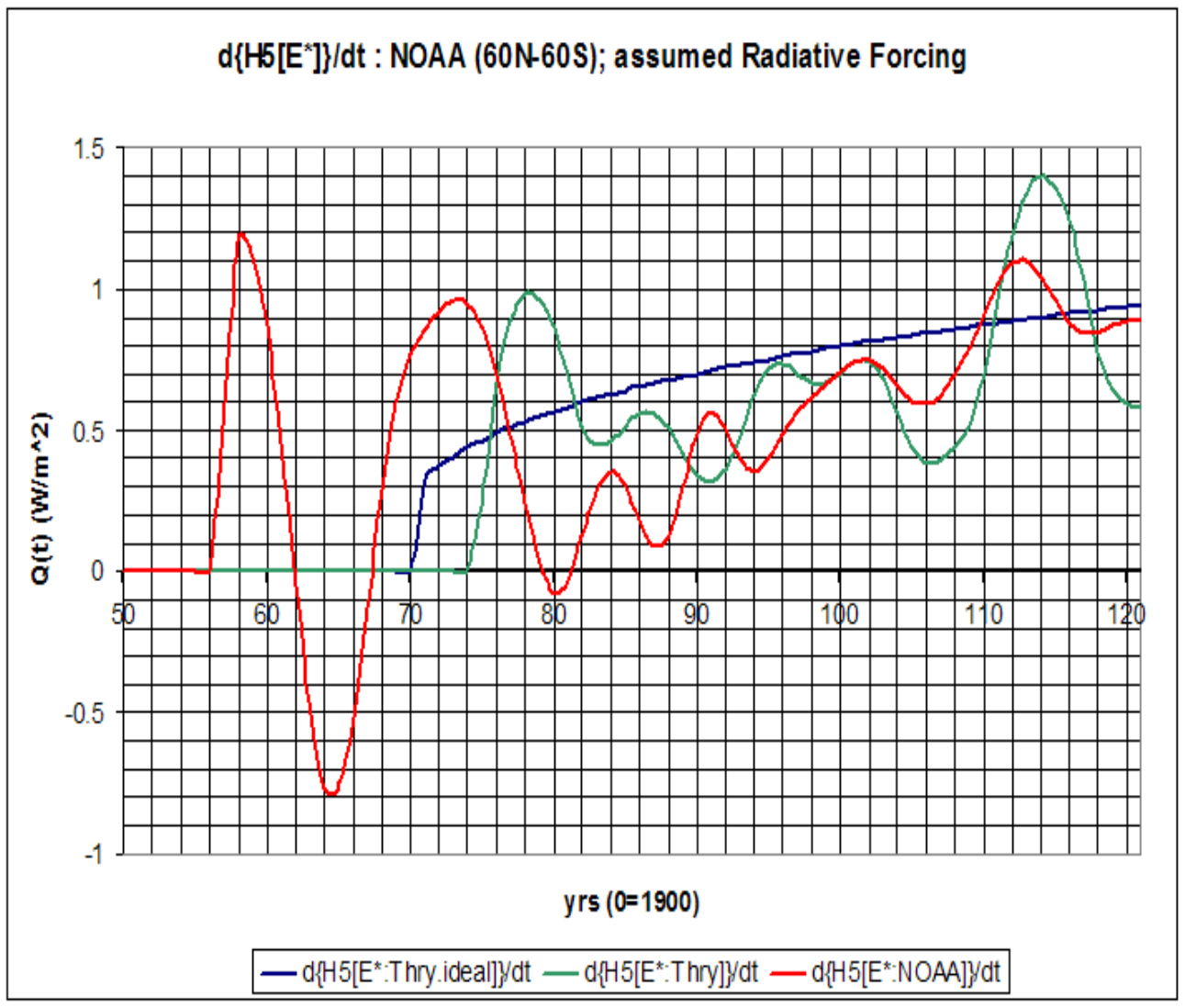

Fig.27B (discrete annual ave. plots; curves for visualization)

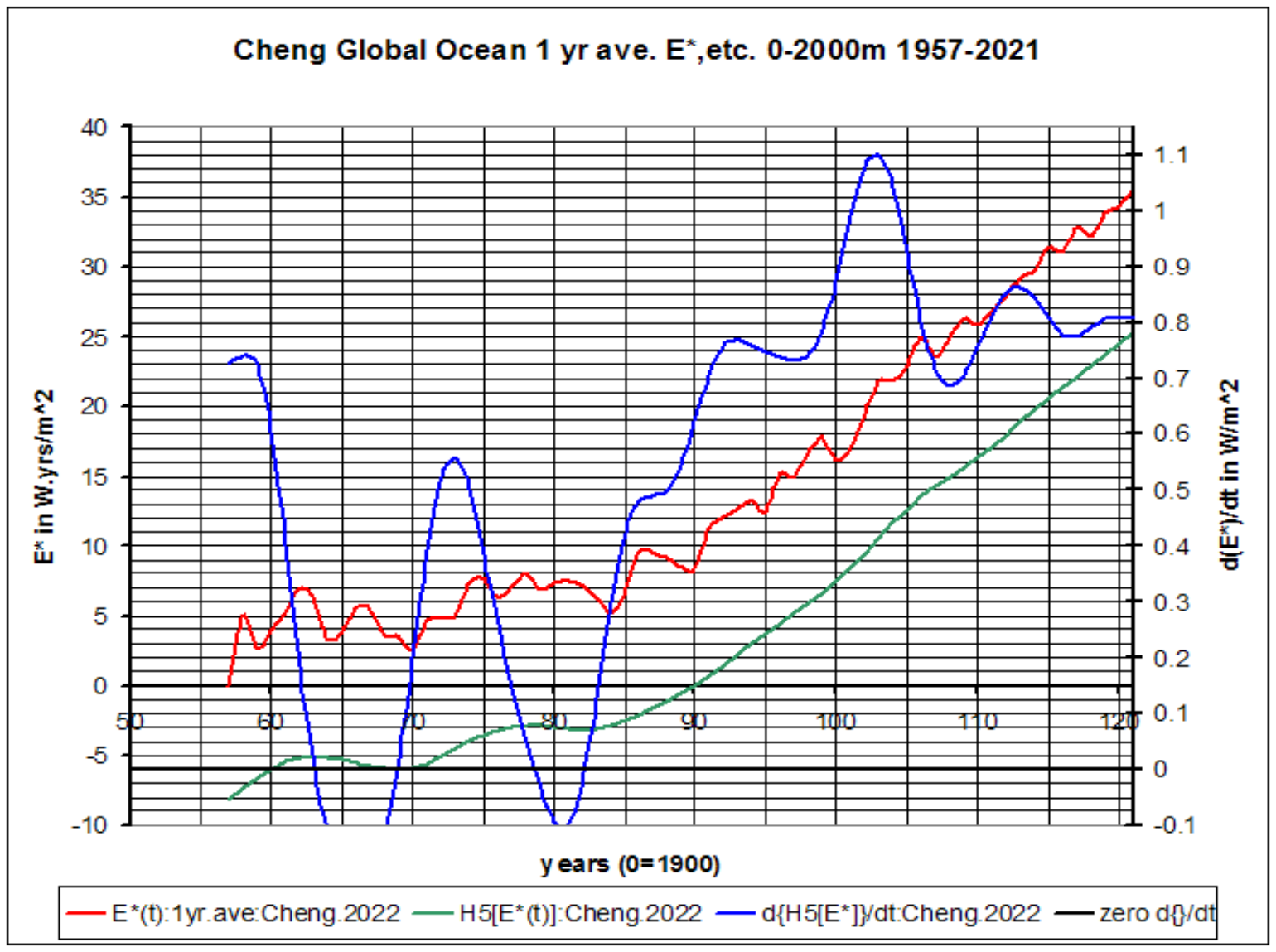

Fig.28 (discrete annual ave. plots; curves for visualization)

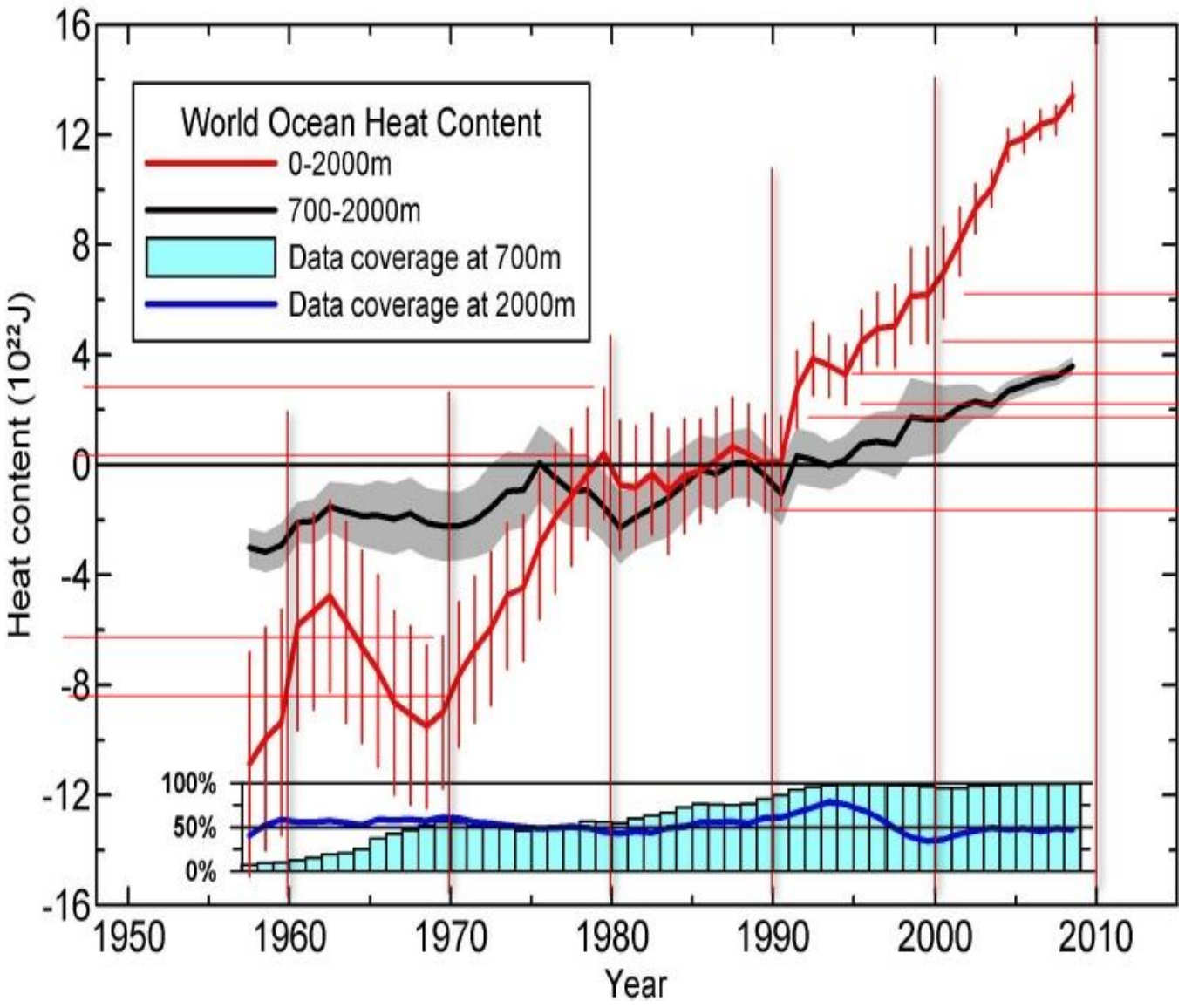

Fig.29 source: ref.[14] 

e.12 $\underline{\Delta}\text{Q}=[d(\text{H5}[\Delta\text{E*}(\text{t}')])/d\text{t}']_{@t'=20,t=2015} - [d(\text{H5}[\Delta\text{E*}(\text{t}')])/d\text{t}']_{@t'=0,t=1995} = 0.81-0.75 = 0.06\ \text{W/m}^2$ .

Even more reliably, the average of $Q_1(t)$ and $Q_2(t)$ (see e.11 above) over 1990-2000 and 2010-2020 are 0.748 and 0.809 W/m$^2$, respectively .These are nearly identical to the specific values of $Q_1$ and $Q_2$ above. The reanalyzed data by Cheng[13] are *purported* more accurate than the direct NOAA data. This "most likely" Cheng data will be used in an alternative orthodox simple direct calculation of ΔQ, $G_{eff}$ and ECS in Section **F**, and simply compared to the "most likely" values determined using the direct NOAA data in Section **E**. Estimation of the uncertainty of $\underline{\Delta}$Q using the Cheng data will not be attempted here.

Certainly, the NOAA profile of $Q(t') \equiv d(\text{H5}[\Delta E^*(t')])/dt'$ (see Fig.24B) after 1980 and through 2020 varies significantly from the Cheng data evaluation (see Fig.28), which asperses the accuracy of either. Further, after H5 filtering, the NOAA value of Q(1972) ≈ 0.8, and Q(1978) ≈ 0.1 W/m$^2$ (see Fig.24B red line). And the power flowing into the oceans should be *roughly* proportional to $d[T_{surface}(t)]/dt$. However, as shown in Fig.24 (or Fig.5.6B...no PIR), the value of the slope of H5[$T_{ocean}$(1972)] is near the minimum for the period from 1970-2010, and the slope of H5[$T_{ocean}$(1978)] is near the maximum; the reverse of Q(1972) and Q(1980) above. Apparently, the possible error of Q(1970 to 1980) is well exceeding 0.7 W/m$^2$ for the NOAA data over this period. And the uncertainty bars for the NOAA ΔE*(t) data in 1990 and 2000 (shown in red bars, see Fig.29 circa 2012 Levitus[14]) is nearly 2/3's that of 1980. So the uncertainty of Q(1990,2000) might also reasonably exceed (2/3)·(0.7) = 0.47 $W/m^2$. Before 1990 the Cheng data suffers from the same limitations.

This amount of uncertainty would be too great to provide any reasonable evaluation of $G_{eff}$ or ECS using the NOAA data directly. As a result a *quasi-theoretical*, *quasi-empirical* (i.e. formal/empirical) method will be employed to determine Q(t) using the NOAA data, as described further below.

**[[ Digression**: note, the clear oscillations of H5[Q(t)] (NOAA Global data: no PIR) in Fig.24B are revealing. The subtraction of the 2'nd order polynomial trend of H5[Q(t)] (≡ *d*(H5[ΔE*(t)])/*d*t) from H5[Q(t)]:1979-2017 leaves the oscillatory component "d(H5[E*])/*dt* - Trend" in Fig.30 . The Fast Fourier Transform of this component in Fig.31 reveals a center frequency cycle of almost exactly 9.3 years, the lunar-earth tidal cycle described in the "PIR" discussion in Section **B**. The best fit of this component to a 9.3 year harmonic yields Q(t)=0.127·cos((t-**1983.05**)·2·π/9.3) .

Therefore, this Q(t) <u>maximum</u> occurs at *precisely* the year (1983) of the lunar-solar-earth tidal maximum (see Fig.22) and is of the same frequency, which is unlikely as a random coincidence. No specific physical mechanism is offered here for the unexpectedly *large* magnitude (and 3x larger *before* filtering), the sign or phase of this *particular* component other than those in section **B**. It is not obviously observable in the Cheng data at all, which might lead us to question the Cheng data validity. There *is* a *slight* possibility this signal is only an error signal *enhanced* by the lunar-earth tidal cycle. None the less, it is an intriguing, unavoidable observation. **The working spreadsheets are found in the "9.3 component" sheet of OCEAN ENERGY RESPONCE-new NOAA-D-Global-T-9.3-comp.xls ]]**

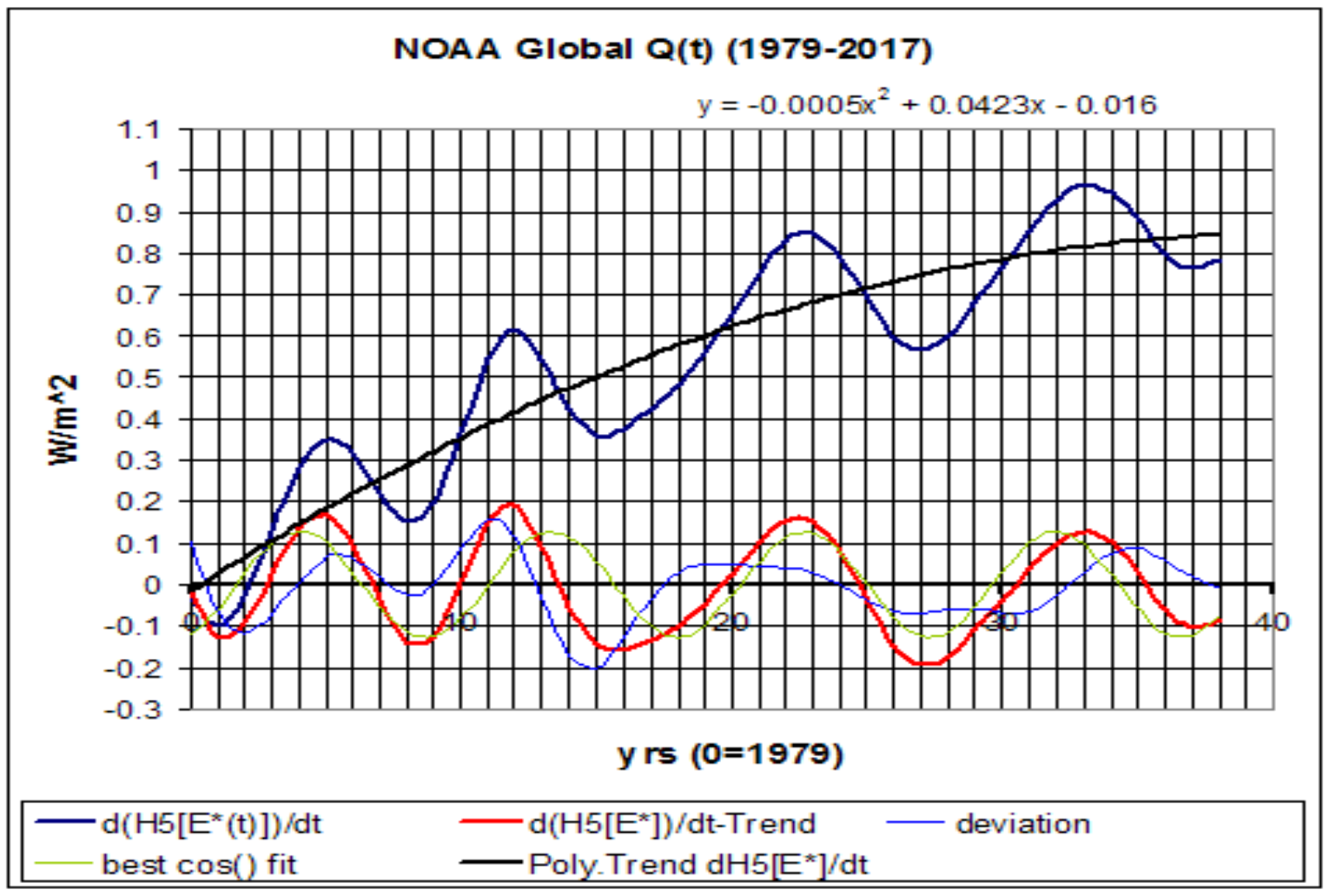

Fig.30 (discrete annual ave. plots; curves for visualization)

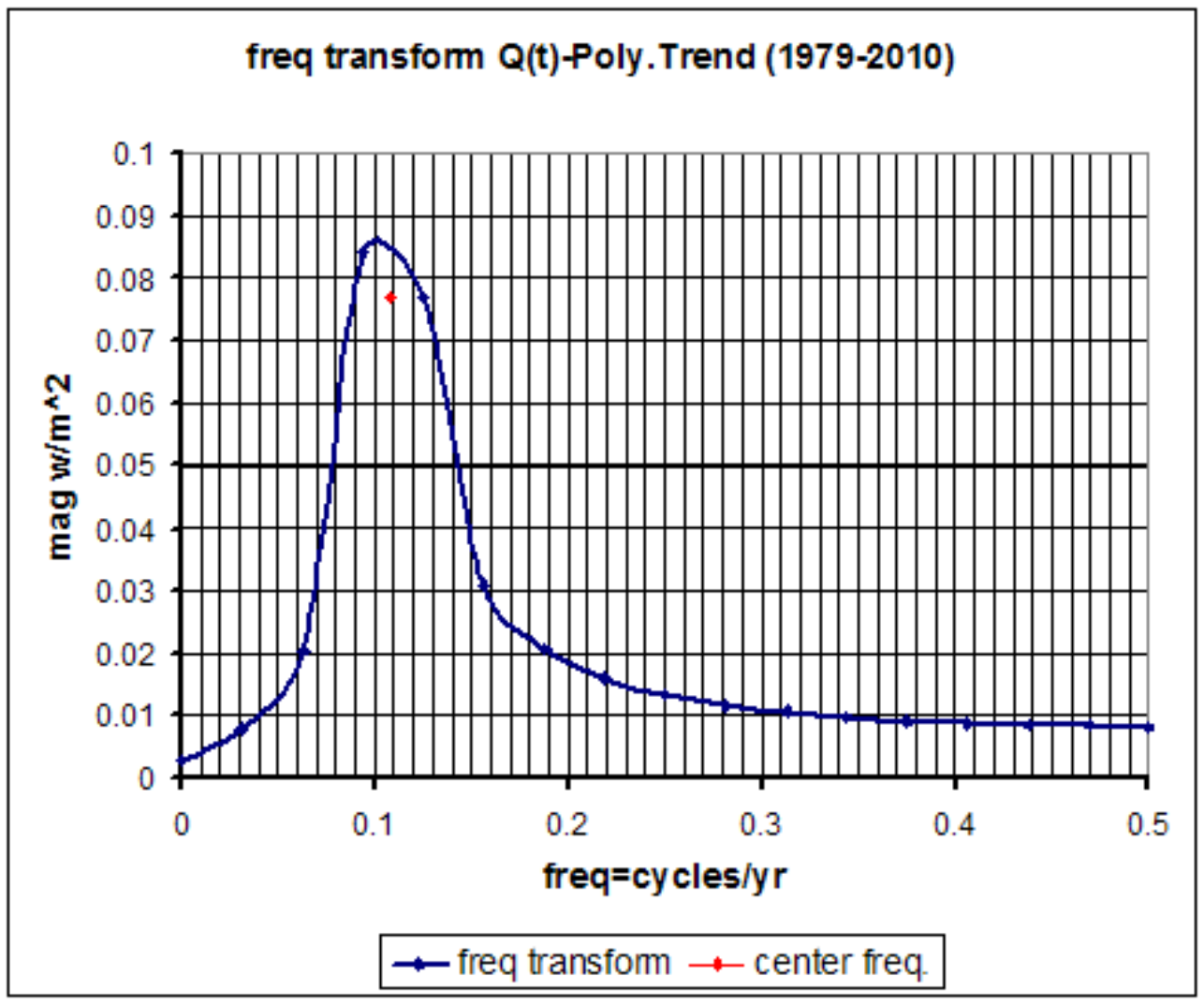

Fig.31 (discrete annual ave. plots; curves for visualization)

Therefore, it is presumed that the *direct* calculation of Q(t)≡*d*{H5[ΔE*(t)]}/*d*t using NOAA data is unreliable at any particular time before 2000. However the value of H5[ΔE*(t)]:NOAA in Fig.24 shows a definitive robust, nearly linear, positive trend. In fact, the best linear fit (or even second order polynomial fit) of H5[ΔE*(t):NOAA] from 2000-2020 exhibits insignificant uncertainty. Therefore, a known theoretical *form* of H5[ΔE*(t)] will be "best fit" to the observed value from 1990-2020, by variation of the coefficients of this *form* of $\Delta E^*(t)_{theoretical}$ . Once found, the value of *d*{H5[ΔE*(t)]}/*dt* ≡ Q(t) can then be precisely calculated. This allows a much more accurate evaluation of the sensitive time derivative terms than by direct observation, ***providing*** the presumed theoretical *form* of ΔE*(t) is correct. This is the "formal/empirical" method.

Q(t) is divided into separable mixed layer *and* deep-ocean parts. The $Q(t)_{mix.layer} = C^*_{eff}(t) \cdot d[\Delta \mathbf{T(t)}_A]/dt$ [see **SS3**] where the bold **T** temperature indicates an area weighted regional average, as is Q. The calculation of effective thermal capacitance $C^*_{eff}(t)$ is essential, and not trivial since mix depth varies widely with location and the seasonal cycle, and ΔT(t) varies widely with location depending on the time or frequency of ΔT(t). Thus, multiplying simple averages is inadequate. This calculation is described and performed in section **SS3** (using a U.S. Naval Research Laboratories mix layer depth data set) for both ramp and harmonic T(t), averaged over the global Oceans and global 60N-60S latitude Oceans (see es3.29,30) [Note: a ramp starting at $t_r$ = 1970, but equaling zero for $t<t_r$, *models* $\Delta T(t)_{Ocean}$ 1948-2021]:

e.13 Global Ocean: $\Delta T(t)$ ramp starting at $t_r$ =1970, $(t-t_r)$>25 yrs...$C^*_{eff}(t) \geq$ **13.8** Watt·yrs/($m^2$·°C)
$\Delta T(t)$ harmonic with >16 yr period.........$C^*_{eff} \geq$ **11.7** Watt·yrs/($m^2$·°C)

e.14 Global Ocean (60N-60S): $\Delta T(t)$ ramp starting at $t_r$ =1970, $(t-t_r)$>25 yrs...$C^*_{eff}(t) \geq$ **14.3** Watt·yrs/($m^2$·°C)
$\Delta T(t)$ harmonic with >16 yr period.......$C^*_{eff} \geq$ **12.3** Watt·yrs/($m^2$·°C) .

In **SS3** it is determined that $C^*_{eff}$ is dependent on the yearly maximum of mix depth (winter), as opposed the yearly average as might be expected. This results in a much larger (nearly double) value of $C^*_{eff}$ . Lateral ocean currents may extend this deeper mix depth to sub-tropic oceans as well, resulting in an even larger $C^*_{eff}$ . *However, the calculations of $G_{eff}$ and ECS are found to be extraordinarily **insensitive** to these alternatives of yearly average values or maximum of yearly values (see section **E** before Fig.34 and Uncertainties section **E.2**).*

For a ramp surface temperature T(t) starting at $t_r$ =1970 with constant time slope "S" (°C/yr), the deep ocean $Q(t)_{deep.ocean} = CC^*\cdot S\cdot(2/\sqrt{\pi})\cdot\sqrt{[t-t_r]} - ½\cdot g_{u.average}\cdot\Delta T(t)$ , as described in section **SS4** for $\mathbf{g_{u.average}=0.41}$ W/(°C·$m^2$) . The calculation of CC* using the formal/empirical method (and requiring $C^*_{eff}(t)$) is described and performed in section **SS5 (see details and figures)** using the NOAA $\Delta E^*(t)$ data, averaged over the global Oceans and global 60N-60S latitude Oceans:

e.15 Global Ocean: $\Delta T(t)$ ramp starting at $t_r$=1970...CC*=**5.1→ (5.51)→6.7** Watt·√[yrs]/($m^2$·°C)
e.16 Global Ocean(60N-60S): $\Delta T(t)$ ramp starting at $t_r$=1970...CC*=**5.47→(6.04)→7.75**Watt·√[yrs]/($m^2$·°C).

*Note, because of the included 5 year averaging of the only available NOAA data, the* **H5[ ]** *filter is achieved by simply applying an H4[] filtering to this data (see es2.3).* ***Also below, "AGW" = Anthropogenic Global Warming.***

The values in e.15,16 (using the NOAA data) include the effects of the much discussed independent transient fluctuations (1995-2015) on the calculation (see end of section **B.1**, and **SS5**), and the Periodic Interference terms (1948-2020). The largest terms in the range are evaluated presuming the temperature deviations are totally due to surface Radiative Ocean Forcing (see Fig.25,25B and Fig.27,27B). The smallest terms are evaluated presuming the temperature deviations *from the linear* are totally due to transient Ocean parametric deviations from the linear AGW Radiative Forcing (Fig.12,16), which produce ***no*** total $\Delta E^*(t)$ variation *in principle* [there is however a variation due to the change of surface outward radiation with temperature, which is accounted and discussed in **SS5** ] (see Fig.24,24B and Fig.26,26B). The true values must be somewhere between these extreme cases. However, as hinted in section **B.1**, the observations are slightly more consistent with the latter case. Note, in Fig.26B,24B (presuming Ocean transient Forcing) the theoretical power flow Q(t) and the NOAA data power flow (red and green lines) are more closely congruent than the similar lines in Fig.27B,25B (presuming Radiative transient Forcing) *during the 2005-2015 "dip" time period* ...but not before 1995. The divergence (loss of correlation) before 1985 is clear, as was predicted previously.

The theoretical and observed values of H5[E*(t)] for 2005-2015 are nearly congruent (thin red and blue lines of Fig.24,26,25,27) after "best fitting". But the theoretical values for Ocean Forcing are slightly above those observed, and the Radiative Forcing values are slightly below, so no preference is indicated. "T:ideal" is the fitting of a surface temperature ramp starting at 1970 to the H5[$\Delta T_{surface}(t)$] after PIR and extrapolation through the "dip" region, i.e. the "ideal" response to the "ideal" AGW radiative linear forcing (see section **C**).

**[[** It is speculated that the Forcing deviations from linear may be Radiative through 2002 due to the volcanic radiative cooling (reasonable), but becomes Ocean parametric Forcing due to the transient "dip" phenomenon (as was suggested in Fig.25B,27B and above). These Plots are shown in **Fig.32,32B,33,33B** below [using **OER-gu-RadaOceanForcing(etc.)-v4.xls**]. The congruence of the theoretical and NOAA observed values of $\Delta E^*(t)$ and $\Delta Q(t)$ are excellent after 1985 (particularly if the puzzling 9.3 year cycle components are ignored), and lend credence to both this speculation and the theoretical and NOAA observed values! The originally questionable NOAA E* data now might be corroborated and verified after 1985! The calculated values of CC* in this case are shown in parentheses in e.15,16 valued between the two fixed cases, as might be expected! **]]**

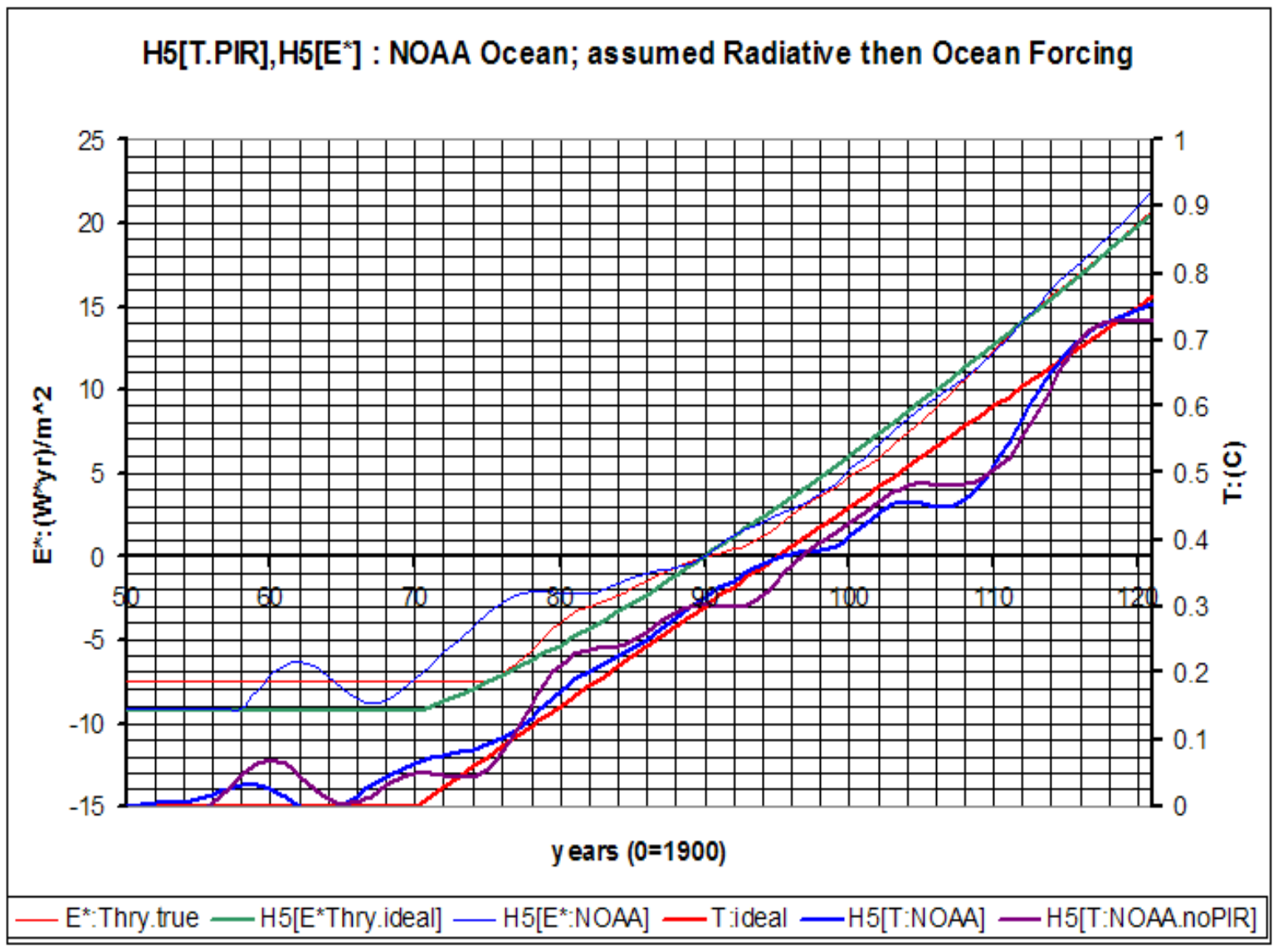

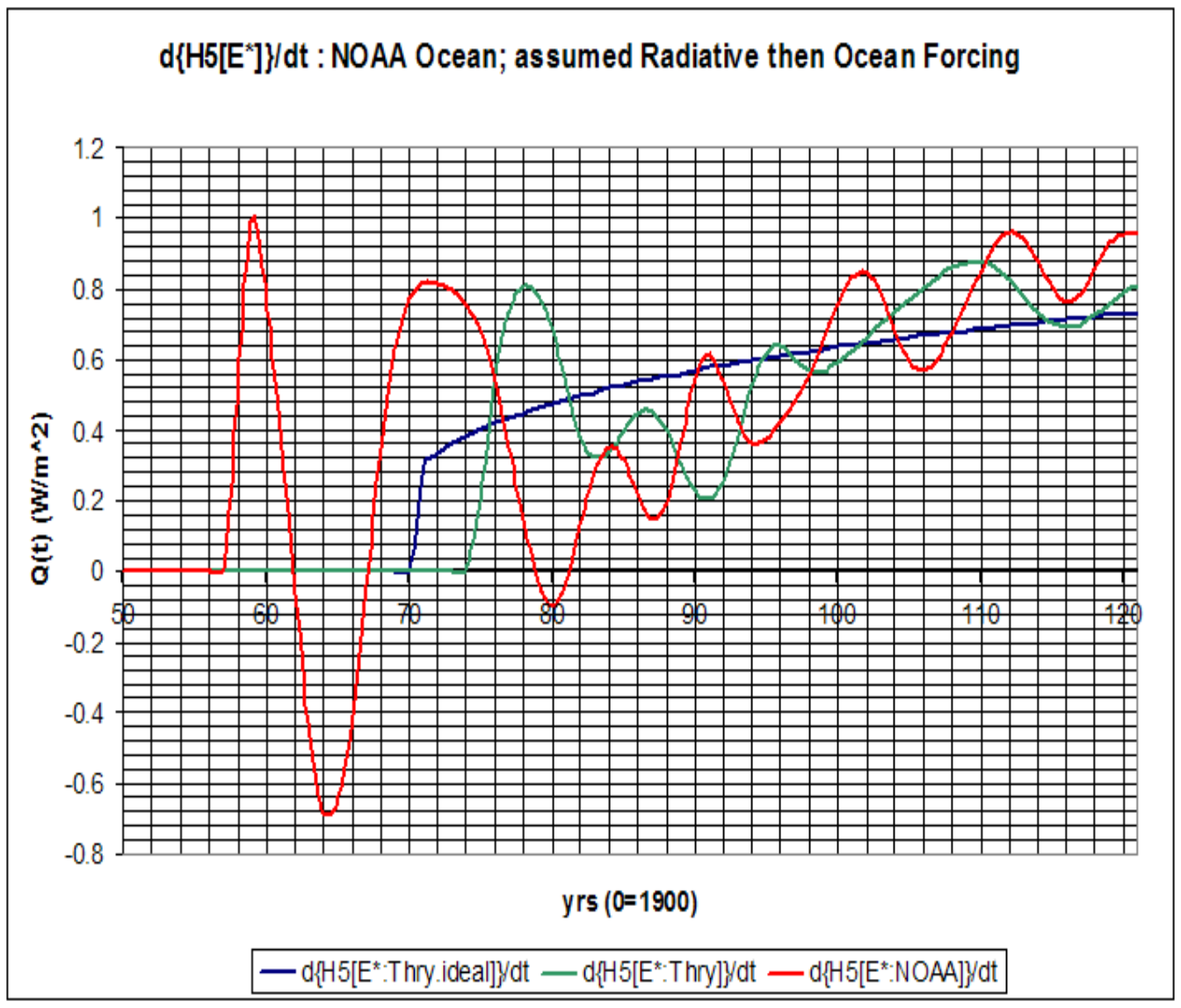

Fig.32 (discrete annual ave. plots; curves for visualization)

Fig.32B (discrete annual ave. plots; curves for visualization)

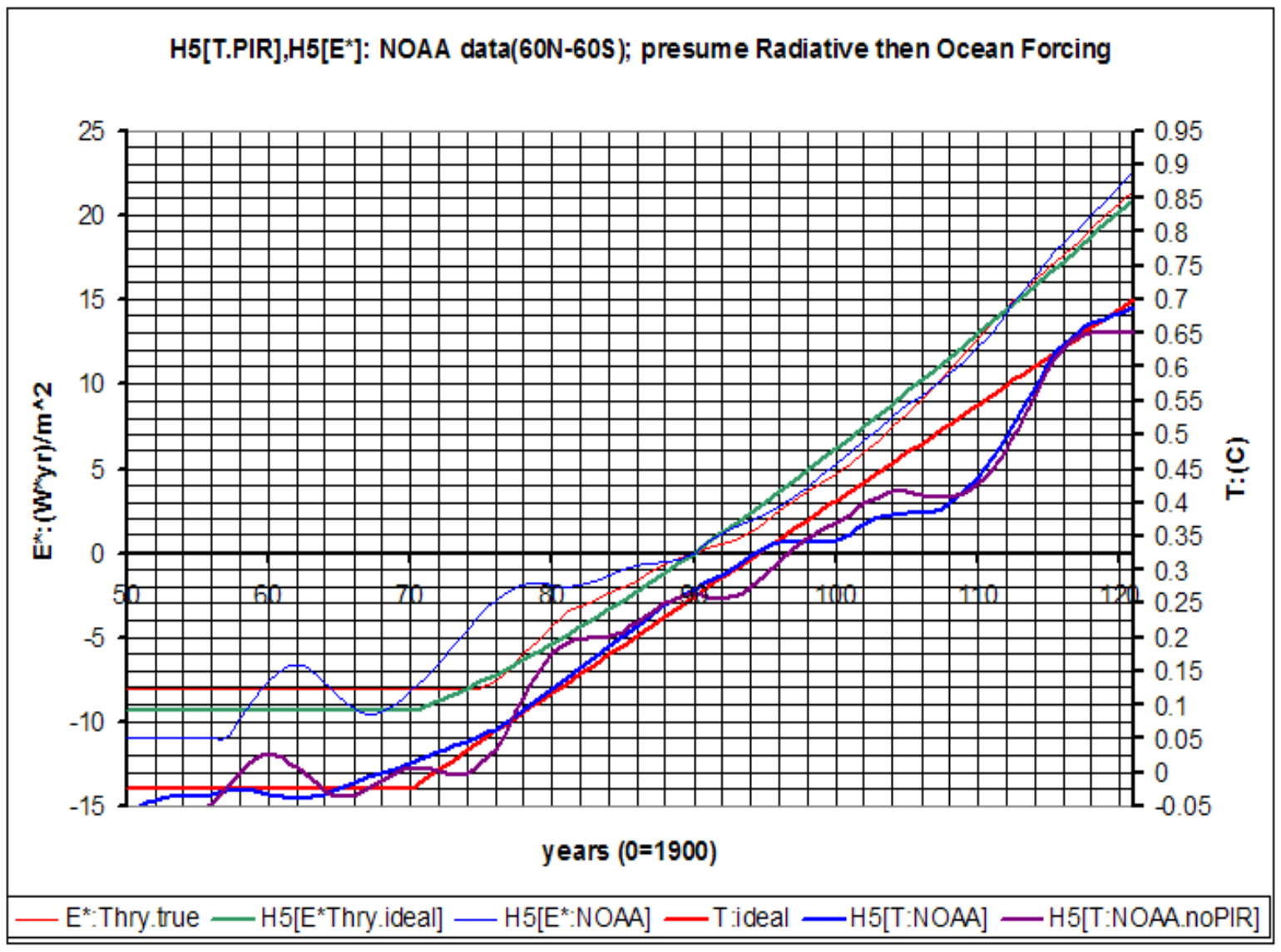

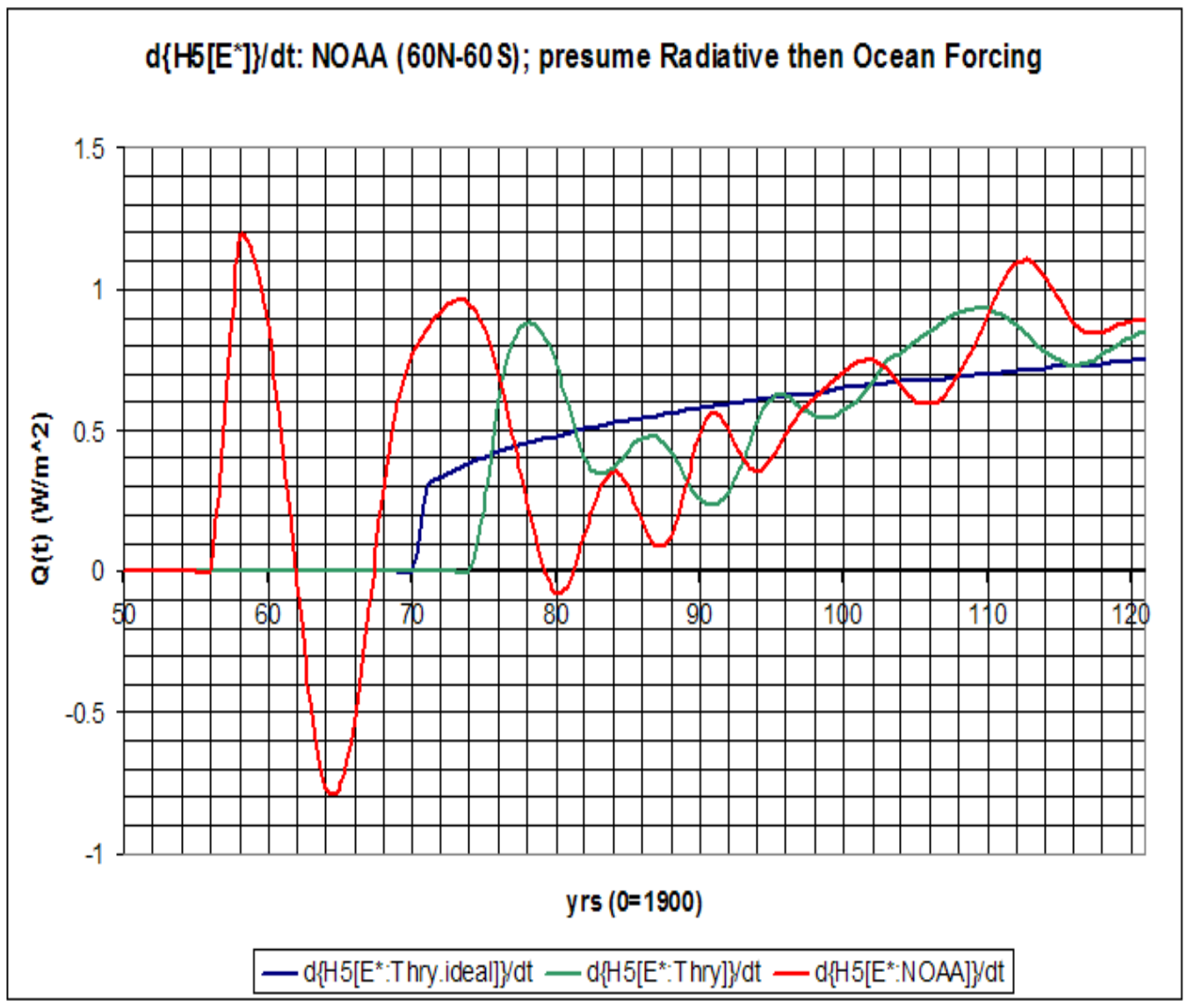

Fig.33 (discrete annual ave. plots; curves for visualization)

Fig.33B (discrete annual ave. plots; curves for visualization)

..................................................

*Alternatively* the value of CC*(60N-60S) can also be estimated using an *independent* "formal/empirical" method using the known Oceans temperature profiles with depth. This is estimated in **SS4**, **SS5, SS9** (for $X_{maximum}$ = 1400m) yielding the CC* ≤ **6.9** Watt·√[yrs]/(m$^2$·°C) (presuming $g_{u.ave}$=**0.41** W/(m$^2$·°C)) **[**and CC*(Global) ≈ (0.9)·CC*(60N-60S) ≤ **6.2** (see last paragraph below)]. This method of CC*(60N-60S) evaluation, by using the direct Global observations of "X" (see es4.15,35), does also yield a theoretical maximum evaluation of $\Delta Q(t)_{Deep.Ocean}$ over a wide range of possible $g_u$ (= 0 to 0.8), and thus a *maximal* value of ECS(Ocean), that is perhaps as accurate as using the Energy Method above (and quite comparable), and does *NOT* depend on the uncertainties of E*(t) or $C^*_{eff}$ or $g_u$ ! Specifically for $X_{eff.maximum}$ = 885m (see **SS5, SS9**):

assuming $g_{u.average}$ = **.2** , **<u>.41</u>**, **.6**, **.8** W/(m$^2$·°C) respectively, then

e.15B Global Ocean : CC* = **4.3**, **<u>6.2</u>**, **7.5**, **8.6** Watt·√[yrs]/(m$^2$·°C),

e.16B Global Ocean (60N-60S): CC* = **4.8**, **<u>6.9</u>**, **8.3**, **9.6** Watt·√[yrs]/(m$^2$·°C),

and correspondingly $ECS_{G.Ocean\,(60N\text{-}60S)}$ = **1.45**, **<u>1.53</u>**, **1.56**, **1.59** °C respectively

; where CC*(Global) ≈ (0.9)·CC*(60N-60S) [see last paragraph below] .

Thus, a nearly *maximum* independent evaluation of ECS (and $\Delta Q(t)$) can be evaluated using the $CC^*_{maximum}$ value (**6.9**) along with the best known value of $g_{u.average}$ = **0.41** in e.15B,16B above, ***and*** this is very compatible with the mid-ranges of e.15,16 above. This independent corroboration tends to generally verify the results of both methods. These lower values of e.15,16 *are thus considered* *most likely herein*. However, conservatively, a *maximal* (ECS upper limit) choice of CC*(60N-60S) ≈ **6.9**(see above) and CC*(Global) ≈ 0.9·6.9=**6.2** will be **used in all following evaluations**. Note, the "ideal" plots of Figures 24-27 indicate the theoretical results for an "ideal" ramp temperature function (radiatively induced by AGW), as shown in the figures.

It is found by trial that if $C^*_{eff}(t)$ is replaced by the simpler *constant* value $C^*_{eff}(t=(t_r+25years))$, then the calculated vales of CC*, Q(t) and $G_{eff}$ are virtually unchanged, and this approximation is used.

Thus, for a nearly linear ramp surface temperature $\Delta \mathbf{T_s}(t)$ of slope "S", for $\mathbf{g_u=0.41}$ W/(m²·°C), starting at $t_r$,

e.17 $$Q(t)_{total} = C^*_{eff} \cdot d[\Delta \mathbf{T(t)_A}]/dt + CC^* \cdot S \cdot (2/\sqrt{\pi}) \cdot \sqrt{[t-t_r]} - \tfrac{1}{2} \cdot g_{u.ave} \cdot \Delta \mathbf{T(t)_A}$$

; for $t > t_r$ , $t_r \approx 1970$ where $\Delta T(t) = 0$ for $t< t_r$ , and for appropriate values of $C^*_{eff}$ and CC* as chosen above. *This* ***form*** *of Q(t) and specifically the value of* $C^*_{eff}$ *for a linear surface T(t) ramp, although derived using reasonable physical principles (see* ***SS3, SS4****), is a simplified model that could be validated or improved using more sophisticated global ocean simulations. The evaluation of CC* could then still be performed as in* ***SS5*** . However the value of Q(t) is not a particularly sensitive component of the ECS evaluation (see section **E.2**).

Note, the value of Q and $\underline{\Delta}$Q is set to zero for land regions as a reasonable and convenient approximation (especially using the trends Methods A and B below, which are largely independent of the effects of *bulk* thermal capacitance, given a linear $\Delta T_{surface}(t)$ !) And the value of Q and $\underline{\Delta}$Q are multiplied by 0.7 for Global averages since the oceans comprise only 70% of the Earths surface. Also the value of ΔE* for the very cold waters of the Polar regions (i.e. higher than 60° N or S Latitude) are self regulated to be very near freezing consistently. Therefore the ΔE* in this region is essentially zero. Thus the values of ΔE*(Global) must be about (0.9)·ΔE*(60N-60S); where the polar oceans (as defined above) are about 10% of the Global Ocean total. And this is used in the calculations of ΔE*(Global) and CC*(Global), as described above and in **SS5**. **[The working spread sheets to produce these values and figures are OER-gu-(etc)-v4.xls, and OE-(etc).xls, /OE-raw-data/, Ocean Energy Response-new Cheng-G-Global.xls; the mix layer depth is found in DMIX-MAX.txt and dmix-source-NRL.txt; $C^*_{eff}$ is calculated in Ceff-calc-(etc).xls]**

---

## E) Modified Analysis and Calculation of $G_{eff}$ and ECS

**E.1)** Basically, this method consists of artificially removing unknown, uncertain or complicating signal components, and then utilizing the resultant linear trend value. Rewrite e.3 using e.4e as

e.18 $$\Delta \mathbf{T_A}(t') \cdot G_{eff} = \Delta F(t') - \Delta Q(t')$$

;where Q(t) is given in e.17 (section **D**, for $\mathbf{g_{u.ave}=0.41}$ W/(m²·°C)) for a nearly linear surface Temperature ramp (and presuming no *transient* Volcanic or Radiative or Ocean parametric fluctuations); where $\Delta \mathbf{T_A}$ is found after PIR (and H5[ ] filtering) for various regions using Fig.4.6,9 and Fig.5.6,9 and Fig.6.3,6 data (section **B**); and where ΔF can be found using Fig.16 [or Fig.17] data in sect. **C,** for near linear AGW Radiative forcing only and omitting volcanic forcing (VS=0 or VS<<<1). **Remember, H5[ ] filtering is first applied to all temporal variables**. Now define an operator "**S[ ]**" that returns the best linear fit slope "S" over some given time range. Apply this operator to both sides of e.18, yielding

e.19 $$S[\Delta \mathbf{T_A}(t') \cdot G_{eff}] = G_{eff} \cdot S[\Delta \mathbf{T_A}(t')] = S[\Delta F(t') - \Delta Q(t')]$$

;where $S[\Delta \mathbf{T_A}(t') \cdot G_{eff}] = G_{eff} \cdot S[\Delta \mathbf{T_A}(t')]$ can be proven by trial or theoretically. **Note: S[ΔQ(t')] = S[ Q(t')]**

METHOD A:

Then rewrite e.19 as

e.20 $G_{eff} = S[\Delta F(t')-\Delta Q(t')] / S[\mathbf{\Delta T_A}(t')]$ .

This method is particularly insensitive to "signal errors" when the variables are roughly linear in time to begin with. Similarly, "S[ ]" might instead return the slope of the best 2'nd order polynomial fit.

METHOD B:

But the ultimate implementation of this method is to solve for the $G_{eff}$ that best equates both sides of e.18 over time. The minimization metric is simply the integration of the absolute value of the difference between both sides over the given time range. Both methods produce nearly identical results, however METHOD B also allows for the additional simultaneous evaluation of the best VS scalar (and *un*likely VS values).

The specific details of $G_{eff}$ calculation using METHOD A and B on NOAA data for the **Global Oceans(60N-60S)** region is presented here. All time data is first H5[ ] filtered. PIR is then used to remove the 21.2 year and 9.3 year cycle T(t) harmonics, etc.; and thus those ***corresponding*** forcing components (which are unknown in any case) can be omitted. Also the transient ΔT(t) component ("dip") seen in Fig.5.9 from 1995 to 2015 is removed and replaced by a linear interpolation/extrapolation shown as line1; and thus the ***corresponding*** ΔForcing and ΔQ(t) ocean parametric fluctuation (which are mysterious in any case) can be and are omitted. Q(t) is then formally/empirically evaluated for a simple linear temperature ramp. This method of PIR and "transient deviation removal" from the underlying T(t) and Q(t) linear trend is a reasoned, reasonable, appropriate liberty, and much superior to incompletely or incorrectly interpreting the effects of these real but otherwise unaccounted phenomena. The selected evaluation range is 1980-2020, and $t_o$=1980. The value of "S" used in the evaluation of Q(t) in e.17, is the best fit slope of T[1980-2020]=0.0137 °C/year, where the ramp begins at about $t_r \approx 1970$ (see Fig.5.9). The value of CC* = **6.9**, using the maximal results preceding e.17 . The value of $C^*_{eff} \approx$ **14.3** (W·yrs/($m^2$°C)), using e.14 . And ΔQ(t) is increased by 7% due to additional extra-Oceanic latent and sensible thermal capacitance/storage [7% is the current[13] orthodox estimation, and is used herein, although greatly exceeding simpler calculations]. As discussed previously the volcanic scalar VS may be set to zero, and then H5[ΔF(t)] is approximated in Fig.16 from the updated IPCC AR5 ERF (see Fig.12, Section **C**).

It should again be noted the values of ΔF(t) (and $\Delta F_{2xCO2}$) above will be multiplied by ~**1.04** for regions defined between 60N – 60S latitude, to account for the increased area weighted green house forcing there. And the value of ΔQ(t) above will be multiplied by 0.70 for true Global (land + ocean) $G_{eff}$ evaluations, since the ocean ΔQ is then averaged over the entire globe area. The curves of ΔT(t), ΔF(t), $\Delta Q(t)_{total}$, $\Delta Q(t)_{deep.ocean}$, $\Delta Q(t)_{mix.layer}$, and ($\Delta F(t)-\Delta Q(t)_{total}$ + offset) so derived are presented in **Fig.34** .

Using METHOD A above, $G_{eff.Oceans(60N-60S)} = S[\Delta F(t')-\Delta Q(t')] / S[\mathbf{\Delta T_A}(t')]$ = .0336/.0137 = **2.45** W/($m^2$·°C).

Using METHOD B, a best fit is achieved by varying $G_{eff}$ and an offset constant added to (ΔF-ΔQ). The offset corrects for any error in the variables at $t_o$=1980, especially in the $\Delta Q_{mix.layer}$ term. It also desensitizes the method to the zero'th order constant, as opposed to the first and second order polynomial terms of the thermal power components, similar to METHOD A. Using METHOD B above, where VS=0, $G_{eff.Oceans(60N-60S)}$= **2.51** W/$m^2$ . When VS is also varied in the minimization, the fit is slightly better, where VS=.025, $G_{eff.Oceans(60N-60S)}$ = **2.56** W/$m^2$ ...very little change. If VS is set to 0.2, then the fit is considerably worse and $G_{eff.Oceans(60N-60S)}$ *increases* to 2.78 W/m2, as previously predicted; VS= zero is the *maximal* ECS choice and quite appropriate. Conditions of a near linear steady ΔT(t) ramp are presumed.

The value of "$ECS_{eff}$" can then be evaluated where $ECS_{eff} \equiv \Delta F_{2xCO2} / G_{eff.near.linear.ramp}$ . The canonical Global value of $\Delta F_{2xCO2}$ = 3.7 W/$m^2$ is as presumed in IPCC AR5. So, $ECS_{eff.Oceans(60N-60S)}$ = 3.7·(1.04)/2.45 = 1.57 °C per doubling of $CO_2$, using the simpler METHOD A, VS=0. [But this selection of $\Delta F_{2xCO2}$ is discussed in section **G.1** and **E.2** .]

These calculations are facilitated using the Excel "Solver" tool, and "SLOPE()" or "LINEST()" functions. **[The working spread sheet used to produce the Fig.34 and values below is "Geff-eval-NOAA-Ocean-60-60-gu-v4.xls"] [The working spread sheets used to produce all such figures and values are "Geff-eval-NOAA-(etc)-gu-v4.xls"]**

All these calculations are repeated for various regions, values of VS, ΔT(t), S[ΔT(t)], C*,CC* using METHOD A and B, and all presented in Table 1 (**Tab.1**) below.

The calculation of effective thermal capacitance $C^*_{eff}(t)$ is essential, and not trivial. And the value of CC* and Q(t) are dependant on $C^*_{eff}(t)$; and Q(t) is an essential component of the $G_{eff}$ and $ECS_{eff}$ evaluation (see section **D** above). However, if the presumed effective mix depth is doubled for latitudes 40N-60N and 40S-60S (i.e. lateral ocean currents might extend the deeper (winter) mix depth to temperate oceans as well), then the value of calculated $G_{eff}$ is increased by only 1.5%. And if the presumed effective mix depth is reduced to the yearly average (which is about 60% the yearly maximum value), then the value of calculated $G_{eff.Ocean}$ is decreased by only 6% . Although, this large change in $C^*_{eff}$ does result in a significant change in calculated CC*. *Fortunately then, any questions of, or uncertainties about the values of $C^*_{eff}(t)$ are largely moot, as the effects on calculated $G_{eff}$ are at most barely significant [see section **E.2**].*

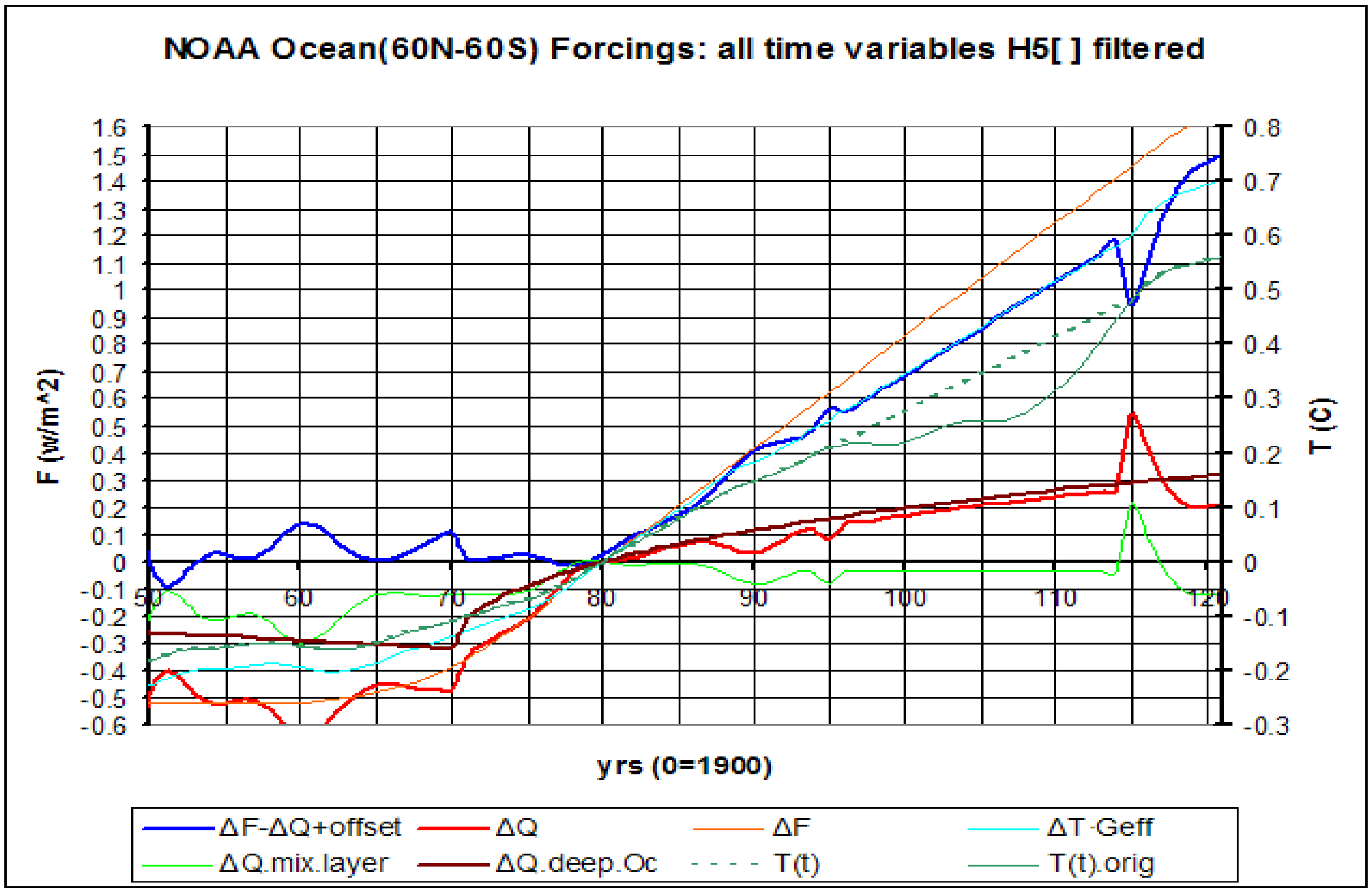

Fig.34 (discrete annual ave. plots; curves for visualization)

| | .nls = near linear steady | C | W·yr | W·√yr | fraction | Geff.nls | | | | ECS.nls (ΔF/2xCO2=3.7*) | | | |
|---|---|---|---|---|---|---|---|---|---|---|---|---|---|
| | .eq = equilibrium | year | C·m² | C·m² | >0 | W/(m²·C) | | | | °C | | | |
| | G. = Global | ↑ | ↑ | ↑ | ↑ | METHOD A | | METHOD B | | METHOD A | | METHOD B | |
| | Region | S | C*eff | CC* | VS=var. | VS=0 | VS=variable | VS=0 | VS=variable | VS=0 | VS=variable | VS=0 | VS=variable |
| 1 | Global .nls | 0.0193 | 13.8 | 6.2 | 0.0172 | 1.65 | 1.66 | 1.71 | 1.72 | 2.242 | 2.229 | 2.164 | 2.151 |
| 2 | Global .eq | | | | | 1.7 | 1.7098 | 1.761 | 1.7716 | 2.177 | 2.164 | 2.101 | 2.089 |
| 3 | G. Ocean .nls | 0.0146 | 13.8 | 6.2 | 0.0377 | 2.2 | 2.25 | 2.175 | 2.19 | 1.682 | 1.644 | 1.701 | 1.689 |
| 4 | G. Land .nls | N.A. | 0 | 0 | 0 | 1.24 | 1.24 | 1.25 | 1.25 | 2.984 | 2.984 | 2.960 | 2.960 |
| 5 | Global (60N-60S) .nls | 0.018 | 14.3 | 6.9 | 0.0148 | 1.84 | 1.86 | 1.87 | 1.9 | 2.091 | 2.069 | 2.058 | 2.025 |
| 6 | Global (60N-60S) .eq | | | | | 1.895 | 1.9158 | 1.926 | 1.957 | 2.030 | 2.009 | 1.998 | 1.966 |
| 7 | G. Ocean (60N-60S) .nls | 0.0138 | 14.3 | 6.9 | 0.0246 | 2.45 | 2.48 | 2.51 | 2.56 | 1.571 | 1.552 | 1.533 | 1.503 |
| 8 | G. Land (60N-60S) .nls | N.A. | 0 | 0 | 0 | 1.37 | 1.37 | 1.36 | 1.36 | 2.809 | 2.809 | 2.829 | 2.829 |

Tab.1

| | ECSx(est.)≈ECS.G.land*X ↓ | | X | ECS | | ECS |
|---|---|---|---|---|---|---|
| 1 | Global Ocean; X=.12/.12 | | 1 | 1.682 | | |
| 2 | Equatorial Ocean; X=.16/.12 | | 1.33 | 2.237 | | |
| 3 | Global Land; X=.32/.32 | | 1 | 2.984 | | |
| 4 | Saudi Arabia; X=.4/.32 | | 1.25 | 3.730 | | |
| 5 | S. America; X=.25/.32 | | 0.78 | 2.327 | | |
| 6 | Africa; X=.26/.32 | | 0.81 | 2.417 | | |
| 7 | Africa exSahara; X=.23/.32 | | 0.72 | 2.148 | | |
| 8 | Thailand: X=.24/.32 | | 0.75 | 2.238 | | |
| 9 | Carribean Area; X=.17/.32 | | 0.53 | 1.581 | → | 2.237 |
| 10 | Malaysia; X=.16/.32 | | 0.5 | 1.492 | → | 2.237 |
| 11 | N. Australia; X=.12/.32 | | 0.375 | 1.119 | → | 2.237 |
| 12 | India; X=.15/.32 | | 0.47 | 1.402 | → | 2.237 |
| 13 | S. Texas; X=.26/.32 | | 0.81 | 2.417 | | |

Tab.2

Because e.2 does not hold exactly true for a Forcing temperature response up to equilibrium , the value of global average $G_{eff}$ does change slightly over time until equilibrium is nearly attained (after many centuries). The value of Global average $\mathbf{G_{eff}}$ , as evaluated above for a near linear steady state condition, is *partly* corrected to the equilibrium value by $\mathbf{G_{eff.equilibrium} \approx 1.03 \cdot G_{eff.near.linear.steady}}$ , as derived in **SS7.1** and tabulated above. [Unfortunately, the equilibrium values of $G_{eff.Ocean}$ and $G_{eff.Land}$ also may vary slightly from their near linear steady evaluations above, though the variation is thought to be small.] This is the difference between the so called *historical* (or near linear steady) versus *theoretical equilibrium* values of $G_{eff}$, also known as the "pattern effect". In section **G.2** this effect, *as simulated by orthodox GCMs*, is shown to be not only highly questionable and unverified but also probably barely significant when using this particular METHOD A or B. In fact, the simplified 3-layer 2-region(Ocean, Land) formal/empirical algebraic Coarse GCM derived in section **SS6** and **SS7.2** indicates the *equilibrium* values for $G_{eff.Land}$ ,and $G_{eff.Global}$ are likely *greater* than the *near linear steady state* (".nls") evaluations obtained using METHOD A or B. However, *maximally*, they are approximated as identical above (see **G.2** for more). Although the $G_{eff.Ocean}$ value could reasonably be 2.5% less (for **CS**=0.9, $k_{mO}$=0). Note, nearly all forcing during the evaluation period is due to increasing GHGs (Green House Gases). Therefore, there is no error in ECS evaluation due to dissimilar forcing sources.

The $\mathbf{ECS_{eff}}$ for the total Globe, global Oceans, and the global Lands are ≤ **2.15, 1.67, 2.96** °C/2x$CO_2$ respectively, *using the method B, VS=variable, and $\Delta F/(2xCO_2)=3.7\ W/m^2$*. [Similarly, for all these same regions from 60N-60S, the values are slightly reduced to **2.03, 1.5, 2.83** °C/2x$CO_2$ respectively, or about 5% smaller.] The value at equilibrium is then estimated to be $ECS_{true.Global}$ = **2.09** °C, which is 70% of the IPCC AR6 ECS[15] estimate of **3.0** °C, similar to the original "influential" evaluation of Otto[16] et al. (**2.0** °C, including all corrections), and 126% of the value reported by Lewis[1] (**1.66** °C/2x$CO_2$). <u>However, the IPCC AR6 estimate of ECS=3.0 °C can be improved (see Lewis[27]) to a more appropriate value of 2.16 °C, which is very close to the value 2.09 °C calculated above.</u>

Also, since global warming of a few °C is only a serious hardship in land regions that are already uncomfortably hot or hot *and* humid <u>consistently</u>, simple ECS estimates are also tabulated in Tab.2 for some continents and countries in the tropic and subtopic zones, using Method A and VS=0 only. Regional temperature linear trends from 1970 to 2020 are evaluated using the NOAA Temperature Global Time Series web-tool[17] and these local trends are fractionally compared to the average Global Land trend. It is assumed the ECS ratios correspond approximately. These $ECS_{eff}$ warming values range from **2.21 to 3.73** °C/2x$CO_2$ , where it is presumed the land ECS is never less than the Equatorial Ocean value. Thus, in tropical/subtropical regions the warming is often about 75% of the global land average, but in some Arabian desert regions it is 45% greater. The European Land average is apparently about 25% greater than the Global Land. A survey of regions with yearly average temperatures over 80 °F indicates the local $ECS_{eff}$ ranges from about **2.21 to 2.33** °C, where it is again assumed the land ECS is never less than the Equatorial Ocean value, and excluding the Arabian desert regions.

Based on GCM simulations it has long been asserted that the greatest greenhouse gas warming generally occurs in the higher latitudes, in winter, and at night. At the current (2020) rates of $CO_2$ increase (~3ppm/year) it would require 140 years before reaching these $CO_2$ doubled levels starting from 2020, and ~140 years before reaching the corresponding ECS temperature increases over land. This evaluation of ECS does not include the very long term rein**forcing** effects of thick ice sheet melting, but this will not much effect the 50N-50S "warm" regions in any case. The direct rein**forcing** effects of ice/snow melt (on albedo and ECS) are included empirically, and the latent energy absorption of ice/snow melt and water evaporation are included in the scaling of $\Delta Q_{Global}$ as discussed above.

The TCR/ECS ratio over the oceans is estimated to be **0.71** as derived in **SS8** [TCR≡Transient Climate Response]. If TCR/ECS over land is assumed to be unity, the global average TCR/ECS is estimated in **SS8** as

$$TCR/ECS_{\ global} = [(0.7)\cdot(TCR/ECS_{\ ocean})\cdot(ECS_{ocean}/ECS_{land}) + 1\cdot(0.3)] \,/\, [(0.7)\cdot(ECS_{ocean}/ECS_{land}) + (0.3)] = 0.83$$

;where 0.3 and 0.7 are the global fractions of land and ocean respectively, and using Tab.1.

Finally, from above, the average $ECS_{eff}$ for land is a factor of 2.96/1.67=**1.77** greater than the ocean average ECS value. [Using HADCRUT T(t) data this becomes 1.77·(.94/1.08)=**1.54**]. This, almost certainly, is mostly due to an enhanced positive cloud reflection "rein**forcing**" that occurs over land. Quite simply, a temperature increase over land only (*all else constant*) has little long term average effect on atmospheric water vapor density over land [land is fundamentally "dry" as opposed to the oceans which evaporate more when warmed...all lands would soon become deserts except for the onshore atmospheric water vapor flow from the oceans], but does increase the local saturation humidity, and thus "evaporates" cloud mist droplets. This decreases solar reflection and is an enhanced positive rein**forcing** *compared to the oceans*. [The misused term "feedback" is replaced by the serendipitously defined "rein**forcing**", herein.] This difference is also partially, but much less, due to enhanced "lift" of latent and sensible water vapor energy to higher altitudes over the ocean surface (e.g. lapse rate variation and ocean evaporation). These two effects are specifically modeled and discussed in **SS6** and **SS7.2**

**[**The "cloud effect" here is purely reflective. It is, and should be considered, completely independent from a water vapor "green house" effect, for first order perturbations. If a cloud slightly evaporates to a reduced reflectivity, the number or density of water molecules in that region does not change, nor does the corresponding green house effect (to first order). Similarly, if water vapor molecules are condensed into a slightly increased cloud reflectivity, the number or density of water molecules in that region does not change, nor does the corresponding green house effect (to first order). It does not matter if the clouds are high or low, warm or cold; the reflection is the same (to first order). The temperature of higher altitude molecules *does* effect the radiation of green house molecules, and this effect is already contained within the independent water vapor rein**forcing** term[15]. But this radiation effect is independent of the form of the molecules (water vapor or micro(~1 micron) water droplet mist) for long wave infrared radiation (~10 micron), for first order perturbations. It is a common and inappropriate modeling formulation to confuse or mix the "cloud effect" with a water vapor "green house effect" for first order "feedback" perturbations.**]**

**E.2) Uncertainty Considerations**

Using Method A , start with e.20, $G_{eff} = ( S[\Delta F(t')] - S[\Delta Q(t')] ) / S[\mathbf{\Delta T_A}(t')]$ ; where $S[\Delta F(t') - \Delta Q(t')] = S[\Delta F(t')] - S[\Delta Q(t')]$ to great accuracy for the functions used here (see **SS7.2** discussion). Also <u>define</u> a fractional change of any variable $A_a$ as $\mathbf{\Delta\%}(A_a) \equiv (A_b/A_a) - 1$ ,where $A_b$ is a modified value of $A_a$. Then the following ratios can be calculated using METHOD A, where the denominator value was usually specified to be +0.2 or -0.2 (i.e. a +20% or –20% variation of that variable):

$\{\Delta\%(ECS_{Global}) / \Delta\%(S[\Delta F(t')_{Global}])\} = [-0.94]_{+20\%}$ , $[-1.52]_{-20\%}$ ;where ΔF is independent of $\Delta F_{2xCO2}$
$\{\Delta\%(ECS_{Global}) / \Delta\%(S[\Delta Q(t')_{Global}])\} = [0.168]_{+20\%}$ , $[0.158]_{-20\%}$
$\{\Delta\%(ECS_{Global}) / \Delta\%(S[\mathbf{\Delta T_A}(t')_{Global}])\} = [1.00]_{+20\%}$ , $[1.00]_{-20\%}$
$\{\Delta\%(ECS_{G.Ocean}) / \Delta\%(C^*_{eff.G.Ocean})\} = [-0.086]_{+20\%}$ , $[-0.09]_{-20\%}$ ;where C* effects the evaluation of CC*
$\{\Delta\%(ECS_{Global}) / \Delta\%(C^*_{eff.G.Ocean})\} = [-0.025]_{+20\%}$ , $[-0.097]_{-20\%}$ ;where C* effects the evaluation of CC*
$\{\Delta\%(ECS_{G.Ocean}) / \Delta\%(CC^*_{G.Ocean})\} = [0.34]_{+20\%}$ , $[0.26]_{-20\%}$

$\{\Delta\%(ECS_{Global}) \;/\; \Delta\%(CC^*_{G.Ocean})\} = [0.28]_{+20\%}$ , $[0.30]_{-20\%}$
$\{\Delta\%(ECS_{G.Ocean}) / \Delta\%(\Delta E^*(t))\} = [0.37]_{+10\%}$ , $[0.335]_{-10\%}$ ;where E*(t) is scaled by a constant fraction
$\{\Delta\%(ECS_{Global}) \;/\; \Delta\%(\Delta E^*(t))\} = [0.24]_{+10\%}$ , $[0.34]_{-10\%}$ ;where E*(t) is scaled by a constant fraction
$\{\Delta\%(ECS_{Global}) / \Delta\%(\Delta F_{2xCO2})\} = [-0.05]_{+10\%}$ , $[-0.06]_{-10\%}$ ;where ΔF is similarly scaled, 90% GHG forcing
$\{\Delta\%(ECS_{Ocean}) / \Delta\%(\Delta F_{2xCO2})\} = [-0.04]_{+10\%}$ , $[-0.154]_{-10\%}$ ;where ΔF is similarly scaled, 90% GHG forcing
$\{\Delta\%(ECS_{Land}) / \Delta\%(\Delta F_{2xCO2})\} = [0.08]_{+10\%}$ , $[0.1]_{-10\%}$ ;where ΔF is similarly scaled, 90% GHG
$\{\Delta\%(G_{eff.Global}) / \Delta\%(\Delta F_{2xCO2})\} = [1.06]_{+20\%}$ , $[1.06]_{-20\%}$ ;where ΔF is similarly scaled, 90% GHG forcing
$\{\Delta\%(G_{eff.Ocean}) / \Delta\%(\Delta F_{2xCO2})\} = [1.14]_{+20\%}$ , $[1.14]_{-20\%}$ ;where ΔF is similarly scaled, 90% GHG forcing
$\{\Delta\%(G_{eff.Land}) / \Delta\%(\Delta F_{2xCO2})\} = [0.91]_{+20\%}$ , $[0.91]_{-20\%}$ ;where ΔF is similarly scaled, 90% GHG forcing

Thus readers may estimate the fractional variation of ECS for any variable they may wish to modify according to their own preference. Uncertainties are so often highly subjective or speculative the readers may decide for themselves. The values given in Tab.1 may be considered the ECS *maximal* median values, according to this analysis.

For the special (but inappropriate?) case where CC* is calculated assuming only Radiative Forcing over the Ocean, i.e. using the largest values of CC* in e.15,16, then $ECS_{G.Oceans(60N\text{-}60S)}$ = 1.59 C (which is 4% greater than reported in Tab.1), $ECS_{G.Oceans}$ = 1.74 C (which is 2.3% greater than in Tab.1), and $ECS_{Global}$ =2.23 C (which is 3% greater). Thus Ocean ECS is either quite small (1.59 °C) and/or insensitive to CC* uncertainty!

For the special (but inappropriate) case where $C^*_{eff}$ is calculated assuming the *effective* mix depth is the yearly average (i.e. 60% of the yearly maximum value used), then CC* for the Global Oceans is calculated to be 16% greater than otherwise, but $ECS_{G.Oceans}$ is only 3.6% greater than otherwise, and $ECS_{Global}$ is 3.4% greater than otherwise. Thus ECS is *extremely* insensitive to C* uncertainty!

Also, a simplified and *maximal* ECS case is to simply omit the uncertain "$g_u$" related terms in all ΔQ(t) [see also **SS4.2**]. A full empirical evaluation of $G_{eff}$ (including the formal/empirical evaluation of CC*) using $g_{u.effective}$ = **0** W/(m$^2$ °C) results in only a 3.5% *decrease* in evaluated $G_{eff}$ for the 60N-60S latitude Ocean region, and thus even less for the Global Ocean average.

None of these special cases effect the Land ECS evaluations.

The calculated values using METHOD A and METHOD B only differ by about 1% on average.

However, it is desirable to at least *coarsely* estimate the Global $ECS_{eff}$ uncertainty upper limit from the uncertainty of the components S[ΔF(t)],S[ΔQ(t)], S[$\mathbf{\Delta T_A}$(t)], $\Delta F_{2xCO2}$ , where $ECS_{eff} = \Delta F_{2xCO2}/G_{eff.nls}$ , and $\Delta F_{total} = \Delta F_{GHG} + \Delta F_{Non\text{-}GHG}$ . This excludes any volcanic component effect, which would only minimize ECS evaluation in any case.

For 1980 through 2020 only:
$\Delta F_{2xCO2}$ : The uncertainty of this is at most +14% or -10%. If the GHG(Green House Gas) component of S[$\Delta F(t)_{total}$] is ~ 90% (which is very likely), then the $\Delta F_{2xCO2}$ and S[$\Delta F(t)_{GHG}$] errors virtually cancel out, and the upper limit $ECS_{eff.Global}$ uncertainty is only +0.6% !

S[$\Delta F(t)_{Non\text{-}GHG}$] : The uncertainty is estimated[7] ±50% [see Fig.13] . Presuming the Non-GHG component of S[$\Delta F_{total}$] is near 10%, then the uncertainty of S[$\Delta F_{total}$] is about ±5% . The upper limit $ECS_{eff.Global}$ uncertainty is then about +7%.

S[$\mathbf{\Delta T_A}$(t)] : The systemic difference of the HADCRUT $\Delta T_s$(t) values from the used NOAA values is about +**4**%, +**8**%, and -**6**% for the Globe, global Ocean, and global Land, respectively. Presuming the premier HADCRUT and NOAA data values are equally likely, then the median value is the average of both, and the error of both is then half the difference. The random measurement data 1σ uncertainty for the PIR and H5[ ] filtered S[$\mathbf{\Delta T_A}$(t)] (after extrapolation through the transient "dip") is about 1% (2% for Land) also. After linear addition, the upper limit NOAA $ECS_{eff}$ error is then about +($1_{rms}$+$\mathbf{2}_{fixed}$)%, +($1_{rms}$+$\mathbf{4}_{fixed}$)%, and +($2_{rms}$+$\mathbf{0}_{fixed}$)% , respectively.

S[ΔQ(t)] : The estimated 1σ error of S[H5[E*(t):NOAA]] (i.e. slope of the thin blue line of Fig.24), is about 4%. Assuming a scalar (systemic) uncertainty of S[ΔE*(t):NOAA] to be ±10%, then the total upper limit $ECS_{eff.Global}$ uncertainty is ≈ {E*(t) error ratio}·$\sqrt{(0.04^2 + 0.1^2)}$ = (0.24)·(0.107) =(0.026)=2.6%. The error ratios above also show sensitivity to $C^*_{eff}$ to be virtually zero. Finally, the alternative "temperature profiled" CC* and $g_u$ evaluations (before and after e.16B in sect. **D**) demonstrate a maximum evaluation of $ECS_{G.Oceans(60N\text{-}60S)}$ < 1.59 °C, which is 4% greater than the tabulated value of 1.53 °C (see Tab.1). This final value of 4% will be used below.

Adding all the *random* uncertainties in quadrature yields a *very* coarse idea of the 1σ $ECS_{eff.Global}$ uncertainty upper limit of [$\sqrt{(0.6^2+7^2+1^2+4^2)}$ +<u>**2**</u>$_{fixed}$] ≈ 10% above the *most likely maximal* value, or <u>$ECS_{eff.Global}$ ≤ (2.15·1.1) ≤ 2.37 °C is better than 84% assured</u>. The uncertainties seem relatively insignificant.

Finally, the "pattern effect" could systemically vary the $ECS_{true.Global}$ value further, but is considered likely to be a reduction or neutral based on the **SS7** Coarse GCM much preferred herein; or likely less than +7% and most likely +3% based on orthodox GCMs. All this is as discussed fully in sections **E.1**, **G.2**, and **H** . An unexplained centuries long positive natural forcing, as seen convincingly in borehole temperature histories, is reasonably possible, as speculated in **G.3**, which probably does *reduce* this upper limit by about 10%. And a larger than expected aerosol cooling after 1980 could reduce this value significantly (see **G.3**)

---

## F) Orthodox Simplified $\underline{\Delta}$Q, $G_{eff}$ , ECS Calculations

In section E many techniques were applied to "artificially" remove various unknown Forcings and Forcing responses from the raw data. All time data is first H5[ ] filtered. PIR is then used to remove the 21.2 year and 9.3 year cycle T(t) harmonics, etc.; and thus the ***corresponding*** forcing components (which are unknown in any case) can be omitted. Also the transient ΔT(t) component seen in Fig.5.6 from 1995 to 2015 is removed and replaced by a linear interpolation/extrapolation shown as line1; and thus the ***corresponding*** ΔF(t) or ΔQ(t) ocean parametric fluctuation (which are mysterious in any case) can be and are omitted. Q(t) is then formally/empirically evaluated for a simple linear ΔT(t) ramp. This method of PIR and "transient deviation removal" from the underlying ΔT(t) linear trend is a reasoned, reasonable, and appropriate liberty, and much superior to incompletely or incorrectly interpreting the effects of these real but otherwise unaccounted phenomena.

However in this section, *for comparison with orthodox methods*, the more orthodox application of the Energy budget method will be employed using the NOAA temperature data, the "authoritative" Cheng[13] E*(t) data, and the updated IPCC AR5 ERF[1] data *excluding* the Volcanic component (which is a *maximal* ECS choice as discussed previously). Only H5[ ] filtering of all yearly averaged time variable raw data will be applied...no PIR, or interpolations, or C* and CC* derivations.

Only the Global $G_{eff}$ will be calculated using e.5, $G_{eff} = (\underline{\Delta}F - \underline{\Delta}Q)/\underline{\Delta}T$ ; and using definitions e.4b,c,f . However it is also useful to first average (integrate) e.3 over the time range (t'-a) to (t'+a), so as to create more reliable time averages of all the variables therein. Therefore $\underline{\Delta}$F, $\underline{\Delta}$Q, and $\underline{\Delta}$T will also be defined in this way for the following evaluation, where 2·a=10 years. The time evaluation range for Δt=20 years, will then be $t'_1$=1995 through $t'_2$=2015; and the averages are then taken from <u>1990 to 2000</u> and <u>2010 to 2020</u>:

b) $\underline{\Delta}T \equiv (\mathbf{\Delta T_A}(t'_2) - \mathbf{\Delta T_A}(t'_1))$ = 0.454-0.068=0.386 °C ;where H5[] values are taken from Fig.4.6B data (no PIR)
c) $\underline{\Delta}F \equiv (\Delta F(t'_2) - \Delta F(t'_1))$ = 1.40-0.60 = 0.80 $W/m^2$ ;where H5[] values are taken from Fig.16 data
f) $\underline{\Delta}Q_{Oceans} \equiv \Delta Q(t'_2) - \Delta Q(t'_1) = \{[d(\Delta E^*(t'))/dt']_{@t'=t'2} - [d(\Delta E^*(t'))/dt']_{@t'=t'1}\}$ = 0.805-0.748 = 0.057 $W/m^2$
; where H5[] values are taken from Fig.28 data (the Cheng data), for the Global Oceans(0-2km) (see e.12).

And then,

e.21 $G_{eff.Global} = (\underline{\Delta}F-(0.7)\cdot(1.07)\cdot(1.08)\cdot\underline{\Delta}Q_{Ocean})/\underline{\Delta}T = (0.80-0.046)/0.386 = \mathbf{1.95\ W/(m^2\cdot°C)}$

e.22 $ECS_{eff.Global} = 3.7/1.95 = \mathbf{1.90\ °C}$

;where $\Delta Q_{Global} = (0.7)\cdot(1.07)\cdot(1.08)\cdot\Delta Q_{Oceans}$ ; where (0.7) is the Global fraction of Ocean area, (1.07) includes the extra-oceanic absorption, and (1.08) approximates the added abyssal absorption, as previously discussed.

Therefore, this more orthodox evaluation of $ECS_{eff}$ (=**1.90** °C) using the E*(t) data of Cheng[13] (circa 2022, done as a comparison between methods) is 12% *smaller* than the value of **2.15** °C using METHOD B. This value is identical to a recent orthodox energy budget evaluation by Spencer[28] (**1.9** °C) using Cheng's data, a similar evaluation range and data sources (plus a volcanic forcing term). These orthodox methods, using the Cheng Ocean Energy data result in smaller ECS values. If the volcanic Forcing component is properly *included* (see Fig.17) the ECS can further reduce over 33%, but is unduly sensitive to the particular evaluation range used, and not considered reliable herein. Exclusion of the volcanic component yields a stable but *maximal* value (=**1.90** °C).

However, for the reasons listed in the first paragraph above, the results using METHOD A or METHOD B are considered to be more reliable, and without using Cheng's highly theoretical reanalyzed data, or requiring the very uncertain volcanic Forcing component.

**[[UPDATE: using the updated Cheng data (circa 2025), and using the averaging range <u>1995-2000</u> and <u>2015-2020</u> (*which avoids most of the Pinatubo volcanic Forcing, and the puzzling Ocean transient Temperature "dip" phenomenon*!!),** then $\underline{\Delta}Q_{Ocean} = 0.097\ W/m^2$ and $\underline{\Delta}T = 0.42$ °C and $\underline{\Delta}F = 0.8\ W/m^2$, and so:

e.21b $G_{eff.Global} = (\underline{\Delta}F-(0.7)\cdot(1.07)\cdot(1.08)\cdot\underline{\Delta}Q_{Ocean})/\underline{\Delta}T = (0.80-0.0785)/0.42 = \mathbf{1.72\ W/(m^2\cdot°C)}$

e.22b $ECS_{eff.Global} = 3.7/1.72 = \underline{\mathbf{2.15\ °C}}$ .

If valid, this would exactly corroborate the value of **2.15** °C using Method B noted above (see **Fig.36** in **G.3**)]]

---

## G) Forcing Uncertainties and Pattern Effects

### G.1) Alternative Forcings

[Note, since methane and other greenhouse gases also approximately follow the logarithmic forcing relationship of $CO_2$, it is conceptually proper and convenient to consider them all as simply additional concentrations of $CO_2$ in simple calculations.]

After independent stratospheric temperature equilibration the effective $\Delta F/2xCO_2$ by MODTRAN calculation is **~4.2** $W/m^2$ (given 20% strato/stratus cumulous cloud cover (.6km-2km)...see **SS6**). The effect of the "sparse" stratosphere is surprisingly large. This is defined herein as the $CO_2$ "**direct forcing**", and allows the effects of the stratosphere on $CO_2$ forcing to be omitted from further modeling; i.e. the *net* upward radiation at the top of **tropo**sphere is the same as at the top of **strato**sphere *after* equilibration. This is the *independent* **direct** forcing (DRF) properly required in the previous <u>empirical</u> $G_{eff}$ and $ECS_{eff}$ calculations involving surface temperature (see end of section **A**). In some theoretical representations the direct forcing is further modified to account for small differences between surface heating versus bulk atmosphere GHG heating. This may be called the *effective* radiative forcing (ERF), but can *only* be theoretically evaluated. By using a representation appropriate for empirical evaluations this theoretical evaluation can be avoided by using the direct radiative forcing concept (DRF or RF) and a $G_{eff.GHG}$ that is *specific* to GHG forcing *only* (see **SS6**). That is the representation used herein, as it provides a more accurate/reliable <u>empirical</u> evaluation, independent of theoretical considerations. Nearly all ΔForcing is due to ΔGHG in the evaluation range, so this is appropriate.

In this case **4.2** $(W/m^2)/2xCO_2$ represents a +13.5% increase from the IPCC AR5 canonical value of ERF = DRF= **3.7** $(W/m^2)/2xCO_2$. And this percentage increase should then also be made to the <u>corresponding</u> anthropogenic ΔForcing component ΔF(t) of Fig.12 (i.e. 1.135 · **0.4** = **0.454** $W/(m^2\cdot$decade) ). Using the error ratio derived in section **E.2**, this would <u>only</u> result in a calculated <u>Ocean</u> ECS change of (+13.5%)·(-0.04) =

-0.54%, and (+13.5%)·(+0.08) = +1.1% over Land, i.e. it is not a problem. If this alternate GHG forcing were to be used then the true values of Ocean $G_{eff}$ would also be about 15.4% larger than all reports above and in Tab.1; the Land $G_{eff}$ would be 12.3% larger. This would increase the Ocean TCR/ECS evaluation by about 4%, and significantly effect the Coarse GCM calculation of cloud reflection rein**forcing** components derived in **SS7**.

However in a complicating twist, it must be noted that in **SS6,7** all the MODTRAN evaluated temperature proportionality radiation parameters $g_{ae}$, $g_{ea}$, $g_{sa}$, $g_{vs}$, $g_{ve}$ were multiplied by a scalar $\boldsymbol{\varepsilon = \tau =}$ **0.85**, which resulted in a close reproduction of the *independent* orthodox IPCC AR6[15] Global Planck (surface) effective radiation "feedback", water vapor "feedback", and cloud reflection "feedback". And for consistency with orthodoxy, this is used herein. But this scaling was not simultaneously performed on the $CO_2$ radiative forcing components $\Delta I_{cs}$ , $\Delta I_{ce}$ ; these remained at the *updated* IPCC AR5 canonical values (see Fig.12). If the $\boldsymbol{\varepsilon}$ scaling is also uniformly applied to the *non-canonical* forcing values above then $\Delta F/2xCO_2$= 0.85· **4.2** = **3.6** $(W/m^2)/2xCO_2$ , and $\Delta F(t)$/decade = **0.454** · 0.85 = **0.39** $W/(m^2$·decade). In other words, the consistent $\boldsymbol{\varepsilon}$ scaling of the *non-canonical* MODTRAN *forcings* above transforms them back into the *canonical* values (within a few percent)!

Therefore, if the $\boldsymbol{\varepsilon}$ scaling is also performed uniformly on the non-canonical MODTRAN calibrated $\Delta F$ values above they are essentially no different from the orthodox/canonical values! If the $\boldsymbol{\varepsilon}$ scaling is not applied to *all* the MODTRAN evaluated parameters above then there will be a slight reduction to the evaluated $ECS_{Global}$, and a considerable change to the $G_{eff}$ and cloud reflection rein**forcing** evaluations. The physical reason or justification for the $\boldsymbol{\varepsilon}$ scaling (if legitimate) is not specifically known to the author. It is certainly possible the MODTRAN values are inexact or have been misapplied to the Coarse GCM in a way that is easily corrected by the simple use of the $\boldsymbol{\varepsilon}$ scaling uniformly on *all* those parameters, and that is assumed herein. Thus, the orthodox/canonical $\Delta F$ values have been and will be used in this treatment, which is again a *maximal* ECS choice.

## G.2) The "Pattern Effect" or $G_{eff}(t)$ variability

Global Climate Models over decadal time spans (especially of cloud cover, rainfall and bulk atmospheric temperature change) are particularly theoretical, unverified, and problematic. This is supported by the fact there is no *standard* model, and the ECS predictions of various CMIP6 models varies by over a factor of two[15,26]. Also, the average bulk global atmospheric warming of an average of CMIP6 models has been shown *incorrect* by a factor of about 2 larger than observations[18]. Further, NOAA RATPAC[19] balloon atmospheric temperature data reveal Global average lower and mid-troposphere temperature trends to be about the same as the surface temperature trends, in contradiction to most CMIP5 and 6 predictions. And CMIP6 GCMs do not generally maintain conservation of energy and/or atmospheric water mass to a significant degree when integrated over centuries[20], resulting in significant non-realistic(?) temperature drift. In particular, atmospheric water conservation errors (which includes cloud formation, precipitation , and evaporation) are significant and cannot be internally corrected in any model.

Subsequently, the IPCC no longer *directly* uses GCMs as the primary predictor of long term climatic temperature change, instead adding "emergent constraints" as a criteria. But, even when only using "emergent constrained" GCMs, most of these proven flaws remain (see section **H** and **ref[18]**). To consider GCMs as effectively unquestionable or verified would be completely unwarranted.

Thus, recently GCM modeled "pattern effects" must be considered purely theoretical, unverified and uncertain. In particular, in a commentary by Lewis[21] (see also Lewis and Mauritsen[22]) "The historical pattern effect is not robust; it varies hugely between models and SST datasets." and "The forced pattern effect is very small in CAM5.3[a model] "...on the order of 2.5% .

Thus "forced pattern effects" are possibly "very small" as stated above using CAM5.3 , and even smaller due to a less harsh, more realistic ramp forcing response (as opposed to step responses). This is an unintended benefit of using METHOD A or B *trend* analyses.

More generally, using LongRunMIP models[23] the median value (from Table 1[23]) of $ECS_{true} / ECS_{eff}$ =**1.027** with a $1\sigma$ upper limit of about **1.07** ; this must only be determined using the year 20 through year 150 option as the $ECS_{eff}$ evaluation range [this *appropriately* avoids the large "pattern effect" variation occurring at the beginning of the Forcing step], and only using a $2xCO_2$ step [a $4xCO_2$ step is definitionally inappropriate and often induces an overly large non-linear response], and removing the one very large "1.22" outlier [a few GCMs are unrealistically hyper-sensitive to Forcings]. This (**1.027**) is a small increase, and several models (i.e. HADCRUT) yielded values *less* than unity. And a less harsh, more realistic ramp forcing, as is used herein, would yield smaller "pattern effect" variations. **[see ECStru-d-ECSeff-C.xls in the repositories]**.

But again, these predictions must be considered purely theoretical, unverified and uncertain. The greatest, if not only, long term variability of $G_{eff}(t)$ (or "pattern effect") is ultimately caused by the large temporal variability of power flow into the Ocean thermal capacitance as compared to the Land...which drops to zero at equilibrium. This is calculated in **SS7.2** using a simplified 3-layer 2-region(Ocean, Land) formal/empirical algebraic Coarse GCM developed in **SS6** . Those results indicate the *equilibrium* values for $G_{eff.Land}$ , and $G_{eff.Global}$ are in fact likely to be *greater* than the *near linear steady state* values (1980-2020) when using METHOD A or B. Although the $G_{eff.Ocean}$ value could reasonably be 2.5% less (for **CS**=0.9, $k_{mO}$=0). In particular $G_{eff.true.Land}$ may be >5% greater, which could be applied to Tab.1,2 $ECS_{Land}$ *reductions.* Any variability (or "pattern effect") between different ocean regions prove to be even less. However, it has been assumed in all evaluations above and in Tab.1,2 that $ECS_{true} = ECS_{eff} \equiv ECS_{near.linear.steady}$ , which is considered herein to be a *maximal* ECS choice..readers may apply whichever value of $ECS_{true}/ECS_{eff}$ they choose to accept.

The near linearity of e.1 has been presumed all along, such that $G_{eff}$ is also independent of $\Delta \mathbf{T_A(t)}$ (excluding "pattern effects") over a small range of several °C (i.e. perturbation theory applies). And this *is* verified empirically and convincingly by the striking linearity of the $\Delta \mathbf{T_A(t)}$ response. See Fig.4.3,6 and Fig.5.3, and the Global Land average of Fig.6.3 which varies over 1.5 °C. Thus, there is no evidence of some extreme temperature dependence of $G_{eff}$ over the evaluation range of several °C . There would be no physically realistic justification for such an extreme non-linearity over so small a temperature range in any case. (Since radiation is a function of the forth power of temperature, any such non-linearity might well enhance cooling and reduce the ECS evaluation.) The ΔForcing variation over the $CO_2$ doubling range is certainly a small perturbation of the total independent Forcing. Therefore, explanations of enhanced ECS evaluations due to the "pattern effect" *or* local $G_{eff}$ temperature dependence are unwarranted.

### G.3) Forcing Uncertainty

The uncertainty of the long term (decadal) trend of ΔF(t) merits further consideration. When a temperature measurement is made there is no fundamental question about its validity, assuming the thermometer is proven accurate. There is no question as to the existence of some unknown "component" of the total temperature that is simply unknown, or unaccounted, or un-acknowledged in the measurement. Temperature is not essentially calculated or determined . It is essentially "measured" by a measuring device.

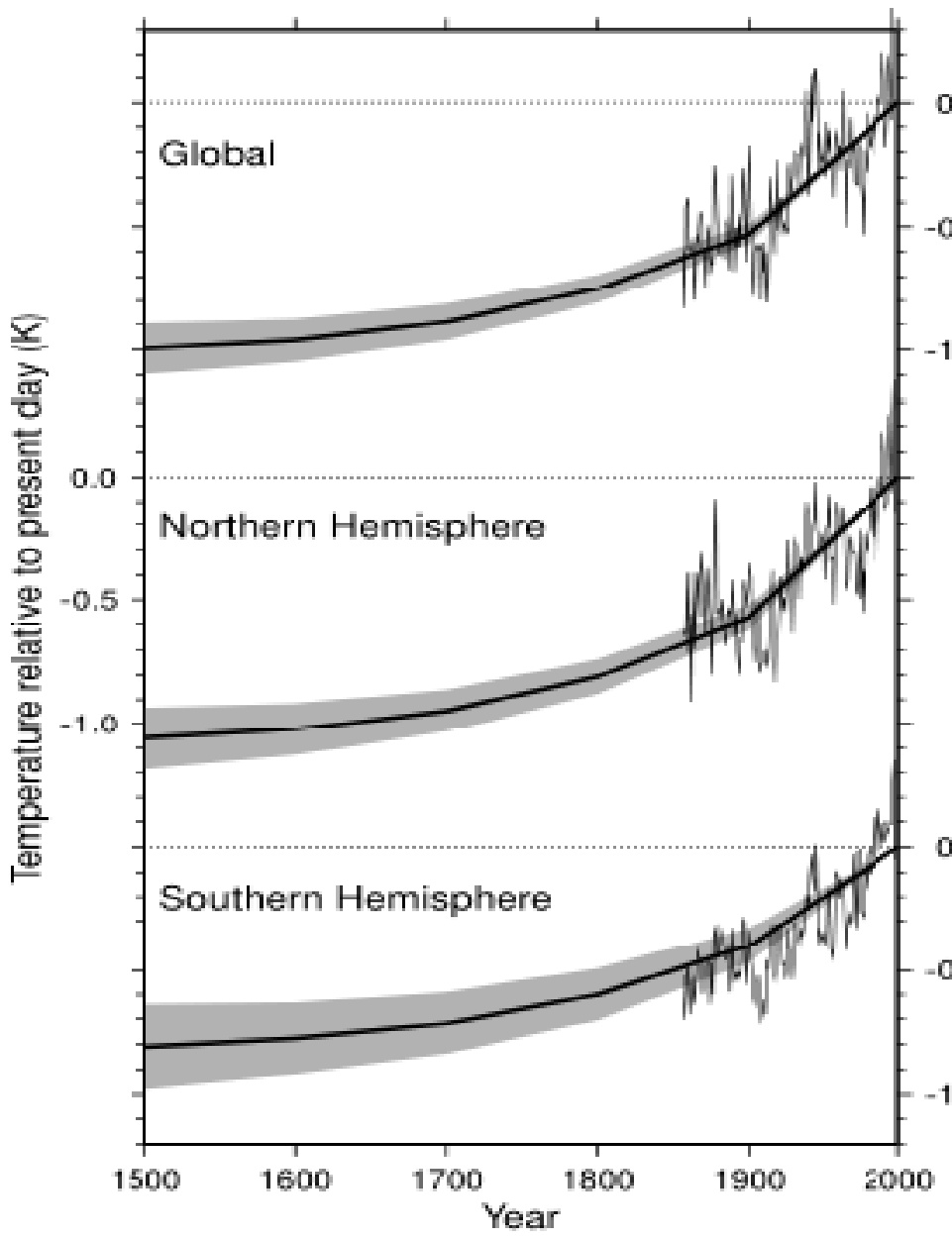

Fig.35 source: ref.[24] CC BY-NC-SA 4

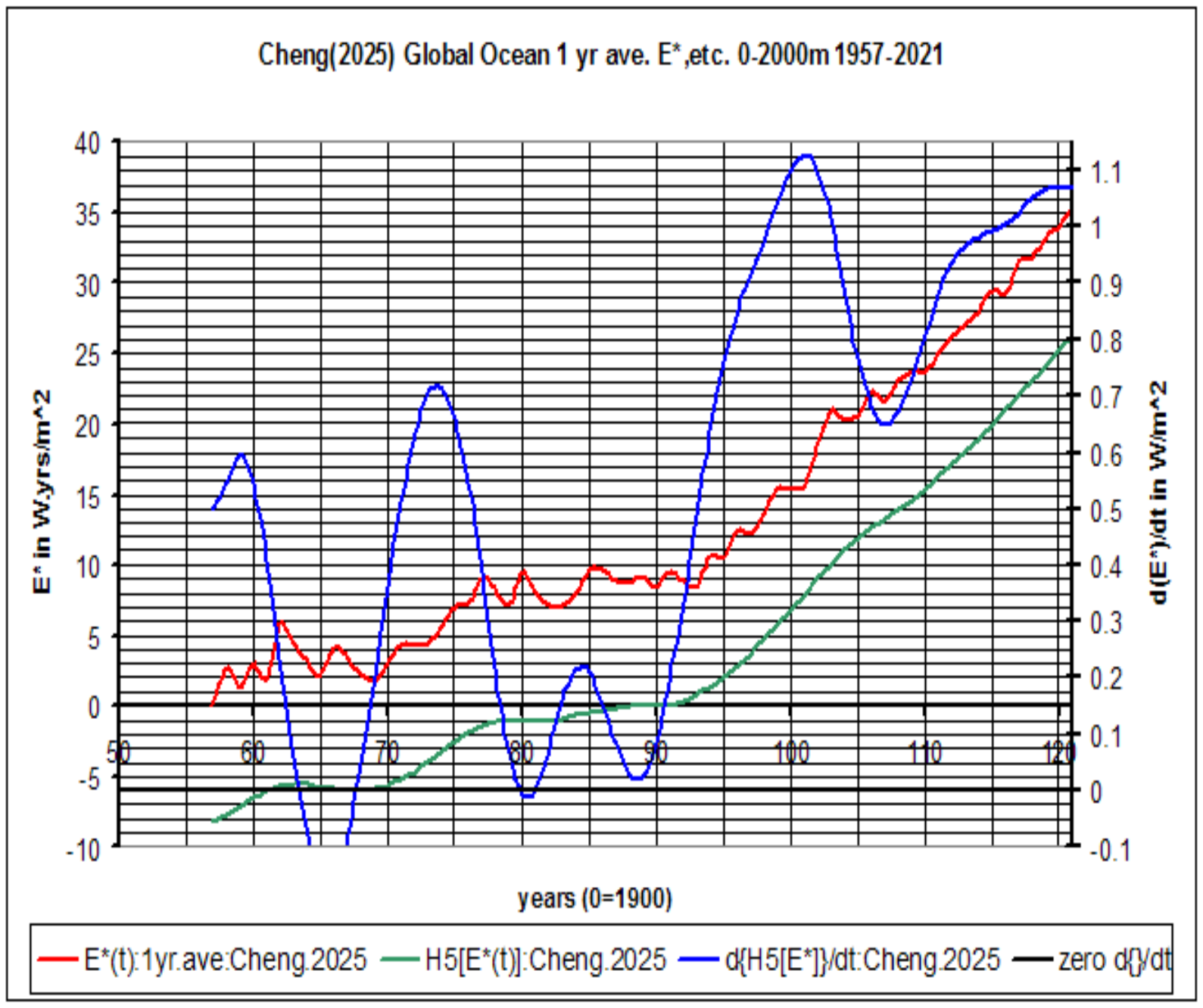

Fig.36 (discrete annual ave. plots; curves for visualization)

All these questions are not so positively or easily answered w.r.t. Forcing (i.e. temperature independent power flow into a region). The possibility of unknown Forcing sources into a region can never be absolutely discounted. And, is a measured Forcing component independent or temperature dependant? This is a critical difference that may be difficult to know. Is a change in cloud reflection due solely to some temperature independent unknown source or fluctuation, or is it a temperature dependant "rein**forcing**" that enhances an independent greenhouse effect?

In the end we must rely on the consistency of these theories or assumptions with **all** known observations and experiments. In that case they are presumed to be valid.

In this treatment that judgment need not be made. It is left to the reader to judge what true long term *independent* natural and anthropogenic forcing trends are in effect 1970 – 2020, and what the uncertainties are. Further, the orthodox updated IPCC AR5 value of 0.4 $(W/m^2)$/yr for anthropogenic sources has been used herein, which is consistent with the canonical IPCC AR5 ERF=DRF of $\Delta F_{2xCO2}$=3.7 $W/m^2$. But is this correct? It is apparently a theoretical consideration as described in **G.1** above, but *fortunately* has virtually no effect on the ECS evaluation (see **E.2** and **G.1**)!

As examples of possible but currently unorthodox natural long term underlying forcing trends, one could cite Shaopeng Huang[24] "Temperature trends over the past five centuries reconstructed from borehole temperatures", where a significant long term and parabolically accelerating Global Land Temperature response to some unknown Forcing was detected starting in the 1500's to the present, as shown in Fig.35 . This evidence is very compelling and would result in a Land ECS reduction of up to 10%. Also, probably less likely, as quoted from the abstract of Ka-Kit Tung[25] , "Quantitatively, the recurrent multi-decadal internal variability, often underestimated in attribution studies, accounts for 40% of the observed recent 50-y warming trend." . In these cases, the evaluated value of ECS would be notably reduced. Also, sulfate and aerosol Forcing (including cloud Forcing) was presumed essentially constant after 1970 in sect. **D**. However sulfate emissions in Fig.21 show the possibility of a sharp drop after 1980, which would significantly increase $S[\Delta F_{total}(t)]$, and then significantly decrease the ECS evaluation. Therefore, the values of $G_{eff}$ and ECS can be modified using the results of section **E.2** by the readers as they see fit.

**[**However...it may be possible to determine the value of the actual independent Forcings by calculation, as opposed to requiring them as an independent theorized quantity. Applying this approach, the value of $G_{eff}$ and Forcings are simultaneously determined, which removes the forcing data uncertainty from the calculation. This approach, using many of the techniques developed herein, may be presented in a subsequent monograph entitled "Earth Radiative Conductivity and Forcings using both Frequency and Trend Analysis".**]**

---

## H) IPCC AR6 ECS Misevaluation

The IPCC AR6 now evaluates ECS using a conglomeration of :
Z) direct GCM evaluations?...No! This is now no longer directly used, being considered unreliable.
**A**) (instrumental) empirical energy balance methods
**B**) atmospheric process understanding
**C**) "emergent (observed) constraints" on GCM simulations (simulations will include the "pattern effect")
**D**) paleo "proxy" data evaluation ,
and conclude a most likely ECS = 3.0 °C .

Remember, in this monograph a simplified 2-D Coarse GCM has been derived which correctly reproduces the individual orthodox IPCC AR6:
1) Global Planck (surface) effective radiation "feedback"
2) Global water vapor "feedback"
3) Global albedo "feedback"
4) Global cloud reflection "feedback"
5) Global $\Delta F(2xCO_2)$ forcing

6) and the observed Surface AND Bulk atmosphere temperature trends for both Land AND Ocean!

It is therefore deemed physically realistic and superior to most alternative GCMs.

Consider:

Results of **A** [(instrumental) empirical energy balance methods]: The Global average $ECS_{eff.1980\text{-}2020}$ is determined herein to have a *maximal* median value of about 2.15 °C , using the energy balance methods. And the Global Land and Global Ocean $ECS_{eff}$ are separately evaluated as well. These results are very nearly certain, with upper limit 1σ uncertainty very approximately +10%...see **E.2** .

And the Coarse GCM verifies that the pattern effect at equilibrium Globally and for global Lands is neutral or slightly "cooling", in the modern era. Therefore $ECS_{true} \leq ECS_{eff}$ above. And explanations of enhanced ECS evaluations due to extreme local $G_{eff}$ temperature dependence are also unwarranted...See **G.2** .

The questionable (see result **C** below) CMIP6 GCMs themselves indicate a median $ECS_{true}/ECS_{eff} = 1.027$ ([ref.23]...using year 20 through year 150 as the $ECS_{eff}$ evaluation range, and only using a $2xCO_2$ step, and removing the one very large "1.22" outlier...see section **G.2** discussion), which is only a barely significant additional "pattern effect" warming.

Results of **B** [atmospheric process understanding]: Although the Coarse GCM reproduces the IPCC orthodox *individual* "feedbacks", the fully evaluated ECS is much smaller. Why?...because the individual feedbacks do NOT add linearly. They interact, and the total positive "feedbacks" are less than the linear sum. This is the error in the IPCC AR6 "process understanding" evaluation. (see **SS7**)

Results of **C**: "Emergent constraints" is a method of more heavily weighting CMIP6 GCMs that better reproduce observed temperature (or other) histories. However, all CMIP6 GCMs significantly over-estimate bulk atmospheric warming [ref.18]. And even those that reproduce Atmospheric and Surface warming *better,* exhibit a wide range of ECS. For example (circa 2020 [ref.18], [ref.26], [ref.23]):

| model | ECS (°C) | $ECS_{true}/ECS_{eff}$ |
|---|---|---|
| CAMS-CSM1-0 (China) | 2.27 | |
| MIROC_ES2L (Japan) | 2.7 | 1.06 ($4xCO_2$) |
| GISS-E2-1-G (NASA) | 2.7 | 1.056 ($4xCO_2$) |
| CNRM_C61r5 (France) | 4.8 | 1.05 ($2xCO_2$) . |

Therefore the CMIP6 GCMs exhibit no particular "expertise" for ECS evaluation, which is not amenable to any *valid* statistical verification. The average of incorrect results is not necessarily the correct result. These GCM ECS evaluations must be considered purely theoretical, unverified, and the large uncertainty leaves them not very useful (see **G.2**). "Stepping" GCM simulators are, in fact, the most error prone evaluators of equilibrium conditions.

Results of **D**: Paleo proxy temperature proportionality constants are theoretical, unproven, and never certain. And the theoretical very long term Paleo Forcing trends in total are also never proven or certain. Paleo ECS evaluations are therefore unreliable in general, requiring several unproven and often unacknowledged "educated" presumptions which are not amenable to any *valid* statistical verification. At best, the uncertainty is too large to be very useful.

Certainly, the empirical Energy Balance method provides the only reliably accurate $ECS_{eff}$ *maximal* evaluation (with upper limit 1σ uncertainty very roughly +10% ...see **E.2**); and additional "pattern effect" equilibrium variations are either negative, neutral, or at most barely significant given Forcing *perturbations* (e.g. $2xCO_2$). Further, Lewis [ref.27] updated and improved the IPCC AR6 $ECS_{true}$ evaluation (using methods **A,B,C,D** above) and estimated an improved value of 2.16 °C , which is very near the empirically derived *maximal* median *value* of 2.09 °C herein! This is consistent with a multitude of similar past scholarly empirical $ECS_{eff}$ evaluations[2].

## I) Summary: Assumptions and Demonstrated Assertions

**Assumptions:**

**1)** The underlying long term Forcing variation between 1970 and 2021 was the same over the Land and Oceans.

**2)** The NOAA [or alternatively the HADCRUT] Global temperature record is reliable between 1948 and 2021.

**3)** The *updated* IPCC AR5[1] anthropogenic ERF record is reliable, as is the *corresponding* $\Delta F/2xCO_2$ forcing.

**4)** There are **NO** *unknown* natural forcing trends, or corresponding warming trend components (Such could change the results significantly!)

**5)** The NOAA or alternatively the Cheng[13] Global Ocean Heat Content general trends (0-2km) are reliable between 1990 and 2020, and the change below 2km is insignificant. The temperature of the abyssal oceans below 2km are self regulated and very close to freezing due to the continual upwelling of cold water sourced from the sinking cold polar waters, and are essentially *independent* of the surface temperature changes directly above. Thus, there is essentially no *dependant* Ocean thermal energy change in this zone and it can be omitted, i.e. the 0-2km zone data is sufficient. [Reports in Cheng[13] (Figure 6) *do* indicate such abyssal ΔEnergy storage is less than 8% of the Ocean total, resulting in an ECS increase of about 1.6%, i.e. insignificant in any case.]

**6)** The energy storage change over and in the Land regions is insignificant *using Method A or B*, and can be neglected. [Method A and B are independent of the effects of bulk thermal capacitance for ramp T(t) , anyway!] The extra-Oceanic storage is estimated[13] at 7% of the Ocean value, results in a 1.5% increase of Land ECS evaluation *at most*, i.e. insignificant in any case.

**Demonstrated Assertions:**

**1)** It is useful to model the climate system as a simplified thermal circuit with interconnected and coupled nodes using concepts such as temperature independent sources (F(t)), temperature independent thermal conductance (G), temperature variable parametric rein**forcing**, thermal capacitance $C_{th}$, and temperature T(t). Thermal (Ocean) Energy "density" is denoted as E*(t) ($W \cdot yrs/m^2$), and $Q(t) \equiv d[E^*(t)]/dt$. The analogy with electrical circuits is exact. [see **SS6,7**]

**2)** Linear time filtering operators (H5[ ],H9[ ], etc.) can be pre-applied to T(t), F(t), and E*(t) without effecting the validity of the "instantaneous" Energy conserved equations to which they belong. [see **SS2**]

**3)** The concept of an effective global average thermal "conductance" ($G_{eff}$) that is a true constant over time is only possible if $\Delta T(x,y,t)=\Delta T_\alpha(x,y) \cdot \Delta T_\beta(t)$ [i.e. see equation e.1,2]. Thus the energy budget methods rely on this assumption as well. In this case there is no "pattern effect". [see **SS1**]

**4)** If e.2 holds, then the "energy budget method" may be applied to Ocean or Land regions *individually* for small signal (perturbation) responses [see **SS1**]. Also, there is no evidence of some extreme temperature dependence of $G_{eff}$ (excluding "pattern effects") over the range of several °C, i.e. perturbation theory applies [see **G.2**].

**5)** Equation e.2 holds to good approximation between 1980 and 2020 for most regions.

**6)** Independent periodic (and generally unacknowledged) Forcing components exist with the periods of 21.2, 9.3, 15.2, and 10.6 years. This is based on temperature frequency analysis spanning a hundred years, and are probably due to the physical effects of the solar magnetic cycle, the solar cycle, lunar precession cycle, etc. . These components can be artificially removed from the temperature record using a technique called PIR (i.e. Periodic Interference Removal). This is accomplished using temperature data from 1948 through 2007.

**7)** By using PIR (preceded by heavy filtering operators on all time dependant variables) an extraordinary underlying linearity of the Ocean and Land temperature response between 1980 and 2003 is revealed (see Fig.4.3,6.3). The linearity extends to 2020 except for a clearly observed transient Ocean Temperature response deviation. This Temperature "dip" (and unknown system fluctuation) can then be removed from the underlying linear trend using linear interpolation/extrapolation. This deviation is likely due to an unknown temperature independent transient Ocean surface energy transport. The validity of the PIR method is proven.

**8)** The energy balance method is applied to evaluation ranges between 1980 and 2020. This avoids the large and very uncertain aerosol cooling forcing component occurring between 1945 and 1970. PIR further reduces the interference errors in the temperature record, corresponding to forcings which are unknown in any case. And the heavy filtering greatly linearizes or regularizes the $\Delta T(t)$, $\Delta F(t)$ and $\Delta E^*(t)$ functions, thereby reducing uncertainty and noise over that range. Thus the evaluation of $G_{eff}$ is made more certain using the most reliable recent Global data (e.g. 1970-2021), and also avoids assuming conditions from the 1800's[1] and after.

**9)** Volcanic Forcing has virtually no effect on the quantity "$\Delta F$-$\Delta Q$" over Lands or certainly Oceans, nor on $\Delta T(t)$. This realization further simplifies and enhances the accuracy of the total Forcing evaluation, and $G_{eff}$ and ECS calculation. And presumption of zero Volcanic Forcing yields a "*maximal*" estimate of ECS (i.e. a larger ECS value).

**10)** The observed values $Q(t) \equiv d[E^*(t)]/dt$ were considered to be too error sensitive and unreliable (especially before 1990) to provide any reasonable evaluation of $G_{eff}$ or ECS using the NOAA data directly. As an alternative, a formal/empirical method is employed to determine Q(t). The *form* of Q(t) [see **SS4**] was derived to be approximately $Q(t)_{total} = C^*_{eff}(t) \cdot d[\Delta \mathbf{T(t)_A}]/dt \ + \ CC^* \cdot S \cdot (2/\sqrt{\pi}) \cdot \sqrt{[t]} - S \cdot t \cdot \mathbf{g_u}/2$ for a ramp surface temperature of slope "S" starting at t = 0; where appropriate values of $C^*_{eff}(t)$ and CC* are determined using Navel Research Laboratory mix depth, NOAA E*(t) and T(t) data, and using a minimization technique [see **SS5**]. An independent robust alternative *maximal* evaluation of $C^*_{eff}$ and CC* is thoroughly demonstrated using Global Ocean Temperature profiles with depth, yielding very similar results! These methods allow a much more accurate evaluation of the sensitive time derivative terms than by direct observation, *providing* the presumed theoretical form is correct. The values of $\Delta Q$ and ECS using the Cheng[13] reanalyzed values of $E^*(t)_{total}$ are provided as an updated and orthodox alternative for comparison, yielding equal or smaller values!

**11)** Using the improvements outlined in 2),6),7),8),9),and 10) above, unknown and unaccounted forcings (and responses) are removed from the data. The results are particularly regularized (linearized) and reliable. $G_{eff}$ is then calculated (presume $\mathbf{g_u}$=0.41) using METHOD A: $G_{eff} = S[\Delta F(t')-\Delta Q(t')] / S[\mathbf{\Delta T_A}(t')]$; where the operator "**S[ ]**" returns the best fit linear slope "S" over 1980-2020 . Method B is more general, but returns nearly identical results. Then $ECS_{eff} \equiv \Delta F_{2xCO2} / G_{eff}$ . The average $\mathbf{ECS_{eff}}$ for the total Globe, global Oceans, and the global Lands are ≤ **2.15, 1.67, 2.96** °C/2x$CO_2$ respectively, using METHOD B, *VS=variable, and* $\Delta F/(2xCO_2)=3.7\ W/m^2$. [Similarly, for all these same regions from 60N-60S, the values are about 5% smaller.] The Global value at true equilibrium then estimated to be $ECS_{true}$ ≤ **2.09** °C/2x$CO_2$ (using METHOD B, VS=.017) is 70% of the IPCC AR6 ECS[15] estimate of **3.0** °C/2x$CO_2$, similar to the original "influential" evaluation of Otto[16] et al. (**2.0** °C, including all corrections), but 126% of the $ECS_{eff}$ value reported by Lewis[1] (**1.66** °C) . The average $ECS_{Global}$ of 36 CMIP6 GCM models[26] is ECS=3.9 °C (1.83-5.67). However, if the models with ECS ≥ 3 °C are omitted as unrealistic, then the average becomes 2.51 °C . This is only about 2.51/2.09=1.20 larger than the ECS evaluated herein.

Further, the IPCC AR6 ECS[15] estimate of ECS=**3.0** °C/2x$CO_2$ can be improved (see Lewis[27]) to a more appropriate value of **2.16 °C**, which is very close to the value **2.09 °C** calculated herein.

The evaluation here of $G_{eff}(t)$ and ECS does not include the very long lag rein**forcing** effects of local ice sheet melting, but this will not much effect the ECS of 50N-50S "warm" regions in any case. The upper limit ECS 1σ uncertainty is estimated to be only about +10%, much less than all orthodox estimates(see section **E.2**).

**12)** The TCR/ECS ratio over the oceans is estimated to be **0.71** as derived in **SS8** . If TCR/ECS over Land is assumed to be unity, the *global* average TCR/ECS is estimated in **SS8** as **0.83** . The average ECS for Land is a factor of 2.96/1.67=**1.77** greater than the Ocean average ECS value [Using HADCRUT T(t) data this becomes 1.77·(.94/1.08)=**1.54**] This, almost certainly, is mostly due to an enhanced positive cloud reflection "rein**forcing**" (or cloud "feedback") that occurs over land. This is also partially, but much less, due to *enhanced* "lift" of latent and sensible water vapor energy to higher altitudes over the ocean surface (e.g. lapse rate variation and ocean evaporation)...see **SS6**, **SS7.2** .

**13)** For comparison with orthodox methods, a more orthodox application of the Energy Budget method is employed using the NOAA Global temperature data, the reanalyzed E*(t) data of Cheng[13] and the updated IPCC AR5 ERF[1] data *excluding* the Volcanic component (which is a *maximal* ECS choice as discussed previously). Only H5[ ] filtering of all yearly averaged time variable raw data was applied to e.5 ... no PIR or other processing was done. Then, the orthodox evaluation of $ECS_{eff}$ = **1.90** °C is 12% *smaller* than the value of **2.15** °C using METHOD B (which is the preferred evaluation). This value (1.90) is identical to a recent orthodox energy budget evaluation by Spencer[28] (**1.9** °C) using the Cheng E*(t) data and similar evaluation range and data sources (plus the Volcanic Forcing term).

**14)** Uncertainties (often speculative or subjective) may be determined separately, based on the readers own preferences. Then, using derived uncertainty ratios of section **E.2**, the values of $G_{eff}$ and ECS can be modified by readers as they see fit. In particular the values of long term independent forcing ΔF(t) is questionable, as well as the "pattern effect"...see section **G**. The true value of ECS could significantly differ from the evaluations herein, depending on those choices. There are some questions. But the ECS and $G_{eff}$ reported herein are the *maximal* ECS most likely values, based on the most orthodox/canonical ΔF estimates.

**15)** The so called "pattern effect" likely results in a barely significant variation of $G_{eff}(t)$ at equilibrium. GCM modeled "pattern effects" are likely a small ~2.5% (up to 7% 1σ variation) maximum warming as stated in section **G.2**. These GCM modeled "pattern effects" must be considered purely theoretical, unverified and uncertain, in any case. In fact these GCMs have proven significant flaws.

However alternatively, $G_{eff}(t)$ is calculated in **SS7.2** using a simplified 3-layer 2-region(Ocean, Land) formal/empirical algebraic Coarse GCM perturbation model developed in **SS6** . Those results indicate the equilibrium values for $G_{eff.Land}$ , and $G_{eff.Global}$ are likely *greater* than the near linear steady state evaluations using METHOD A or B. Although the $G_{eff.Ocean}$ value could reasonably be 2.5% less (for **CS**=0.9, $k_{mO}$=0). In particular $G_{eff.Land}$ may be more than 5% greater, which might also be applied to Tab.1,2 $ECS_{Land}$ *reductions*. However, $G_{eff.equilibrium} = G_{eff\,(1980-2020)}$, and $ECS_{true}=ECS_{eff(1980-2020)}$ is the *maximal* ECS choice used herein.

The Coarse GCM derived above (where scalar **ε=τ = 0.85)** duplicates the *individual* orthodox IPCC AR6[15]: 1) *Global* Planck (surface) effective radiation "feedback", 2) *Global* water vapor "feedback", 3) *Global* cloud reflection "feedback", 4) *Global* $\Delta F_{2xCO2}$ forcing, 5) and the observed Surface ***and*** Bulk atmosphere temperature trends for both Land ***and*** Ocean! **It is therefore deemed physically realistic and superior to most alternative GCMs.**

**16)** There have been recent speculations that $\Delta CH_4$ (a strong GHG) concentration is naturally proportional to ΔT, as a positive rein**forcing** mechanism. This would decrease the *independent* historical ΔCH4 Forcing and increase the ECS calculation. This is one proposed ECS "enhancing" speculation that *might* actually be physically realistic, and must be resolved independently. Also, speculations by Soon[29] et al. propose modifications to the NOAA land temperature measurements and additional solar forcing variations. However, the land temperature trend proposed therein over the evaluation period (1980-2020) are virtually identical with the NOAA values. And the solar forcing variations proposed are insignificant compared to the direct anthropogenic ΔRF over that period. Therefore even if these speculations were correct, the evaluated values of ECS herein would remain unchanged!

ECS results using HADCRUT temperature data instead are similar, but 6% smaller over land, and 8% larger over oceans. If there **are** *unknown* natural positive forcing trend components, the ECS evaluation could be notably smaller, and this evidence is compelling[24] (see sect. **G.3**). The ECS evaluations herein are considered *maximal*.

..................................................................................................................................................................

**17) Conclusion**...Societal Implications:

Global warming is only a *serious* hardship in Land regions that are already uncomfortably hot or hot *and* humid consistently. Using *maximal* estimations of ECS (see Tab.1,2) it was determined that Temperature increases in Land regions having yearly average temperatures over 80 °F would be on the order of ≤ **2.2 to 2.33 °C** (~4.0 °F) at equilibrium after 140 years of $CO_2$ emissions at the current rate of $CO_2$ increase (3ppm/year). This excludes the Arabian desert regions, which warm more. This *is* of **concern** over century long time frames, but certainly no immediately alarming or catastrophic or existential threat; especially when considering how energy production technologies will likely improve over 140 years. And expected warming over Oceans is *not* extreme.

IPCC and NASA reports also indicate no change in extreme weather events over the last 50 years, outside of natural variation, which should be expected to continue, and not change suddenly. This excludes heat wave maxima, which *should* increase corresponding to the general global warming. However, 9 times as many persons die of extreme cold exposure than extreme heat exposure, so this concern is likely moot in any case. Similarly, tide gauge measurements on geologically stable coastlines (or using geologically compensated data) with 100 year records indicate moderate sea-level rise (≤ 2 mm/year) and no significant long term acceleration outside of natural variation over the last 50 years of $CO_2$ and temperature increase, which should be expected to continue, and not change suddenly. All of these situations will of course be monitored for change into the future. The next 20 years will prove much, at which time reasoned policy changes could be made if and when "alarming" accelerations do actually evolve.

It is particularly ironic, that while the so called "green" movement claims the earth is on a fast path to calamity, the earth is now much more literally green (i.e. increased flora as seen from orbit) than over the last 50 years, due to $CO_2$ fertilization and warming. And one must then also balance the benefits of carbon based fuels for human societal happiness, and prosperity, which are almost beyond our ability to appreciate.

Therefore, achieving a "NetZero" carbon emission agenda within several decades is not only operationally unobtainable, impoverishing and debilitating, but unnecessary. A "NetConstant" or "NetReduction" global carbon emission agenda would be a more realistic and effective approach over many decades, with a "NetZero" in global $CO_2$ emissions realized on the order of centuries, concurrent with realistic (and affordable) advances in technology and climatological forecasting.

---

This Monograph, Sub-Sections SS1 through SS9, additional figures, references and papers, referenced working spread sheets, and all data are available at the following publicly accessible repositories:

1) https://drive.google.com/folderview?id=1-fqvvDYj2Kr4WrHKYXN_1e7Yf7nNPAwd

---

Acknowledgements:

Thanks to Max Reason for providing the XBASIC compiler public and free.
Thanks to Nic Lewis for advising the importance of addressing the "pattern effect" issue.
Thanks to the University Of Chicago for providing the MODTRAN analysis application public and free.
Thanks to all referenced authors and organizations for access to their work, figures, and data.
Thanks to Microsoft for developing their indispensable Excel application.

---

Meteorology/Climatology specialists who find this monograph "Worthy of Consideration" might contact the Author as references. It would be appreciated. And please do share with any interested parties, Thank You.

---

No Conflicts of Interest are hereby declared. 

---

[1] N. Lewis & J. Curry, "The Impact of Recent Forcing and Ocean Heat Uptake Data on Estimates of Climate Sensitivity", Journal of Climate, Aug. 2018

[2] P. M. Forster, "Inference of Climate Sensitivity from Earth's Energy Budget", Annu. Rev. Earth Planet. Sci., June 2016

[3] S. Sherwood, "An Assessment of Earth's Climate Sensitivity Using Multiple Lines of Evidence", Review of Geophysics, July 2020

[4] C. D. Keeling & T. Whorf, "Possible Forcing of Global Temperature by the Oceanic Tides", Proc. Natl. Acad. Sci. USA, Aug. 1997

[5] B. Copeland & A. Watts, "Evidence of a Lunisolar Influence on Decadal AND Bidecadal Oscillations in Globally Averaged Temperature Trends", wattsupwiththat.com, May 2009

[6] Ian R. Wilson, "Lunar Tides and the Long-Term Variation of the Peak Latitude Anomaly of the Summer Sub-Tropical High Pressure Ridge over Eastern Australia", The Open Atmospheric Science Journal, 2012

[7] IPCC WG1AR5 Chapter 8 - Anthropogenic and natural radiative forcing, page 667, "...we assess the ratio ERF/RF =1..."

[8] Schmidt et al., "Volcanic Radiative Forcing From 1979 to 2015", J. of Geophysical Research, Nov. 2018,Fig2,Table1

[9] Gregory et al., "Small global mean cooling due to volcanic radiative forcing", Climate Dynamics, Mar. 2016, Fig1a

[10] S. J. Smith et al, "Historical Sulfur Dioxide Emissions1850-2000: Methods and Results", PNNL-14537, Jan. 2004

[11] https://tamino.wordpress.com/2010/08/23/antrhopogenic-global-cooling/
https://www.ncdc.noaa.gov/paleo/icecore/greenland/summit/document/gispdata.htm

[12] https://www.ncei.noaa.gov/access/global-ocean-heat-content/index.html

[13] L. Cheng et al, "Improved estimates of ocean heat content from 1960 to 2015", Science Advances, Mar. 2017

[14] Levitus et al , "World ocean heat content and thermosteric sea level change (0–2000 m), 1955–2010", Geophysical Research Letters, May 2012

[15] IPCC_AR6_WGI_TS.pdf, TS.3.2 : including Fig TS.17,16

[16] A. Otto et al , "Energy Budget Restraints on Climate Response", Nature Geoscience 6, 415–416 (2013)

[17] https://www.ncei.noaa.gov/access/monitoring/climate-at-a-glance/global/time-series/globe/land_ocean/ann/12/

[18] R. McKitrick & J. Christy, "Pervasive Warming Bias in CMIP6 Tropospheric Layers", Earth and Space Science – Research Letter, 07/2020

[19] https://www.ncei.noaa.gov/pub/data/ratpac/ratpac-a/

[20] D. Irving, et al, "A Mass and Energy Conservation Analysis of Drift in the CMIP6 Ensemble", Journal of Climate, 4/2021

[21] N. Lewis, https://nicholaslewis.org/committed-warming-and-the-pattern-effect/ , commentary, 1/2021

[22] N. Lewis & T. Mauritsen, "Negligible Unforced Historical Pattern Effect on Climate Feedback Strength Found in HadISST-Based AMIP Simulations", Journal of Climate, 1/2021

[23] M. Rugenstein et al, "Equilibrium Climate Sensitivity Estimated by Equilibrating Climate Models", Geophysical Research Letters,11/2019

[24] Shaopeng Huang et al, "Temperature trends over the past five centuries reconstructed from borehole temperatures", NATURE, vol. 403, 17 Feb. 2000

[25] Ka-Kit Tung et al, "Using data to attribute episodes of warming and cooling in instrumental records", PNAS (Early Edition), Jan., 2013

[26] N. Scafetta, "CMIP6 Ensemble Members vs. Global Surface Temperatures", Climate Dynamics, Sept. 2022: Figure 8

[27] N. Lewis, "Objectively combining climate sensitivity evidence", Climate Dynamics, Sept 2022

[28] R. Spencer & J. Christy, "Effective climate sensitivity distributions from a 1D model of global ocean and land temperature trends, 1970–2021", Theoretical and Applied Climatology Aug. 2023

[29] W. Soon et al, "The Detection and Attribution of Northern Hemisphere Land Surface Warming (1850-2018) in Terms of Human and Natural Factors: Challenges of Inadequate Data", Climate (MDPI), Aug. 2023

**Sub-Sections**

## SS1 Thermal $G_{eff}$ Theory v.4

Michael D. Mill...Mar. 2024 [Contact: m.d.mill.climate@gmail.com]

For a given contiguous region "A" on the globe the instantaneous total *temperature independent* power flow "in" must equal the rate of total Energy Storage (i.e. regional power storage) plus the rate of total power flow "out" at any and all times "t", as is required by the law of energy conservation, i.e.

es1.11 $\int^A F(x,t)\cdot da = \int^A [d(E^*(x,t))/dt]\cdot da + \int^A I(x,t)\cdot da$

where:

1) $\int^A f(x)\cdot da$ indicates the areal integration of f(x) over the region "A", "*d*a" is the incremental area, "x" represents the enumeration of any and every location in the area "A" (i.e. "x" is **not** a one dimensional spatial variable), A≡ the area value of region "A", and "t" is the time variable
2) F(x,t) ≡temperature independent power flow "in" per unit area ="Forcing" (= $W/m^2$); Note, Forcing can be defined for an incremental area or large area equally.
3) E*(x,t)≡Energy storage per unit area (= $W\cdot years/m^2$); Note, E* can be defined for an incremental area or large area equally.
4) I(x,t) ≡Power "radiated out" per unit area vertically to space ***plus*** power transferred "out" horizontally via mass flow, mixing or conduction to adjacent areas (but *NOT* the ocean depths) (= $W/m^2$). Note, I can be defined for an incremental area or large area equally.

But let

es1.12
$F(x,t) \equiv F(x,t_o) + \Delta F(x,(t-t_o))$
$I(x,t) \equiv I(x,t_o) + \Delta I(x,(t-t_o))$
$E^*(x,t) \equiv E^*(x,t_o) + \Delta E^*(x,(t-t_o))$
$t' \equiv (t- t_o)$
$\Delta F(x, (t-t_o)) = \Delta F(x,t') \equiv F(x,t)-F(x,t_o)$
$\Delta I(x, (t-t_o)) = \Delta I(x,t') \equiv I(x,t)-I(x,t_o)$
$\Delta E^*(x, (t-t_o)) = \Delta E^*(x,t') \equiv E^*(x,t)-E^*(x,t_o)$
$\Delta T(x, (t-t_o)) = \Delta T(x,t') \equiv T(x,t)-T(x,t_o)$

Note, $\Delta F=\Delta I=\Delta E^*=\Delta T=0$ at $t=t_o$ or $t'=0$ .

Then rewrite es1.11 as

es1.13 $\int^A [F(x,t_o)+\Delta F(x,t')]\cdot da = \int^A [d( E^*(x,t_o) + \Delta E^*(x,t') )/dt']\cdot da + \int^A [I(x,t_o)+\Delta I(x,t')]\cdot da$

Note above , $E^*(x,t_o)$ is not variable in time, and so $d(E^*(x,t_o))/dt' = 0$ .

But since es1.11 is true for $t=t_o$ ,we can subtract es1.11 at $t=t_o$ from es1.13 yielding

es1.21 $\int^{A} \Delta F(x,t')\cdot da = \int^{A} \{d(\Delta E^*(x,t'))/dt' - [d(\Delta E^*(x,t'))/dt']_{at\ t'=0}\}\cdot da + \int^{A} \Delta I(x,t)\cdot da$

Now divide es1.21 by area A yielding

es1.22 $\mathbf{\Delta F}(t') = \{d(\mathbf{\Delta E^*}(t'))/dt' - [d(\mathbf{\Delta E^*}(t'))/dt']_{at\ t'=0}\} + \mathbf{\Delta I}(t')$

;where the **bold** delta terms indicate areal averages over region "A" only, which are functions of time only.

We can and will make the assertion that for perturbations the $\Delta I(x,t)$ term can be synthesized as a very general function of global temperature $\Delta T(x,t)$ as follows:

es1.31 $\Delta I(x,t) = \int^{Globe} G(x,y)\cdot \Delta T(y,t)\cdot da_y$

;where "y" is an enumeration of global location that is independent from "x", and the areal integration is over the entire globe. In other words $\Delta I(x)$ is a function of the $\Delta T$ at not only "x", but also at other locations due to fluid flow, mixing, conduction, or other coupling mechanisms.

Also require that $\Delta T(y,t)$ is of the following form over the entire globe:

es1.32 $\Delta T(y,t) = \Delta T_{\alpha}(y)\cdot \Delta T_{\beta}(t)$ thus $\mathbf{\Delta T}(t) = \mathbf{\Delta T_{\alpha}}\cdot \Delta T_{\beta}(t)$ and $\Delta T_{\beta}(t) = [\mathbf{\Delta T}(t)/\mathbf{\Delta T_{\alpha}}]$

;where the **bold** delta terms indicate areal averages over region "A" only, and es1.32 has been so averaged. In other words, the relative proportional value of temperatures across the global does not change over time. Now integrate es1.31 over area "A", divide by A, and using es1.32 will yield:

es1.33 $(1/A)\cdot\int^{A} \Delta I(x,t)\cdot da_x = (1/A)\cdot\int^{A}\int^{Globe} G(x,y)\cdot \Delta T_{\alpha}(y)\cdot \Delta T_{\beta}(t)\cdot da_y\cdot da_x$

$\mathbf{\Delta I}(t) = (1/A)\cdot\int^{A}\int^{Globe} G(x,y)\cdot \Delta T_{\alpha}(y)\cdot[\mathbf{\Delta T}(t)/\mathbf{\Delta T_{\alpha}}]\cdot da_y\cdot da_x$

$\mathbf{\Delta I}(t) = \mathbf{\Delta T}(t)\cdot(1/A)\cdot(1/\mathbf{\Delta T_{\alpha}})\cdot\int^{A}\int^{Globe} G(x,y)\cdot \Delta T_{\alpha}(y)\cdot da_y\cdot da_x$

es1.34 $\mathbf{\Delta I}(t) = \mathbf{\Delta T}(t)\cdot G_{eff.A}$ ;where

es1.35 $G_{eff.A} \equiv (1/A)\cdot(1/\mathbf{\Delta T_{\alpha}})\cdot\int^{A}\int^{Globe} G(x,y)\cdot \Delta T_{\alpha}(y)\cdot da_y\cdot da_x$ .

Note $G_{eff.A}$ is a true constant <u>independent</u> of time, under empirically verifiable conditions of es1.32 . Thus $G_{eff}$ is definable for an area "A" which is any contiguous subsection of the globe, or the entire globe. Now rewrite es1.22 using es1.34

es1.36 $\mathbf{\Delta T}(t')\cdot G_{eff.A} = \mathbf{\Delta F}(t') - \{d(\mathbf{\Delta E^*}(t'))/dt' - [d(\mathbf{\Delta E^*}(t'))/dt']_{at\ t'=0}\}$ or

es1.37 $\mathbf{\Delta T}(t')\cdot G_{eff.A} = \mathbf{\Delta F}(t') - \{\mathbf{Q}(t') - \mathbf{Q}(t')_{at\ t'=0}\}$

$\mathbf{\Delta T}(t')\cdot G_{eff.A} = \mathbf{\Delta F}(t') - \mathbf{\Delta Q}(t')$ ;where

es1.38 $\mathbf{Q}(t') \equiv d(\mathbf{\Delta E^*}(t'))/dt'$ and $\mathbf{\Delta Q}(t') \equiv \mathbf{Q}(t') - \mathbf{Q}(t'=0)$

<u>Note that all the effects of horizontal thermal power flow (i.e. mixing), water vapor reinforcing, cloud reflection reinforcing, etc. (due to local or adjacent heating) is all contained within the single constant $G_{eff.A}$ term</u>! All these effects are proportional to $\mathbf{\Delta T}(t)$ only because es1.32 holds. Thus Energy budget methods may be applied to Land or Ocean subsections as well as the entire Globe during certain time periods where es1.32 holds to good approximation.

In **SS2** it is shown that all these linear power equations and results will hold for linearly time averaged and time filtered variables as well as for the "instantaneous" values. Thus $\mathbf{\Delta F}(t')$, $\mathbf{\Delta Q}(t')$, and $\mathbf{\Delta T}(t')$ in **bold** can be considered to <u>be *both* area averaged and time averaged or filtered variables</u>, provided the time averaging is applied identically to all the time dependant variables.

**SS2 Filter/Operator theory v.4** Michael D. Mill...Mar. 2024 [Contact: m.d.mill.climate@gmail.com]

Let $H_x[f(t)]$ be the following functional operator over t:

es2.1 $H_x[f(t)] \equiv f(t+\Delta\cdot x)$ .

Note then also,

es2.1b $H_x[a\cdot f(t)] = a\cdot f(t+\Delta\cdot x) = a\cdot H_x[f(t)]$ .

And specifically let $H^1[\ ]$ be the simple 3 point moving average operator

es2.2 $H^1[f(t)] \equiv (\ f(t-\Delta) + f(t) + f(t+\Delta)\ )/3$ , and

es2.3 $H^4[f(t)] \equiv H^1[H^1[H^1[H^1[f(t)]\ ]\ ]\ ]$ ,

which is the simple 3 point moving operator applied 4 times in succession.

And let $H^5[f(t)]$ be <u>defined</u> as the $H^4[\ ]$ operator followed by a simple 5 point moving average. Similarly, let $H^9[f(t)]$ be <u>defined</u> as the $H^4[\ ]$ operator followed by a simple 9 point moving average. Then note that :

es2.4
$$\begin{aligned} H^1[H^1[f(t)]] = \{ \quad & f(t+\Delta+\Delta)+f(t+\Delta)+f(t-\Delta+\Delta) \\ & + f(t+\Delta)+f(t)+f(t-\Delta) \\ & + f(t+\Delta-\Delta)+f(t-\Delta)+f(t-\Delta-\Delta) \quad \}/9 \ . \end{aligned}$$

Thus it may be evident that all these operators $H^n[\ ]$ are simply a scaled sum of $H_x[\ ]$ operators, i.e.

es2.5 $H^n[\ ] = \sum_x \alpha_x\cdot H_x[\ ]$ .

Then it is directly proven:

es2.6

a)
$$\begin{aligned} H^n[a\cdot f_1(t)+b\cdot f_2(t)] = \sum_x \{\alpha_x\cdot H_x[a\cdot f_1(t)+b\cdot f_2(t)]\} &= \sum_x \{\alpha_x\cdot[\ a\cdot H_x[f_1(t)] + b\cdot H_x[f_2(t)]\ ]\} \\ &= a\cdot\sum_x \{\alpha_x\cdot H_x[f_1(t)]\}+ b\cdot\sum_x \{\alpha_x\cdot H_x[f_2(t)]\} \\ &= a\cdot H^n[f_1(t)] + b\cdot H^n[f_2(t)] \qquad \text{, or} \end{aligned}$$
$$H^n[a\cdot f_1(t)+b\cdot f_2(t)] = a\cdot H^n[f_1(t)] + b\cdot H^n[f_2(t)]$$

b)
$$\begin{aligned} H^n[df(t)/dt] = \sum_x \alpha_x\cdot H_x[df(t)/dt] = \sum_x \alpha_x\cdot[df(t)/dt]_{t\to t+\Delta\cdot x} = \sum_x \alpha_x\cdot[df(t+\Delta\cdot x)/dt] &= \sum_x \alpha_x\cdot[d(H_x[f(t)])/dt] \\ &= d\{\sum_x \alpha_x\cdot H_x[f(t)]\}/dt \\ &= d\{H^n[f(t)]\}/dt \qquad \text{, or} \end{aligned}$$
$$H^n[df(t)/dt] = d\{H^n[f(t)]\}/dt$$

c)
$$\begin{aligned} H^n[\int f(t)\cdot dt] = \sum_x \alpha_x\cdot H_x[\int f(t)\cdot dt] = \sum_x \alpha_x\cdot[\int f(t)\cdot dt]_{t\to t+\Delta\cdot x} = \sum_x \alpha_x\cdot[\int f(t+\Delta\cdot x)\cdot dt] &= \sum_x \alpha_x\cdot[\int H_x[f(t)]\cdot dt] \\ &= \int\{\sum_x \alpha_x\cdot H_x[f(t)]\}\cdot dt \\ &= \int H^n[f(t)]\cdot dt \qquad \text{, or} \end{aligned}$$
$$H^n[\int f(t)\cdot dt] = \int H^n[f(t)]\cdot dt \ .$$

Now we can apply $H^n[\ ]$ to both sides of es1.11 in **SS1**, and using es2.6a,b yields:

es2.7 $H^n[\int^A F(x,t)\cdot da] = H^n\{\int^A [d(E^*(x,t))/dt]\cdot da\} + H^n[\int^A I(x,t)\cdot da]$ ,or
$\int^A H^n[F(x,t)]\cdot da = \int^A [d(H^n[E^*(x,t)])/dt]\cdot da + \int^A H^n[I(x,t)]\cdot da$

Thus, the "instantaneous" time dependant variables of es1.11 can be replaced with their time filtered forms, and that equation and *all* following results must hold true. More specifically application of $H^n[\ ]$ to both sides of es1.22,36,37,38 , and using es2.6a,b yields respectively:

es2.8 $H^n[\Delta\mathbf{F}(t')] =\{ H^n[d(\Delta\mathbf{E}^*(t'))/dt'] - H^n[d(\Delta\mathbf{E}^*(t'))/dt']_{at\ t'=0} \} + H^n[\Delta\mathbf{I}(t')]$
$=\{ d(H^n[\Delta\mathbf{E}^*(t')])/dt' - [d(H^n[\Delta\mathbf{E}^*(t')])/dt']_{at\ t'=0} \} + H^n[\Delta\mathbf{I}(t')]$

es2.9 $H^n[\mathbf{\Delta T}(t')]\cdot G_{eff.A} = H^n[\mathbf{\Delta F}(t')] - \{ H^n[d(\Delta\mathbf{E}^*(t'))/dt'] - H^n[d(\Delta\mathbf{E}^*(t'))/dt']_{at\ t'=0} \}$
$= H^n[\mathbf{\Delta F}(t')] - \{ d(H^n[\Delta\mathbf{E}^*(t')])/dt' - [d(H^n[\Delta\mathbf{E}^*(t')])/dt']_{at\ t'=0} \}$

es2.10 $H^n[\mathbf{\Delta T}(t')]\cdot G_{eff.A} = H^n[\mathbf{\Delta F}(t')] - \{ H^n[\mathbf{Q}(t')] - H^n[\mathbf{Q}(t')]_{at\ t'=0} \}$
$= H^n[\mathbf{\Delta F}(t')] - \mathbf{\Delta}\{H^n[\mathbf{Q}(t')]\}$ ;where

es2.11 $\mathbf{Q}(t') \equiv d(\mathbf{\Delta E}^*(t'))/dt$ , and so
$H^n[\mathbf{Q}(t')] = H^n[d(\mathbf{\Delta E}^*(t'))/dt]$
$H^n[\mathbf{Q}(t')] = d(H^n[\mathbf{\Delta E}^*(t')])/dt'$ ;and where

es2.11b $\mathbf{\Delta\{}H^n[\mathbf{Q}(t')]\} \equiv \{ H^n[\mathbf{Q}(t')] - H^n[\mathbf{Q}(t')]_{at\ t'=0} \}$

;and where the **bold** delta terms indicate areal averages over region "A" only.

Now, if it is understood that all time variables are both area averaged over area "A" *and* time filtered in some specific way then es2.8,9,10,11,11b can be written more simply as, respectively:

es2.22 $\Delta F(t') =\{d(\Delta E^*(t'))/dt' - [d(\Delta E^*(t'))/dt']_{at\ t'=0} \} + \Delta I(t')$

es2.36 $\Delta T(t')\cdot G_{eff.A} = \Delta F(t') - \{d(\Delta E^*(t'))/dt' - [d(\Delta E^*(t'))/dt']_{at\ t'=0} \}$ or

es2.37 $\Delta T(t')\cdot G_{eff.A} = \Delta F(t') - \{Q(t') - Q(t')_{at\ t'=0} \}$
$= \Delta F(t') - \Delta Q(t')$ ;where

es2.38 $Q(t') \equiv d(\Delta E^*(t'))/dt'$ ;and where

es2.38b $\Delta Q(t') \equiv \{Q(t') - Q(t')_{at\ t'=0} \}$ .

Note, this is simply the area and time filtered versions of es1.22,36,37,38 .

Using time filtered variables will reduce high frequency variability and uncertainty, and may reveal underlying linear dependencies.

Applying these operators as defined may not be possible near the end points. At these points the operator is linearly defined so as to simply extend the function slope linearly through the end point. This is a small liberty taken to maximize the available data, but as it happens is a very good approximation using these variable data sets. Regardless, the endpoint altered operators are still linear, and all results above hold true exactly, in any case!

## SS3 Mix Layer Thermal Capacitance Theory v.4

### SS3.1 Ramp Forcing Case

The surface mix layer of the oceans is a region wherein the temperature of the entire surface layer is the same as the surface due to wave and current mixing mechanisms. The equalization of the layer occurs on the order of a few months or less, and therefore can be considered as instantaneous for yearly averages or greater. It can therefore be considered as a "bulk" thermal capacitance.

The depth of this layer varies with Global position , and is given by "$d_{mix}(x)$", where "x" represents the enumeration of any and every location in the area "A" (i.e. "x" is **not** a one dimensional spatial variable). Let ΔE*(x) be the Δenergy of the ocean mix layer per unit surface area at "x" (see **SS1** and es1.11 definitions). The thermal capacitance per unit *area* (C*) for sea water is then calculated using the specific heat of sea water :

es3.1 $C^* \equiv [\Delta E^*(x)] / \Delta T(x) = [d_{mix}(x)\cdot\Delta T(x)\cdot 3950\cdot(kW\cdot sec/(m^3\cdot°C))\cdot(year/(3.15\cdot 10^7\cdot sec)) ]/\Delta T(x)$

es3.2 $C^*(x)=d_{mix}(x)\cdot(\sim 0.13)\cdot(W\cdot years/(°C\cdot m^3))$; where $d_{mix}$ is in meters .

Not only does this depth vary with Global position , but also with time during the seasonal cycle. Yearly averaged data is used in this analysis, so what "average" depth is appropriate? For purposes herein, the yearly maximum depth is the appropriate yearly value. The depth reaches a maximum in winter, and this *total* volume reaches the surface temperature $T(t_{winter1})$. As the mix depth decreases to its summer depth and temperature, the cooler deeper water remains in place (does not rise), and the thermal energy therein remains constant in place also. As the seasonal cycle continues to the next winter, the temperature of the maximum depth returns to $T(t_{winter2})$. *The additional warming and cooling energy flow of the normal yearly seasonal cycle adds and averages to zero over the year*. The total long term average change in water thermal energy from winter1 to winter2 is then given by (using es.1,2:

es3.3 $\Delta E^*=d_{mix.maximum}\cdot(0.13)\cdot(T(t_{winter2}) - T(t_{winter1}))\cdot(W\cdot years/(°C\cdot m^3))$, or for long time averages simply

es3.4 $\Delta E^*(x,t)=C^*(x)\cdot\Delta T(x,t)$ ;<u>where now</u>

es3.5 $C^*(x)=d_{mix.max}(x)\cdot(0.13)\cdot(W\cdot years/(°C\cdot m^3))$ .

Taking the time derivative of es3.4, and using e.4d, yields (for the mix layer only)

es3.6 $Q_{mix.layer}(x,t) \equiv d(\Delta E^*(x,t))/dt = C^*(x)\cdot d(\Delta T(x,t))/dt$ .

Similarly, using es4.31 for the deep ocean energy (see **SS4**)

es3.7 $Q(t)_{deep.ocean} = CC^*\cdot S\cdot(2/\sqrt{\pi})\cdot\sqrt{[t-t_r]} - (½)\cdot g_{u.ave}\cdot S\cdot(t-t_r)$ , or
$= CC^*\cdot S\cdot(2/\sqrt{\pi})\cdot\sqrt{[t]} - (½)\cdot g_{u.ave}\cdot\Delta T(t)$ ,

where $t_r=0$, S is the surface temperature slope *presumed* roughly invariant over time t > 0, and CC* is *approximated* as spatially independent over the entire large ocean region "A". Then let

es3.8 $Q(x,t) \equiv Q(x,t)_{total} = Q_{mix.layer}(x,t) + Q(t)_{deep.ocean}$ .

The form of es1.37 or es2.37 for small local areas, positionally denoted by "x", is

es3.9 $\Delta T(x,t)\cdot G_{eff.A} = \Delta F(x,t) - \{Q(x,t) - Q(x,t)_{at\ t=0}\}$ .

But asserting time derivatives are all zero for $t \le 0$, then es3.9 may be rewritten using es3.6,7,8 as

es3.10 $\Delta T(x,t)\cdot(G_{eff.A} - (½)\cdot g_u) + C^*(x)\cdot d(\Delta T(x,t))/dt + CC^*\cdot S(x)\cdot(2/\sqrt{\pi})\cdot\sqrt{[t]} = \Delta F(x,t)$

If we specify that $\Delta F(x,t)$ is a ramp function of constant slope for $t > 0$, where $\underline{d(\Delta F(x,t))/dt \equiv \Delta F'(x)}$, then

es3.11 $\Delta F(x,t) = t \cdot \Delta F'(x)$ .

And similarly *assume* $\Delta T(x,t)$ is a ramp function of roughly invariant slope for $t > 0$, where $\underline{d(\Delta T(x,t))/dt \equiv \Delta T'(x,t)}$, then

es3.12 $\Delta T(x,t) \approx t \cdot \Delta T'(x,t)$, approximately.

Now es3.10 can be rewritten:

es3.13 $\Delta T'(x,t) = (t\cdot\Delta F'(x)) / [t\cdot(G_{eff.A} - (½)\cdot g_{u.ave}) + C^*(x) + CC^*\cdot(2/\sqrt{\pi})\cdot\sqrt{[t]}\ ]$ .

However, due to the errors of the approximation of es3.12, it is found by numerical trial in **SS8,** that the nearly exact solution for realistic parameters is actually

es3.13b $\Delta T'(x,t) \cong (t\cdot\Delta F'(x)) / [t\cdot(G_{eff.A} - (½)\cdot g_{u.ave}) + C^*(x) + \mathbf{(0.4)}\cdot CC^*\cdot(2/\sqrt{\pi})\cdot\sqrt{[t]}\ ]$ !

Define an "area weighted average" operator as $\mathbf{A[}f(x)\mathbf{]} \equiv (1/A)\cdot\int^A f(x)\cdot da$ [see **es1.11 1)** ]. Then *define* the effective mix layer thermal capacitance (per area) by

es3.14 $C^*_{eff} \cdot \mathbf{A[}\Delta T'(x,t)\mathbf{]} = \mathbf{A[}\ C^*(x) \cdot \Delta T'(x,t)\ \mathbf{]}$ , or

es3.15 $C^*_{eff} = \mathbf{A[}\ C^*(x) \cdot \Delta T'(x,t)\ \mathbf{]} / \mathbf{A[}\Delta T'(x,t)\mathbf{]}$

;where $\Delta T'(x,t)$ is given by es3.13b . It can be determined then for $\Delta F'(x)$ spatially invariant

es3.16 $C^*_{eff}\ \big|_{t\to\infty} = \mathbf{A[}\ C^*(x)\ \mathbf{]}$ and $C^*_{eff}\ \big|_{t\to 0} = 1/(\ \mathbf{A[}\ 1/C^*(x)\ \mathbf{]}\ )$ ,

and the value for any time between is between these 2 limits.

Now apply the area weighted average to es3.6, and using es3.14 yields

es3.17 $\mathbf{A[}\ Q_{mix.layer}(x,t)\ \mathbf{]} = \mathbf{A[}\ C^*(x)\cdot d(\Delta T(x,t))/dt\ \mathbf{]} = C^*_{eff} \cdot \mathbf{A[}\Delta T'(x,t)\mathbf{]} =$ , or

es3.18 $\mathbf{Q_{A.mix.layer}}(t) = C^*_{eff} \cdot \Delta\mathbf{T'_A}(t) = C^*_{eff} \cdot d(\Delta\mathbf{T_A}(t))/dt$

;where $\Delta\mathbf{T_A}$ and $\mathbf{Q_A}$ in **bold** indicates an **Area** weighted average of $\Delta T$ and $\Delta Q$ over area A.

This is used in e.17, and a useful replacement in e.3,4d; where $C^*_{eff}$ is defined in es3.14,15 for the Ramp forcing case. $C^*_{eff}$ (and the area weighted averages) are calculated for various values of t = time in working spreadsheets **Ceff-Calc-ramp.xls**, **Ceff-Calc-ramp-work.xls**, and **Ceff-Calc-ramp-60N-60S.xls** .

## SS3.2 Harmonic Forcing Case

This analysis follows that of SS3.1 up to es3.6,7,9 , where the harmonic phasor form of these (review **SS4**) become (note "j" indicates the imaginary unity):

es3.19 $\Delta Q_{mix.layer}(x) = C^*(x)\cdot j\omega\cdot \Delta T(x)$
es3.19b $\Delta Q_{deep.ocean}(x) = [CC^*\cdot\sqrt{(j\omega)} - (½)\cdot g_u]\cdot\Delta T(x) = [\ CC^*\cdot\sqrt{\omega}/\sqrt{2} + j\cdot CC^*\cdot\sqrt{\omega}/\sqrt{2} - (½)\cdot g_u]\cdot\ \Delta T(x)$
es3.20 $\Delta T(x)\cdot G_{eff.A} = \Delta F(x) - \{\ \Delta T(x)\cdot\ [CC^*\cdot\sqrt{\omega}/\sqrt{2} + j\cdot CC^*\cdot\sqrt{\omega}/\sqrt{2} - (½)\cdot g_u + C^*(x)\cdot j\omega]\}$

;where the harmonic component of a constant is zero. Then rewrite es3.20 as

es3.21 $\Delta T(x) = \Delta F(x) / [X+ jY]$ ;where

es3.22 $X \equiv [G_{eff.A} - (½)\cdot g_u + CC^*\cdot\sqrt{(\omega/2)}]$ and $Y \equiv [CC^*\cdot\sqrt{(\omega/2)} + C^*(x)\cdot\omega]$ .

Rewrite es3.21 as

es3.23 $\Delta T(x) = [\Delta F(x)\cdot X/(X^2+Y^2)] - j[\Delta F(x)\cdot Y/(X^2+Y^2)] = DD-j\cdot EE$ ;where

es3.24 $DD \equiv [\Delta F(x)\cdot X/(X^2+Y^2)]$ and $EE \equiv [\Delta F(x)\cdot Y/(X^2+Y^2)]$ .

Similarly:

es3.25 $C^*(x)\cdot\Delta T(x) = [C^*(x)\cdot\Delta F(x)\cdot X/(X^2+Y^2)] - j[C^*(x)\cdot\Delta F(x)\cdot Y/(X^2+Y^2)] = AA-j\cdot BB$ ;where

es3.26 $AA \equiv [C^*(x)\cdot\Delta F(x)\cdot X/(X^2+Y^2)]$ and $BB \equiv [C^*(x)\cdot\Delta F(x)\cdot Y/(X^2+Y^2)]$ .

Now define $C^*_{eff}$ as in es3.14 :

es3.27 $C^*_{eff} = \mathbf{A[}C^*(x)\cdot\Delta T'(x,t)\ \mathbf{]/A[}\Delta T'(x,t)\mathbf{]} = j\omega\cdot\mathbf{A[}\ C^*(x)\cdot\Delta T(x,t)\ \mathbf{]}/(j\omega\cdot\mathbf{A[}\Delta T(x,t)\mathbf{]})$ .

Or using es3.23 and es3.25 ,

es3.28 $$C^*_{eff} = \frac{\mathbf{A}[AA] - j\cdot\mathbf{A}[BB]}{\mathbf{A}[DD] - j\cdot\mathbf{A}[EE]} = \frac{\{\mathbf{A}[AA]\cdot\ \mathbf{A}[DD] + \mathbf{A}[BB]\cdot\ \mathbf{A}[EE]\}}{\mathbf{A}[DD]^2 + \mathbf{A}[EE]^2} + j\frac{\{\ \mathbf{A}[AA]\cdot\mathbf{A}[EE] - \mathbf{A}[DD]\cdot\ \mathbf{A}[BB]\}}{\mathbf{A}[DD]^2 + \mathbf{A}[EE]^2}$$ .

This $C^*_{eff}$ evaluation can then be used as in es3.18 . $C^*_{eff}$ (and the area weighted averages) here is calculated for the harmonic case, and various values of $\omega$ = angular frequency, in working spreadsheets **Ceff-Calc-6-freq.xls** and **Ceff-Calc-6-freq-60N-60S.xls** . It is thus determined that the harmonic component distinction and magnitude is insignificant and not used.

---

The calculated results of **SS3.1** and **SS3.2** are:

es3.29 Global Ocean: $\Delta T(t)$ ramp starting at $t_r$, $(t-t_r)$>25 yrs...$C^*_{eff}(t) \geq$ **13.8** Watt·yrs/(m²·°C)
$\Delta T(t)$ harmonic with >16 yr period.......$C^*_{eff} \geq$ **11.7** Watt·yrs/(m²·°C)
es3.30 Global Ocean (60N-60S): $\Delta T(t)$ ramp starting at $t_r$, $(t-t_r)$>25 yrs...$C^*_{eff}(t) \geq$ **14.3** Watt·yrs/(m²·°C)
$\Delta T(t)$ harmonic with >16 yr period.......$C^*_{eff} \geq$ **12.3** Watt·yrs/(m²·°C) .

## SS4 Deep Ocean Thermal Admittance Theory v.4

Unlike the mix-layer, which can be considered a simple bulk thermal capacitance (over long time averages...see **SS3**), all the ocean below this layer (i.e. the "deep ocean") is composed of a continuous distribution of true thermal capacitance and thermal conductance. If the surface temperature increases by a small step ($\Delta T(t)$), power will flow into the capacitance, but dropping to zero over time at equilibrium. This is the $\Delta Q(t)_{deep.ocean}$ term discussed in sections **A** and **D**.

Further, the temperature of the abyssal oceans (excluding the polar regions) are self regulated below some depth, very close to freezing due to the continual upwelling of cold water sourced from the sinking cold polar waters, and are essentially independent of the surface temperature changes directly above. And this upwelling is the only reason the ocean temperatures have not equilibrated due to conduction vertically over the millennia.

This combination of upwelling and conduction all leads to a quite complex evolution of water thermal energy density and temperature with depth and time. Two cases will be considered where: (I) the thermal conductance parameter is constant with depth, *or* (II) decreases from the surface linearly to zero at some depth $-X_o$ . The reality is reasonably somewhere between these two extremes. The $\Delta Q(t)_{deep.ocean}$ term can then be, and is, directly formalized.

---

## Power density due to upwelling

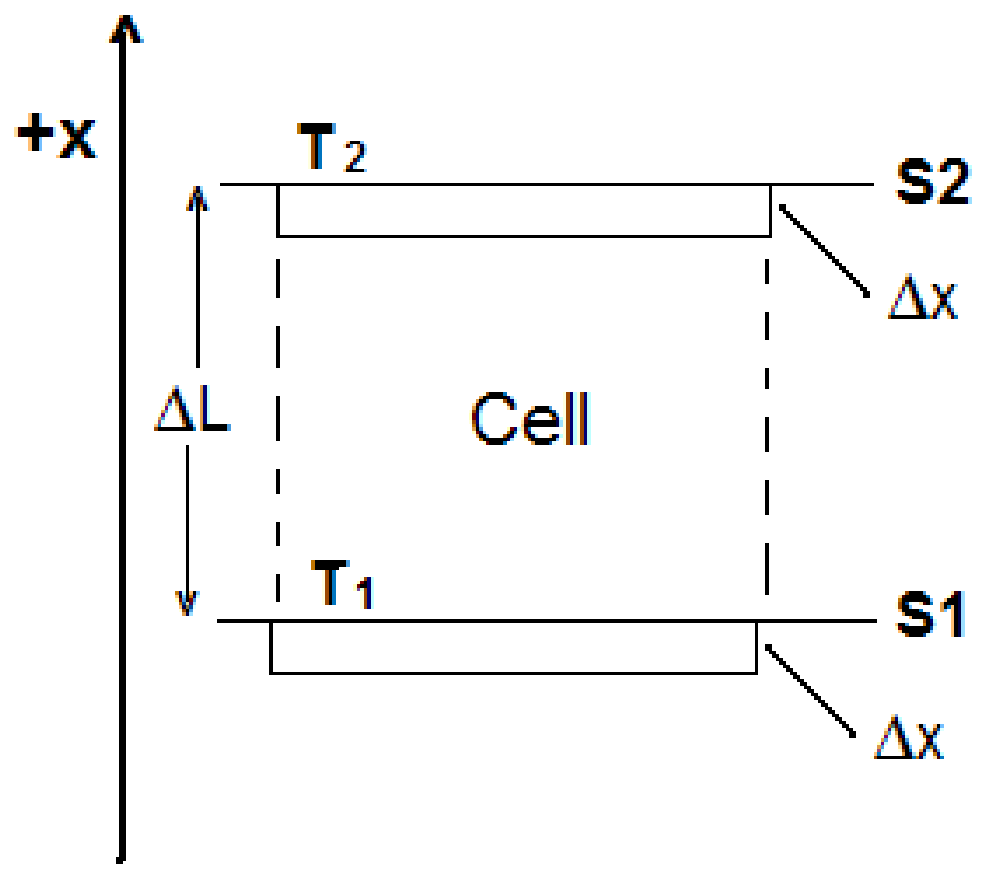

Fig.SS4.1

Figure SS4.1 shows a cross section of a cell of deep ocean water of area **ΔA** where +x is a position and direction increasing from the ocean floor to the surface. The upwelling of water is occurring at a positive velocity "**v**" in the +x direction. Consider this upwelling velocity approximately constant with depth, except very near the surface where it must reduce to zero and transition into purely lateral flow. Over a short time **Δt** a small slab of thickness **Δx** flows across the surface $S_1$ into the cell of volume **ΔV**. The power flow *into* this cell across $S_1$ is given by:

es4.1 $\Delta E/\Delta t = (E_o+\Delta T_1\cdot C\cdot \Delta x\cdot \Delta A)/\Delta t$ ;where

a) C is the change of thermal energy density of ocean water per Δ°C (~0.13 watt·yrs/(m$^3$·°C)...see es3.2)
b) $\Delta x_1\cdot\Delta A$ is the volume of the slab
b) $E_o$ is the thermal energy of the water slab at a temperature $T_o$ (where $T \equiv T_o + \Delta T$)
c) E is the total thermal energy of the water slab at a temperature $T = T_o + \Delta T$
d) $\Delta T_n$ is the change in temperature (from $T_o$) at surface $S_n$ .

The power flow *out* of this cell across $S_2$ is given by es4.1 except $\Delta T_1$ is replaced by $\Delta T_2$. The **total** power flow *into* this cell per *volume*, i.e. power density "**P***", is then given by

es4.2 $P^* = [(\Delta T_1 - \Delta T_2)\cdot C\cdot \Delta x\cdot \Delta A]/[\Delta t\cdot \Delta V]$ .

Using the relations $\Delta t=\Delta x/v$, $\Delta T_2 - \Delta T_1 \cong \Delta L\cdot d(\Delta T(x))/d(x)$, and $\Delta V=\Delta L\cdot \Delta A$ in es4.2 yields

es4.3 $P^* = -v\cdot C\cdot d(\Delta T(x))/d(x) = -g_u \cdot d(\Delta T(x))/d(x)$ ;where

es4.4 $g_u(x) \equiv v(x)\cdot C$ .

[Note, P* is <u>not</u> more generally a function of $dv(x)/dx$ if the *lateral* fluid velocity is zero everywhere. Then $dv(x)/dx = 0$ since the *net* thermal mass flow into the cell is always zero (i.e. $\nabla\cdot\mathbf{v} = 0$).]

Studies[1,2] have found the deep Oceans (for 60N-60S, i.e. *excluding the near freezing polar Oceans* higher than 60N or 60S latitude) average upwelling velocity $v_{average} \approx 1\cdot 10^{-7}$ m/sec. = 3.15 m/year. With uncertainty[1] about ± 33%. Therefore $\mathbf{g_{u.average} \approx 3.15\cdot(0.13) = 0.41\ (W/(m^2\cdot{}^\circ C))}$ is a rough estimate of magnitude.

Using Fig.SS4.1 and es4.1, where positive power flow is in the +x direction, the power flow per area (i.e. "**I(x,t)**") across surface 1, and for $T_1=T_o + \Delta T_1$, is given by (see es4.1)

es4.5a $I(T_o + \Delta T_1) = \Delta E/(\Delta t\cdot \Delta A) = (E_o + \Delta T_1\cdot C\cdot \Delta x\cdot \Delta A)/(\Delta t\cdot \Delta A)$
$= E_o/(\Delta t\cdot \Delta A) + \Delta T_1\cdot C\cdot v$ .

And for $\Delta T_1=0$

es4.5b $I(T_o) = E_o/(\Delta t\cdot \Delta A)$ .

Therefore the *change* in power flow per area across surface 1 in the +x direction, for a given change in temperature at surface 1, is given by

es4.5c $\Delta I(\Delta T_1) = I(T_o + \Delta T_1) - I(T_o) = \Delta T_1\cdot C\cdot v$
$= \Delta T_1(x,t)\cdot C\cdot v(x)$
$= \Delta T_1(x,t)\cdot g_u(x)$

; using es4.5a,b and es4.4 .

**<u>Power density due to conduction</u>**

Thermal conduction through water can be of many types: dispersion, convection, radiative, etc. . But in all cases the basic form is the same. Again using figure SS4.1, where positive power flow is in the +x direction, the power flow per area (i.e. "**I(x,t)**") of surface 1 or 2 is given by

es4.6a $I(x,t) = -g_c(x)\cdot dT(x,t)/dx$

;where $g_c$ is the effective conduction coefficient for all forms of conduction with units W/(m °C) .

The **total** power flow *into* this cell per volume, i.e. power density "P*", is then given by:

---

1 A. Ganachaud et al, "Improved estimates of global ocean circulation, heat transport and mixing from hydrographic data", Nature Nov 2000 p454

2 X. Liang et al, "Global Ocean Vertical Velocity From a Dynamically Consistent Ocean State Estimate ", Journal of Geophysical Research: Oceans, Oct 2017; sect. 3.1

es4.6b $P^*(x,t) = \Delta A\cdot(-I_{out} + I_{in})/\Delta V = [g_c(x_2)\cdot dT_2(x_2,t)/dx - g_c(x_1)\cdot dT_1(x_1,t)/dx]\cdot\Delta A/(\Delta L\cdot\Delta A)$
$\cong \{\Delta A\cdot\Delta L\cdot d[g_c(x)\cdot dT(x,t)/dx]/dx\}/(\Delta L\cdot\Delta A)$
$= d[g_c(x)\cdot dT(x,t)/dx]/dx$

es4.6c $P^*(x,t) = dg_c(x)/dx \cdot dT(x,t)/dx + g_c(x) \cdot d^2T(x,t)/dx^2$ ;where the properties of derivatives are used .

---

The power density flow into the cells thermal capacitance is given by

es4.7 $P^*(x,t) = C \cdot dT(x,t)/dt$

Then the power density flow into the cell thermal capacitance must equal the total conductive and upwelling power flow onto the cell. So, using es4.3,6c,7 yields

es4.8 $C \cdot dT(x,t)/dt = dg_c(x)/dx \cdot dT(x,t)/dx + g_c(x)\cdot d^2T(x,t)/dx^2 - g_u\cdot d(T(x,t))/d(x)$ .

But let $T(x,t) \equiv T_o(x) + \Delta T(x,t)$ . Let $\Delta T(x,t)=0$. It follows that

es4.9 $C \cdot dT_o(x)/dt = dg_c(x)/dx \cdot dT_o(x)/dx + g_c(x)\cdot d^2T_o(x)/dx^2 - g_u\cdot d(T_o(x))/d(x)$ .

Then rewrite es4.8 as

es4.10 $C \cdot d[T_o(x)+\Delta T(x,t)]/dt = dg_c(x)/dx \cdot d[T_o(x)+\Delta T(x,t)]/dx + g_c(x)\cdot d^2[T_o(x)+\Delta T(x,t)]/dx^2 +$
$- g_u\cdot d[T_o(x)+\Delta T(x,t)]/d(x)$

Now subtract es4.9 from es4.10 yielding

es4.11 $C \cdot d\Delta T(x,t)/dt = dg_c(x)/dx \cdot d\Delta T(x,t)/dx + g_c(x)\cdot d^2\Delta T(x,t)/dx^2 - g_u\cdot d(\Delta T(x,t))/d(x)$ .

---

**Case I**

Assume $g_c(x)=g_{co}$ is a constant with depth. And $g_u$ is considered approximately constant with depth below the bottom of the mix layer, which is defined as **x = 0** , and is the region of analysis. And assume an equilibrium solution for T(x,t) (where $d(Tx,t)/dt = 0$) of

es4.12 $T(x) = a\cdot\exp[\beta\cdot x] + b$ ;where $\exp[z] \equiv e^z$ .

Then es4.8 simplifies to

es4.13 $0 = 0 + g_{co}\cdot a\cdot\beta^2\cdot\exp[\beta\cdot x] - g_u\cdot a\cdot\beta\cdot\exp[\beta\cdot x]$ , or
$= g_{co}\cdot\beta^2 - g_u\cdot\beta$ , or
$\beta = g_u/g_{co}$ .

Let $T_s$ be the "surface" temperature at x=0 , $T_b$ is the ocean bottom temperature (or very nearly), $T_s > T_b$, and x < 0 beneath the surface. The +x direction is upward toward the surface in the direction of the upwelling. The solution must then be

es4.14 $T(x) = (T_s-T_b)\cdot \exp[\beta\cdot x] + T_b$ , where the magnitude of T(x) *decays* into the depths.

Note x = $-X_{1/e}$ is depth at which the temperature changes by (1-1/e) of is full variation of ($T_s$-$T_b$) . So $\beta=1/X_{1/e}$ . This depth on *average* is known from observations of this temperature profile to be on the order of $\mathbf{X_{1/e} \sim 700m}$. Note that below 2000m the variation of T(x) is very small (relatively). And so using es4.13 yields

es4.15 $X_{1/e} = 1/\beta = g_{co}/g_u$ , or

es4.16 $g_{co} = g_u \cdot X_{1/e}$

Using es4.4 (and the subsequent evaluation) in es4.16 yields $g_{co}$ on the order of

es4.17 $g_{co} = g_u \cdot X_{1/e} \approx (0.41)\cdot 700\ \text{m}\cdot(\text{W}/(\text{m}^2\cdot{}^\circ\text{C})) = 287\ \text{W}/(\text{m}\cdot{}^\circ\text{C})$ .

............................

Now *assume* a time variable form of ΔT(x,t) using harmonic **phasor** variables and mathematics (as per the methods of electrical engineering, <u>where "**j**" is the unit imaginary number</u>), where " **T** " is a phasor value, and " $a\cdot\exp[j\omega\cdot t]$ " is the independent input surface $\Delta T(t)_{surface}$ phasor [where $\exp[z] \equiv e^z$]. Then the assumed form is

es4.18 $\mathbf{T} \equiv \Delta T(x,t) = a\cdot\exp[\beta\cdot x]\cdot\exp[j\omega\cdot t]$ .

Then use es4.18 in es4.11 yielding:

es4.19 $C\cdot j\omega\cdot\mathbf{T} = 0 + g_{co}\cdot\beta^2\cdot\mathbf{T} - g_u\cdot\beta\cdot\mathbf{T}$

$C\cdot j\omega = g_{co}\cdot\beta^2 - g_u\cdot\beta$

$0= g_{co}\cdot\beta^2 - g_u\cdot\beta - C\cdot j\omega$

Therefore , using the quadratic equation

es4.20 $\beta = \{g_u \pm \sqrt{[g_u^2 + 4g_{co}\cdot C\cdot j\omega]}\}/(2\cdot g_{co})$ .

Using es4.17,4,1a and $\omega \geq 0.02$ radians/year [i.e. 1/2 period = 157 years] yields $4g_{co}\cdot C\cdot\omega \approx 2.98$, $g_u^2 = 0.17$. So $4g_{co}\cdot C\cdot\omega/g_u^2 = 17.5$, and the bracket terms can be approximated :

es4.21 $\beta = \{g_u \pm\sqrt{[\, g_u^2 + 4g_{co}\cdot C\cdot j\omega]}\}/(2\cdot g_{co})$

$\cong \{g_u \pm \sqrt{[4g_{co}\cdot C\cdot j\omega]}\}/(2\cdot g_{co})$

$= \pm\sqrt{[g_{co}\cdot C\cdot j\omega]}/(g_{co}) + g_u/(2\cdot g_{co})$

es4.22 $\beta = \pm\sqrt{[j\omega\cdot C/g_{co}]} + g_u/(2\cdot g_{co}) = \pm\sqrt{[j]}\cdot\sqrt{[\omega\cdot C/g_{co}]} + g_u/(2\cdot g_{co})$ .

Now rewrite es4.18 using es4.22

es4.23 $\mathbf{T} \equiv \Delta T(x,t) = a\cdot\exp\{[\pm\sqrt{[j]}\cdot\sqrt{[\omega\cdot C/g_{co}]} + g_u/(2\cdot g_{co})]\cdot x\}\cdot\exp[j\omega\cdot t]$ .

Note: since ΔT must decay into the depths, and since x<0 below the surface, only the $+\sqrt{[j]}$ option is physically appropriate.

...................................

Finally, solve for the thermal admittance of the ocean at the bottom of the ocean mix layer (x=0). Power flow per area *into* the ocean surface by heat *conduction* (downward...see Fig.SS4.1) is given in es4.6a as

es4.24 $\Delta I(0,t) = g_{co}\cdot\, d\Delta T(x,t)/dx\,|_{x=0}$ . Using es4.23 this becomes

es4.25 $\Delta I(0,t) = g_{co}\cdot a\cdot\{\sqrt{[j]}\cdot\sqrt{[\omega\cdot C/g_{co}]}+g_u/(2\cdot g_{co})\}\cdot \exp\{[\sqrt{[j]}\cdot\sqrt{[\omega\cdot C/g_{co}]} + g_u/(2\cdot g_{co})]\cdot x\}\cdot \exp[j\omega\cdot t]\Big|_{x=0}$

$= a\cdot\{\sqrt{[j]}\cdot\sqrt{[\omega\cdot C\cdot g_{co}]}+g_u/2\}\cdot \exp[j\omega\cdot t]$ .

The deep ocean thermal phasor admittance (i.e. Y(ω)) , using es4.23,25 , is then given by

es4.26a $Y(\omega)_{conduction} \equiv \Delta I(0,t)/\Delta T(0,t)$

$= \{a\cdot(\sqrt{[j]}\cdot\sqrt{[\omega\cdot C\cdot g_{co}]} +g_u/2)\cdot \exp[j\omega\cdot t]\}/\{a\cdot \exp[j\omega\cdot t]\}$

$= \sqrt{[j]}\cdot\sqrt{[\omega\cdot C\cdot g_{co}]} + g_u/2$

$= \sqrt{[j]}\cdot\sqrt{[\omega\cdot C\cdot g_{co}]} + g_u/2$ .

Finally, determine an admittance component due to the upwelling mechanism (using es4.5c) where power flow I is in the -x direction; specifically

es4.26b $Y(\omega)_{upwelling} \equiv \Delta I_{upwelling}(0,t)/\Delta T(0,t) = -\Delta T(0,t)\cdot(g_u(0))/ \Delta T(0,t)$

$= -g_u(0)$ .

The total admittance is the just the sum of es4.26a and es4.26b,

es4.26c $Y(\omega)_{total} = \sqrt{[j]}\cdot\sqrt{[\omega\cdot C\cdot g_{co}]} + g_u/2 - g_u$

$= \sqrt{[j]}\cdot\sqrt{[\omega\cdot C\cdot g_{co}]} - g_u/2$

.......................................

**It is important to clarify** that for ω→0 (i.e. equilibrium solutions of finite non-zero amplitude) in es4.21 the exact solution of es4.21 becomes $\boldsymbol{\beta = g_u/g_{co}}$ . In this case, following through the calculations above, es4.26c becomes

es4.26d $Y(\omega\to 0)_{total} = g_u - g_u = 0$ !!

Thus, after a step temperature change ΔT the Δpower flow into the deep ocean ultimately reduces to zero, and $\Delta Q_{deep.ocean}(t = \infty) \equiv d(\Delta E^*(t))/dt = \Delta I(t = \infty) = \Delta T\cdot Y(\omega\to 0)_{total} = 0$ , <u>as is expected and required of $\Delta Q_{deep.ocean}$</u> (where $\Delta E^*(t) \equiv \Delta E(t)/\Delta Area$).

[The dependence on √ω is also common in passive *distributed* "RC" electronic devices.]

..........................................

Conversion to a Laplace s-space transform simply requires replacing "jω" by "s" , so es4.26c becomes

es4.27 $Y(s) = \sqrt{s}\cdot CC^* - g_u/2$ , where $CC^* \equiv \sqrt{[C\cdot g_{co}]}$ .

Using es4.17 and es4.1a in es4.27 we can estimate

es4.28 $CC^* = \sqrt{[287\ W/(m\cdot{}^\circ C)\cdot 0.13\ watt\cdot yrs/(m^3\cdot{}^\circ C)]} = 6.1\ (W\sqrt{yrs})/(m^2\cdot{}^\circ C)$ .

This is identical to the value of CC*(60N-60S) calculated in **SS5** ( = 6.04, presuming hybrid forcing, where $g_u$ =0.41) <u>using a completely independent Energy method and data</u> (see e.16 or es5.13). The *form* of Y(s) is the more important result , but this corroboration does tend to validate the results of both SS4 and SS5!

**[It is noted, this method of CC*(60N-60S) evaluation using the direct Global observations of "X" (see es4.15,35), does yield a theoretical *maximal* evaluation of $\Delta Q(T_s(t))_{Deep.Ocean}$ over a wide range of possible $g_u$ (=0 to .8) and thus a *maximal* value of ECS(Ocean) that is likely as accurate as using the Energy methods of sect. D (and quite comparable), and does *NOT* rely on the uncertain values of E*(t) or $C^*_{eff}$ or**

**$g_u$ ! This alternative evaluation will be also be performed in SS5 and compared to the Energy method values.]**

For a surface temperature ramp function of slope "S" ($\Delta T$/year) starting at t=0, the power flow per area (I(t)) into the deep ocean thermal *capacitance* (i.e. excluding the mix layer bulk capacitance) can be calculated using the Laplace transformed variables to be:

es4.29 $\Delta I(s) = Y(s)\cdot\Delta T(s)$
$= S\cdot CC^*\cdot(\sqrt{s}/s^2) - (½)\cdot S\cdot g_u/s^2 = S\cdot CC^*\cdot s^{-(3/2)} - (½)\cdot S\cdot g_u/s^2$

; where S· <u>$1/s^2$</u> is the Laplace transform of <u>T(t)=S·t</u>, and using es4.27. Now take the Inverse Laplace Transform[3] of es4.29 yielding

es4.30 $\Delta I(t) = S\cdot CC^*\cdot(2/\sqrt{\pi})\cdot\sqrt{t} - (½)\cdot S\cdot g_u\cdot t$, or

es4.31 $\Delta Q_{deep.ocean}(t) \equiv d(\Delta E^*(t))/dt = \Delta I(t) = S\cdot CC^*\cdot(2/\sqrt{\pi})\cdot\sqrt{t} - (½)\cdot S\cdot g_u\cdot t$

; <u>where $\Delta E^*(t) \equiv \Delta E(t)/\Delta area$</u> .

---

**<u>Case II</u>**

Assume an equilibrium solution for T(x,t) (where $d(Tx,t)/dt = 0$) such that T(x) decreases linearly with depth to a value of $\mathbf{T_b}$ at some depth $-X_o$, and remains constant at this value to the bottom. Again, the bottom of the mix layer is defined as elevation x=zero, x becoming more negative with depth. Let $\mathbf{T_s}$ be the "surface" temperature at x=0. And $g_u$ is again considered approximately constant with depth below the bottom of the mix layer. Then in this region of decrease T(x) is assumed of the form

es4.32 $T(x) = a\cdot x + b$

Then es4.8 simplifies to

es4.33 $0 = a\cdot dg_c(x)/dx + g_c(x)\cdot 0 - g_u\cdot a$ , or
$g_u = dg_c(x)/dx$ .

However, for $x < -X_o$ , $g_c(x)$ must equal a constant value of zero! Otherwise a constant value $T_b$ could not be a valid solution to the bottom, as is indicated in **Case I** above. The solution for $g_c(x)$ in the region of temperature decrease must then be

es4.34 $g_c(x) = g_u\cdot(x + X_o)$ , but
$g_c(x) = 0$ for $x < -Xo$ .

Let $g_{co} \equiv g_c(x=0)$ . Again, the average value of $X_o$ is known from observations of this temperature profile to be on the order of ~700m. Using es4.4 in es4.34 at x=0 yields $g_{co}$ on the order of

es4.35 $g_{co} = g_u\cdot X_o \approx (0.41)\cdot 700\ m\cdot(W/(m^2\cdot{}^\circ C)) = 287\ W/(m\cdot{}^\circ C)$ . So,

es4.36a $g_c(x) = g_{co}\cdot(1 + x/X_o)$ , for $x >= -X_o$

---

[3] https://www.wolframalpha.com/input/?i=inverse+Laplace+transform+1%2F%28s%5E2%2B1%29 ;online application

..................................................

At underline equilibrium the total $\Delta$power flow per area (i.e. $\Delta I(x,t)$) **into** the ocean surface at x=0 is given by

es4.36b $\Delta I(x{=}0)_{total} = \Delta I(x{=}0)_{conduction} + \Delta I(\Delta T(x{=}0))_{upwelling}$

Using es4.24,5c then yields

es4.36c $\Delta I(x{=}0)_{total} = g_{co} \cdot d(\Delta T(x))/dx\big|_{x=0} - \Delta T(x{=}0) \cdot g_u(x)$

But rewrite es4.32 for the described temperature profile as $T(x)=(T_s-T_b)\cdot(1+x/X_o) + T_b$ . Since $T_b$ is constant this can be rewritten $T(x)_o + \Delta T(x) = (T_{so}+\Delta T_s - T_b)\cdot(1+x/X_o) + T_b$ ; where $X_o$ is essentially unchanged for a small $\Delta T_{so}$ . Subtracting out the constant large signal terms yields

es4.36d $\Delta T(x) = \Delta T_s \cdot (1+x/X_o)$ . And then
es4.36e $d\Delta T(x)/dx\big|_{(x=0)} = \Delta T_s/X_o$ .

Using es4.35,36e in es4.36c then yields

es4.36f $\Delta I(x{=}0)_{total} = g_{co} \cdot d\Delta T(x)/dx\big|_{x=0} - \Delta T(x{=}0) \cdot g_u(x)$
$= g_{co} \cdot \Delta T_s/X_o - \Delta T_s \cdot g_u(x)$
$= g_u \cdot \Delta T_s - \Delta T_s \cdot g_u(x)$
$= 0$ !!

Note, this is the expected and required result when equilibrium is attained after some disturbance $\Delta T_s$ from the steady state condition.

......................................

Now as in **Case I** above, assume a phasor form solution for T(x,t)

es4.37 $\mathbf{T} \equiv \Delta T(x,t) = a \cdot \exp[\beta(x) \cdot x] \cdot \exp[j\omega \cdot t]$ .

Using calculus it can be shown that $d(\exp[\beta(x)\cdot x])/dx = \beta(x)\cdot\exp[\beta(x)\cdot x]$ as $x\to 0$ , and also $d^2(\exp[\beta(x)\cdot x])/dx^2 = (2\cdot d\beta(x)/dx + \beta^2(x))\cdot\exp[\beta(x)\cdot x]$ as $x\to 0$ . All results below are for the case $x\to 0$ ! Then use es4.37,36a in es4.11 yielding:

es4.38 $C\cdot j\omega\cdot\mathbf{T} = (g_{co}/X_o)\cdot(\beta\cdot\mathbf{T}) + g_c(x)\cdot[\,2\cdot\beta'(x) + \beta(x)^2]\cdot\mathbf{T} - g_u\cdot(\beta\cdot\mathbf{T})$ ; where $\beta'(x) \equiv d\beta(x)/dx$ .

Assume

es4.39 $\beta'(x) = -\alpha\cdot(\beta(x)-\alpha)$ for $x\to 0$

;where $\alpha$ is a positive constant (to be proven below). Then rewrite es4.38 as

es4.40 $C\cdot j\omega\cdot\mathbf{T} = (g_{co}/X_o)\cdot(\beta\cdot\mathbf{T}) + g_c(x)\cdot[\,-2\cdot\alpha\cdot\beta(x)+2\cdot\alpha^2 + \beta(x)^2]\cdot\mathbf{T} - g_u\cdot(\beta\cdot\mathbf{T})$
$0 = \beta^2\cdot g_c(x) + \beta\cdot[(g_{co}/X_o) - g_u - 2\cdot\alpha\cdot g_c(x)] - C\cdot j\omega + 2\cdot\alpha^2\cdot g_c(x)$
$0 = \beta^2\cdot g_c(x) + \beta\cdot[\,0 - 2\cdot\alpha\cdot g_c(x)] - (C\cdot j\omega - 2\cdot\alpha^2\cdot g_c(x))$

;where, using es4.35, $(g_{co}/X_o) - g_u = g_u - g_u = 0$ . Therefore, using the quadratic equation

es4.41 $\beta = \{2\cdot\alpha\cdot g_c(x) + \sqrt{[\,\alpha^2\cdot(2\cdot g_c(x))^2 + 4\cdot g_c(x)\cdot(C\cdot j\omega - 2\cdot\alpha^2\cdot g_c(x))\,]}\}/(2\cdot g_c(x))$
$\beta = \alpha + \sqrt{[\,\alpha^2 + (C\cdot j\omega - 2\cdot\alpha^2\cdot g_c(x))/g_c(x)]}$ , as $x\to 0$ .

Assume

es4.42 $\alpha^2 << C\cdot\omega/g_{co}$ (to be proven below) .

Then, using es4.36a, es4.41 can be approximated as

es4.43 $\beta \cong \alpha + \sqrt{[C\cdot j\omega/g_c(x)]}$
$\beta \cong \alpha + \sqrt{[C\cdot j\omega/g_{co}]}\cdot(1 + x/X_o)^{-(1/2)}$ , as $x\rightarrow 0$ .

Calculate the derivative

es4.44 $\beta'(x) \equiv d\beta(x)/dx = \sqrt{[C\cdot j\omega/g_{co}]}\cdot(-1/2)\cdot(1/X_o)\cdot(1 + x/X_o)^{-(3/2)}$ , as $x\rightarrow 0$ .

Rewriting es4.39 and using es4.43,44 yields

es4.45 $\alpha \equiv -\beta'(x)/(\beta(x) - \alpha) \mid_{x\rightarrow 0} = (1/2)\cdot(1/X_o)$ .

So es4.39 is verified.

Now verify es4.42 for $\omega \geq .02$ radians/year [i.e. 1/2 period = 157 years], $X_o \approx 700$m, $g_{co} \approx 287$ W/(m·°C), C ≈ 0.13 W·yrs/(m³·°C), $\alpha \approx 0.00071$ (1/m) [see es4.1a,34,35,45] used in

es4.46 $\alpha^2\cdot g_{co} << C\cdot\omega$ , or
$1 << C\cdot\omega/[\alpha^2\cdot g_{co}]$
$1 << 18.0$

;which is adequate verification.
.............................................

Again, solve for the thermal admittance of the ocean at the bottom of the ocean mix layer. Power flow per area *into* he ocean surface by heat *conduction* (downward...see Fig.SS4.1) is given in es4.6a as

es4.47 $\Delta I(0,t) = g_{co}\cdot d\Delta T(x,t)/dx \mid_{x=0}$ . Using es4.37,43 this becomes

es4.48 $\Delta I(0,t) = g_{co}\cdot\{\alpha + \sqrt{[j]}\cdot\sqrt{[\omega\cdot C/g_c(x)]}\}\cdot\mathbf{T}\mid_{x=0}$
$= \{\alpha\cdot g_{co} + \sqrt{[j]}\cdot\sqrt{[\omega\cdot C\cdot g_{co}]}\}\cdot \mathbf{T}(0,t)$

The deep ocean thermal phasor admittance (i.e. Y(ω)) , using es4.37,48,45,35 , is then given by

es4.48b $Y(\omega)_{conduction} \equiv \Delta I(0,t)/\Delta T(0,t)$
$= \{\alpha\cdot g_{co} + \sqrt{[j]}\cdot\sqrt{[\omega\cdot C\cdot g_{co}]}\}\cdot\mathbf{T}(0,t)/\mathbf{T}(0,t)$
$= \{\alpha\cdot g_{co} + \sqrt{[j]}\cdot\sqrt{[\omega\cdot C\cdot g_{co}]}\}$
$= \sqrt{[j]}\cdot\sqrt{[\omega\cdot C\cdot g_{co}]} + g_{co}/(2\cdot X_o)$
$= \sqrt{[j]}\cdot\sqrt{[\omega\cdot C\cdot g_{co}]} + g_u/2$

Use the same methods as in **Case I** from es4.26b through es4.31 .

es4.48c $Y(\omega)_{total} = Y(\omega)_{conduction} + Y(\omega)_{upwelling} = \sqrt{[j]}\cdot\sqrt{[\omega\cdot C\cdot g_{co}]} + g_u/2 - g_u$
$= \sqrt{[j]}\cdot\sqrt{[\omega\cdot C\cdot g_{co}]} - g_u/2$

But as $\omega\rightarrow 0$ the above formula does not hold. **It is then important to clarify**, as shown in es4.36f, that after a step surface temperature change ΔT, the Δpower flow into the deep ocean ultimately reduces to zero, and $\Delta Q_{deep.ocean}(t = \infty) \equiv d(\Delta E^*(t))/dt = \Delta I(t = \infty) = 0$ , as is expected and required of $\Delta Q_{deep.ocean}$ (where $\Delta E^*(t) \equiv \Delta E(t)/\Delta area$).

We can find, as in **Case I**, for a surface temperature ramp function of slope "S" (ΔT/year) starting at t=0, the power flow per area (I(t)) into the deep ocean thermal *capacitance* (i.e. excluding the mix layer bulk capacitance) can be calculated as

es4.49 $\Delta Q_{deep.ocean}(t) \equiv d(\Delta E^*(t)/dt = \Delta I(t) = S\cdot CC^*\cdot(2/\sqrt{\pi})\cdot\sqrt{t} - (½)\cdot g_u\cdot S\cdot t$

; where $CC^* \equiv \sqrt{[C\cdot g_{co}]}$, and where

es4.50 $Y(s) = \sqrt{s}\cdot CC^* - g_u/2$ .

The only difference in **Case II** is that $X_{1/e}$ is replaced by $X_o$ . It seems the *form* for $\Delta Q_{deep.ocean}(t)$ is essentially independent of the *form* of the conduction coefficient profile $g_c(x)$ !

---

## SS4.2 $g_u$ effective

As in es1.37,38 the area weighted average of $\mathbf{\Delta Q_{deep.ocean}}(t)$ is of primary interest. The **bold** delta terms indicate areal averages over some given region A, and **A[**f(y)**]** indicates the areal average operator on some function of "y" over A; where "y" here represents the enumeration of any and every location in the area "A" (i.e. "y" is **not** a one dimensional spatial variable). Taking the average of es4.49 yields

es4.51a $\mathbf{\Delta Q_{deep.ocean}}(t) = \mathbf{\Delta I}(t) = \mathbf{A[}S(y)\cdot CC^*(y)\mathbf{]}\cdot(2/\sqrt{\pi})\cdot\sqrt{t} - \mathbf{A[}g_u(y)\cdot S(y)\mathbf{]}\cdot(½)\cdot t$
b $\equiv \mathbf{A[}S(y)\mathbf{]}\cdot CC^*_{eff}\cdot(2/\sqrt{\pi})\cdot\sqrt{t} - \mathbf{A[}S(y)\mathbf{]}\cdot g_{u.effective}\cdot(½)\cdot t$

; where $CC^*(y) \equiv \sqrt{[C\cdot g_{co}(y)]}$ . A rough approximation of $g_u(y) \equiv v(y)\cdot C \approx +0.41$ (W/(m$^2\cdot$°C)) [see es4.4] has been used for the temperate and tropical oceans (roughly 60N-60S), but it becomes smaller and negative in the polar oceans [see **SS9** for a more exact evaluation]. *In fact the Global Ocean areal average of* v*(y) and $g_u(y)$ must be zero since there is no net change of volume of the deep ocean (below the bottom of the mix layer)*. However in polar locations where the "upwelling" velocity becomes zero or negative the temperature change is very small, i.e. the near surface waters are self regulating to be near freezing. Therefore the Global Ocean average term **A[**$g_u(y)\cdot S(y)$**]** **(**and the value of $g_{u.effective}$**)** in es4.51b remains positive for a positive temperature change "S", but must be somewhat less (~0.9) of the 60N-60S latitude averages.

Also, as shown in es1.37

es4.52 $\mathbf{\Delta T}(t')\cdot G_{eff.A} = \mathbf{\Delta F}(t') - \mathbf{\Delta Q}(t')$ ,or
$G_{eff.A} = [\mathbf{\Delta F}(t') - \mathbf{\Delta Q}(t')]/\mathbf{\Delta T}(t')$ .

The subtracted second term of es4.51 results in a reduced value of $\mathbf{\Delta Q}(t')$, an increased value of $G_{eff.A}$, and a decreased value of $ECS_{eff}$ .

Therefore a simplified, and *maximal* ECS option *could* be to simply omit the "$g_u$" related terms in all ΔQ(t). In fact, a full empirical evaluation of $G_{eff}$ (including the formal/empirical evaluation of CC*) using $g_{u.effective}$ = **0** W/(m$^2\cdot$°C) results in only a 3.5% *decrease* in evaluated $G_{eff}$ for the 60N-60S latitude Ocean region, and thus even less for the Global Ocean average.

## SS5 Deep Ocean Coefficient Determination v.4

**[Note: the working spreadsheets for all these figures and calculations is available in "OER-gu-(etc.)-v4.xls, ". The definitions of all terms and operators is fully explained in the Main section D and A.]**

The evaluation of CC* is of critical importance, and so it will be described in considerable detail for the case of the Global Oceans (60N-60S) using the Ocean Energy formal/empirical method presuming:
(1) All Forcing components are simple surface Radiative Forcing,
(2) Deviations from the linear AGW (Anthropogenic Global Warming) Radiative Forcing are only caused by Ocean parametric Forcing (described in **D** after e.16),
(2B) Hybrid Forcing...Radiative Forcing, followed by some Ocean parametric Forcing deviations.
This is then followed by:
(3) An alternate formal/empirical method as described in **SS4** (following es4.27,28).

**1)**

A known theoretical *form* of H5[ΔE*(t)] will be "best fit" to the observed value from 1990-2020, by variation of the coefficients of this *form*. This is the "formal/empirical" method. Thus using es4.31 and es3.18, for a nearly linear ramp surface temperature $\Delta\mathbf{T}_s(t)$ of slope "S" [where the **bold T** indicates an area weighted average], starting at $t_r$ =1970,

es5.1 $$Q(t)_{total} = C^*_{eff}\cdot d[\Delta\mathbf{T}_s(t)]/dt \;+\; CC^*\cdot S\cdot(2/\sqrt{\pi})\cdot\sqrt{[t-t_r]} \;-\; S\cdot[t-t_r]\cdot g_u/2$$

; for $t>t_r$ . The appropriate values of $C^*_{eff}$ (see **SS3**) are:

es5.2a Global Ocean: ΔT(t) ramp starting at $t_r$, $(t-t_r)$ = 25 yrs...$C^*_{eff}$ = **13.8** Watt·yrs/(m²·°C)
ΔT(t) harmonic with >16 yr period.......$C^*_{eff}$ ≥ **11.7** Watt·yrs/(m²·°C)
es5.3a Global Ocean (60N-60S): ΔT(t) ramp starting at $t_r$, $(t-t_r)$ = 25 yrs...$C^*_{eff}$ = **14.3** Watt·yrs/(m²·°C)
ΔT(t) harmonic with >16 yr period........$C^*_{eff}$ ≥ **12.3** Watt·yrs/(m²·°C)

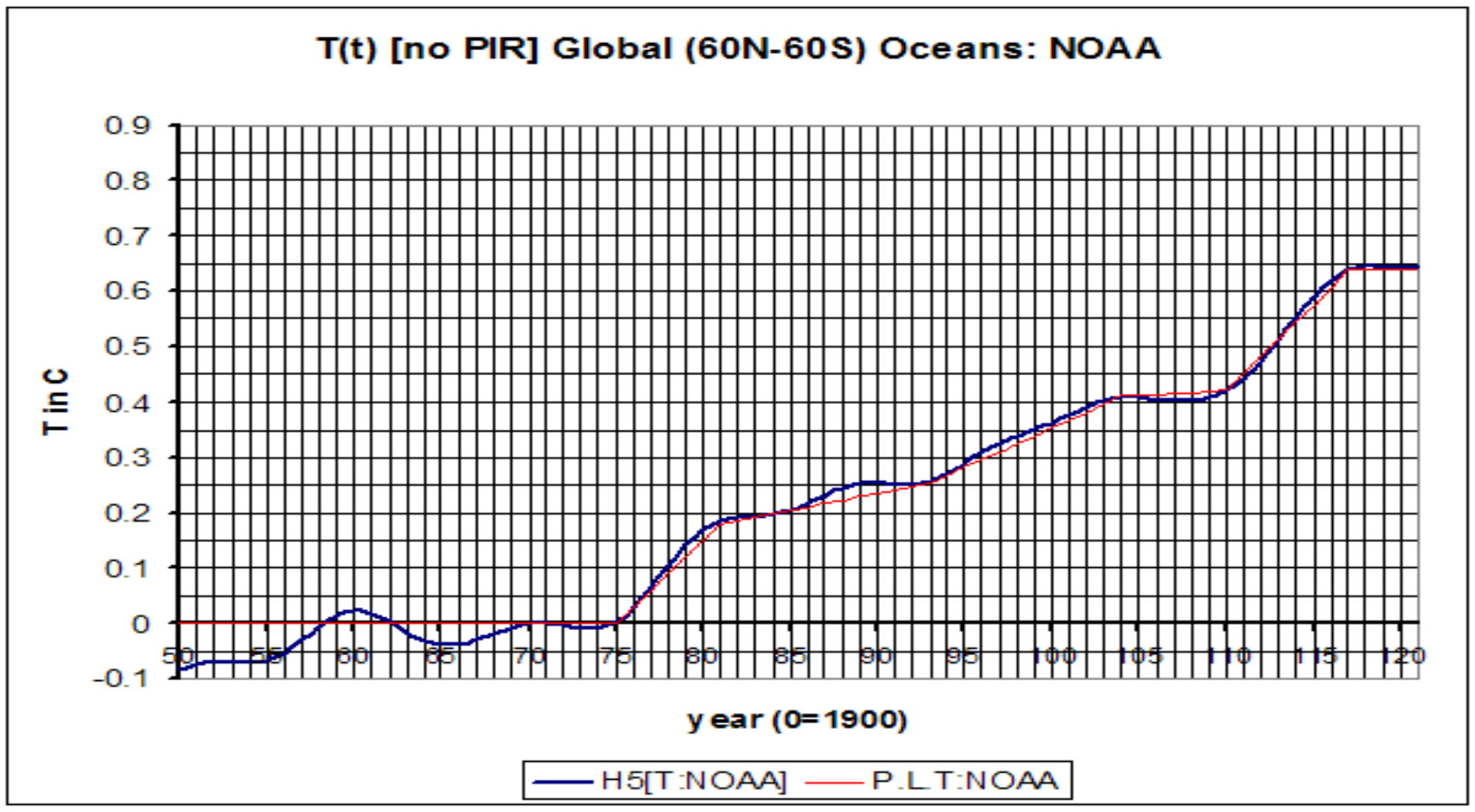

Fig.SS5.1

A piecewise linear (P.L.) approximation is made of H5[**T**(t):NOAA.Ocean(60N-60S)] as plotted in Fig.SS5.1 (see "P.L. T:NOAA"). This P.L. temperature can be represented as a sum of delayed ramp functions of various delay and slope "S" [using a technique common in Electrical Engineering]. The response to this temperature function is then just the sum of the responses to these individual ramp functions, such as given in

es.5.1 above.

Integrate es5.1 over time (let t=0 at the ramp start time) to get

es5.4 $\Delta E^*(t)_{total} = C^*_{eff}\cdot\Delta \mathbf{T}_s(t) + \Sigma[CC^*\cdot S\cdot(2/\sqrt{\pi})\cdot(2/3)\cdot t^{(3/2)} - S\cdot t^2\cdot g_u/4 ]\,|_{\text{summed over all ramps}}$.

Presuming $\mathbf{g_u=0.41}$ , the sum of the deep ocean later two terms in es5.4 for *each* such individual delayed ramp function, plus the first mix layer term, is the theoretical " E*: Thry.true " , as is plotted in Fig.27 (see the thin red line, sect. **D**) . Then " *d*(E*:Thry)/*d*t " ≡ $\Delta Q(t)_{theory}$ is plotted in Fig27B. Note, CC* is an unknown parameter at this point.

The corresponding total value of observed NOAA Ocean energy (0-2km) per area is H5[ΔE*(t):NOAA] and is plotted in Main Fig.27 (the thin blue line "H5[E*:NOAA]") **[see raw data in "OE-NOAA60N-60S-list-1959-2019.xls"]**

Therefore the only unknown theoretical parameter to be evaluated is "CC*" . A difference metric is defined as the *difference* between the slopes (S) of the best linear fit to the (1)theoretical and (2)observational values of H5[ΔE*(t)] over the evaluation range (in this case, 1990-2020). The Excel Solver tool is used to solve for the best fit "CC*" parameters that minimizes this metric. [Using "best linear fitting" enhances the importance of trends in the analysis, and diminishes the importance of specific values, which is a philosophy used throughout this monograph.]

Presuming $\mathbf{g_u=0.41}$, the best fit values of CC* for Global Oceans and Global Oceans 60N-60S latitude are:

es5.5 Global Ocean [$\mathbf{g_u}$=0.41·(0.9)] : for a ΔT(t) ramp starting at $t_r$...CC* = **6.7** Watt·√[yrs]/($m^2$·°C)
es5.6 Global Ocean (60N-60S) : for a ΔT(t) ramp starting at $t_r$...CC* = **7.75** Watt·√[yrs]/($m^2$·°C) .

**[see working spreadsheets "OER-gu-RadForcing(etc.)-v4"]**

**2)**

Similarly, CC* can be evaluated presuming the Ocean temperature deviations from the AGW linear (Fig.12 ,16) are totally due to transient internal Ocean parametric fluctuations, which can produce ***no*** total ΔE*(t) variation within the Ocean *in principle*. There is however a variation due to the change of surface outward radiation with temperature, which is accounted below. The ideal linear response is shown as the best fit linear ramp starting at 1970 (in thick red Fig.26) to H5[$T(t)_{.after.PIR}$:NOAA(60N-60S)] (in thick blue Fig.26) after PIR *and* after the "dip" fluctuation has been artificially removed by interpolation/extrapolation. This deviation of the true surface temperature H5[$T_s(t)_{no.PIR}$:60N-60S] from the ideal linear response is denoted as "$\delta\mathbf{T_s}(t)$". This changes the upward radiation by $G_{eff.Ocean}\cdot\delta\mathbf{T_s}(t)$ , and must result in an change of power flow into the Oceans by the negative of this amount (by the conservation of power) . Taking a numerical time integration yields

es5.7 $H5[\Delta E^*(t)] = - \int^t G_{eff.Ocean}\cdot\delta\mathbf{T_s}(t)\cdot dt$ .

Therefore the total H5[E*:NOAA(60N-60S)] for an ideal temperature ramp plus Ocean parametric variations is given by

es5.8 $\Delta E^*(t)_{total} = C^*_{eff}\cdot S\cdot t + CC^*\cdot S\cdot(2/\sqrt{\pi})\cdot(2/3)\cdot t^{(3/2)} - S\cdot t^2\cdot g_u/4 - \int^t G_{eff.Ocean}\cdot\delta\mathbf{T_s}(t)\cdot dt$

;where we have used the single ramp version of es5.4, $G_{eff} \approx 2$ W/($m^2$·°C) , "S" is the slope of the ideal ramp, and let t=0 at the ramp start.

CC* is then evaluated as in **1)** above. Presuming $\mathbf{g_u=0.41}$, the best fit values of CC* for Global Oceans and Global Oceans 60N-60S latitude are:

es5.9 Global Ocean [$\mathbf{g_u}$=0.41·(0.9)] : ΔT(t) ramp starting at $t_r$...CC* = **5.1** Watt·√[yrs]/($m^2$·°C)
es5.10 Global Ocean (60N-60S) : ΔT(t) ramp starting at $t_r$...CC* = **5.47** Watt·√[yrs]/($m^2$·°C) .

**[see working spreadsheets "OER-gu-OceanForcing(etc.)-v4.xls" ]**

**2B)**

**[[** It is speculated that the Forcing deviation from linear may be Radiative through 2002 due to the volcanic radiative cooling (reasonable), but becomes Ocean parametric Forcing due to the transient "dip" phenomenon (as was suggested in Fig.25B,27B and following text). These Plots are shown in **Fig.32,32B,33,33B** [and using **OER-gu-RadaOceanForcing(etc.)-v4.xls**]. The congruence of the theoretical and NOAA observed values of $\Delta E^*(t)$ and $\Delta Q(t)$ are *excellent* after 1985 (particularly if the puzzling 9.3 year cycle components are ignored), and lend credence to both the speculation, and the theoretical and NOAA observed values! The originally questionable NOAA E* data now may be corroborated and verified after 1985! The calculated values of CC* in this case are shown in parentheses in e.15,16 valued between the two fixed cases, as might be expected!**]]**

**3)**

Using es4.16,27 gives $g_{co} = g_u \cdot X$ and $CC^* \equiv \sqrt{[C \cdot g_{co}]} = \sqrt{[C \cdot g_u \cdot X]}$ ; where $X = X_{1/e}$ or $X_o$ . We can calculate the Global 60N-60S average of $X_{eff.maximum}$=885m [see **SS9**]. For some regions (e.g. the Indian Ocean or higher latitudes) "X" is ≈1400m; but in the Tropics "X" is less than 300m (see various Ocean temperature profiles in the repositories "/OceanTprofiles/"). Further, "X" is larger in some regions primarily because the upwelling velocity is much reduced (see es4.4 $g_u(x) \equiv v(x) \cdot C$); particularly in the higher latitudes nearer the polar oceans where upwelling must reduce to zero and turn negative (see **SS4.2**). Thus the term $g_u(x) \cdot X(x)$ , and values of $g_{co}$ and then CC* are expected to be relatively constant spatially; where the Global Ocean (60N-60S) average of $g_{co} \approx g_{u.average} \cdot X_{eff.maximum} \approx 0.41(W/(m^2 \cdot °C)) \cdot 885m = 363\ (W/(m \cdot °C))$ is a reasonable *maximum* value **[see SS9].**

In this case the Global Ocean(60N-60S) average of CC* (and a corresponding maximum ECS (see sect. **E**)) can then be easily calculated using the equations es9.7,8 :

for $g_{u.average}$ = **.2** , <u>**.41**</u>, **.6**, **.8** W/(m²·°C) respectively, then

es5.11 Global Ocean : CC* = **4.3**, <u>**6.2**</u>, **7.5**, **8.6** Watt·√[yrs]/(m²·°C),

es5.12 Global Ocean (60N-60S): CC* = **4.8**, <u>**6.9**</u>, **8.3**, **9.6** Watt·√[yrs]/(m²·°C),

and corresponding $ECS_{G.Ocean\ (60N-60S)}$ = **1.45**, <u>**1.53**</u>, **1.56**, **1.59** °C respectively

; where CC*(Global) ≈ (0.9)·CC*(60N-60S) [see Main sect. **D**, last paragraph] .

Note, the calculated value of ECS for the Global Ocean (60N-60S) varies by only + 2% over a 50% increase of $g_u$'s most likely value (=0.41), and these are certainly *maximal* values calculated using *maximal* "X" values, using this Ocean profile method of evaluation.

.................................................................

Tabulating the results of methods **1), 2), 2B), 3)** above, and presuming $g_{u.ave}$ =0.41, yields

| | method: | **(1)** | **(3)** | **(2B)** | **(2)** | |
|---|---|---|---|---|---|---|
| es5.13 | CC*[Global Ocean (60N-60S)] = | 7.75 | <u>6.9</u> | (6.04) | 5.47 | Watt·√[yrs]/(m²·°C) |
| es5.14 | CC*[Global Ocean(**$g_u$**=0.41·(0.9))] = | 6.7 | <u>6.2</u> | (5.51) | 5.1 | |

The results above of method **1)** are not considered as realistic as method **2)** (see discussion after e.16 sect. **D**). Results of the speculated hybrid Forcing **2B)** (in brackets) seem to most closely reproduce NOAA observed values of Ocean Energy and Power flow over time, and may be the most likely. And the results of method **3)** are considered particularly robust and independent *maximum* values. Therefore the results of **(2), (2B)** and **(3)** above do seem to corroborate each other; where the lower values of method **(2B)** are then considered most likely herein, but not definitively. The values of method **(3)** are used as the *maximal* choice in calculations reported herein.

**SS6 Coarse GCM v.4** Michael D. Mill...Mar. 2024 [Contact: m.d.mill.climate@gmail.com]

A simplified coarse 3-layer 2-region(Ocean, Land) formal/empirical algebraic small signal GCM is developed using atmospheric radiative parameters evaluated using the MODTRAN[4] [see [ref.4] in **SS6.3**] application, using others estimated, and others evaluated by matching model temperature results to observations. Figure Fig.SS6.1 diagrams the Thermal Circuit for the Oceans region as used in this development. The MODTRAN evaluations are placed at the end of this subsection.

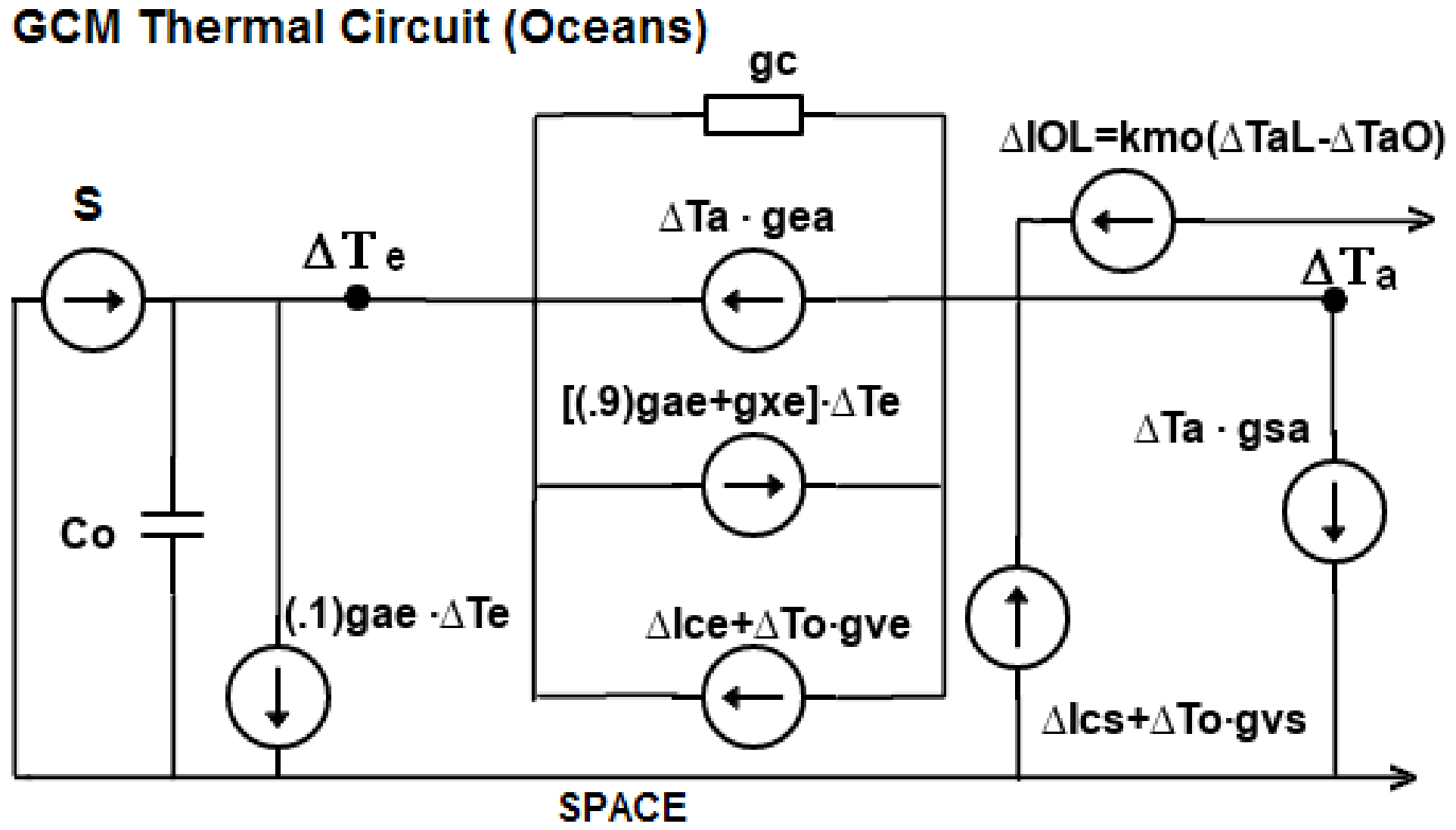

Fig.SS6.1 (Note: all crossing lines are connected)

In this model the earth is modeled as an Ocean region adjacent to a Land region, since the difference between "wet" water and "dry" land (i.e. the thermal capacitive and evaporative properties) is by far the most important to model. The Ocean region is explicitly separated into 3 layers: 1) the ocean body, 2) the lowest 100 meters of the atmosphere above the ocean (the surface atmosphere), 3) and the bulk atmosphere above this. **[**The Ocean is also implicitly separated into a mix layer and a deep layer. The bulk atmosphere is also implicitly separated into the troposphere and the stratosphere.**]** There are then 3 associated nodes: 1) the earth surface (i.e. the Ocean surface or Land surface) which represents the Δtemperature (i.e. $\mathbf{\Delta T_e}$ Fig.SS6.1) of the surface atmosphere, 2) some average ΔT of the bulk atmosphere (i.e. $\mathbf{\Delta T_a}$ Fig.SS6.1), 3) and **space** with $\Delta T \equiv 0$. Using 2 layers for the atmosphere is certainly a coarse approximation but will be sufficient to allow for any significant difference between $\mathbf{\Delta T_e}$ and $\mathbf{\Delta T_a}$ , or between *surface* forcing and *GHG atmospheric* forcing at equilibrium. The Ocean thermal circuit above is coupled to a similar mirrored Land circuit connected by the "→" lines shown. This is the exact analog of an electrical circuit where current flow is analogous to thermal power flow per unit area ($W/m^2$), voltage is analogous to Temperature, electrical conductivity is analogous to thermal conductivity ($(W/m^2)/°C$), electrical capacitance is analogous to thermal capacitance* ($(W{\cdot}yrs/m^2)/°C$), and where power flow is conserved at the nodes. Note, in this circuit these subscripts have the following meanings:

L ≡ "Land"
O ≡ "Ocean"
e ≡ "earth surface" or earth surface atmosphere (Land *or* Ocean)
s ≡ "space"
su ≡ "sun"
v ≡ "water vapor"

a ≡ "bulk atmosphere"

Further, "$\Delta T$" is the change in temperature at the subscripted location, and "$\Delta I$" is the change in power flow per unit area of the subscripted type, and specifically

$\Delta T_O(t)$ ≡ change of Ocean surface temperature , $\Delta T_{aO}(t)$ ≡ change of Ocean bulk atmosphere temperature
$\Delta T_L(t)$ ≡ change of Land surface temperature , $\Delta T_{aL}(t)$ ≡ change of Land bulk atmosphere temperature .

In this model *only* the Ocean thermal capacitance element is a source of temperature response time delay to an independent source, when time averaged over long periods (> 1 year).

**[[** It is proposed here *fundamentally* that a surface perturbation temperature increase over land only (*all else constant*) has little long term average effect on atmospheric water vapor density over land. Land is fundamentally "dry" as opposed to the oceans which evaporate more when warmed...all lands would soon become deserts except for the onshore atmospheric water vapor flow from the oceans. Land warming only produces a secondary effect, where 80% of all land rainfall re-evaporates to the atmosphere regardless of any perturbation Land temperature increase. Such warming *does* increase the local saturation humidity, and thus "evaporates" cloud mist droplets. This decreases solar reflection and is an enhanced Land positive rein**forcing** *compared to the Oceans*.

Therefore, *only* Ocean surface perturbation temperature changes significantly effect global atmospheric water vapor density in this model. This model proposal is either an *extremely* economical and insightful characterization of the hydrological process on the regional scale, or it is not. The reader may decide. But this relation, beyond any other, establishes the fundamental character of regional temperature response to Forcing that follows. See the application to **1)**, **9)**, **10)** below. **However, the alternative hydrological supposition is also considered at the end of SS7**, so the "correct" model result can be bracketed between these two extremes!

Further, the "cloud effect" here is purely reflective. It is, and should be, completely independent from a water vapor "green house" effect, for first order perturbations. If a cloud slightly evaporates to a reduced density, the number or density of water molecules in that region does not change, nor does the corresponding green house effect (to first order). Similarly, if water vapor molecules are condensed into a slightly increased cloud density, the number or density of water molecules in that region does not change, nor does the corresponding green house effect (to first order). It does not matter if the clouds are high or low, warm or cold; the reflection is the same (to first order). The temperature of higher altitude molecules *does* effect the radiation of green house molecules, and this effect is already contained within the independent water vapor rein**forcing** term. But radiation absorption/emission is independent of the form of the molecules (gas vapor or micro (~1 micron) water droplet mist) for long wave infrared radiation (~10 micron), for first order perturbations. It is a common and inappropriate modeling formulation to confuse or equate the "cloud effect" with a "green house effect" for first order "feedback" perturbations.**]]**

All the thermal circuit components of Fig.SS6.1 will now be described:

**1)** The S source is of the form $S=\Delta I_{su} + \beta^{*}\cdot\Delta T_O + \gamma^{*}\cdot\Delta T_a + g_{sn}\cdot\Delta T_e$
;where $\Delta I_{su}$ is the direct *independent* solar forcing variation at the surface only; $\beta^{*}\cdot\Delta T_O$ is the solar forcing reduction due to increased cloud formation (i.e. cloud condensation) caused by increased water evaporation at the ocean surface proportional to the increase in Ocean (only) surface $\Delta T_O$, and where **$\beta^{*}$ ((W/m$^2$)/°C)** should only be negative **[**For the **Land** region the results are similar except the change in water vapor density is multiplied by a scaler "**$\mu$**" (<1) , so the change of absorbed Δpower.flow/area is given by $\mu\cdot\beta^{*}\cdot\Delta T_O(t)$**]**; $\gamma^{*}\cdot\Delta T_a$ is the solar forcing increase due to decreased cloud formation caused by increased saturation humidity (i.e. cloud evaporation) proportional to the increase in local atmospheric $\Delta T_a$ , and where **$\gamma^{*}$ ((W/m$^2$)/°C)** should only be positive; and $g_{sn}\cdot\Delta T_e$ is the solar forcing increase due to decreased solar snow and ice reflection caused by increased melt proportional to the increase local surface temperature $\Delta T_e$ and where **$g_{sn}$ ((W/m$^2$)/°C)** should only be positive. Note, $\beta^*$ and $\gamma^*$ are evaluated by matching temperature model results to observations. The value of $g_{sn}$ is taken from IPCC AR6.

**2)** $C_O$ represents the total Ocean thermal capacitance/area. However the change in total ocean thermal energy/area, given by $\Delta E^*(t)$, is an observed quantity. Thus the total power flow/area is given by $d(\Delta E^*(t))/dt \equiv Q(t)$. And the change in power flow into Ocean E* storage as measured from $t_o$ is $\mathbf{\Delta Q(t)=Q(t)-Q(t_o)}$ , and this is the form used in the thermal circuit equations.

**3)** The radiated Δpower flow/area out of the surface atmosphere (and surface) into the bulk atmosphere due to temperature increase at the surface is given by $g_{ae}\cdot\Delta T_e(t)$ , where $\mathbf{g_{ae}}$ **((W/m$^2$)/°C ) >0** is evaluated using MODTRAN[4]. However 10% of this radiation (see Fig.SS6.2) flows unabsorbed directly to space. This is given by $(0.1)\cdot g_{ae}\cdot\Delta T_e(t)$ , and the portion absorbed into the bulk atmosphere is given by $(0.9)\cdot g_{ae}\cdot\Delta T_e(t)$ .

**4)** The radiated Δpower flow/area out of the bulk atmosphere into the surface atmosphere (or surface) due to bulk atmosphere temperature increase is given by $g_{ea}\cdot\Delta T_a(t)$ , where $\mathbf{g_{ea}}$ **((W/m$^2$)/°C) >0** is evaluated using MODTRAN.

**5)** An increasing surface temperature also *enhances* the "lift" rate of latent and sensible water vapor energy from the surface layer to the higher bulk layer altitudes over the Ocean surface (due to lapse rate variation and ocean evaporation). This occurs to a lesser extent over the Land regions also. So the general relation is given by Δpower flow/area = $g_{xe}\cdot\Delta T_e(t)$ ; where $\mathbf{g_{xO}}$ **and** $\mathbf{g_{xL}}$ **((W/m$^2$)/°C) >0** are individually evaluated by matching temperature model results to observations.

**6)** An increasing surface temperature also *enhances* the "lift" of thermal energy by convection from the surface layer to the higher bulk layer altitudes. The general relation is given by Δpower flow/area = $g_c\cdot(\Delta T_e(t) - \Delta T_a(t))$ ; where $\mathbf{g_c}$ **((W/m2)/°C) >0** is evaluated by educated estimate (see Fig.SS6.2)... $g_c \sim(17\ W/m^2)/(294K\text{-}267K) \sim 0.6((W/m^2)/°C)$ [294K and 267K are the atmospheric temperatures at the surface and 5 km, respectively.]

**7)** The decrease of radiated Δpower flow/area out of the bulk atmosphere into space, due to an increase of bulk atmosphere $\Delta CO_2$ direct independent forcing, is given by $\Delta I_{cs}$ ; where $\mathbf{\Delta I_{cs}}$ **((W/m$^2$)) >0 if** $\mathbf{\Delta CO_2}$ **>0**, and is evaluated using MODTRAN....see Fig.SS6.1. Note, this $\Delta I_{cs}$ is defined as **Direct** $CO_2$ Radiative Forcing (DRF or RF) that occurs only after equilibration of the stratospheric temperature (see section **G.1** and **SS6.3.2**)

**8)** The increase of radiated Δpower.flow/area out of the bulk atmosphere into the surface atmosphere (or surface) due to bulk atmosphere $\Delta CO_2$ direct independent forcing increase is given by $\Delta I_{ce}$ , where $\mathbf{\Delta I_{ce}}$ **((W/m$^2$)) >0 if** $\mathbf{\Delta CO_2>0}$ is evaluated using MODTRAN.

**9)** For the **Ocean** region only, the decrease of radiated Δpower flow/area out of the bulk atmosphere into space due to the increase of bulk atmosphere water vapor (a GHG) is given by $g_{vs}\cdot\Delta T_O(t)$ ; where the increase in bulk water vapor density is *only* caused by the increase in Ocean surface temperature and its evaporation rate, and where $\mathbf{g_{vs}}$ **((W/m$^2$)/°C) >0** is estimated using MODTRAN...see Fig.SS6.1 . For the **Land** region the results are similar except the change in water vapor density is multiplied by a scaler "**μ**" (**<1**) , so the decrease of radiated Δpower flow/area is given by $\mu\cdot g_{vs}\cdot\Delta T_O(t)$ {see **1)** [bracketed comments] above}.The value of μ is an educated and formal/emperically modeled estimate, and is limited and tested by various trial cases.

**10)** The increase of radiated Δpower.flow/area out of the bulk atmosphere into the surface atmosphere (or surface) due to the increase of bulk atmosphere water vapor (a GHG) is given by $g_{ve}\cdot\Delta T_O(t)$ ; where the increase in bulk water vapor density is *only* caused by the increase in Ocean surface temperature and its evaporation rate, and where $\mathbf{g_{ve}}$ **((W/m$^2$)/°C) >0** is estimated using MODTRAN. For the **Land** region the results are similar except the change in water vapor density is multiplied by a scaler "**μ**" (<1) , so the change of radiated Δpower flow/area is given by $\mu\cdot g_{ve}\cdot\Delta T_O(t)$

**11)** The radiated Δpower flow/area out of the bulk atmosphere into space due to bulk atmosphere temperature increase is given by $g_{sa}\cdot\Delta T_a(t)$ , where $\mathbf{g_{sa}}$ **((W/m$^2$)/°C) >0** is evaluated using MODTRAN.

**12)** The increase of thermal Δpower.flow/area into the Ocean regions from the Land regions due to atmospheric currents (or mixing) is approximated by $\Delta I_{OL}=k_{mO}\cdot(\Delta T_{aL}-\Delta T_{aO})$ ; where $\mathbf{k_{mO}}$ **((W/m$^2$)/°C) >0** , and whose value is unknown, but *is* limited and tested by various trial cases, and Occam's Razor. The warmer region injects thermal energy into the cooler one. Similarly for Land, $\Delta I_{LO}=k_{mL}\cdot(\Delta T_{aO}-\Delta T_{aL})$ ; where $\mathbf{k_{mL}}$ **((W/m$^2$)/°C) >0** . Conservation of power flow requires that $A_L\cdot\mathbf{k_{mL}} = \mathbf{k_{mO}}\cdot A_O$ , or $\mathbf{k_{mL}} = \mathbf{k_{mO}}\cdot(A_O/A_L)$, where $A_O$ and $A_L$ are the total Ocean and Land areas, respectively.

---

## SS6.2 Circuit Solutions

Conservation of power *into* node $\Delta T_e$ (see Fig.SS6.1) yields:

es6.1 $$0 = \Delta I_{su} + \beta^*\cdot\Delta T_O + \gamma^*\cdot\Delta T_a + g_{sn}\cdot\Delta T_e - \Delta Q(t) - (.1)g_{ae}\cdot\Delta T_e + (\Delta T_a-\Delta T_e)\cdot g_c + g_{ea}\cdot\Delta T_a + \Delta I_{ce} + \Delta T_O\cdot g_{ve} - ((.9)g_{ae}+g_{xe})\cdot\Delta T_e$$

es6.2 $$0 = \Delta I_{su} + \Delta I_{ce} + \Delta T_O\cdot(\beta^* + g_{ve}) + \Delta T_e\cdot[g_{sn} - g_c - (.1)g_{ae} - g_{xe} - (.9)g_{ae}] - \Delta Q(t) + \Delta T_a\cdot[\gamma^* + g_{ea} + g_c]$$

Define the ratios:

es6.3 $$\boldsymbol{\alpha \equiv \Delta I_{ce}/\Delta I_{cs}} \;;\; \boldsymbol{\nu \equiv g_{ve}/g_{vs}} \;;\; \boldsymbol{\sigma \equiv \Delta T_O/\Delta T_L}$$

; where e.2 and these proportionalities hold over the entire time evaluation range (except for σ), or specifically at t'=zero or t'=infinity.
**[**Note "σ" is not necessarily known at this point, but can be evaluated easily by iterations of the final solution using the Excel "Solver" tool (see **SS7.2**). This technique greatly simplifies the algebra.**]**

For the Ocean region $\mathbf{\Delta T_e=\Delta T_O}$ . Then es6.2 can be rewritten:

es6.4 $$\Delta T_O = [\Delta I_{su} + \Delta I_{cs}\cdot\alpha + \Delta T_{aO}\cdot\mathbf{E} - \Delta Q(t)]/\mathbf{D}$$

;where $\mathbf{E} \equiv [\gamma^* + g_{ea} + g_c]$ , and
$\mathbf{D} \equiv [g_c + (1)g_{ae} + g_{xO} - g_{sn} - (\beta^* + \nu\cdot g_{vs})]$ .

For the Land region $g_{vs}$ is replaced by $\mu\cdot g_{vs}$ (see **9)** and **10)** above) , and so using "σ" above $g_{vs}\cdot\Delta T_O(t)$ is replaced by $\mu\cdot\sigma\cdot g_{vs}\cdot\Delta T_L(t)$. <u>And similarly $\beta^*\cdot\Delta T_O$ is replaced by $\mu\cdot\sigma\cdot\beta^*\cdot\Delta T_L$</u> (see **1)** above). Also $\Delta Q(t)$ is presumed zero over the Land region, and $g_{xO}$ is replaced by $g_{xL}$. All other radiative characteristics are presumed the same. And the independent forcing over Land is presumed the same as over the Oceans.
Using these replacements, es6.2 can then be rewritten for the Land region, where $\mathbf{\Delta T_e=\Delta T_L}$, as:

es6.5 $$\Delta T_L = [\Delta I_{su} + \Delta I_{cs}\cdot\alpha + \Delta T_{aL}\cdot\mathbf{E}]/\mathbf{DL}$$

;where $\mathbf{E} \equiv [\gamma^* + g_{ea} + g_c]$ , and
$\mathbf{DL} \equiv [g_c + (1)g_{ae} + g_{xL} - g_{sn} - \mu\cdot\sigma\cdot(\beta^* + \nu\cdot g_{vs})]$ .

---

Conservation of power into node $\Delta T_a$ for the Ocean (see Fig S6.1) yields:

es6.6 $0=\Delta I_{OL} + \Delta I_{cs} + \Delta T_O \cdot g_{vs} - \Delta T_a \cdot g_{sa} + (\Delta T_e - \Delta T_a) g_c - \Delta T_a \cdot g_{ea} + ((.9) g_{ae} + g_{xe}) \cdot \Delta T_e - \Delta I_{ce} - \Delta T_O \cdot g_{ve}$

;where $\Delta I_{OL}=k_{mO} \cdot (\Delta T_{aL} - \Delta T_{aO})$ (see **12)** above).

Define the ratios

es6.7 $\boldsymbol{\lambda \equiv \Delta T_{aO}/\Delta T_{aL}}$ , $\boldsymbol{\rho \equiv (A_O/A_L)}$

[Note "λ" is not necessarily known at this point, but can be evaluated easily by iterations of the final solution using the Excel "Solver" tool (see **SS7.2**). This technique greatly simplifies the algebra.]

Then rewrite es6.6 as,

es6.8 $0=\Delta I_{cs} + \Delta T_O (g_{vs} - g_{ve}) + \Delta T_e ( (.9) g_{ae} + g_c + g_{xe}) + \Delta T_{aO} (1/\lambda - 1) k_{mO} + \Delta T_a ( - g_{sa} - g_c - g_{ea}) - \Delta I_{ce}$

For the Ocean region $\boldsymbol{\Delta T_e = \Delta T_O}$ , $\boldsymbol{\Delta T_a = \Delta T_{aO}}$ . Then, using es6.3, es6.8 can be rewritten:

es6.9 $\Delta T_{aO} = [\Delta I_{cs} \cdot (1-\alpha) + \Delta T_O \cdot \mathbf{A}]/\mathbf{B}$

;where $\mathbf{A} \equiv [g_{vs}(1-\nu) + (.9) \cdot g_{ae} + g_c + g_{xO}]$ , and
$\mathbf{B} \equiv [ (1-1/\lambda) k_{mO} + g_{sa} + g_c + g_{ea}]$ .

.....................................

For the Land region es6.8 is rewritten

es6.10 $0=\Delta I_{cs} + \Delta T_O \cdot \mu \cdot (g_{vs} - g_{ve}) + \Delta T_e ( (.9) g_{ae} + g_c + g_{xe}) + \Delta T_{aL} (\lambda - 1) \cdot \rho \cdot k_{mO} + \Delta T_a (- g_{sa} - g_c - g_{ea}) - I_{ce}$

;where $\Delta I_{LO} = k_{mO} \cdot \rho (\Delta T_{aO} - \Delta T_{aL})$ (see **12)** above)
$= k_{mO} \cdot \rho \cdot \Delta T_{aL} (\lambda - 1)$

Again, for the Land region $\boldsymbol{\Delta T_e = \Delta T_L}$ , $\boldsymbol{\Delta T_a = \Delta T_{aL}}$ . Then, using es6.3, es6.10 can be rewritten:

es6.11 $\Delta T_{aL} = [\Delta I_{cs} \cdot (1-\alpha) + \Delta T_L \cdot \mathbf{AL}]/\mathbf{BL}$

;where $\mathbf{AL} \equiv [\mu \cdot \sigma \cdot g_{vs}(1-\nu) + (.9) g_{ae} + g_c + g_{xL}]$ , and
$\mathbf{BL} \equiv [ (1-\lambda) \cdot \rho \cdot k_{mO} + g_{sa} + g_c + g_{ea} ]$ .

---

Therefore, the full solution for the Ocean surface, using es6.4 and es6.9, is:

es6.12 $\Delta T_O = [\Delta I_{su} + \Delta I_{cs} \cdot \alpha + \Delta T_{aO} \cdot \mathbf{E} - \Delta Q(t)]/\mathbf{D}$
$= \{ \Delta I_{su} + \Delta I_{cs} \cdot \alpha + \mathbf{E}[\Delta I_{cs} \cdot (1-\alpha) + \Delta T_O \cdot \mathbf{A}]/\mathbf{B} - \Delta Q(t) \}/\mathbf{D}$ or,

$\Delta T_O [\mathbf{D} - \mathbf{E} \cdot \mathbf{A}/\mathbf{B}] = \Delta I_{su} + \Delta I_{cs} \cdot \alpha + \Delta I_{cs} \cdot \mathbf{E} \cdot (1-\alpha)/\mathbf{B} - \Delta Q(t)$ .

Let

es6.12.b $\mathbf{F} \equiv \mathbf{E}/\mathbf{B}$ , or $\mathbf{F} \equiv [\gamma^* + g_{ea} + g_c] / [(1-1/\lambda)\cdot k_{mO} + g_{sa} + g_c + g_{ea}]$ .

Then, es6.12 is rewritten:

es6.13 $\Delta T_O = \{\Delta I_{su} + \Delta I_{cs}\cdot\mathbf{CS} - \Delta Q(t)\}/\{\mathbf{D}\text{-}\mathbf{F}\cdot\mathbf{A}\}$

;where $\mathbf{CS} \equiv [\alpha + \mathbf{F}(1-\alpha)]$ , or

es6.14 $\Delta T_O(t) = [\Delta I_{su}(t) + \Delta I_{cs}(t)\cdot\mathbf{CS} - \Delta Q(t)] / G^*_O$

;where $G^*_O \equiv \mathbf{D}\text{-}\mathbf{F}\cdot\mathbf{A}$ .

It is also convenient to rewrite $G^*_O$ as:

es6.14.b $G^*_O = g_c[1\text{-}\mathbf{F}] + g_{ae}[1\text{-}(.9)\mathbf{F}] + g_{xO}[1\text{-}\mathbf{F}] - g_{sn} - \beta^* - g_{vs}[\nu + \mathbf{F}(1\text{-}\nu)]$

; which allows us to easily determine the water vapor rein**forcing** component ($g_{vs}[\nu + \mathbf{F}(1\text{-}\nu)]$), and the snow albedo rein**forcing** ($g_{sn}$). Evaluating the Cloud Reflection Rein**forcing** is more involved since $\gamma^*$ is contained within **F**, so es6.14 is more appropriate.

However, for GHG forcing (where $\Delta I_{su}(t)$=0) $G_{eff.O.GHG}$ must be defined as in e.1 and es1.37, i.e.

es6.15 $\Delta T_O(t)\cdot G_{eff.O.GHG} \equiv \Delta I_{out}(t) = \Delta I_{cs}(t) - \Delta Q(t)$ ; where $\Delta I_{cs}(t)$ is the only external independent forcing term into the earth system (from space).

Using es6.14 to replace $\Delta I_{cs}(t)$ in es6.15 then yields

es6.15b $G_{eff.O.GHG} = G^*_O/\mathbf{CS} + [\Delta Q(t) / \Delta T(t)]\cdot[1/\mathbf{CS} - 1]$

After parameter values are determined by comparison to observed values (see SS**7.2**), we find realistically that $1 \geq \mathbf{CS} \geq .9$ , and realistically the last term of es6.15b is only ≤ +2.5% of the total. Here $G_{eff.O.GHG}$ is the effective thermal conductance for GHG (Green House Gas...e.g. $CO_2$) forcing only. And $\Delta I_{cs}$ is the direct $CO_2$ external radiative forcing (DRF or RF) into the earth. Note alternatively, in es6.14, "$\Delta I_{cs}\cdot\mathbf{CS}$" can be thought of as the *effective* radiative forcing (ERF), which is used in most of the recent literature.
.............................................

Similarly, the full solution for the Land surface, using es6.5 and es6.11, is:

es6.16 $\Delta T_L = [\Delta I_{su} + \Delta I_{cs}\cdot\alpha + \Delta T_{aL}\cdot\mathbf{E}]/\mathbf{DL}$
$= \{\Delta I_{su} + \Delta I_{cs}\cdot\alpha + \mathbf{E}\cdot[\Delta I_{cs}\cdot(1-\alpha) + \Delta T_L\cdot\mathbf{AL}]/\mathbf{BL}\}/\mathbf{DL}$ or,

$\Delta T_L[\mathbf{DL} - \mathbf{E}\cdot\mathbf{AL}/\mathbf{BL}] = \Delta I_{su} + \Delta I_{cs}\cdot\alpha + \Delta I_{cs}\cdot\mathbf{E}\cdot(1-\alpha)/\mathbf{BL}$ .

Let $\mathbf{FL} \equiv \mathbf{E}/\mathbf{BL}$ , or $\mathbf{FL} \equiv [\gamma^* + g_{ea} + g_c] / [(1-\lambda)\cdot\rho\cdot k_{mO} + g_{sa} + g_c + g_{ea}]$ .

Then, es6.16 is rewritten:

es6.17 $\Delta T_L = \{\Delta I_{su} + \Delta I_{cs}\cdot\mathbf{CSL}\}/\{\mathbf{DL}\text{-}\mathbf{FL}\cdot\mathbf{AL}\}$

;where $\mathbf{CSL} \equiv [\alpha + \mathbf{FL}(1-\alpha)]$ , or

es6.18 $\Delta T_L(t) = [\Delta I_{su}(t) + \Delta I_{cs}(t)\cdot \mathbf{CSL}] / G^*_L$

;where $G^*_L \equiv \mathbf{DL\text{-}FL\cdot AL}$ .

As above, according to the definition of $G_{eff}$ in e.1 and es1.37, $G_{eff.L.GHG}$ can be evaluated rewriting es6.18:

es6.19 $\Delta T_L(t) = \Delta I_{su}(t) / G^*_L + \Delta I_{cs}(t) / G_{eff.L.GHG}$

;where $G_{eff.L.GHG} \equiv G^*_L / \mathbf{CSL}$ , and $\Delta I_{su}$ is thought to be essentially zero.

Again, $G_{eff.L.GHG}$ is the effective thermal conductance for GHG (Green House Gas...e.g. $CO_2$) forcing only. And $\Delta I_{cs}$ is the only *direct* external $CO_2$ radiative forcing (DRF or RF) into the earth.

[ In actuality 32% (see Fig. S6.2) of the short wave solar radiation ($\Delta I_{su}$) is first absorbed in the bulk atmosphere. The general solution for both solar and $CO_2$ forcing is then approximated by:

es6.14b $\Delta T_O = [(0.68)\Delta I_{su} + (0.32)\Delta I_{su}\cdot\mathbf{F} + \Delta I_{cs}\cdot\mathbf{CS} - \Delta Q(t)] / G^*_O$

es6.18b $\Delta T_L = [(0.68)\Delta I_{su} + (0.32)\Delta I_{su}\cdot\mathbf{FL} + \Delta I_{cs}\cdot\mathbf{CSL}] / G^*_L$ ]

In section **SS7** a spread sheet evaluation of this model is used (using observed variable histories) to predict equilibrium $ECS_{true}$ as compared to the steady state linear empirical $ECS_{eff}$ calculated herein, and more.

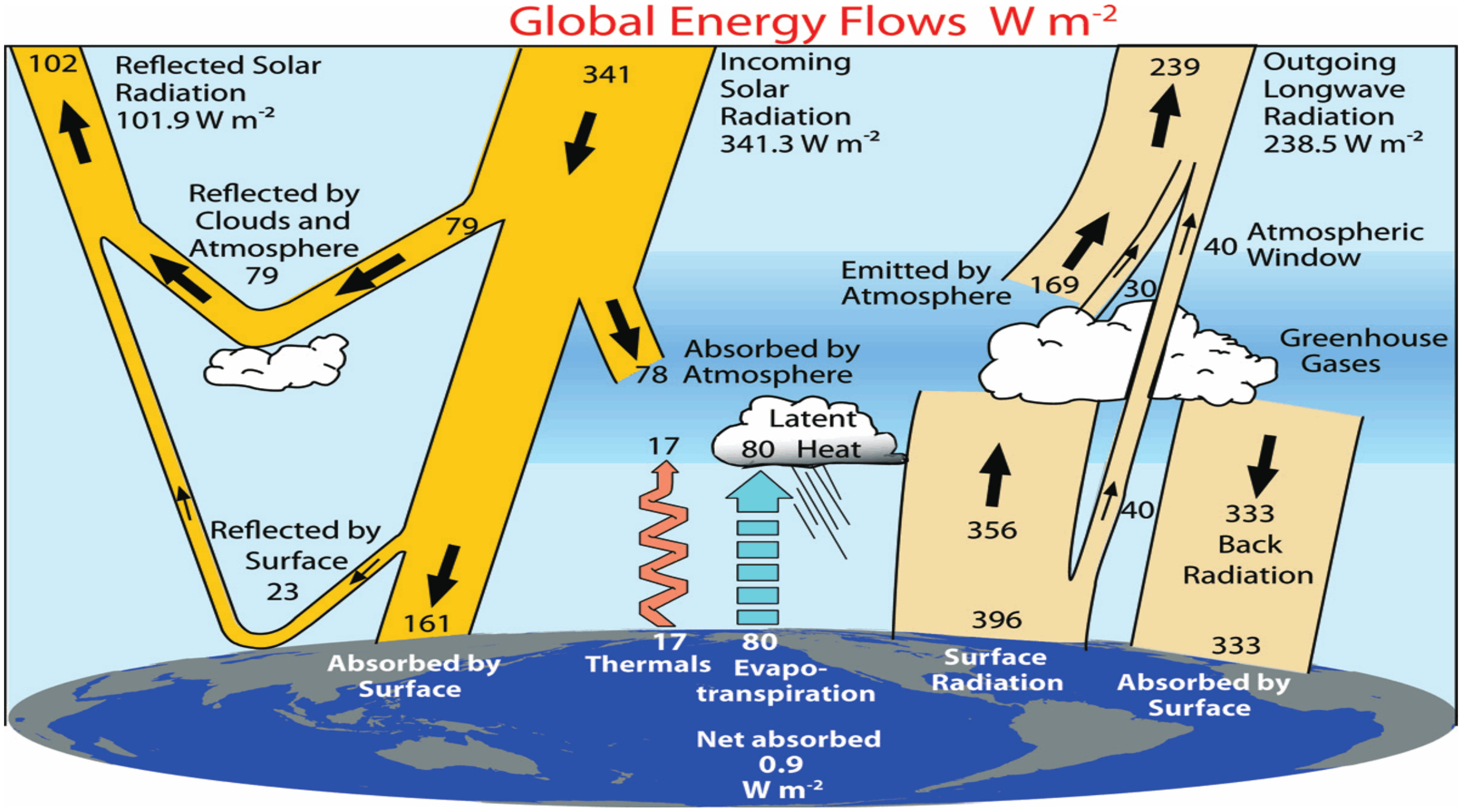

Fig.SS6.2 Trenberth et al, 2009 [CC BY-NC-SA 4]

---

## <u>SS6.3 MODTRAN parameter evaluations</u>

Evaluations of $g_{sa}$, $g_{ae}$, $g_{ea}$, $g_{vs}$, $g_{ve}$, and $\alpha \equiv \Delta I_{ce} / \Delta I_{cs}$ are possible using various options of the versatile MODTRAN[4] application. It is found by trial that the "Mid-latitude Summer" atmospheric model represents the

[4] https://climatemodels.uchicago.edu/modtran/

Global average atmospheric model very well. Further, the cloud cover is modeled by a heavy Strato/Stratus CU (0.66 km base to 2 km top) in effect 20% of the time. Clear skies are assumed for the other 80% . A 30% coverage of thinner clouds is also used as described, where appropriate. These are all, of course, coarse approximations. The change of "up" and "down" radiation is calculated at various altitudes for the following change of conditions:

**.1**) $\Delta T_{surface}$ = + 1 °C; where water vapor density remains constant, and $CO_2$ = 400 ppm. This permits evaluation of the $g_{sa}$, $g_{ae}$, $g_{ea}$ parameters. [Note: the entire atmospheric column changes by the same amount.]
**.2**) CO2 is doubled from 400 ppm to 800 ppm ; where $\Delta T_{surface}$ = 0 °C , and vapor density remains constant. This permits evaluation of the $\Delta I_{ce}/(2xCO_2)$, $\Delta I_{cs}/(2xCO_2)$, and $\alpha \equiv \Delta I_{ce} / \Delta I_{cs}$ parameters.
**.3**) a $\Delta T_{surface}$ = + 1 °C where water vapor density remains constant *is changed* to a $\Delta T_{surface}$ = + 1 °C where water vapor Relative Humidity remains constant. This approximately models the effects of increasing surface temperature on atmospheric water vapor density, and that water vapor on radiation. This permits evaluation of the $g_{vs}$ and $g_{ve}$ parameters.

**ΔI units are W/m²**, ↑ indicates radiation up (towards space), ↓ indicates radiation down (towards earth surface).

---

**SS6.3 .1a** Calculated with cloud cover

surface ↑ : ΔI=422-416.4=5.6
100m ↑ : ΔI=422.96-417.3=5.66
100m ↓ : ΔI=420.4-414.8=5.6
12km ↑ : ΔI=284.5-280.3=4.2
12km ↓ : ΔI=2.85-23.40=0.45
70km ↑ : ΔI=277.7-273.6=4.1

The non-absorbing radiation window feed-through Δpower (i.e. 10% of surface ↑ = (0.1)(5.60)=0.56) is not due to tropospheric warming, so remove this effect artificially:

12km ↑ : ΔI=4.2-0.56=3.64
70km ↑ : ΔI=4.1-0.56=3.54

Stratospheric + 1 °C warming causes 0.45 W/m² of 12km ↓ (see above). We will assume it also causes 0.45 W/m² of 70km ↑ . For sparse atmospheres the radiation is nearly the same in both directions. Since this is not due to tropospheric warming, this effect is also artificially removed:

12km ↓ : ΔI=0.45-0.45=0
70km ↑ : ΔI=3.54-0.45=3.09 .

So, the net Δpower *into* the stratosphere due to tropospheric warming is given by:

ΔI=(12km↑)-(12km↓)-(70km↑)=3.64-0.0-3.09=0.55 .

After stratospheric equilibration (i.e. warming) this $\Delta I_{stratosphere}$ is radiated ½ upward and ½ downward. So:

12km ↓ : ΔI=0.0+(0.55/2)=0.275
70km ↑ : ΔI=3.09+(0.55/2)=3.365 .

Therefore:

12km ↑(total)=(12km↑)-(12km↓)=3.64-0.275=3.365
70km ↑(total)=3.365 .

Note that 12km↑(total) and 70km↑(total) are identical, as is expected after stratospheric equilibration.

The following parameters can then be evaluated from above:

$g_{sa}$=12km↑(total)=3.365 (W/$m^2$)/°C
$g_{ae}$=100m↑ =5.66 (W/$m^2$)/°C
$g_{ea}$=100m↓ =5.60 (W/$m^2$)/°C

**SS6.3 .1b** Calculated for clear sky

surface ↑ : ΔI=422.0-416.34=5.64
100m ↑ : ΔI=422.96-417.31=5.65
100m ↓ : ΔI=325.93-322.16=3.77
12km ↑ : ΔI=301.38-396.98=4.40
12km ↓ : ΔI=23.15-22.71=0.44
70km ↑ : ΔI=293.87-289.60=4.27

The non-absorbing radiation window feed through Δpower (i.e. 10% of surface ↑ = (0.1)(5.60)=0.56) is not due to tropospheric warming, so remove this effect artificially:

12km ↑ : ΔI=4.4-0.56=3.84
70km ↑ : ΔI=4.27-0.56=3.71

Stratospheric + 1 °C warming causes 0.44 W/$m^2$ of 12km ↓ (see above). We will assume it also causes 0.44 W/$m^2$ of 70km ↑ . For sparse atmospheres the radiation is nearly the same in both directions. Since this is not due to tropospheric warming, this effect is also artificially removed:

12km ↓ : ΔI=0.44-0.44=0.0
70km ↑ : ΔI=3.71-0.44=3.27 .

So, the net Δpower *into* the stratosphere is given by:

ΔI=(12km↑)-(12km↓)-(70km↑)=3.84-(0.0)-3.27=0.57 .

After stratospheric equilibration (i.e. warming) this $\Delta I_{stratosphere}$ is radiated ½ upward and ½ downward. So:

12km ↓ : ΔI=0.0+(0.57/2)=0.285
70km ↑ : ΔI=3.27+(0.57/2)=3.555 .

Therefore:

12km ↑(total)=(12km↑)-(12km↓)=3.84-0.285=3.555
70km ↑(total)=3.555 .

Note that 12km↑(total) and 70km↑(total) are identical, as is expected after stratospheric equilibration.

The following parameters can then be evaluated from above:

$g_{sa}$=12km↑(total)=3.555 (W/m$^2$)/°C
$g_{ae}$=100m↑ =5.65 (W/m$^2$)/°C
$g_{ea}$=100m↓ =3.77 (W/m$^2$)/°C .
......................................................

The average for all sky conditions is then:
**$g_{sa}$**=12km↑(total)=(0.2)(3.365) + (0.8)(3.555)=3.52 (W/m$^2$)/°C
**$g_{ae}$**=100m↑ =(0.2)(5.66) + (0.8)(5.65) =5.65 (W/m$^2$)/°C
**$g_{ea}$**=100m↓ =(0.2)(5.60) + (0.8)(3.77) =4.14 (W/m$^2$)/°C .

---

**SS6.3 .2a** Calculated with cloud cover

surface ↑ : ΔI=416.36-416.36=0
100m ↑ : ΔI=417.31-417.31=0
100m ↓ : ΔI=414.8-414.8=0
12km ↑ : ΔI=277-280.3=-3.3
12km ↓ : ΔI=25.2-23.4=1.8
70km ↑ : ΔI=270-273.6=-2.6

So, the net Δpower *into* the stratosphere is given by:

ΔI=(12km↑)-(12km↓)-(70km↑)=-3.3-1.8-(-2.6)= -2.5 .

After stratospheric equilibration (i.e. warming) this $\Delta I_{stratosphere}$ is radiated ½ upward and ½ downward. So:

12km ↓ : ΔI=1.8+(-2.5/2)=0.55
70km ↑ : ΔI=-2.6+(-2.5/2)=-3.85 .

Therefore:

12km ↑(total)=(12km↑)-(12km↓)=-3.30-0.55= -3.85
70km ↑(total)=-3.85 .

Note that 12km↑(total) and 70km↑(total) are identical, as is expected after stratospheric equilibration.

The following parameters can then be evaluated from above:

$\Delta I_{cs}$=-12km↑(total)=3.85 (W/m$^2$)/2x$CO_2$
$\Delta I_{ce}$=100m↓ =0 (W/m$^2$)/2x$CO_2$

**SS6.3 .2b** Calculated for clear sky

surface ↑ : ΔI=416.36-416.36=0
100m ↑ : ΔI=417.31-417.31=0
100m ↓ : ΔI=324.68-322.16=2.52
12km ↑ : ΔI=293.25-296.98=-3.73
12km ↓ : ΔI=24.53-22.71=1.82
70km ↑ : ΔI=286.59-289.60=-3.01

So, the net Δpower *into* the stratosphere is given by:

ΔI=(12km↑)-(12km↓)-(70km↑)=-3.73-1.82- (-3.01)=-2.54 .

After stratospheric equilibration (i.e. warming) this $\Delta I_{stratosphere}$ is radiated ½ upward and ½ downward. So:

12km ↓ : ΔI=1.82+(-2.54/2)=0.55
70km ↑ : ΔI=-3.01+(-2.54/2)=-4.28 .

Therefore:

12km ↑(total)=(12km↑)-(12km↓)=-3.73-0.55= -4.28
70km ↑(total)=-4.28 .

Note that 12km↑(total) and 70km↑(total) are identical, as is expected after stratospheric equilibration.

The following parameters can then be evaluated from above:

$\Delta I_{cs}$=-12km↑(total)=4.28 (W/m$^2$)/2x$CO_2$
$\Delta I_{ce}$=100m↓ =2.52 (W/m$^2$)/2x$CO_2$

..........................................................................................................................................

The average for all sky conditions is then:

$\mathbf{\Delta I_{cs}}$=(0.2)(3.85) + (0.8)*(4.28)=4.19 ($W/m^2$)/2x$CO_2$
$\mathbf{\Delta I_{ce}}$=($0.3^+$)(0) + ($0.7^-$)*(2.52)=$1.76^-$ ($W/m^2$)/2x$CO_2$ [i.e. $30^+$% "light" cloud cover].

$\boldsymbol{\alpha} \equiv \Delta I_{ce}/\Delta I_{cs} = 1.76^-/4.19 = 0.42^-$ .

---

<u>**SS6.3 .3a** Calculated with cloud cover</u>

surface $\uparrow$ : $\Delta$I=422.016-422.016=0
100m $\uparrow$ : $\Delta$I=422.96-422.96=0
100m $\downarrow$ : $\Delta$I=420.45-420.45=0
12km $\uparrow$ : $\Delta$I=283.51-284.52=-1.01
12km $\downarrow$ : $\Delta$I=24.25-23.85=0.40
70km $\uparrow$ : $\Delta$I=276.70-277.70=-1.00

So, the net $\Delta$power *into* the stratosphere is given by:

$\Delta$I=(12km$\uparrow$)-(12km$\downarrow$)-(70km$\uparrow$)=-1.01-0.4-(-1.00)=-0.41 .

After stratospheric equilibration (i.e. warming) this $\Delta I_{stratosphere}$ is radiated ½ upward and ½ downward. So:

12km $\downarrow$ : $\Delta$I=.40 +(-.41/2)=0.195
70km $\uparrow$ : $\Delta$I=-1.00+(-.41/2)=-1.205 .

Therefore:

12km $\uparrow$(total)=(12km$\uparrow$)-(12km$\downarrow$)=-1.01-0.195=-1.205
70km $\uparrow$(total)=-1.205 .

Note that 12km$\uparrow$(total) and 70km$\uparrow$(total) are identical, as is expected after stratospheric equilibration.

The following parameters can then be evaluated from above:

$g_{vs}$=-12km$\uparrow$(total)=1.205 ($W/m^2$)/˚C
$g_{ve}$= 100m$\downarrow$ = 0 ($W/m^2$)/˚C .

<u>**SS6.3 .3b** Calculated for clear sky</u>

surface $\uparrow$ : $\Delta$I=422.016-422.016=0
100m $\uparrow$ : $\Delta$I=22.96-422.96=0
100m $\downarrow$ : $\Delta$I=329.39-325.93=3.46
12km $\uparrow$ : $\Delta$I=300.03-301.38=-1.35
12km $\downarrow$ : $\Delta$I=23.53-23.15=0.38
70km $\uparrow$ : $\Delta$I=292.52-293.87=-1.35

So, the net $\Delta$power *into* the stratosphere is given by:

$\Delta$I=(12km$\uparrow$)-(12km$\downarrow$)-(70km$\uparrow$)=-1.35-0.38-(-1.35)=-0.38 .

After stratospheric equilibration (i.e. warming) this $\Delta I_{stratosphere}$ is radiated ½ upward and ½ downward. So:

12km $\downarrow$ : $\Delta$I= 0.38+(-0.38/2)=0.19
70km $\uparrow$ : $\Delta$I=-1.35+(-0.38/2)=-1.54 .

Therefore:

12km $\uparrow$(total)=(12km$\uparrow$)-(12km$\downarrow$)=-1.35-0.19=-1.54
70km $\uparrow$(total)=-1.54 .

Note that 12km$\uparrow$(total) and 70km$\uparrow$(total) are identical, as is expected after stratospheric equilibration.

The following parameters can then be evaluated from above:

$g_{vs}$=-12km$\uparrow$(total)=1.54 ($W/m^2$)/˚C
$g_{ve}$= 100m$\downarrow$ = 3.46 ($W/m^2$)/˚C
.....................................................................
The average for all sky conditions is then:

$\mathbf{g_{vs}}$ =(0.2)(1.205) + (0.8)(1.54)=1.47 ($W/m^2$)/˚C
$\mathbf{g_{ve}}$ =($0.3^+$)(0) + ($0.7^-$)(3.46)=$2.42^-$ ($W/m^2$)/˚C
....[i.e. $30^+$% "light" cloud cover]

## SS7 Equilibrium vs. Near Linear Steady ECS

The concept of an effective global average thermal conductance ($G_{eff}$) that is a true constant over time is only possible if $\Delta T(x,y,t)=\Delta T_{\alpha}(x,y)\cdot\Delta T_{\beta}(t)$ [see equation e.2]. Thus the energy budget methods rely on this presumption as well. In this case there is no "pattern effect" [see **SS.1**]. However, because e.2 does not hold exactly true for a forced temperature response up to equilibrium , the value of global average $G_{eff}$ does change slightly over time until equilibrium is nearly attained (after many centuries). The value of global average $\mathbf{G_{eff}}$, as evaluated for a near linear steady state condition, is *partly* corrected to the equilibrium value in section **SS7.1**. A more sophisticated analysis of Equilibrium vs. Near Linear Steady ECS utilizing the Coarse GCM (model) of **SS6** is undertaken in section **SS7.2** . [see definition of "ECS" in e.7c]

### SS7.1

Using e.5

es7.1 $G_{eff} =(\underline{\Delta}F-\underline{\Delta}Q)/\underline{\Delta}\mathbf{T}$

; using definitions e.4b,c,e,f for any given region. Under near linear steady conditions (".nls"...e.g. 1980-2020) the Global ("G") value of $\underline{\Delta}T$ is then given by

es7.2 $\underline{\Delta}\mathbf{T}_G = (\underline{\Delta}F_G-(.7)\cdot\underline{\Delta}Q_O) / G_{eff.G.nls}$

;where $\underline{\Delta}Q_O$ represents the Global Ocean average value, (.7) is the Ocean fraction of the entire Globe, and $(.7)\cdot\underline{\Delta}Q$ is then the Global average of $\underline{\Delta}Q$ . Similarly for the Ocean and Land regions respectively:

es7.3 $\underline{\Delta}\mathbf{T}_O = (\underline{\Delta}F_O-\underline{\Delta}Q_O) / G_{eff.O.nls}$

es7.4 $\underline{\Delta}\mathbf{T}_L = \underline{\Delta}F_L / G_{eff.L.nls}$

;where $\underline{\Delta}Q_L$ is assumed zero, and all $\Delta\mathbf{T}$ values are area weighted averages for that region. By definition

es7.5 $\underline{\Delta}\mathbf{T}_G = (.7)\underline{\Delta}\mathbf{T}_O + (.3)\underline{\Delta}\mathbf{T}_L$

;where (.7) and (.3) are the Global fractions of the Ocean and Land regions respectively. Rewrite es7.5 using es7.2,3,4 yielding:

es7.6 $(\underline{\Delta}F_G-(.7)\cdot\underline{\Delta}Q_O)/G_{eff.G.nls} = (.7)(\underline{\Delta}F_O-\underline{\Delta}Q_O)/G_{eff.O.nls} + (.3)\underline{\Delta}F_L/G_{eff.L.nls}$ , or

es7.7 $\underline{\Delta}F_G/G_{eff.G.nls} = (.7)\underline{\Delta}F_O/G_{eff.O.nls} - (.7)\underline{\Delta}Q_O/G_{eff.O.nls} + (.3)\underline{\Delta}F_L/G_{eff.L.nls} + (.7)\cdot\underline{\Delta}Q_O/G_{eff.G.nls}$ .

If we assert that $\underline{\Delta}F_G = \underline{\Delta}F_O = \underline{\Delta}F_L$ , and then dividing es7.7 by $\underline{\Delta}F_G$ yields:

es7.8 $1/G_{eff.G.nls} = (.7)/G_{eff.O.nls} + (.3)/G_{eff.L.nls} + (.7)(\underline{\Delta}Q_O/\underline{\Delta}F_O)(1/G_{eff.G.nls} - 1/G_{eff.O.nls})$ , or

es7.9 $(1/G_{eff.G.nls})[1 - (.7)(\chi)(1- G_{eff.G.nls}/G_{eff.O.nls})] = [(.7)/G_{eff.O.nls} + (.3)/G_{eff.L.nls}]$ , or

es7.10 $(1/G_{eff.G.nls})(0.969) = [(.7)/G_{eff.O.nls} + (.3)/G_{eff.L.nls}]$

;where $\chi \equiv \underline{\Delta}Q_O/\underline{\Delta}F_O \approx 0.2$ , and $(G_{eff.G.nls}/G_{eff.O.nls}) \approx 1.71/2.19 = 0.78$ .

But define $G_{eff.G.equilibrium}$ only for a system that starts and ends at equilibrium (see e.6,7), and then $\underline{\Delta}Q_O=0$, and then $\chi=0$. So at equilibrium es7.9 becomes:

es7.11 $(1/G_{eff.G.equilibrium}) = [(.7)/G_{eff.O.equilibrium} + (.3)/G_{eff.L.equilirium}]$ .

Finally, if we could **assume**

es7.12 $G_{eff.O.nls} = G_{eff.O.equilibrium}$ , and
es7.13 $G_{eff.L.nls} = G_{eff.L.equilirium}$ ,

then using es7.11,12,13 in es7.10 yields:

es7.14 $(1/G_{eff.G.nls})(0.969) = 1/G_{eff.G.equilibrium}$ , or
es7.15 $[G_{eff.G.equilibrium} / G_{eff.G.nls}] = 1.032$ .

Therefore, under this last assumption, we see a partial "pattern effect" resultant of Ocean thermal capacitance is to *increase* $G_{eff.Global}$ by 3% at equilibrium; and thus $ECS_{true.Global} = (0.97)\cdot ESC_{nls.Global}$ ; where $ESC_{eff.Global} \equiv ESC_{nls.Global}$. This is then an expected "tendency". However using the more sophisticated modeling and analysis of the next section we need not make this assumption to determine $ECS_{true.Global}$ , and which also reveals many other important properties of the Earth Climate System.

---

**SS7.2**

A solution for $\Delta T_O$, $\Delta T_{aO}$, $\Delta T_L$, $\Delta T_{aL}$ using the Coarse 3-layer 2-region(Ocean, Land) formal/empirical algebraic GCM small signal model is developed in **SS6** (review **SS6**). Now define an operator "**S[ ]**" that returns the best linear fit slope "S" over some given time range. Apply this operator to both sides of es6.14, yielding

es7.16 $S[\Delta T_O(t)] = S[\ \Delta I_{su}(t) + \Delta I_{cs}(t)\cdot \mathbf{CS} - \Delta Q(t)\ ] / G^*_O$
es7.17 $S[\Delta T_O(t)] \cong \{\ S[\Delta I_{su}(t)] + S[\Delta I_{cs}(t)]\cdot \mathbf{CS} - S[\Delta Q(t)]\ \} / G^*_O$

;where the near linearity of S[ ] *in this case* is used and can be proven (and tested) since $\Delta I_{cs}(t)$ is nearly linear and $\Delta I_{su}(t)=0$; and **CS** and $G^*_O$ are essentially true constants over the evaluation range (1980-2020). Similarly, **all** time variables in all **SS6** equations can and will now be considered to be the "best linear fit slope" values without change or error, which improves reliability of the empirical values and simplifies matching of data to model parameters.

The working spread sheet **GCM8-altGeffOcs.xls** implements the GMC and provides a solution for $\Delta T_O$ and $\Delta T_L$ using selected model parameters and given independent direct Forcings. And more importantly, provides a solution for the parameters $\beta^*$, $\gamma^*$, $g_{xO}$ and $g_{xL}$ (and $\lambda$ and $\sigma$), by matching the theoretical and empirical values of $\Delta T_O$, $\Delta T_{aO}$, $\Delta T_L$, $\Delta T_{aL}$ within the Excel Solver tool.

**[[** Evaluating $\mathbf{\Delta T_{aO}/\Delta T_O}$ and $\mathbf{\Delta T_{aL}/\Delta T_L}$ :

The empirical values of $\Delta T_O(t)$, $\Delta T_L(t)$, $\Delta Q(t)$, $\Delta I_{cs}(t)$, (and $\Delta I_{su}(t)\equiv 0$), are determined in the main body of the monograph. The value of $\mathbf{\Delta T_{aO}/\Delta T_O}$ and $\mathbf{\Delta T_{aL}/\Delta T_L}$ are determined as follows. The NOAA RATPAC (radiosonde balloon data) best fit ΔT(t) slope over 1970 to 2020 for the Global Surface and Global Bulk Troposphere ("BT"...1.5 to 5.5km) is given[5] as 0.21 °C/decade and 0.21 °C/decade, respectively [see working spread-sheet **ratpac.xls**]. Therefore,

$\mathbf{\Delta T_{surface.Global}} / \mathbf{\Delta T_{BT.Global}} = 0.21/0.21 = 1.00$ .

Similarly, satellite data[6] using *both* RSS and UAH versions (1979-2014) indicate

---

[5] https://www.ncei.noaa.gov/data/ratpac/access/

[6] https://judithcurry.com/2015/03/04/differential-temperature-trends-at-the-surface-and-in-the-lower-atmosphere/

$\mathbf{\Delta T_{LT.Land}}$ / $\mathbf{\Delta T_{LT.Ocean}}$ = 0.167/0.1 =1.67 and
= 0.179/0.107=1.67 , respectively;
[note, "LT" ≡ "Lower Troposphere"].

And the NOAA values (with PIR) for 1975-2020 taken from Main section **B** indicate

$\mathbf{\Delta T_{Land}}$ / $\mathbf{\Delta T_{Ocean}}$ = 1.37/.62 = 2.20 .

Using the subscripts a ≡ atmosphere (LT≈BT) , O ≡ Ocean surface, L ≡ Land surface, G ≡ Global surface , aO≡ Ocean atmosphere (LT≈BT), etc., and using the above ratios :

$\Delta T_G = (1.00)\Delta T_{aG}$
$( (.7)\Delta T_O + (.3)\Delta T_L ) = 1.00( (.7)\Delta T_{aO} +(.3)\Delta T_{aL} )$
$( (.7)\Delta T_O + (.3)(2.2)\Delta T_O ) = 1.00( (.7)\Delta T_{aO} +(.3)(1.67)\Delta T_{aO} )$
$\Delta T_O(1.36) = \Delta T_{aO}(1.20)$ , so
$\underline{\mathbf{\Delta T_{aO} / \Delta T_O} = 1.36/1.20 = 1.13}$ .

Further,
$\Delta T_{aO} = (1.13)\Delta T_O$
$\Delta T_{aL}/1.67 = (1.13)\Delta T_L/2.2$ , so
$\underline{\mathbf{\Delta T_{aL} / \Delta T_L} = (1.13)(1.67)/2.2 = 0.86}$ . **]]**

The other independent ΔForcings (slopes), ΔTemperatures (slopes), and *predetermined* parameters used in this evaluation [see **SS6** and Main sections **B)**, **C)** ,**D)** ] are:

$\Delta I_{cs}$ = 0.4 (W/m$^2$)/decade
$\Delta I_{ce}$ = 0.169 (W/m$^2$)/decade
$\Delta I_{su}$ = 0.0 (W/m$^2$)/decade
$\Delta Q$ = 0.075 (W/m$^2$)/decade

$\Delta T_O$ = 0.147 °C/decade
$\Delta T_L$ = 0.323 °C/decade
$\Delta T_{aO} / \Delta T_O$ = 1.13
$\Delta T_{aL} / \Delta T_L$ = 0.86

$g_{sn}$ = 0.35 (W/m$^2$)/°C
$g_{ae}$ = 5.66 (W/m$^2$)/°C
$g_{ea}$ = 4.11 (W/m$^2$)/°C
$g_{sa}$ = 3.53 (W/m$^2$)/°C
$g_c$ = 0.6 (W/m$^2$)/°C
$g_{vs}$ = 1.47 (W/m$^2$)/°C
$g_{ve}$ = 2.42 (W/m$^2$)/°C
$k_{mO}$ = 0.5 (W/m$^2$)/°C
$\mu$ = 0.5

; and the most uncertain or unknown parameters are evaluated by matching theoretical and empirical values of $\Delta T_O$, $\Delta T_{aO}$, $\Delta T_L$, $\Delta T_{aL}$ using spread sheet **GCM8-altGeffOcs.xls** . These are then found to be:

$\beta^* = -3.7$ (W/m$^2$)/°C
$\gamma^* = +3.4$ (W/m$^2$)/°C
$g_{xO} = 2.6$ (W/m$^2$)/°C
$g_{xL} = 1.3$ (W/m$^2$)/°C , and also

**CS** = 1.015 .

The values of $k_{mO}$ and $\mu$ are also uncertain (and might be evaluated from other sources) , but a wide range of possibilities will be tested below that yield similar results.

However, IPCC AR6[7] indicates that for *Global surface Forcing* the independent and separately evaluated water vapor, cloud reflection, and snow/ice Global equilibrium "feedbacks" are +1.3, +0.42, and +0.35 (W/m$^2$)/°C , respectively. And the equilibrium Global fundamental surface Planck effective radiative conductance is 3.22 (W/m$^2$)/°C (excluding any rein**forcings**, i.e. "feedbacks"). Trial solutions (using surface forcing) of this Coarse model using **GCM8-altGeffOcs.xls** yield nearly identical equilibrium results for all the above if radiation parameters $g_{ea}$, $g_{ae}$, $g_{sa}$ are multiplied by a scalar **$\varepsilon$** $\equiv 0.85$, and water vapor forcing parameters $g_{vs}$, $g_{ve}$ are multiplied by a similar scalar **$\tau$** $\equiv 0.85$ . In other words, if the MODTRAN evaluated parameters are slightly and simply modified, this Coarse GCM reproduces the **individual** IPCC AR6 "feedback" and effective Planck surface radiation results! This duplication is unlikely from random coincidence, and thus tends to corroborate both this model and the AR6 results. And for consistency with orthodoxy, this is done henceforth. The physical reason for the **$\varepsilon = \tau$** scaling (if legitimate) is not specifically know to the author.

If these IPCC AR6 "feedbacks" are simply added linearly then the resulting equilibrium Global effective "conductance" for surface forcing should be $G_{eff.Global}$ = 3.22-1.3-0.42-0.35 = 1.15 (W/m$^2$)/°C . And this is the value presented in IPCC AR6[7] . However this Coarse model indicates the separate rein**forcing** mechanisms do in fact interact significantly in a non-linear manner when acting in concert concurrently. Equations es6.12b and es6.14b show how $G_{eff}$ is dependant on a product of $\gamma^*$ and $g_{vs}$ , indicating a non-linear interaction between the cloud reflection and water vapor rein**forcing** mechanisms. *Proper* solutions (using surface Forcings) at equilibrium of this model for *Global surface Forcings* (using **GCM8-altGeffOcs.xls**) yield equilibrium $G_{eff.Global}$ = 1.82 (W/m$^2$)/°C (presuming $k_{mO}$ =0.5, $\mu$ = 0.5, and $F_{2xCO2}$ =3.7(W/m$^2$)), which is a 58% increase over the AR6 result. For $k_{mO}$ =0, $\mu$ = 0.5 then equilibrium $G_{eff.Global}$ = 1.64, which is a 43% increase over the AR6 result. This indicates a likely source of error in the simplified **linear** AR6 formulation of "feedbacks"[7] and $G_{eff.Global}$ .

When the model radiation and water vapor forcing parameters are slightly modified *using $\varepsilon$ and $\tau$ as above*, then the following parameters are again evaluated by matching theoretical and empirical values of $\Delta T_O$, $\Delta T_{aO}$, $\Delta T_L$, $\Delta T_{aL}$ within **GCM8-altGeffOcs.xls** . Assuming **$\mu$=0.5**, **$k_{mO}$=0.5** (W/m$^2$)/°C These are found to be:

$\beta^* = -3.44$ (W/m$^2$)/°C
$\gamma^* = +2.76$ (W/m$^2$)/°C
$g_{xO} = 1.94$ (W/m$^2$)/°C
$g_{xL} = 1.04$ (W/m$^2$)/°C , and also

**CS** = 1.008

However if **$\mu$=0.5**, **$k_{mO}$ = 0** (i.e. zero atmospheric mixing), then the best matched parameters become:

$\beta^* = -2.14$ (W/m$^2$)/°C
$\gamma^* = +1.95$ (W/m$^2$)/°C

[7] IPCC_AR6_WGI_TS.pdf, TS.3.2 : including Fig TS.17,16

$g_{xO}$ = 2.32 (W/m$^2$)/°C
$g_{xL}$ = 0.64 (W/m$^2$)/°C , and also
**CS** = 0.914 .
But the final calculated values of equilibrium Global $G_{eff}$ for both $k_{mO}$'s differ by only about 3 percent, i.e. $G_{G.eff}$ = 1.85 and 1.80 (W/m$^2$)/°C, respectively.

In fact, extensive trial evaluations of all calculated $G_{eff}$'s using wide variations of all unknown or uncertain model parameters indicates Land and Global $G_{eff}$ at equilibrium is usually 2% to 5% *greater* than $G_{eff}$ calculated for the near linear steady conditions from 1980-2020 (see **GCM8-altGeffOcs.xls** and **GCM8-work1-altGeffOcs.xls**) . Under these conditions the calculated "pattern effect" is thus a small *cooling* effect at equilibrium (2 to 5%). The exception is in cases of μ ~ 0.2 (unlikely) *concurrent with* $k_{mO}$ ~ 1 (W/m$^2$)/°C (again, unlikely), in which case $G_{eff.Land}$ at equilibrium is slightly smaller (a few percent) [and **CS** = 1.055]. This exception is *not* true for the Global average. For the global Oceans, the $G_{eff.Ocean}$ value could also reasonably be 2.5% *less* (for **CS**=0.9, $k_{mO}$=0). In any case, the differences are only a few percent such that $ECS_{true} \leq ECS_{eff}$ is a very good approximation. Explanations of enhanced ECS evaluations due to the "pattern effect" are unwarranted.

In **SS6** it was argued (realistically?) that "... *only* Ocean perturbation temperature changes significantly effect global Land atmospheric water vapor density in this model." However if we were to consider the opposite extreme, namely that *only* Land perturbation temperature changes significantly effect global Land atmospheric water vapor density, this could be easily modeled herein by simply replacing "σ" with unity in the equations for **AL** and **DL** (see **SS6.2**). Then "μ" becomes the scalar for this Land effect.

Under *this* presumption (not realistic?) the calculated "pattern effect" is instead a possible small *warming* effect at equilibrium over Land, Oceans or Global total of about 4%, 2%, and < 0% respectively, using realistic parameters. Even if we were to suppose that the reality lies halfway between these two extremes (which can also be modeled), then the "pattern effect" is still essentially neutral or negative at equilibrium. And this is the presumption *already* made in Main section **E** and tables Tab.1 , Tab.2 , in any case! Again, explanations of enhanced ECS evaluations due to the "pattern effect" are unwarranted.
....................................................

The value of $k_{mO}$ (or $k_{mL}$) is not explicitly evaluated herein, but an estimate might be made indirectly as follows. Occam's Razor is a philosophical statement that the simplest models or answers that match observations are more likely to be true than those more complex. Likewise models using the fewest independent parameters, or even the *smallest* magnitude parameters are more likely true. This has a statistical basis, given no "a priori" knowledge. It is found by trials that the *average* magnitude of the best fit parameters $k_{mO}$, β*, γ* , $g_{xO}$ , $g_{xL}$ reach a significant minimum as $k_{mO}$ is reduced to zero (see values above).

Further, if $k_{mO}$ = 1.0 (W/m$^2$)/°C then $g_{xO} \approx g_{xL}$ , which seems physically unrealistic [remember, $g_{xO}$ represents the temperature dependant power flow from the surface into the atmosphere due to the lift of sensible and latent water vapor energy, which would reasonably be much larger over the Oceans than "dry" Land]. For $k_{mO}$ = 0.5 (W/m$^2$)/°C then it is found $g_{xO} \approx 1.87 \cdot g_{xL}$, and for $k_{mO}$ = 0 (W/m$^2$)/°C then it is found $g_{xO} \approx 3.6 \cdot g_{xL}$ , which both seem more realistic. Therefore it may be expected the true value of $k_{mO} \leq 0.5$ (W/m$^2$)/°C, with even smaller values somewhat more likely.
..................................................

The near linear steady value of $G_{eff.Global}$ can be derived from the Land and Ocean values by remembering

es7.20 $\Delta T_{Global} = 0.7 \cdot \Delta T_{Ocean} + 0.3 \cdot \Delta T_{Land}$ .

Now, presuming the forcing is globally uniform, and applying e.20 yields

es7.21 $(\Delta F - 0.7 \cdot \Delta Q_{Ocean})/G_{eff.Global} = 0.7 \cdot (\Delta F - \Delta Q_{Ocean})/G_{eff.Ocean} + 0.3 \cdot \Delta F/G_{eff.Land}$

;where $\Delta Q_{Ocean}$ represents the Global Ocean average value, (.7) is the Ocean fraction of the entire Globe, and (.7)·ΔQ is then the average of $\Delta Q_{Global}$ . Let $\chi \equiv \Delta Q_{Ocean}/\Delta F$ , then es7.21 is reformed as

es7.22 $G_{eff.Global} = (1-0.7\cdot\chi)/[0.7\cdot(1-\chi)/G_{eff.Ocean} + 0.3/G_{eff.Land}]$ .

..................................................................

The equilibrium value of $G_{eff.Global}$ can be obtained by simply setting ΔQ=0, then es7.22 yields

es7.23 $G_{eff.Global} = [0.7/G_{eff.Ocean} + 0.3/G_{eff.Land}]^{-1}$ .

..................................................................

After independent stratospheric temperature equilibration the effective $\Delta F/2xCO_2$ by MODTRAN calculation is **~4.2** $W/m^2$ (given 20% strato/stratus cumulous cloud cover (.6km-2km)...see **SS6.3.2** end page) . The effect of the "sparse" stratosphere is surprisingly large. In this case **4.2** $(W/m^2)/2xCO_2$ represents a +13.5% increase from the IPCC AR5 canonical value of ERF = DRF= **3.7** $(W/m^2)/2xCO_2$. And this percentage increase should then also be made to the corresponding anthropogenic GHG(Green House Gas) ΔForcing component ΔF(t) of the Main section Fig.12 (i.e. 1.135 · **0.4** = **0.454** $W/(m^2\cdot decade)$ ) .

However in a complicating twist, it must be noted that in **SS6,7** all the MODTRAN temperature proportionality radiation parameters $g_{ae}$, $g_{ea}$, $g_{sa}$, $g_{vs}$, $g_{ve}$ were multiplied by a scalar **ε = τ = 0.85**, which resulted in a close reproduction of the *independent* orthodox IPCC AR6[7] equilibrium Global Planck surface radiation "feedback", water vapor "feedback", and cloud reflection "feedback". And for consistency with orthodoxy, this is used herein. But this scaling was *not* simultaneously performed on the $CO_2$ radiative forcing components $\Delta I_{cs}$ , $\Delta I_{ce}$; these remained at the *updated* IPCC AR5 canonical values (see Main Fig.12). If the **ε** scaling *is* also uniformly applied to the *non-canonical* Forcing values above then $\Delta F/2xCO_2$= 0.85· **4.2** = **3.6** $(W/m^2)/2xCO_2$ , and ΔF(t)/decade = **0.454** · 0.85 = **0.39** $W/(m^2\cdot decade)$. In other words, the consistent **ε** scaling of the *non-canonical* MODTRAN calibrated *Forcings* above transforms them back into the *canonical* values (within a few percent)!

Therefore, if the **ε** scaling is also performed uniformly on the non-canonical MODTRAN calibrated ΔF values above there is essentially no change from the orthodox/canonical values. If the **ε** scaling is not applied to all the MODTRAN evaluated parameters above then there will be a reduction in the evaluated ECS, and a considerable change to the $G_{eff}$ and cloud reflection rein**forcing** evaluations. The physical reason or justification for the **ε** scaling (if legitimate) is not specifically known to the author. It is certainly possible the MODTRAN values are inexact or have been misapplied to the Coarse GCM in a way that is easily corrected by the simple use of the **ε** scaling uniformly on *all* of those values, and that is assumed herein. Thus, the **ε** scaling and orthodox/canonical ΔF values have been and will be used in this treatment, which is again a *maximal* ECS choice.

.......................................................

The Coarse GCM derived in **SS6** and above (where scalar **ε=τ = 0.85)** duplicates the *individual* orthodox IPCC AR6[7]:

1) *Global* Planck (surface) effective radiation "feedback",
2) *Global* water vapor "feedback",
3) *Global* albedo "feedback",
4) *Global* cloud reflection "feedback",
5) *Global* $\Delta F_{2xCO2}$ forcing,
6) and the observed Surface ***and*** Bulk atmosphere Global.ave temperature trends for both Land ***and*** Ocean!

**It is therefore deemed physically realistic and superior to most alternative GCMs.**

## SS8 Ramp forcing response and TCR v.4

A general solution for **Ocean** surface temperature response to a ramp forcing function starting at t=0 is derived. Using e.3,4e and es3.8,6 , and given all $d(\,)/dt = 0$ for $t \le 0$, the form of $\Delta T(t)$ is given by

es8.1 $\Delta \mathbf{T}(t)\cdot G_{eff.O} = \Delta F(t) - \Delta Q_{total}(t) = \Delta F(t) - \Delta Q_{deep.Ocean}(t) - \Delta Q_{mix.layer}(t)$
$= \Delta F(t) - \Delta Q_{deep.Ocean}(t) - C^*_{eff}\cdot d(\Delta T(t))/dt$ .

Now take the Laplace transform of es8.1, and using es4.27,29 rewrite it as

es8.2 $\Delta \mathbf{T}(s)\cdot G_{eff.O} = \Delta F(s) - Y(s)\cdot\Delta \mathbf{T}(s) - s\cdot C^*_{eff}\cdot\Delta \mathbf{T}(s)$
$= \Delta F(s) - (\sqrt{s}\cdot CC^* - g_u/2)\cdot\Delta \mathbf{T}(s) - s\cdot C^*_{eff}\cdot\Delta \mathbf{T}(s)$ , or

es8.3 $\Delta \mathbf{T}(s) = \Delta F(s)/[\, G_{eff.O} + \sqrt{s}\cdot CC^* + s\cdot C^*_{eff} - g_u/2 \,]$ .

Remember, at equilibrium the $g_u$ term actually is reduced to zero (see **SS4**). For a ramp forcing function $\Delta F(s)=F_o\cdot(1/s^2)$, so es8.3 becomes

es8.4 $\Delta \mathbf{T}(s) = F_o/[s^2\cdot(\, G_{eff.O} + \sqrt{s}\cdot CC^* + s\cdot C^*_{eff} - g_u/2 \,)]$ .

If Ocean thermal admittance were zero (i.e. $CC^*=0$ and $C^*_{eff}=0$, and $g_u=0$), the solution would simply be

es8.5 $\Delta \mathbf{T}(s) = F_o/[s^2\cdot G_{eff.O}]$ .

If we set $F_o=G_{eff.O}$ then the solution is simply a unity ramp function. Normalize es8.4 to this case by setting $F_o=G_{eff.O}$ , yielding

es8.6 $\Delta \mathbf{T}(s)_N = G_{eff.O}/[s^2\cdot(\, G_{eff.O} + \sqrt{s}\cdot CC^* + s\cdot C^*_{eff} - g_u/2 \,)]$ .

The realistic parameters used (as determined in **SS6,7** for **$\mu$=0.5**, **$k_{mO}$=.5** W/(m$^2\cdot$°C)) are
$G_{eff.O}$ = 2.2 W/(m$^2\cdot$°C), $CC^*\cong$ 5.7 W$\cdot\sqrt{[yrs]}$/(m$^2\cdot$°C), $C^*_{eff}$=14 W$\cdot$yrs/(°C$\cdot$m$^2$), $g_u$=.41 W/(m$^2\cdot$°C) .

Taking the Numerical Inverse Laplace Transform (NILT) of es8.6 yields $\Delta \mathbf{T}(t)_N$ . This is plotted in Fig.SS8.1 as " T(t):nilt:g=2.2-.2 " . The time derivative is also plotted as " T′(t):nilt ". The value of **TCR/ECS$_{eff}$** is then simply $\Delta \mathbf{T}(70)_N/70 = 50/70 =$ **0.71** (review definitions of ECS and TCR elsewhere). [The working spreadsheets are "**TCRdECS(etc).xls**"]. For the Global Oceans 60N-60S Latitude the calculated value of **TCR/ECS$_{eff}$** is $\Delta \mathbf{T}(70)_N/70 = 47.7/70 =$ **0.68** . [If $\Delta F_{2xCO2}$ =4.2 W/m$^2$ (as opposed to 3.7), then $G_{eff.Ocean}$=2.57 W/(m$^2\cdot$°C) (as opposed to 2.2), and it can be determined **TCR/ECS$_{eff}$ = 0.75** .] For **$k_{mO}$=0** W/(m$^2\cdot$°C), and $G_{eff.O}$ = 2.15 W/(m$^2\cdot$°C), then **TCR/ECS$_{eff}$ = 0.705** .

The calculation of $\Delta T'(t)$ as estimated in es3.13 (see **SS3**) is a simplistic formula of $\Delta T'(t)$ and not generally correct, but is very nearly exact for the specific parameters used *only* if CC* is replaced by CC*$\cdot$(0.4). It is a convenient compact form required for the estimation of $C^*_{eff}$ in **SS3**. So $\Delta T'(t)$ (=es3.13) and $\Delta T(t)=\int\Delta T'(t)\cdot dt$ , are plotted as " T1' " and " T1 " respectively in Fig.SS8.1 .

Interestingly, using es3.12,13 , the estimation

es8.7 $\Delta T(t) \approx t\cdot\Delta T'(t) \approx (t\cdot t\cdot\Delta F')\,/\{\, t\cdot G_{eff} + C^*(x) + CC^*\cdot(2/\sqrt{\pi})\cdot\sqrt{[t]} \,\}$

is shown to be a good *general* approximation for most parameters. The values of T(t) and $T'(t)=dT(t)/dt$ using this form are also plotted as " T2 " and " T2' " respectively in that figure. The comparison to the exact NILT $\Delta T(t)$ solution is respectable with an extraordinary reduction of mathematical complexity!

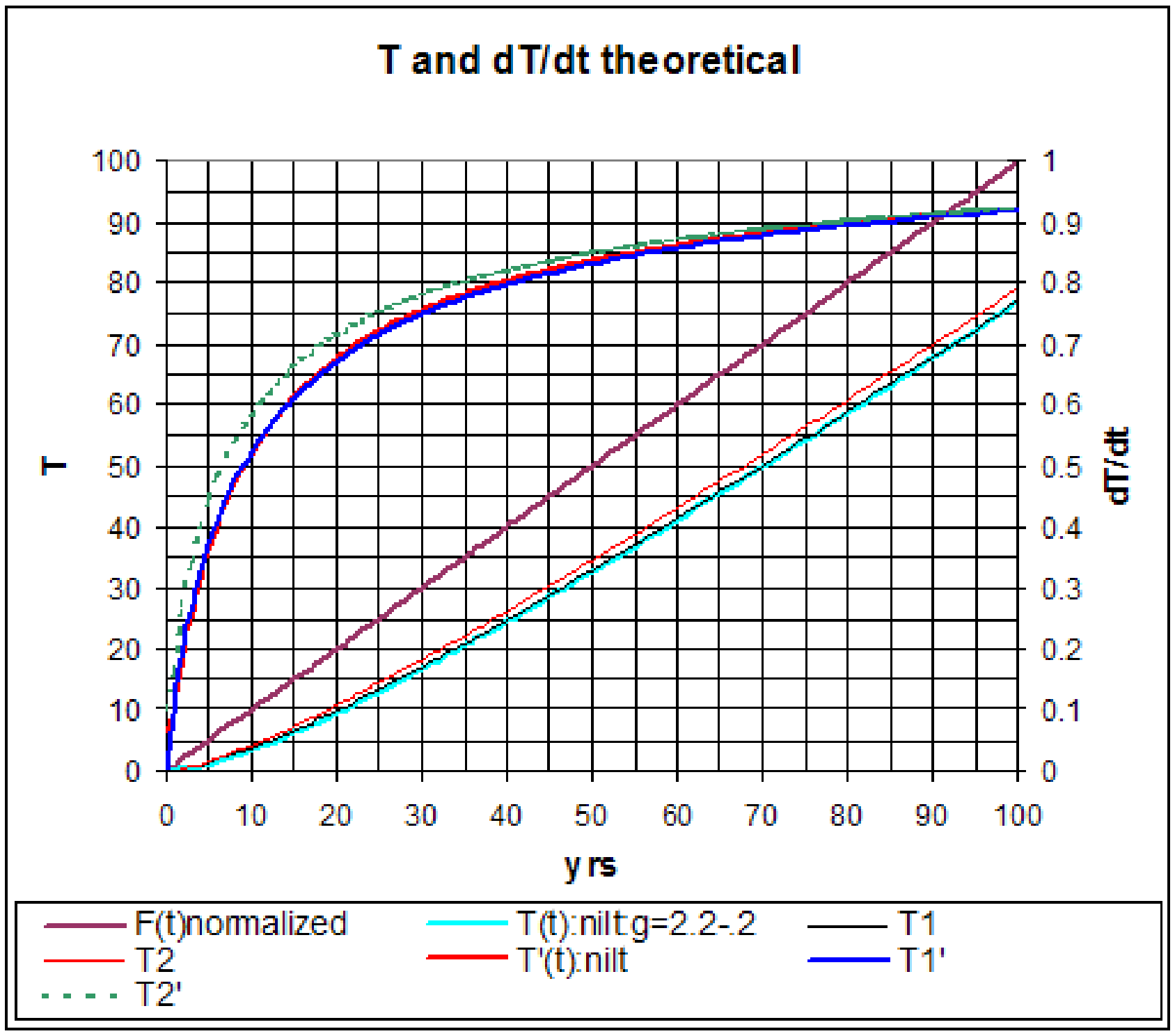

Fig.SS8.1

---

The general solution for **Land** surface temperature response to a ramp forcing function is not quite so straight forward. Start with es6.16 and replace all $\sigma \cdot \Delta T_L$ with $\Delta T_O$ (see es6.3). Specifically, if $I_{su}=0$,

es8.8 $\Delta T_L \cdot (g_c+g_{ae}+g_{xL}-g_{sn}) - \Delta T_O \cdot \mu \cdot (\beta^*+\nu \cdot g_{vs}) = \Delta I_{cs} \cdot \alpha + \mathbf{FL} \cdot \Delta I_{cs} \cdot (1-\alpha) +$
$+ \mathbf{FL} \cdot \{\Delta T_L \cdot (g_c+0.9 \cdot g_{ae}+g_{xL})+\Delta T_O \cdot \mu \cdot g_{vs} \cdot (1-\nu)\}$ , or

$\Delta T_L \cdot [g_c+g_{ae}+g_{xL}-g_{sn} - \mathbf{FL} \cdot (g_c+0.9 \cdot g_{ae}+g_{xL})] = \Delta I_{cs} \cdot (\alpha+\mathbf{FL} \cdot (1-\alpha)) + \Delta T_O \cdot [\mathbf{FL} \cdot \mu \cdot g_{vs} \cdot (1-\nu) + \mu \cdot (\beta^*+\nu \cdot g_{vs})]$ , or

es8.9 $\Delta T_L = \{\Delta I_{cs} \cdot \mathbf{CSL} + \Delta T_O \cdot \mathbf{AA}\}/\mathbf{DD}$

;where $\mathbf{CSL} \equiv (\alpha+\mathbf{FL} \cdot (1-\alpha))$, $\mathbf{AA} \equiv \mathbf{FL} \cdot \mu \cdot g_{vs} \cdot (1-\nu) + \mu \cdot (\beta^*+\nu \cdot g_{vs})$, $\mathbf{DD} \equiv [g_c+g_{ae}+g_{xL}-g_{sn}-\mathbf{FL} \cdot (g_c+0.9 \cdot g_{ae}+g_{xL})]$.

Take the Laplace transform of es6.14, and using es4.27,29 and es3.6 yields

es8.10a $\Delta T_O(s) = [\Delta I_{cs}(s) \cdot \mathbf{CS} - \Delta Q_{total}(s)]/G^*_O$
es8.10b $\Delta Q_{total}(s) = (\sqrt{s} \cdot CC^* - g_u/2) \cdot \Delta \mathbf{T}_O(s) + s \cdot C^*_{eff} \cdot \Delta T_O(s)$ .

Rewrite es8.10a using es8.10b yields

es8.11 $\Delta T_O(s) = \Delta I_{cs}(s) \cdot \mathbf{CS}/\{ G^*_O + \sqrt{s} \cdot CC^* + s \cdot C^*_{eff} - g_u/2\}$ .

Now use es8.11 in es8.9 yielding

es8.12 $\Delta T_L(s) = (\Delta I_{cs}(s)/\mathbf{DD}) \cdot \{ \mathbf{CSL} + \mathbf{CS} \cdot \mathbf{AA}/[ G^*_O + \sqrt{s} \cdot CC^* + s \cdot C^*_{eff} - g_u/2] \}$ .

For a ramp function $\Delta I_{cs}(s) = \Delta I_{cso}/s^2$ . If Ocean thermal admittance were zero (i.e. $CC^*=0$, $g_u=0$ and

$C^*_{eff}$ =0), the solution would simply be

es8.13 $\Delta T_L(s) = (\Delta I_{cso}/(s^2 \cdot \mathbf{DD})) \cdot \{ \mathbf{CSL} + \mathbf{CS{\cdot}AA}/G^*_O \}$ .

Let $\Delta I_{cso} = \mathbf{DD}/\{ \mathbf{CSL} + \mathbf{CS{\cdot}AA}/G^*_O \}$. Then es8.13 $\Delta T_L(s)$ becomes the unity ramp function $1/s^2$. Normalize es8.12 to this case by using this value of $\Delta I_{cso}$ . Then es8.12 becomes

es8.14 $\Delta T_L(s)_N = \mathbf{M} \cdot \{ \mathbf{CSL} + \mathbf{CS{\cdot}AA}/[ G^*_O + \sqrt{s} \cdot CC^* + s \cdot C^*_{eff} - g_u/2] \}/s^2$ ;where

$\mathbf{M} \equiv 1/[\mathbf{CSL} + \mathbf{CS{\cdot}AA}/G^*_O]$ .

The solution is quite sensitive to the value of $k_{mO}$ assumed in the evaluation of parameters (see **SS7**). For $k_{mO}$ = 0 and μ=0.5, then **CS·AA** = -0.35 , **CSL** = 0.91, $G^*_O \cong 1.98$ , **M** = 1.36 . The other parameters are as in es8.6 . Taking the Numerical Inverse Laplace Transform (NILT) of es8.14 yields $\Delta T_L(t)_N$ . This is plotted in Fig.SS8.2 as " T(t):nilt,kmo=0 " . The value of **TCR/ECS$_{eff}$** is then simply $\Delta\mathbf{T}(70)_N/70$ = 75.3/70 = **1.075** (review definitions of ECS and TCR elsewhere).

Similarly, for $k_{mO}$ = 0.5 and μ=0.5, then **CS·AA** = -1.07 , **CSL** = 0.95, $G^*_O \cong 2.23$ , **M** = 2.13 . The other parameters are as in es8.6 . Taking the Numerical Inverse Laplace Transform (NILT) of es8.14 yields $\Delta T_L(t)_N$ . This is plotted in Fig.SS8.2 as " T(t):nilt,kmo=.5 " . The value of **TCR/ECS$_{eff}$** is then simply $\Delta\mathbf{T}(70)_N/70$ = 88.5/70 = **1.26** .

These unexpected Land values >1 are due to the lagging effects of Ocean warming on increased Land atmosphere water vapor density and cloud formation. If ramp forcing were to stop suddenly at a constant value then the warming over the oceans would continue to some equilibrium value, but the temperature over the Land would immediately begin to cool to some equilibrium value due to increasing cloud cover. The true magnitude of this effect over Land is not determinable herein unless $k_{mO}$ is well determined by some other means. Because of this uncertainty, the land value of TCR/ECS$_{eff}$ might be assumed to simply be unity.

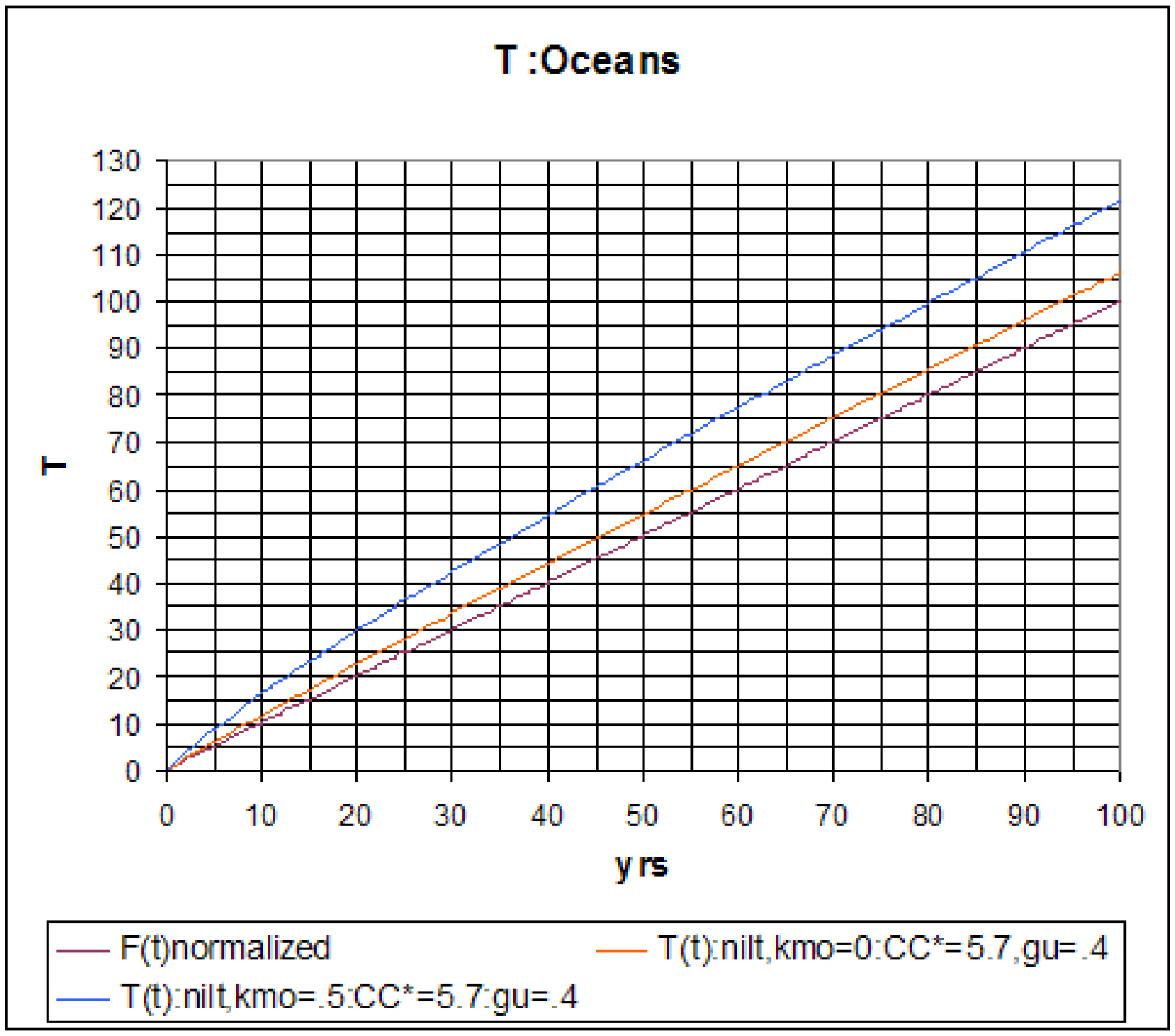

Fig.SS8.2
...............................................

The Global average TCR/ECS is calculated from the Ocean and Land values as:

$TCR/ECS_{Global} = \{(.7)\cdot TCR_O + (.3)TCR_L\} / \{(.7)\cdot ECS_O + (.3)ECS_L\}$
$= \{(.7)\cdot TCR_O/ECS_L + (.3)TCR_L/ECS_L\} / \{(.7)\cdot ECS_O/ECS_L + (.3)\}$
$= \{(.7)\cdot(TCR_O/ECS_O)\cdot(ECS_O/ECS_L) + (.3)\cdot(TCR_L/ECS_L)\} / \{(.7)\cdot(ECS_O/ECS_L) + (.3)\}$

;where 0.3 and 0.7 are the global fractions of land and ocean.

———————— Michael D. Mill...Mar. 2025 [Contact: m.d.mill.climate@gmail.com]—[copyright: CC BY 4]—

## SS9 Average CC*, $g_u$, and $X_{eff}$ Calculations...Upper Limits v.4

Using the theory of Energy absorption into the deep ocean (excluding the mix layer) as developed in **SS4** and reiterated in **SS5 3)** then

es9.1 $g_{co}(\theta) = g_u(\theta)\cdot X(\theta)$ and
es9.2 $CC^*(\theta) \equiv \sqrt{[C\cdot g_{co}(\theta)]} = \sqrt{[C\cdot g_u(\theta)\cdot X(\theta)]}$

; where X= $X_{1/e}$ or $X_o$, and "θ" represents the latitude angle of a position on the Globe (see es4.16 and 4.27), and "C" ≡ $(0.13)\cdot watt\cdot yrs/(m^3\cdot{}^\circ C)$ . We will make two simplifying assumptions, i.e. the Ocean encompasses the entire Globe (i.e. no Land), and that $g_u(\theta)$ is of the simple form

es9.3 $g_u(\theta) = a\cdot\cos(\theta\cdot 3/2)$

;where $g_u(\theta)$ is maximum at the equator (θ= 0°) and drops to zero at about 60° latitude N and S. The necessity of this basic form is discussed in **SS4.2** , The precise form of these simplifying approximations produce very little variation in the final evaluations of ECS. Also we presume the 60N to 60S area weighted average value of $g_u(\theta) \approx 0.41\ W/(m^2\cdot{}^\circ C)$ , as was introduced following es4.4 . The 60N to 60S areal average operator is represented as **A[**f(θ)**]** in **bold**, and all such averaged variables are also printed as **bold** (e.g. **CC***). The value of X(θ) does vary, or is uncertain, but it is certainly ≤ 1400 m (see various Ocean temperature profiles in the repositories "/OceanTprofiles/" , and particularly the Indian Ocean maximum) , and so we will simplify X(θ) to be a constant maximum of $X_{max}$=1400 m. This will guarantee that the **CC*** evaluation , and the resulting ECS calculations, will be *maximal* values. We can now calculate the Ocean Global 60N-60S averages.

Note that generally, for symmetrical functions in θ, we can write

es9.4 $\mathbf{A[}f(\theta)\mathbf{]} = \{P\cdot {}_0\int^{\pi/3} f(\theta)\cdot\cos(\theta)\cdot d\theta\}/\{P\cdot {}_0\int^{\pi/3}\cdot\cos(\theta)\cdot d\theta\}$
$= \{{}_0\int^{\pi/3} f(\theta)\cdot\cos(\theta)\cdot d\theta\}/\{{}_0\int^{\pi/3}\cdot\cos(\theta)\cdot d\theta\}$

;where P is an appropriate constant multiplier of the Global 60N-60S integration, and always cancels out!
Specifically, using es9.3, $\mathbf{g_u}$ is calculated as

es9.5 $\mathbf{A[}a\cdot\cos(\theta\cdot 3/2)\mathbf{]} = \{{}_0\int^{\pi/3} a\cdot\cos(\theta\cdot 3/2)\cdot\cos(\theta)\cdot d\theta\}/\{{}_0\int^{\pi/3}\cdot\cos(\theta)\cdot d\theta\}$ , or
$\mathbf{g_u} = a\cdot\{0.6\}/\{.866\} = a\cdot 0.693$ , and so
$a = \mathbf{g_u}\cdot 1.44$ ; and then rewrite es9.3 as

es9.6 $g_u(\theta) = 1.44\cdot \mathbf{g_u}\cdot\cos(\theta\cdot 3/2)$ .

The Tropical deep-ocean Δtemperature profile must be modeled approximately as a combination of an upper ocean (excluding the mix layer!) where $g_c$ is very small (due to the large temperature gradient stabilization of convection processes) and a lower region where $g_c$ is larger due to normal convection processes.

So this Tropical profile consists approximately of an upper exponential Δtemperature decay for $X_{1/e} \approx$ 280m and covering 2/3 of the total temperature drop, *followed* by a lower separate constant decay over an $X_o$ = 1400 m and covering the remaining 1/3 of the total temperature drop. Elsewhere the profile is a simple exponential or constant decay to the bottom fixed temperature, where $X_{max}$=1400m (see Fig.SS9.1). The Tropical range is estimated to be 20S to 20N, and the rest is 60S to 60N but excluding the Tropical range.

Therefore, using es9.2,4,6, and where $X=X_{max}$=1400m, and X/5= 280 m:

es9.7 **A[**CC*(θ)**] = CC*** =

$$= \{ \tfrac{2}{3}\cdot {}_0\!\int^{\pi/9} \sqrt{[1.44\cdot \mathbf{g_u}\cdot C\cdot (X/5)\cdot \cos(\theta\cdot 3/2)]}\cdot\cos(\theta)\cdot d\theta + \tfrac{1}{3}\cdot {}_0\!\int^{\pi/9} \sqrt{[1.44\cdot \mathbf{g_u}\cdot C\cdot X\cdot \cos(\theta\cdot 3/2)]}\cdot\cos(\theta)\cdot d\theta +$$

^------------------upper Tropical--------------------------^ ^------------------------lower Tropical----------------^

$$+ 1\cdot {}_{\pi/9}\!\int^{\pi/3} \sqrt{[1.44\cdot \mathbf{g_u}\cdot C\cdot X\cdot \cos(\theta\cdot 3/2)]}\cdot\cos(\theta)\cdot d\theta \} / \{0.866\}$$

^--------------------extra Tropical-----------------------^

$$= \{ [2/(3\cdot\sqrt{5}) + \tfrac{1}{3}]\cdot \sqrt{[1.44\cdot C\cdot X\cdot \mathbf{g_u}]}\cdot {}_0\!\int^{\pi/9} \sqrt{[\cos(\theta\cdot 3/2)]}\cdot\cos(\theta)\cdot d\theta +$$
$$+ \sqrt{[1.44\cdot C\cdot X\cdot \mathbf{g_u}]}\cdot {}_{\pi/9}\!\int^{\pi/3} \sqrt{[\cos(\theta\cdot 3/2)]}\cdot\cos(\theta)\cdot d\theta \} / \{0.866\}$$

$$= \{ [2/(3\cdot\sqrt{5}) + \tfrac{1}{3}]\cdot\sqrt{[1.44\cdot C\cdot X\cdot \mathbf{g_u}]}\cdot(0.334) + \sqrt{[1.44\cdot C\cdot X\cdot \mathbf{g_u}]}\cdot(0.363) \} / \{0.866\}$$

$$= \{ (0.631)\cdot(10.37)\cdot(0.334) + (1)\cdot(10.37)\cdot(0.363) \} / \{0.866)$$
$$= \underline{6.87}$$ , so

es9.7b **CC*** = <u>6.87</u> $W\cdot\sqrt{yr} /(m^2\cdot°C)$

Then define $X_{eff}$ by (see es9.2):

es9.8 **CC*** = $\sqrt{[C\cdot \mathbf{g_u}\cdot X_{eff}]}$ , or

es9.8b $X_{eff} = (6.87)^2 / (0.13\cdot 0.41) = 885$ m

Remember, since we have assumed $X(\theta) = X_{max}$ everywhere, these are upper limits or *maximal* values of **CC*** and $X_{eff}$ ! These are "first order estimates" for the Oceans 60N to 60S .

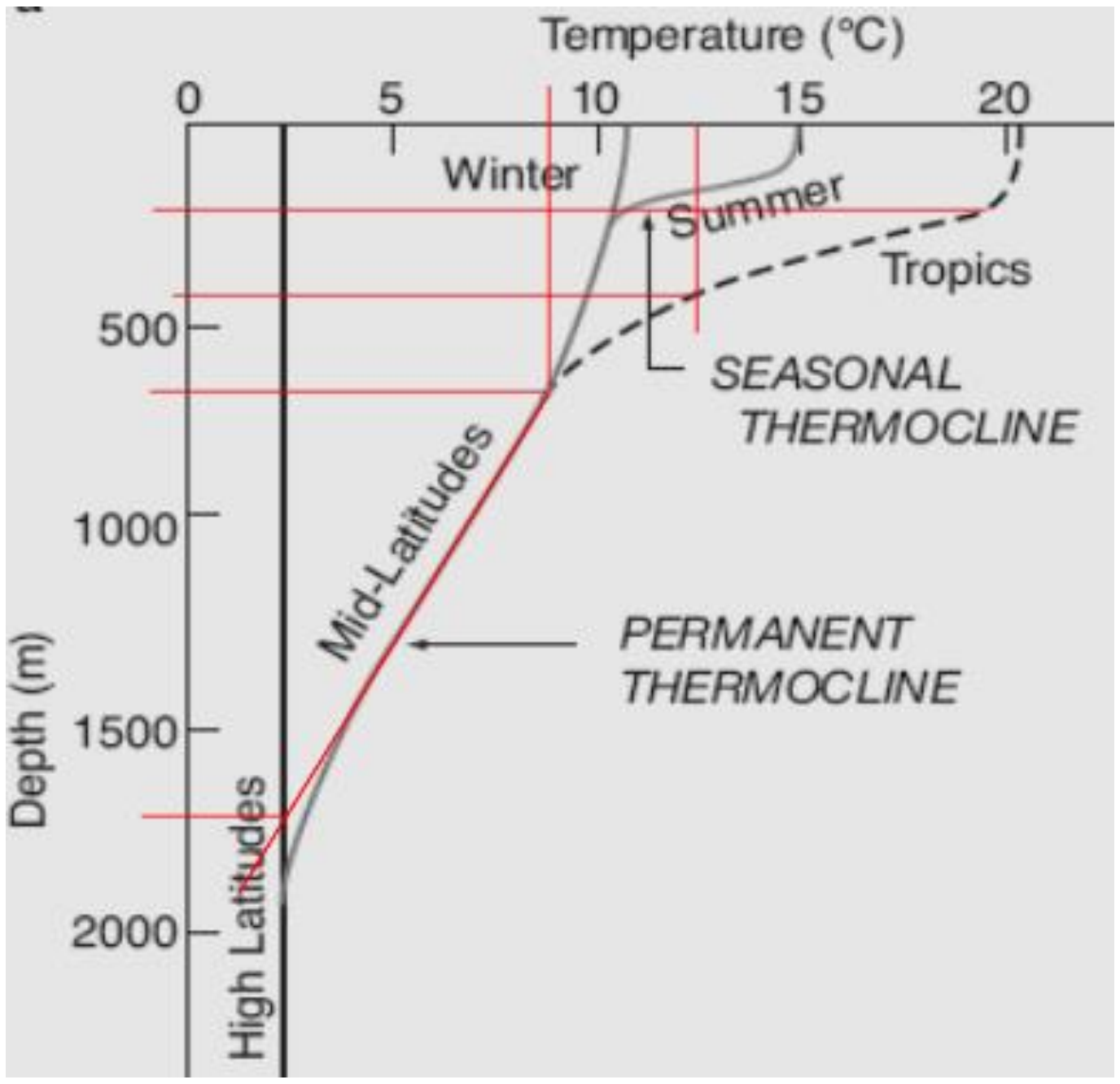

Fig.SS9.1 Characteristic (maximum depth) Ocean Temperature profiles with depth. [CC BY-NC-SA 4]